\documentclass[10pt]{article}

\usepackage{amsmath}
\usepackage{amssymb}

\usepackage{graphicx}

\usepackage{cite}

\usepackage{color} 

\usepackage[labelfont=bf,labelsep=period,justification=raggedright]{caption}

\makeatletter
\renewcommand{\@biblabel}[1]{\quad#1.}
\makeatother

\date{}

\newcommand{\Red}[1]{\textcolor{red}{#1}}
\newcommand{\script}[1]{{\mbox{\scriptsize #1}}}

\newcommand{\CITE}[1]{ \cite{#1}}
\newcommand{\VEC}[1]{\mathbf{#1}}
\newcommand{\Eq}[1]{Eq. {#1}}
\newcommand{\Eqs}[1]{Eqs. {#1}}
\newcommand{\SEq}[1]{Eq. {#1}}
\newcommand{\SEqs}[1]{Eqs. {#1}}
\newcommand{\EqPunc}[1]{}
\newcommand{\EqPeriod}[1]{}
\newcommand{\Fig}[1]{Fig. {#1}}
\newcommand{\Figure}[1]{Fig. {#1}}
\newcommand{\Figs}[1]{Figs. {#1}}
\newcommand{\SFig}[1]{Fig. {#1}}
\newcommand{\SFigs}[1]{Figs. {#1}}
\newcommand{\Table}[1]{Table {#1}}
\newcommand{\Tables}[1]{Tables {#1}}
\newcommand{\STable}[1]{Table {#1}}
\newcommand{\STables}[1]{Tables {#1}}

\newcommand{\FigureInText}[1]{}

\newcommand{\FigureInLegends}[1]{}
\newcommand{\TableInLegends}[1]{#1}
\renewcommand{\VEC}[1]{\mbox{\boldmath{$#1$}}} 

\renewcommand{\FigureInText}[1]{}

\renewcommand{\FigureInLegends}[1]{}
\renewcommand{\TableInLegends}[1]{#1}

\newcommand{\BF}[1]{\textbf{#1}}

\renewcommand{\SEq}[1]{Eq. S1-{#1} in Supporting Information, Text S1}
\renewcommand{\SEqs}[1]{Eqs. S1-{#1} in Supporting Information, Text S1}
\renewcommand{\SFig}[1]{Fig. {#1}}
\renewcommand{\SFigs}[1]{Figs. {#1}}
\renewcommand{\STable}[1]{Table {#1}}
\renewcommand{\STables}[1]{Tables {#1}}

\begin{document}

\begin{flushleft}
{\Large
\textbf{Selective Constraints on Amino Acids
Estimated by a Mechanistic Codon Substitution Model
with Multiple Nucleotide Changes}
}
\\
\vspace*{1em}
Sanzo Miyazawa
\\
Graduate School of Engineering, Gunma University, Kiryu, Gunma 376-8515, Japan
\\
E-mail: sanzo.miyazawa@gmail.com
\end{flushleft}

\section*{Abstract}

\paragraph*{Background}
Empirical substitution matrices represent the average tendencies 
of substitutions over various protein families 
by sacrificing gene-level resolution. 
We develop a codon-based model, 
in which mutational tendencies of codon, a genetic code,
and the strength of selective constraints against amino acid 
replacements can be tailored to a given gene.
First, selective constraints averaged over proteins 
are estimated by maximizing the likelihood
of each 1-PAM matrix of empirical amino acid (JTT, WAG, and LG)
and codon (KHG) substitution matrices.
Then, selective constraints specific to given proteins 
are approximated as a linear function of those estimated from
the empirical substitution matrices.

\paragraph*{Results}
Akaike information criterion (AIC) values indicate 
that a model allowing multiple nucleotide changes fits
the empirical substitution matrices significantly better.
Also, the ML estimates of transition-transversion bias
obtained from these empirical matrices are  
not so large as previously estimated.
The selective constraints are characteristic of proteins 
rather than species. 
However, their relative strengths
among amino acid pairs can be approximated 
not to depend very much on 
protein families but amino acid pairs,
because
the present model, 
in which selective constraints are approximated to be 
a linear function of those estimated from the JTT/WAG/LG/KHG matrices,
can provide a good fit to 
other empirical substitution matrices including
cpREV for chloroplast proteins and
mtREV for vertebrate mitochondrial proteins.

\paragraph*{Conclusions/Significance}
The present codon-based model with the ML estimates of
selective constraints and with 
adjustable mutation rates of nucleotide
would be useful as a simple substitution model in
ML and Bayesian inferences of molecular phylogenetic trees,
and enables us to obtain  biologically meaningful
information at both nucleotide and amino acid levels
from codon and protein sequences.

\section*{Introduction}
Any method for inferring molecular phylogeny is 
implicitly or explicitly based on
the evolutionary mechanism of nucleotide or amino acid 
substitutions, and
the reliability of phylogenetic analyses strongly depends on models
assumed for the substitution processes of nucleotide and amino acid.
Mutational events occur at the individual nucleotide level, but
selective pressure primarily operates at the amino acid level.
Thus, a codon-based model of amino acid substitutions
has a potential to be preferable to 
both mononucleotide substitution models
\CITE{K:80,HKY:85,TN:93}
and amino acid substitution models
\CITE{DSO:78,JTT:92,AH:96,YNH:98,AWMH:00,DMG:00,WG:01,LG:08,HJLR:08},
because it can take into account
both mutational tendencies at the nucleotide level
and selective pressure on amino acid replacements as well as
the knowledge of a genetic code. 
Schneider et al. 
\CITE{SCG:05}
and Kosiol et al. 
\CITE{KHG:07}
empirically estimated a codon substitution matrix from a large number of
coding sequence alignments.  
However, 
the tendencies of substitutions differ among 
nuclear, mitochondrial\CITE{AH:96},
and chloroplast genes
\CITE{AWMH:00}.
Delport et al. 
\CITE{DSGMP:10,DSBGMP:10}
pointed out that
empirical substitution matrices 
represent the average tendencies
of substitutions over various protein families 
by sacrificing gene-level resolution.
A mechanistic codon substitution model,
in which one can change a genetic code, 
and adjust mutational tendencies at the codon level and
selectional preferences on amino acid replacements,
is potentially more superior than empirical codon substitution matrices.

A main difference between the current mechanistic codon substitution models
\CITE{MJ:93,GY:94,MG:94,YNH:98,WG:04,DP:07,YN:08,SK:08,SK:09,DSGMP:10,DSBGMP:10}
resides in the estimation of 
selective constraints against amino acid replacements.
(1) In \CITE{MG:94,WG:04,YN:08},
the difference between nonsynonymous and synonymous substitution rates 
was taken into account but 
the amino acid dependences of selective constraints were not 
taken into account; i.e., single selective constraints.
(2) In \CITE{MJ:93,GY:94,YNH:98}, 
selective constraints against amino acid replacements  were
evaluated from physico-chemical properties of amino acids.
(3) In \CITE{DP:07,SK:08,SK:09}, codon exchangeabilities for nonsynonymous changes 
were evaluated from
those in empirical amino acid substitution matrices.
(4) In \CITE{DSGMP:10,DSBGMP:10},
selective constraints were grouped, and the number of groups and
the strength of selective constraint of each group were optimized
for a given protein phylogeny.
The fourth method has the highest resolution of selective constraints
employing as many substitution groups as necessary.
However, it seems to be a very computer-intensive calculation\CITE{DSBGMP:10}.
Here, we try to estimate selective constraint for each type of amino acid
replacement by maximizing the likelihood of individual empirical substitution matrices.  
Unlike the present method, in the previous methods of this third category
codon exchangeabilities for nonsynonymous changes were
assumed to be proportional to the corresponding amino acid exchangeability
\CITE{SK:08}, or a codon substitution matrix was restricted to
yield amino acid exchangeabilities equal to empirically-derived ones
\CITE{DP:07}. 
The empirical substitution matrices fitted are
1-PAM amino acid substitution frequency matrices, 
the JTT matrix\CITE{JTT:92},
the WAG matrix\CITE{WG:01},
and
the LG matrix\CITE{LG:08},
evaluated from relatively large data of nuclear-encoded proteins,
the mtREV matrix\CITE{AH:96}
from vertebrate mitochondrial proteins, and
the cpREV matrix\CITE{AWMH:00}
from chloroplast-encoded proteins, 
and also a 1-PAM codon substitution frequency matrix (KHG)\CITE{KHG:07}.
In the following, these empirical substitution frequency matrices 
corresponding to 1 PAM will be simply referred to 
by their common acronyms, JTT, WAG, LG, KHG, mtREV, and cpREV.

In most of the reversible Markov models for codon substitutions, 
instantaneous rates for codon substitutions that require multiple nucleotide
changes were assumed to be equal to $0$.
\CITE{MJ:93,GY:94,MG:94,DSGMP:10}.
However, in all empirical substitution matrices 
unnegligible amounts of rates are assigned 
to amino acid replacements
that require multiple nucleotide changes.
Variations in substitution rates or time intervals
would yield significant amounts of probabilities for
the multi-step substitutions. 
Alternative explanation is that the significant fraction of these substitutions 
occurred with multiple nucleotide changes.
Thus, both of them are taken into account in the present work. 
It is assumed that substitution rates are
distributed with a $\Gamma$ distribution.
The use of $\Gamma$ distribution for rate variation 
has been attempted in many studies
\CITE{JN:90,Y:93}.
Multiple nucleotide changes are assumed to occur
in the same order of time as single nucleotide changes do.

Interdependence of nucleotide substitutions at three codon positions
\CITE{YNH:98}
and also spanning codon boundaries
\CITE{WG:04}
have been pointed out.
Evidences for a high frequency, 
which is the order of 0.1 per site per billion years,
of double-nucleotide substitutions
were found in diverse organisms 
by Averof et al. 
\CITE{ARWS:00},
although
there is a report 
\CITE{SWE:03}
indicating a low rate of 
double-nucleotide mutations in primates.
Bazykin et al. 
\CITE{BKOSK:04}
pointed out a 
possibility of successive single compensatory substitutions 
for multiple nucleotide changes. 
Recently, many codon models relaxing mathematical assumptions 
in a more sophisticated way than the models
of Goldman and Yang 
\CITE{GY:94}
and Muse and Gaut 
\CITE{MG:94}
are devised to study and to detect evidence of positive selection
in codon evolutionary processes;
see Anisimova and Kosiol 
\CITE{AK:09}
for a review.

In the Singlet-Doublet-Triplet (SDT) mutation model\CITE{WG:04},
single-nucleotide, doublet and triplet mutations spanning 
codon boundaries are taken into account, but
double nucleotide mutations at the first and the third positions in a codon
were not taken into account. 
The dependences of selective constraints on amino acid pairs 
were not taken into account.
In the present model,
it is assumed that
nucleotide mutations occur independently at each codon position
and so any double nucleotide mutation occurs as frequently as doublet mutations.
The codon substitution rate matrix of KHG\CITE{KHG:07} indicates that some types of
double nucleotide mutations at the first and the third positions
frequently occur.

Close relationships between selective constraints on amino acids and
physico-chemical properties of amino acids and protein structures
have been pointed out
\CITE{DSO:78,G:74,MMY:79,MJ:93,DMG:00,CHRKT:07,CWS:07}.
We suppose
that the relative strengths of selective constraints 
among amino acid pairs
do not strongly depend on species, organelles, and even protein families but
amino acid pairs.
Then, we examine the performance of the present codon-based model, in which 
selective constraints 
are approximated to be a linear function of those estimated from
JTT, WAG, LG, or KHG,
in respect of how well other empirical substitution matrices 
including cpREV and mtREV can be fitted
by adjusting parameters such as mutational tendencies and 
the strength of selective constraints.
It is shown that these maximum likelihood (ML) estimators 
of the selective constraints
perform better than any physico-chemical estimation.
It is also indicated that
the present model yields good values of 
Akaike information criterion (AIC)
for a phylogenetic tree
of mitochondrial coding sequences in comparison with 
the codon model almost equivalent to mtREV.
If the present model is applied to the ML inference of phylogenetic trees,
it will allow us to estimate
mutational tendencies at the nucleotide level, which are
specific to each species and organelle, such as 
transition-transversion bias and
the ratio of nonsynonymous to synonymous rate.
One of the interesting results revealed by the present model 
is that the ML estimators of transition to transversion bias
calculated from the empirical substitution matrices
are not so large as previously estimated.
Also, AIC values 
indicate that 
a model allowing multiple nucleotide changes fits the empirical
substitution matrices 
and the phylogeny of vertebrate mitochondrial proteins significantly better.

The
present codon-based model with the new estimates
for selective constraints on amino acids
is useful as
a simple evolutionary model for phylogenetic
estimation,
and also useful to generate log-odds for codon substitutions
in protein-coding sequences with any genetic code.

\section*{Methods}

\noindent
\subsection*{A mechanistic codon substitution model with multiple nucleotide changes}
\vspace*{1em}

In early codon substitution models\CITE{MJ:93,GY:94},
the probabilities of multiple nucleotide replacements in the infinitesimal time difference
$\Delta t$ were completely neglected by assuming them to be $O(\Delta t^2)$,
when the probabilities of single nucleotide replacements are taken to be $O(\Delta t)$.
In other words,
the instantaneous mutation rate $M_{\mu\nu}$ from codon $\mu$ to $\nu$
was assumed to be equal to zero for codon pairs 
requiring multiple nucleotide replacements.
However, multiple nucleotide mutations may not be neglected in real protein evolution 
\CITE{T:87,YNH:98,ARWS:00,WG:04,BKOSK:04,KHG:07}.
Here, multiple nucleotide changes are assumed to occur
with the same order of time as single nucleotide changes occur, but
unlike the SDT model\CITE{WG:04}
a mutation process is simplified in such a way that
mutations independently occur at each position 
of a codon.
Thus, the mutation rate matrix for a codon is defined here as
\begin{eqnarray}
M_{\mu \nu}
	&\equiv& 
	\prod_{i=1}^3 [\delta_{\mu_i\nu_i} + (1-\delta_{\mu_i\nu_i}) (B_i)_{\mu_i\nu_i} ]
	\mbox{ for } \mu \neq \nu
		\label{eq: def_mutation_rate_matrix}
\end{eqnarray}
where $B_i$ is a mutation rate matrix between the four types of nucleotides at the $i$th codon position, 
$\delta_{\mu_i\nu_i}$ is the Kronecker's $\delta$,
and the index $\mu_i$
means the $i$th nucleotide in the codon $\mu$; 
$\mu = (\mu_1, \mu_2, \mu_3)$ where $\mu_i \in \mbox{\{ a, t, c, g \} }$.
Assuming that the rate matrix $B_i$ satisfies the detailed balance condition, 
it is represented as
\begin{eqnarray}
(B_{i})_{\mu_i \nu_i} &=&
		(m_i)_{\mu_i \nu_i} f_{i,\nu_i}^{\script{mut}}
		\hspace*{1em} \mbox{ for } i = 1, 2, 3
		\label{eq: def_base_mutation_rate_matrix}
	\\
(m_i)_{\mu_i \nu_i} &=& (m_i)_{\nu_i \mu_i}
		\label{eq: def_m_matrix}
	\\
f^{\script{mut}}_{\nu = (\nu_1, \nu_2, \nu_3)} &=& f_{1, \nu_1}^{\script{mut}} f_{2, \nu_2}^{\script{mut}} f_{3, \nu_3}^{\script{mut}}
		\label{eq: def_f_mut}
\end{eqnarray}
where $f_{i, \nu_i}^{\script{mut}}$ is the equilibrium composition of nucleotide $\nu_i$ 
at the $i$th codon position, and 
$(m_i)_{\mu_i \nu_i}$ 
is 
the exchangeability between nucleotides $\mu_i$ and $\nu_i$ at the $i$th codon position.
As a result of the detailed balance condition assumed for the $B_{i}$,
the $M$ also satisfies the detailed balance condition;
\begin{eqnarray}
        f_{\mu}^{\script{mut}} M_{\mu \nu} &=& f_{\nu}^{\script{mut}} M_{\nu \mu}
		\label{eq: detailed-balance-M}
\end{eqnarray}

The instantaneous substitution rate $R_{\mu\nu}$
from codon $\mu$ to $\nu$
can be represented as the product of
the mutation rate $M_{\mu\nu}$ and 
the average rate of fixation $F_{\mu\nu}$,
which is defined to be the average fixation probability multiplied by the 
chromosomal population size, 
for mutations from codon $\mu$ to $\nu$ under selection pressure;
$
R_{\mu\nu} 
	\propto
	\; M_{\mu\nu} F_{\mu\nu}  
	\mbox{ for } \mu \neq \nu
$.
Let us assume that the $R$ also satisfies
the detailed balance condition; that is,
\begin{eqnarray}
	f_{\mu} R_{\mu \nu} &=& f_{\nu} R_{\nu \mu}
		\label{eq: detailed-balance-R}
		\label{eq: detailed-balance}
\end{eqnarray}
where
$f_{\mu}$ is the equilibrium codon composition of the substitution rate matrix $R$.
The detailed balance condition \Eq{\ref{eq: detailed-balance-R}} for the $R$
is equivalent with a condition that 
$R_{\mu\nu}$ can be
expressed to be a product of the $(\mu, \nu)$ element of a symmetric matrix and
the equilibrium composition $f_{\nu}$.  
Similarly, the detailed balance condition \Eq{\ref{eq: detailed-balance-M}} for the $M$ 
is equivalent with
a condition that the matrix whose ($\mu, \nu$) element 
is equal to $M_{\mu\nu} / f_{\nu}^{\script{mut}}$ is symmetric.
Thus, the detailed balance conditions for the $M$ and the $R$ require that
the average fixation rate
$F_{\mu\nu}$
must be represented as the product of 
the two terms, $f_{\nu} / f_{\nu}^{\script{mut}}$ and $e^{w_{\mu\nu}}$,
where $w_{\mu\nu} = w_{\nu\mu}$;
$
	F_{\mu\nu} = (f_{\nu} / f_{\nu}^{\script{mut}}) e^{w_{\mu\nu}} 
		\; \mbox{ for } \mu \neq \nu
$.
Then, the codon substitution rate 
$R_{\mu\nu}$ can be represented as
\begin{eqnarray}
R_{\mu\nu} 
	&=& C_{\script{onst}}
	\; M_{\mu\nu} \frac{f_{\nu}}{f_{\nu}^{\script{mut}}} e^{w_{\mu\nu}} 
		\; \mbox{ for } \mu \neq \nu
	\label{eq: def_substitution_rate_matrix}
\end{eqnarray}
where 
$C_{\script{onst}}$ is an arbitrary scaling constant.
By taking the frequencies of stop codons to be zero,
the probability flow from any codon to a termination codon 
and its inverse flow are set to zero.
The unit of time is chosen by determining
the arbitrary scaling constant $C_{\script{onst}}$ in 
\Eq{\ref{eq: def_substitution_rate_matrix}}
in such a way that
the total rate of the rate matrix $R$ is equal to one;
\begin{eqnarray}
- \sum_{\mu} f_{\mu} R_{\mu\mu}  &=& 1
	\label{eq: def_of_unit_time}
\end{eqnarray}
Therefore, only the relative values among $M_{\mu\nu}$ are meaningful.
The frequency-dependent term $f_{\nu} / f_{\nu}^{\script{mut}}$ represents
the effects of selection pressures at the DNA level as well as at the amino acid level, which
change the codon frequency from
the mutational equilibrium frequency $f_{\nu}^{\script{mut}}$
to the frequency $f_{\nu}$ specific to a gene.
The fixation rate is obviously equal to 0 for lethal mutations
and 1 for neutral mutations.
Here, we approximate the average quantity $e^{w_{\mu\nu}}$ over mutants
to be independent of codon frequencies.
The quantity $e^{w_{\mu\nu}}$ is the same as the one 
called the rate of acceptance by Miyata et al.
\CITE{MMY:79}.
We assume that
selection pressure against 
codon replacements principally appears on an amino acid sequence
encoded by a nucleotide sequence; 
$w_{\mu \nu}$ for the codon pair $(\mu, \nu)$ is equal to 
the selective constraint $w_{a b}$ for the encoded amino acid pair $(a, b)$.
\begin{eqnarray}
e^{w_{\mu\nu}} &\equiv&
\left\{ 
	\begin{array}{l}
	\sum_{a} \sum_{b \in \mbox{\{ amino acids \} } } C_{\mu a} C_{\nu b} e^{w_{ab}} 
		\\
	\  \hspace*{2em}
		\mbox{for } 
	\hspace*{1em} \mu, \nu \notin \mbox{\{ stop codons \} and } \mu \neq \nu
		\\
	0 \hspace*{2em} 
		\mbox{for } \hspace*{1em} \mu \mbox{ or } \nu \in \mbox{\{ stop codons \} and } \mu \neq \nu
	\end{array}
	\right.
	\label{eq: ratio_of_neutral_mutations}
\end{eqnarray}
where $C_{\mu a}$ is a genetic code table and 
takes the value one if codon $\mu$ encodes amino acid $a$, 
otherwise zero.  
At the amino acid level, there should be no selection pressure 
against synonymous mutations.
Thus, the $w_{ab}$ 
satisfies
\begin{eqnarray}
w_{ab} &=& w_{ba} 
\hspace*{1em} , \hspace*{1em}
w_{aa} = 0
	\label{eq: condition_of_fitness}
\end{eqnarray}

The matrix $w$ will be directly estimated by maximizing the likelihood of 
an empirical substitution matrix, or 
it will be evaluated for a specific protein family
as a linear function of such an estimate of $w_{ab}$;
\begin{eqnarray}
w_{ab} &\equiv& \beta w^{\script{estimate}}_{ab} + w_0 (1 - \delta_{ab})
	\label{eq: estimation_of_fitness}
\end{eqnarray}
In \Eq{\ref{eq: estimation_of_fitness}},
$\delta_{a b}$ is the Kronecker's $\delta$,
and $w^{\script{estimate}}_{ab}$ means the 
estimate of $w_{ab}$, which is either a physico-chemical estimate
or a ML estimate calculated from a specific substitution matrix,
and satisfies \Eq{\ref{eq: condition_of_fitness}}.
The parameter $\beta$, which is non-negative, 
adjusts the strength of selective constraints for a protein family.
The parameter $w_0$ controls the ratio of nonsynonymous to synonymous substitution rate, 
but it will be ineffective
and may be assumed to be equal to 0 if amino acid sequences rather than codon sequences are analyzed.

Then,  the substitution probability matrix $S(t)$ at time t
in a time-homogeneous Markov process can be calculated as
\begin{eqnarray}
	S(t) = \exp ( R t)
\end{eqnarray}
Because the rate matrix $R$ satisfies the detailed balance condition,
the $S(t)$ also satisfies it. 
Therefore, a substitution process is modeled as a reversible Markov process.
The $S(t)$ and the $R$ that satisfy the detailed balance condition 
can be easily diagonalized with real eigenvalues and eigenvectors
\CITE{MJ:93};
the eigenvalues of $R$ are the same as those of a symmetric matrix
whose $(\mu, \nu)$ element is equal to $(f_{\mu}/f_{\nu})^{1/2}R_{\mu \nu}$.

If multiple nucleotide changes were completely ignored, then
\Eq{\ref{eq: def_mutation_rate_matrix}} would  be simplified as
$M_{\mu\nu} = ((1-\delta_{\mu_1\nu_1}) (B_1)_{\mu_1\nu_1}\delta_{\mu_2\nu_2}\delta_{\mu_3\nu_3} )
 +(\delta_{\mu_1\nu_1}(1-\delta_{\mu_2\nu_2}) (B_2)_{\mu_2\nu_2}\delta_{\mu_3\nu_3} )
 + (\delta_{\mu_1\nu_1}\delta_{\mu_2\nu_2}(1-\delta_{\mu_3\nu_3}) (B_3)_{\mu_3\nu_3} )$, 
whose formulation for a codon mutation rate matrix 
with \Eq{\ref{eq: def_base_mutation_rate_matrix}}
is essentially the same as the one proposed by 
Muse and Gault 
\CITE{MG:94}.
Here, it should be noted that 
$(B_{i})_{\mu_i \nu_i}$ in \Eq{\ref{eq: def_base_mutation_rate_matrix}}
is defined to be proportional to the equilibrium nucleotide composition
$f^{\script{mut}}_{i, \nu_i}$.
Alternatively, 
one may define $M_{\mu\nu}$ as
$M_{\mu\nu} = \prod_{i=1}^3 [\delta_{\mu_i\nu_i} + (1-\delta_{\mu_i\nu_i}) (m_i)_{\mu_i\nu_i} ] 
f^{\script{mut}}_{\nu}$ 
in the same way as
Miyazawa and Jernigan
\CITE{MJ:93}
and others 
\CITE{GY:94,YNH:98}
defined it to be
proportional explicitly to the composition of the base triplet, $f^{\script{mut}}_{\nu}$.
This alternative definition 
with \Eqs{\ref{eq: def_substitution_rate_matrix}} and \ref{eq: def_of_unit_time}
is equivalent to
\Eqs{\ref{eq: def_mutation_rate_matrix}} and \ref{eq: def_base_mutation_rate_matrix} 
with $f_{\nu_i}^{\script{mut}} = 0.25$,
and thus it is a special case in the present formulation;
see 
\CITE{RLP:08} 
for justifications of this alternative definition.

In the present analyses, we assume for simplicity 
that $(m_i)_{\mu_i \nu_i}$ and $f_{i, \nu_i}^{\script{mut}}$
do not depend on codon position $i$; that is,
$(m_i)_{\xi \eta} = m_{\xi \eta}$ and 
$f_{i, \xi}^{\script{mut}} = f_{\xi}^{\script{mut}}$, where $\xi, \eta \in \{ a, t, c, g\}$.
This assumption is reasonable because mutational tendencies may not
depend on a nucleotide position in a codon.
Let us define $m_{[tc][ag]}$ to represent the average of 
the exchangeabilities of the transversion type,
$m_{ta}$, $m_{tg}$, $m_{ca}$, and $m_{cg}$, 
and likewise $m_{tc|ag}$ to represent the average of 
the exchangeabilities of the transition type,
$m_{tc}$ and $m_{ag}$. 
We use the ratios $\{ m_{\xi\eta}/m_{[tc][ag]} \}$ 
as parameters for exchangeabilities, and
$m_{[tc][ag]}$ to represent the ratio of the exchangeability of double 
nucleotide change to that of single nucleotide change 
and also the ratio of the exchangeability of triple nucleotide change to that of
double nucleotide change;
note that 
the exchangeabilities of single, double, and triple nucleotide changes
are of $O(m_{[tc][ag]}), O(m_{[tc][ag]}^2)$, and $O(m_{[tc][ag]}^3)$
in \Eq{\ref{eq: def_mutation_rate_matrix}}, respectively,
and that \Eq{\ref{eq: def_of_unit_time}} must be satisfied.
Then, multiple nucleotide changes in a codon can be completely
neglected by making the parameter $m_{[tc][ag]}$ approach zero
with keeping $\{ m_{\xi\eta}/m_{[tc][ag]} \}$ constant 
in \Eq{\ref{eq: def_of_unit_time}}.
Also, it is noted that double nucleotide changes 
at the first and the third positions in a codon are assumed to 
occur as frequently as doublet changes.

\noindent

\vspace*{2em}
\noindent
\subsection*{Empirical substitution matrices used for model fitting}
\vspace*{1em}

Maximum likelihood (ML) values
are calculated 
for each 1-PAM substitution frequency matrix, which corresponds to
the time duration of 1 amino acid substitution per 100 amino acids,
of the JTT
\CITE{JTT:92}, 
the WAG
\CITE{WG:01},
the LG
\CITE{LG:08},
the cpREV
\CITE{AWMH:00},
and
the mtREV
\CITE{AH:96}
amino acid substitution matrices,
and of
the KHG codon substitution matrix
\CITE{KHG:07}.
We have arbitrarily chosen
the transition matrices of 1-PAM, whose time interval is
long enough for the significant number of substitutions to occur and also
too short for multi-step substitutions to cover multiple nucleotide changes.
JTT is an accepted point mutation matrix compiled from
the pairs of closely related proteins encoded in nuclear DNA.
WAG, LG, cpREV, and mtREV are
amino acid substitution matrices estimated 
by maximizing the likelihood of a given set of optimum phylogenetic trees.
The KHG matrix used is the one named
ECMunrest in the supplement of their paper, for which
multiple nucleotide changes are allowed.
JTT, WAG, LG, and KHG 
were all calculated from nuclear-encoded proteins,
although JTT was calculated by a different method from the others. 
The matrices of cpREV and mtREV were
calculated from proteins encoded 
in chloroplast DNA,
and in vertebrate mitochondrial DNA,
respectively.
It should be noted here that a non-universal genetic code 
is used in the mitochondrial DNA.

\vspace*{2em}

\noindent
\subsection*{Average of a transition matrix over time or over rate}
\vspace*{1em}

In the present study, model parameters are estimated by
maximizing the likelihood of each 1-PAM substitution frequency matrix of
JTT, WAG, LG, cpREV, mtREV, and KHG.
In the case of JTT,
the pairs of 
closely related sequences were used to count substitutions
and the transition matrix was calculated by completely neglecting 
multiple substitutions at a site in a parsimony method.  
Thus, JTT should be considered to consist of
substitutions that occurred in various time intervals (various branch lengths).
The substitution rate matrices of WAG, LG, mtREV, cpREV and KHG
were estimated by the ML method 
for a given set of protein phylogenetic trees.
Each site of protein families may have evolved with a different rate.
As a result, these substitution matrices
may be regarded as an average over different substitution rates.
Here we assume that evolutionary time intervals 
or substitution rates 
for each substitution matrix
are distributed in a $\Gamma$ distribution.
There have been many attempts
\CITE{JN:90,Y:93} 
of using a $\Gamma$ distribution for rate variation.

If the substitution rate matrix $R$ is assumed to vary only by a scalar factor,
the mean of a substitution matrix irrespective of over-time and over-rate 
will be calculated as 
\begin{eqnarray}
\lefteqn{
\langle S \rangle(\tau, \sigma) \equiv \int_{0}^{\infty} S(t) \Gamma(t ; \tau, \sigma) dt 
}
\nonumber 
\\
&=& \int_{0}^{\infty} \frac{1}{\Gamma(\tau)} \exp \{ - ( I - \sigma R) \frac{t}{\sigma} \} (\frac{t}{\sigma})^{\tau-1} \frac{d t}{\sigma}
= [ ( I - \sigma R )^{-1} ]^{\tau}
	\label{eq: gamma_distributed}
\end{eqnarray}
where $\Gamma(t ; \tau, \sigma)$ is the probability density function of a $\Gamma$ distribution with
a scale parameter $\sigma$ and a shape parameter $\tau$, $\Gamma(\tau)$ is the $\Gamma$ function,
and $I$ is the identity matrix.
The mean and the variance of the $\Gamma$ distribution $\Gamma(t ; \tau, \sigma)$ 
are equal to $\tau \sigma$ and  $\tau \sigma^2$, respectively.
Here we should recall that the rate matrix $R$ is normalized such that
the total rate per unit time is equal to one; see \Eq{\ref{eq: def_of_unit_time}}.

\vspace*{2em}

\vspace*{2em}

\noindent
\subsection*{Evaluation of the log-likelihood of an empirical substitution matrix}
\vspace*{1em}
The log-likelihood of the empirical frequency, 
$A_{\kappa\lambda} = N f^{\script{obs}}_{\kappa} S^{\script{obs}}_{\kappa\lambda} $, 
of substitutions from $\kappa$ to $\lambda$
in the present model can be calculated as
\begin{eqnarray}
\ell(\VEC{\theta})
&=& N \sum_{\kappa} \sum_{\lambda} f^{\script{obs}}_{\kappa} S^{\script{obs}}_{\kappa\lambda} \log (f_{\kappa} \langle S \rangle(\tau,\sigma)_{\kappa\lambda})
	\label{eq: log-likelihood}
\end{eqnarray}
where $\kappa$ and $\lambda$ mean one of the amino acid types for amino acid substitution matrices 
or one of the codon types for codon substitution matrices,
$S^{\script{obs}}$ is an observed transition probability matrix
corresponding to the accepted point mutation matrix $A$,
$f^{\script{obs}}_{\kappa}$ is the observed composition of 
amino acid or codon $\kappa$,
and 
$N$ is the total number of amino acid or codon sites 
compared to count substitutions. 
The observed composition $f^{\script{obs}}_{\kappa}$ 
is assumed to be the equilibrium composition of
$S^{\script{obs}}$.
$\VEC{\theta}$ is a set of parameters and 
$\hat{\VEC{\theta}} = \arg \max_{\VEC{\theta}} \ell(\VEC{\theta})$ 
is a set of the maximum likelihood (ML) estimators.
Similarly, the estimate $\hat{I}_{\script{KL}}$ of the 
Kullback-Leibler (K-L) information by replacing the real distribution to the observed
frequency distribution is calculated as
\begin{eqnarray}
\lefteqn{
\hat{I}_{\script{KL}}(\VEC{\theta})
}
\nonumber \\
&=& \sum_{\kappa} \sum_{\lambda} f^{\script{obs}}_{\kappa} S^{\script{obs}}_{\kappa\lambda} [ \log (f^{\script{obs}}_{\kappa} S^{\script{obs}}_{\kappa\lambda})
				 - \log (f_{\kappa} \langle S \rangle(\tau,\sigma)_{\kappa\lambda}) ]
		\\
&=& - \ell(\VEC{\theta}) / N + \sum_{\kappa} \sum_{\lambda} f^{\script{obs}}_{\kappa} S^{\script{obs}}_{\kappa\lambda} \log (f^{\script{obs}}_{\kappa} S^{\script{obs}}_{\kappa\lambda})
	\label{eq: KL-information}
\end{eqnarray}
Maximum log-likelihood $\ell(\hat{\VEC{\theta}})$ corresponds to the minimum 
of the estimate of K-L information, $\hat{I}_{\script{KL}}(\hat{\VEC{\theta}})$.

The transition probability, $S(t)_{ab}$, 
between amino acids $a$ and $b$ and
the composition, $f_a$, of amino acid $a$
are related to those for codons as follows.
\begin{eqnarray}
f_{a} S(t)_{ab} &\equiv&
\sum_{\mu} \sum_{\nu} C_{\mu a} f_{\mu} S(t)_{\mu \nu} C_{\nu b}
	\\
f_{a} &\equiv&
	\sum_{\mu} C_{\mu a} f_{\mu}
\end{eqnarray}

The goodness of a model and the significance of parameters 
can be indicated by Akaike Information Criterion (AIC).  The AIC value is defined as
\begin{eqnarray}
\lefteqn{
\mbox{AIC} 
}
\nonumber \\
&\equiv&
-2 \ell(\hat{\VEC{\theta}}) + 2 \cdot (\mbox{number of adjustable parameters})
		\label{eq: def_AIC}
	\\
\lefteqn{
\Delta \mbox{AIC} 
}
\nonumber \\
&\equiv& \mbox{AIC} + 2  N \sum_{\kappa} \sum_{\lambda} f^{\script{obs}}_{\kappa} S^{\script{obs}}_{\kappa\lambda} \log (f^{\script{obs}}_{\kappa} S^{\script{obs}}_{\kappa\lambda})
	\label{eq: orig_def_dAIC}
	\\
	&=& 2 N \hat{I}_{\script{KL}}(\hat{\VEC{\theta}}) + 2 \cdot (\mbox{number of adjustable parameters})
		\label{eq: def_dAIC}
\end{eqnarray}
For convenience, $\Delta \mbox{AIC}$, which is equal to 
a constant value
added to the AIC value,
is also defined above. 
The AIC and $\Delta \mbox{AIC}$ always take a non-negative value. 
Models with smaller AIC and $\Delta \mbox{AIC}$ can be considered 
to be more appropriate 
\CITE{A:74}.

Parameters in the present model are
$\beta$, $m_{\xi \eta}$, $f_{\eta}^{\script{mut}}$, 
$f_{\eta}$, 
$\tau$, and $\sigma$. 
Assuming
that the observed process of substitutions is in the stationary state,
the estimates of the equilibrium codon and the equilibrium amino acid compositions,
$\hat{f}_{\mu}$ and $\hat{f}_{a}$,
are taken to be the observed composition of the codon and of the amino acid:
\begin{eqnarray}
\hat{f}_{\mu} &=& f^{\script{obs}}_{\mu}  \; , \hspace*{2em} \hat{f}_a = f^{\script{obs}}_a
		\label{eq: estimate_of_amino acid frequency}
\end{eqnarray}
In the case of amino acid sequences, for which 
their coding sequences are not available, 
codon compositions may be parameterized by
\begin{eqnarray}
	\hat{f}_{\mu} &=& \frac{\sum_{a} C_{\mu a} \hat{f}_{a} f_{\mu}^{\script{usage}}}{\sum_{a} C_{\mu a} \sum_{\nu} C_{\nu a} f_{\nu}^{\script{usage}}}
	\\
	f^{\script{usage}}_{\nu = (\nu_1, \nu_2,\nu_3)} &=& f_{\nu_1}^{\script{usage}} f_{\nu_2}^{\script{usage}} f_{\nu_3}^{\script{usage}}
	\label{eq: estimator_of_codon_frequencies}
\end{eqnarray}
In the present analyses, this parameterization is used 
for the equilibrium codon compositions in amino acid sequences.

Then, the shape parameter $\tau$ of a $\Gamma$ distribution for variations in 
mutation rates or evolutionary time intervals
for observed codon or amino acid substitutions is estimated by equating 
the ratio of the expected number of substitutions in the model to its observed value.
\begin{eqnarray}
	\sum_{\kappa} \hat{f}_{\kappa} \langle S \rangle(\hat{\tau},\sigma)_{\kappa\kappa} 
	&=& 
	 \sum_{\kappa} f^{\script{obs}}_{\kappa} S^{\script{obs}}_{\kappa\kappa}
	\label{eq: estimator_of_tau}
\end{eqnarray}
Other parameters
$\beta$, $m_{\xi \eta}$, $f_{\eta}^{\script{mut}}$, 
$f^{\script{usage}}_{\eta}$, and $\sigma$ are evaluated as ML estimators
or fixed to a proper value.
The observed transition matrix $S^{\script{obs}}_{\kappa\lambda}$ corresponding to
1-PAM is used here; 
PAM means accepted point mutations per 100 amino acids.
\begin{eqnarray}
	\sum_{a} f^{\script{obs}}_{a} S^{\script{obs}}_{aa} &=& 0.99
\end{eqnarray}

\noindent

\vspace*{2em}
\noindent
\subsection*{The total number of site comparisons ($N$) for each empirical substitution matrix}
\vspace*{1em}

In the case of JTT, 
59190 accepted point mutations found in 16130 protein sequences were used
to build a substitution probability matrix of 1-PAM
\CITE{JTT:92}.
Thus, the total number $N$ of amino acid comparisons
for JTT is assumed to be equal to $N = 59190/0.01$.
On the other hand,
a phylogenetic tree for cpREV
is based on $9957$ amino acid sites of 45 proteins
encoded in chloroplast DNAs of 9 species
\CITE{AWMH:00},
and the one for mtREV is based on 
$3357$ amino acid sites of the complete mitochondrial DNA
from 20 vertebrate species (3 individuals from human)
\CITE{AH:96}.
Thus, the total number of site comparisons $N$ for them may be approximated to be equal to
the number of amino acid sites multiplied by
the number of branches in the phylogenetic tree used to evaluate
the transition matrices; that is, 
$N \approx 9957 \cdot (2 \cdot 10 - 3) = 169269$ for cpREV, and 
$N \approx 3357 \cdot (2 \cdot 22 - 3) = 137637$ for mtREV.
The BRKALN database consisting of 50867 sites and 895132 residues 
was used to estimate WAG. Thus, 
$N \approx 895132 \cdot 2 - 50867 \cdot 3 = 1637663$
is used for WAG
\CITE{WG:01,LG:08}.
To evaluate LG, 3412 of 3912 alignments consisting of 49637 sequences, 599692 sites,
and 6697813 residues are used
\CITE{LG:08}.
Therefore,
$N \approx (6697813 \cdot 2 - 599692 \cdot 3) \cdot 3412 / 3912 = $ 
$10114373$
is assumed for LG.
These crude estimates of $N$ are used to evaluate
the AICs of JTT, WAG, LG, cpREV and mtREV.

In the case of KHG, 
which was estimated
by maximizing
a likelihood of a set of phylogenetic trees of coding sequences
of 7332 nuclear protein families taken from Pandit database
\CITE{WBQRG:06},
the total numbers of residues and sites are not written 
in Kosiol et al. 
\CITE{KHG:07},
so that an AIC value is not given for KHG in the following.

\section*{Results}

Models,
each of which includes a different number of parameters
and is a special case of models including more parameters,
are fitted by a maximum likelihood method
to each of the 1-PAM amino acid substitution frequency matrices,
JTT\CITE{JTT:92},
WAG\CITE{WG:01},
and 
LG\CITE{LG:08}
for proteins encoded in nuclear DNA,
cpREV\CITE{AWMH:00}
for chloroplast DNA, and mtREV\CITE{AH:96}
for mitochondrial DNA.
Also, the models are fitted to the 1-PAM codon substitution
frequency matrix of KHG\CITE{KHG:07} for nuclear DNA.
The selective constraints $w_{ab}$ are either directly estimated
by ML
or evaluated from a known estimate $w^{\script{estimate}}_{ab}$ by 
\Eq{\ref{eq: estimation_of_fitness}} that
includes two parameters $\beta$ and $w_0$.
The parameter $w_0$ is fixed here to $0$ for amino acid substitution matrices
because the likelihood of an amino acid substitution matrix does not strongly depend on $w_0$;
codon substitution data are required to
reliably estimate the value of $w_0$, which significantly
affects the ratio of nonsynonymous to synonymous substitution rate.
Each model is named to indicate either the method to estimate $w_{ab}$
or the name of $w_{ab}^{\script{estimate}}$ with a suffix 
meaning the number of ML parameters.
Each model is briefly described in \Table{\ref{tbl: model_names}}.
The Nelder-Mead Simplex algorithm has been used for the maximization of
likelihoods.

\vspace*{2em}
\noindent
\subsection*{
The effects of selective constraints
}
\vspace*{1em}

First, the No-Constraints models, 
in which selective constraints do not depend on amino acid pairs, 
$\beta = 0$ 
in \Eq{\ref{eq: estimation_of_fitness}},
were examined to see how well nucleotide mutation rates, codon frequencies 
and a genetic code
can explain the observed frequencies of amino acid substitutions in
JTT, WAG, cpREV, and mtREV; 
the No-Constraints models disallowing multiple nucleotide changes
are equivalent to mononucleotide substitution models, 
because $w_0 = 0$ is used here.
The $\Delta \mbox{AIC}$ value and the ML estimates 
for each parameter set are listed in
\Table{\ref{tbl: optimizations_AIC}} and
\STable{\ref{tbl: optimizations_no-selection}}, 
respectively.
Please refer to Supporting Information, Text S1, for details.
These No-Constraints models serve as a reference to
measure how selection models can improve the likelihoods.
Then,  
we examine various estimations of selective constraints on amino acids
based on
the physico-chemical distances of amino acids evaluated by Grantham
\CITE{G:74} 
and by
Miyata et al.
\CITE{MMY:79} 
and mean energy increments due to an amino acid substitution.
These models are called Grantham, Miyata, 
and Energy-Increment-based (EI) models, respectively.
Please refer to Supporting Information, Text S1, for the definition of the mean energy increment
and for the details of each model.
The $\Delta \mbox{AIC}$ values and the ML estimates
for these models with various sets of parameters are
also listed in 
\Table{\ref{tbl: optimizations_AIC}}, and
\STables{\ref{tbl: optimizations_EI_JTT_WAG_CpRev_MtRevRmtx}} and
\ref{tbl: optimizations_selection_JTT_WAG},
respectively.
Comparisons of $\Delta \mbox{AIC}$ values between the models 
in \Table{\ref{tbl: optimizations_EI_AIC}} indicate that
the selective constraints on amino acids
representing conservative selection against amino acid substitutions
significantly improve the $\Delta \mbox{AIC}$ values of all substitution matrices.
It is also indicated that 
the Miyata's physico-chemical distance performs better
in all parameter sets than the Grantham's distance,
This result is 
consistent with
that of Yang et al.\CITE{YNH:98}
for mitochondrial proteins.
The present physico-chemical evaluation of selective constraints 
(EI models)
fits JTT and WAG even better than the 
Miyata's distance scale, although
the performances of both the methods are almost same for cpREV and mtREV.
One of the important facts in these results is that
allowing multiple nucleotide changes in a codon
significantly improve the AIC irrespective of
the estimations of selective constraints; compare 
the $\Delta\mbox{AIC}$ values
between the Grantham-10 and the Grantham-11,
between the Miyata-10 and the Miyata-11,
and
between the EI-10 and the EI-11.

\vspace*{2em}
\noindent
\subsection*{
The effects of multiple nucleotide changes 
on ML estimations}
\vspace*{1em}

In principle, all parameters $\{ w_{ab} \}$ for selective constraints 
can be optimized in the case of codon sequences. 
In the case of protein sequences,
all 190 non-diagonal elements of $w$ in addition to the parameters for
mutational tendencies at the nucleotide level and others
cannot simultaneously be optimized; 
the number of freedoms in a general reversible model for an amino acid transition matrix
is equal to 209.

In order to see how well amino acid substitution matrices can be
explained with the assumption of successive single nucleotide substitutions,
let us optimize $w_{ab}$ corresponding to single-step amino acid pairs
by assuming that only single nucleotide mutations are possible, i.e.,
by $m_{[tc][ag]} \rightarrow 0$ with $m_{\xi\eta} /m_{[tc][ag]} =\mbox{constant}$
in \Eq{\ref{eq: def_of_unit_time}}.
The number of $w_{ab}$ for the single-step amino acid pairs is equal to 75
in the case of the universal genetic code.
All 75 $w_{ab}$ for the single-step amino acid pairs have been optimized 
for each of JTT and WAG
together with the nucleotide exchangeabilities $\{ m_{\xi\eta} \}$, 
the equilibrium nucleotide composition $\{ f^{\script{mut}}_{\xi} \}$, 
the codon usage parameters $\{ f^{\script{usage}}_{\xi} \}$
and the scale parameter $\sigma$;
the total number of the parameters is equal to 87 in addition to the
19 amino acid frequencies and the shape parameter $\tau$.
This maximum likelihood model to estimate the matrix $w$ is called ML 
with a suffix meaning the number of ML parameters;
see \Table{\ref{tbl: model_names}}.
The ML estimates of these parameters except $\hat{w}_{ab}$
for the ML-87 are listed in 
\Table{\ref{tbl: optimizations_wij_JTT_WAG}} for JTT and WAG.

In the lowest rows of this table, 
the ratio of the total nucleotide substitution rate
per codon to the codon substitution rate, 
which represents 
the average number of nucleotide changes for substituting a codon,
the ratio of the total transition to the total transversion rate per codon, 
and the ratio of nonsynonymous to synonymous substitution rate per codon are listed for the models.
The sum of the total transition and the total transversion rates per codon
is equal to the total nucleotide substitution rate per codon.
The lowest three rows list their values in the case of 
$\sigma \rightarrow 0$ and $w_{ab} = 0$, and the second
lowest three rows for the case of $\sigma \rightarrow 0$.
Thus, the differences of their values 
between the lowest and second lowest three rows
represent the effects of selective constraints on amino acids ($w_{ab}$),
and those between the second lowest and the third lowest three rows
describe the effects of rate/time variations 
on the substitution matrix.
If codon substitutions proceed by successive single nucleotide changes,
i.e., $m_{[tc][ag]} \rightarrow 0$,
then the ratio of the total nucleotide to the codon substitution rate 
will be equal to 1 
in the case of $\sigma \rightarrow 0$.

Here it should be noticed that the nonsynonymous and the synonymous substitution rates 
are defined not to be rate per site but simply rate per codon.
The sum of the nonsynonymous and the synonymous substitution rates
is equal to the codon substitution rate. 
The ratio of the nonsynonymous to the synonymous substitution rate per codon
does not corresponds to the ratio of
nonsynonymous to synonymous substitutions per site,
$K_A/K_S$\CITE{MY:80},
but the ratio of nonsynonymous to synonymous substitutions per codon, $M_A/M_S$\CITE{MY:80}.
The ratio ($N_A/N_S$\CITE{MY:80}) of the effective number of nonsynonymous sites 
to that of synonymous sites per codon corresponds to
the ratio of nonsynonymous to synonymous rate in the case of 
no selective constraints ($w_{ab}=0$).
In the present models,
$K_A/K_S$ indicating the effects of selection on amino acid replacements
corresponds to
the nonsynonymous to synonymous substitution rate ratio
in the case of $\sigma \rightarrow 0$ 
divided by that in the case of $w_{ab}=0$ and $\sigma \rightarrow 0$.
\Table{\ref{tbl: optimizations_wij_JTT_WAG}} indicates 
that selection on amino acids is conservative, because
the ratio of nonsynonymous to synonymous rate per codon 
is much smaller in the case of $\sigma \rightarrow 0$ 
than in the case of $w_{ab}=0$ and $\sigma \rightarrow 0$.

As expected, the AIC value drastically decreases from that of the EI-14 in both cases of JTT and WAG,
indicating that the introduction of many parameters may be still 
appropriate.
However, there are large discrepancies between the observed transition matrix and the one
estimated by the ML-87.  Let us see the discrepancies between them in terms of
log-odds.

A log-odds matrix introduced by Dayhoff et al.
\CITE{DSO:78}
is one of the representations of amino acid substitution propensities.
The $(\kappa, \lambda)$ element of the log-odds matrix
is defined to be the logarithm of odds 
to find an amino acid pair $(\kappa, \lambda)$ in comparison with random sequences.
The odds $O_{\kappa\lambda}$ is equal to the $(\kappa, \lambda)$ element of 
transition matrix divided by the amino acid composition $f_{\lambda}$.
\begin{eqnarray}
	O(S(t))_{\kappa\lambda} &\equiv& S(t)_{\kappa\lambda} / f_{\lambda}
		\\
	\mbox{log-}O(S(t))_{\kappa\lambda} &\equiv& \frac{10}{\log 10} \log O(S(t))_{\kappa\lambda}
		\label{eq: def-log-odds}
\end{eqnarray}
The proportional constant in \Eq{\ref{eq: def-log-odds}} is the one 
originally used by Dayhoff et al.
\CITE{DSO:78}.

In \Fig{\ref{fig: ML-87_log-odds_1pam_JTT}}, the log-odds
$\mbox{log-}O(\langle S \rangle(t))_{ab}$
corresponding to the 1 PAM transition matrix of the ML-87 model
fitted to JTT are plotted against those calculated from JTT.
Plus, circle and cross marks show the log-odds for
one-, two-, and three-step amino acid pairs,
respectively.  
Although the estimated values of log-odds 
for one-step amino acid pairs are almost exactly equal to those of the JTT matrix, 
there are still large discrepancies between the log-odds values for 
two- and three-step amino acid pairs, indicating a non-stepwise manner of codon substitutions.
Similar discrepancies are also found 
in 
\SFig{\ref{fig: ML-87_log-odds_1pam_WAG}} 
for WAG.

We have examined how the AIC is improved by enabling
multiple nucleotide changes in a codon.
The selective constraints $\{w_{ab}\}$ for multiple nucleotide changes
are classified into 6 groups 
according to the amounts of discrepancies between the observed 
and the estimated values
of the log-odds as shown in \Fig{\ref{fig: ML-87_log-odds_1pam_JTT}}.
Then, the ML estimates of 94 parameters including
7 additional parameters, $w_{ab}$ for the 6 groups of 
multiple nucleotide changes and the parameter 
$m_{[tc][ag]}$ for the rate of multiple nucleotide change, are calculated. 
This model is called ML-94. 
Also, 
the values of $\{ w_{ab} \}$ for multi-step amino acid pairs 
are calculated
by maximizing the likelihood with fixing the values
of all other parameters including $w_{ab}$ for the single-step
amino acid pairs; this model is called here ML-94+ by appending the "+" mark.
It should be noted that these values of $\hat{w}_{ab}$ 
for the multi-step amino acid pairs in the ML-94+ 
are not ML estimates at all.
The ML estimates $\hat{w}_{ab}$ for single-step amino acid pairs, 
the classification of multi-step amino acid pairs into the 6 groups,
and the ML estimates for those categories of $w_{ab}$
are provided in Supporting Information, Data S1.
As shown in
\Table{\ref{tbl: optimizations_wij_JTT_WAG}},
the ML estimates of $m_{\xi\eta}$,
$f^{\script{mut}}_{\xi}$, and $f^{\script{usage}}_{\eta}$
for the ML-87 model 
are very different from those for the ML-94,
and some of them for the ML-87 seem to be unrealistic.
For example,
$\hat{m}_{ta}/\hat{m}_{[tc][ag]}$ is evaluated to be smaller than $0.1$.
Also, the small value of $\hat{f}^{\script{usage}}_{t+a}$
indicates the extremely biased usage of codons.
The ML estimate $\hat{\sigma}$ of a $\Gamma$ distribution is too large.
These parameters are forced in the ML-87 
to take such values to reduce the discrepancies between
the observed and the estimated counts for multi-step amino acid pairs.
In the ML-94 model, the ML estimators of these parameters take
more reasonable values.
However, it may also yield unreasonable estimates 
for
codon usage parameters, $\{ f^{\script{usage}}_{\xi} \}$;
for example, $\hat{f}^{\script{usage}}_{t+a} = 0.221$ in the ML-94 for WAG,
and $\hat{f}^{\script{usage}}_{c} = 0.249 \cdot \hat{f}^{\script{usage}}_{c+g} =  0.14$ 
in the ML-94 for LG.
Thus, the ML-91 model with $f^{\script{usage}}_{\xi} = 0.25$, which means equal codon usage, may be better than the ML-94.
The ML-91 model was applied for JTT, WAG, and LG, 
and the ML estimates for them in the ML-91 are also listed in
\Table{\ref{tbl: optimizations_wij_JTT_WAG_LG}}.

The ML estimators $\hat{m}_{\xi\eta}$, $\hat{f}^{\script{mut}}_{\xi}$, and $\hat{\sigma}$
show a similar tendency between the ML-91 models for all the amino acid
substitution matrices, i.e., JTT, WAG, and LG.
The parameter $m_{[tc][ag]}$ for multiple nucleotide changes 
and the scale parameter $\sigma$ for rate variation are both significant 
for all the matrices.
The values of $\hat{m}_{tc|ag} / \hat{m}_{[tc][ag]} > 1$
for JTT, WAG, and LG
indicate that the mean exchangeability of the transition type
is larger than that of the transversion type 
in all the matrices.

As shown in \Fig{\ref{fig: ML-91_log-odds_1pam_JTT}} for JTT and 
in 
\SFig{\ref{fig: ML-91_log-odds_1pam_WAG}} 
for WAG, 
the large discrepancies of the log-odds for the multi-step amino acid pairs disappear
in the ML-91, in which multiple nucleotide changes
are taken into account.
The AIC values of JTT and WAG are significantly improved 
by enabling multiple nucleotide changes
in the ML-91.
This fact confirms that
multiple nucleotide changes are statistically significant
and should be taken into account to build a codon substitution model. 

\vspace*{2em}
\noindent
\subsection*{
ML estimation for the KHG codon substitution matrix
}
\vspace*{1em}

If a codon substitution matrix is used for model fitting with
the assumption of multiple nucleotide changes,
all 190 parameters of selective constraints $\{ w_{ab} \}$ will be able to be optimized.
The ML-200 model has been fitted to the 1-PAM codon substitution frequency 
matrix of KHG, which
was empirically estimated without any restriction 
on multiple nucleotide changes\CITE{KHG:07}.

The log-odds values for the codon pairs requiring single, double, and triple nucleotide changes
are shown in 
\Figs{\ref{fig: ML200_logodds_codon_1pam_KHG}A, \ref{fig: ML200_logodds_codon_1pam_KHG}B, 
and \ref{fig: ML200_logodds_codon_1pam_KHG}C}, respectively.
In  these figures, upper triangle, plus, circle, and cross marks
show the log-odds values for synonymous pairs and one-, two-, and three-step
amino acid pairs, respectively.
The dotted line shows the line of values 
where the observed and the estimated values of log-odds
are equal to each other.
The log-odds of the codon pairs requiring single/double/triple nucleotide changes
for one/two/three-step amino acid pairs respectively
tend to fall along the dotted line in comparison with the log-odds
of the other codon pairs.
In other words,  the log-odds of the
codon pairs for which any nucleotide change is accompanied by an amino
acid change are correctly estimated.
On the other hand, the estimated log-odds values do not well agree with
the observed ones for synonymous codon pairs shown by the upper triangles.
These estimated log-odds can be adjusted only by changing
nucleotide mutation rates, i.e., $m_{\xi\eta}$ and $f^{\script{mut}}_{\xi}$.
Thus, 
the approximations of the independence and of no difference of 
nucleotide exchangeabilities between nucleotide positions
may be limited; see \Eq{\ref{eq: def_mutation_rate_matrix}}.

The codon pairs, whose log-odds values are less than $-30$ and 
which require more nucleotide changes than
the least nucleotide changes required for the corresponding amino acid pair,
tend to be located
in the upper region than in the lower region of the dotted line;
see plus marks in \Fig{\ref{fig: ML200_logodds_codon_1pam_KHG}B} and 
plus and circle marks in \Fig{\ref{fig: ML200_logodds_codon_1pam_KHG}C}.
Such a tendency is more clear in 
\Fig{\ref{fig: ML200_logodds_codon_1pam_KHG}C},
in which plus and circle marks corresponding to one- and two-step amino acid pairs
are mostly located far from and almost in parallel to the dotted line.
The estimated values of the log-odds for
these one- and two-step amino acid pairs are greater by 10 -- 15
than the observed values.

In \Fig{\ref{fig: logodds_vs_exchange_scaled_1pam_observed_KHG_3x}D},
the log-exchangeabilities of the codon pairs 
requiring 
triple nucleotide changes 
in the 1-PAM KHG matrix are plotted against their log-odds of the 1-PAM KHG matrix.
The log-exchangeability is defined here to be 
$(10 / \log 10) \log [R^{\script{KHG}}_{\mu\nu} \cdot t_{\script{1-PAM}} / f_{\nu} ]$.
The log-exchangeabilities of the codon pairs corresponding to three-step amino acid 
pairs are all nearly equal to their log-odds.
The smallest log-exchangeabilities of these codon pairs reach almost $-40$.
However, there are many codon pairs whose log-exchangeabilities are smaller than
$-40$, and all of them correspond to one- or two-step amino acid pairs.
The log-exchangeabilities of these codon pairs
are significantly smaller than their log-odds,
indicating that
almost all substitutions of these codon pairs were estimated 
in KHG
not to occur by triple nucleotide changes
but rather by successive single or double nucleotide changes.

In the present model, codon exchangeabilities are approximated by the product of
nucleotide exchangeabilities; see \Eq{\ref{eq: def_mutation_rate_matrix}}
for the exact expression. 
Therefore, all codon exchangeabilities for
triple nucleotide changes are in the same order of magnitude, 
and specific codon pairs cannot be significantly less exchangeable.
Thus, 
the present approximation for codon exchangeabilities may have a limitation,
unless those exchangeabilities of KHG are underestimated.
Estimation of the exchangeabilities for those codon pairs,
which require more nucleotide changes than
the least nucleotide changes required for the corresponding amino acid pair,
may be less reliable than for the others.

The ML estimates $\hat{m}_{\xi\eta}$, $\hat{f}^{\script{mut}}_{\xi}$ and $\hat{\sigma}$ 
for KHG are listed in 
\Table{\ref{tbl: optimizations_wij_JTT_WAG_LG_KHG}}.
The scale parameter $\sigma$ of the $\Gamma$ distribution is 
estimated to be $0.0$ for KHG, meaning that variations in rates 
need not be taken into account for KHG.
There is a different tendency in the $\{ \hat{m}_{\xi\eta} \}$ between KHG and 
the amino acid substitution matrices.
One remarkable difference between them
is that the parameter $m_{tc|ag}/m_{[tc][ag]}$ for transition-transversion bias
is estimated to be greater than one in the ML-91 for JTT, WAG, and LG
but to be less than one in the ML-200 for KHG. 
This estimation of transition to transversion bias
for KHG results from a fact that the ratio of 
the total transition to the total transversion substitution rate 
is actually equal to $0.765$ 
in KHG,
although this fact is contrary to the common understanding of transition-transversion bias.
Because selective constraints on amino acids more favor transitions than transversions,
transition-transversion bias in nucleotide mutation rates for KHG
must be much less than $0.765$.
Actually the ratio of the total transition to the total transversion mutation rate 
is estimated to be 0.427; 
see \Table{\ref{tbl: optimizations_wij_JTT_WAG_LG_KHG}}.

\vspace*{2em}
\noindent
\subsection*{
Comparison of ML estimates $\hat{w}_{ab}$ among the present models}
\vspace*{1em}

In \Table{\ref{tbl: correlation_of_wij_between_EI_JTT_WAG_LG_KHG}},
the correlation coefficients of $\hat{w}_{ab}$ between the present models
are listed. 
The lower half of the table lists those for
single-step amino acid pairs, and the upper half lists those for
multi-step amino acid pairs by excluding 
the amino acid pairs that belong to the least exchangeable class
at least in one of the models.
Each model name of JTT/WAG/LG-ML91+ and
KHG-ML200 means the empirical substitution matrix and the method
used to estimate selective constraints, $w_{ab}$.
In the following, these ML estimates of $w_{ab}$ 
will be specified as $\hat{w}_{ab}^{\script{JTT/WAG/LG-ML91+}}$
and $\hat{w}_{ab}^{\script{KHG-ML200}}$.
In the EI method, 
selective constraints are approximated by a linear function of
the energy increment due to an amino acid substitution,
$\Delta \hat{\varepsilon}^{\script{c}}_{ab} + \Delta \hat{\varepsilon}^{\script{v}}_{ab}$,
which is defined by 
\SEqs{4, S1-5, and S1-6};
therefore,
$\hat{w}_{ab}^{\script{EI}} \equiv - (\Delta \hat{\varepsilon}^{\script{c}}_{ab} + \Delta \hat{\varepsilon}^{\script{v}}_{ab})$.

The correlations of the ML estimates $\{ \hat{w}_{ab} \}$ 
between the JTT-ML91+, the WAG-ML91+, and the LG-ML91+ 
are very strong even for the multi-step amino acid pairs.
Comparisons of the ML estimates of selective constraints between various models
are shown in \SFig{\ref{fig: wij_JTT-ML91+_vs_KHG-ML200}}.
The $\{ \hat{w}_{ab}^{\script{KHG-ML200}} \}$ estimated 
from the KHG codon substitution matrix
are less correlated with $\{ \hat{w}_{ab}^{\script{JTT/WAG/LG-ML91+}} \}$
from the other amino acid substitution matrices, especially less 
for the multi-step amino acid pairs.
The ML estimates 
$\{ - \hat{w}_{ab} \}$
for the multi-step amino acid pairs 
are relatively smaller in the KHG-ML200 than in the JTT/WAG/LG-ML91+ models;
see \SFig{\ref{fig: wij_JTT-ML91+_vs_KHG-ML200}}.

The correlations of $\{ \hat{w}_{ab} \}$
between the EI and others are not as good as
those between the other estimates, but
they are significant especially between the EI and the KHG-ML200
even for the multi-step amino acid pairs.
In \Fig{\ref{fig: wij_JTT-ML91+_vs_EI}}A,
the ML estimates $\{ - \hat{w}_{ab}^{\script{JTT-ML91+}} \}$
in the JTT-ML91+
are plotted against 
the energy increments $\{ - \hat{w}_{ab}^{\script{EI}} \}$
due to an amino acid substitution;
the least exchangeable category of multi-step amino acid pairs 
are not shown in this figure.
Similar plots for the WAG-ML91+ and for the LG-ML91+ are shown in 
\SFig{\ref{fig: wij_EI_vs_WAG-ML91+}}.
The ML estimates $\{ - \hat{w}_{ab}^{\script{KHG-ML200}} \}$
for all amino acid pairs in the KHG-ML200
are plotted against the energy increments $\{ - \hat{w}_{ab}^{\script{EI}} \}$ in 
\Fig{\ref{fig: wij_KHG-ML200_vs_EI}}B.
No drastic difference in the correlation between these two quantities 
is found among one-, two-, and three-step amino acid pairs.
The correlations of $\{ \hat{w}_{ab} \}$ 
between the EI and the other models
are better for the ML-91 than for the ML-87;
the correlation coefficient between them
for the single step amino acid pairs
is equal to $0.19$ for the JTT-ML87 but $0.66$ for the JTT-ML91
and $0.30$ for the WAG-ML87 but $0.68$ for the WAG-ML91.
The ML estimates $\{ - \hat{w}_{ab} \}$ 
for the single step amino acid pairs
are compared 
between the ML-87 and the ML-91 models
in \SFig{\ref{fig: wij_JTT-ML87_vs_JTT-ML91}}.

In the next section, we will examine whether the differences 
among these estimates of
$w_{ab}$
are significant in representing selective constraints on amino acids.

\vspace*{2em}
\noindent
\subsection*{
Performance of the ML estimates $\{ \hat{w}_{ab} \}$
and the characteristics of nucleotide mutations estimated
}
\vspace*{1em}

The present model for codon substitutions is designed to separate
selective pressures at the amino acid level from mutational events at the nucleotide level.
Both unequal usage of degenerate codons and different rates of transition and transversion
are characteristic of a genetic system specific to each species and each organelle.
On the other hand, the relative strengths of selective constraints 
on amino acids would be
far less specific to each species and each protein than each type of amino acid,
although the mean strength of the selective constraints
is specific to each protein family.
Thus, we tried to approximate selective constraints ($w_{ab}$)
for empirical substitution matrices including cpREV and mtREV
by a linear function of those ($\hat{w}_{ab}$) 
estimated from each of JTT, WAG, LG, and KHG; $\hat{w}_{ab}^{\script{JTT/WAG/LG-ML91+}}$
and $\hat{w}_{ab}^{\script{KHG-ML200}}$ are used as 
$w^{\mbox{estimate}}_{ab}$ in \Eq{\ref{eq: estimation_of_fitness}}.
We call these models JTT/WAG/LG-ML91+ or KHG-ML200,
which mean the empirical substitution
matrix and the model used to estimate $w^{\mbox{estimate}}_{ab}$,
with a suffix meaning the number of ML parameters; 
see \Table{\ref{tbl: model_names}}.

In \Table{\ref{tbl: optimizations_wij_AIC}},
the ML values for these models with the various sets of parameters are listed
for all empirical substitution matrices.
The ML estimates in the JTT/WAG/LG-ML91+-11 and the KHG-ML200-11 models 
are listed 
in \Tables{\ref{tbl: optimizations_wij_JTT_WAG_LG_X-ML}, 
\ref{tbl: optimizations_wij_CpRev_MtRevRmtx_X-ML},
and
\ref{tbl: optimizations_wij_KHGasm_KHG_X-ML}}.
The JTT-ML91+-0, the WAG-ML91+-0 and the LG-ML91+-0 models
are the codon-based models corresponding 
to the JTT-F, the WAG-F and the LG-F amino-acid-based model, respectively,
in which the JTT, the WAG and the LG rate matrices 
with an adjustment for the equilibrium frequencies of amino acids 
are used as a substitution rate matrix,
because all 11 parameters of $m_{\xi\eta}$, $f^{\script{mut}}_{\xi}$, 
and $\sigma$ 
are fixed to the values of their ML estimators in the ML-91+ 
for JTT, WAG, and LG;
$\beta = 1$ and $w_0 = 0$ are assumed,
However, a critical difference is that a genetic code cannot be taken into account in the JTT/WAG/LG-F 
but in the JTT/WAG/LG-ML94+-0.
This difference between both models can been clearly seen in the present models 
applied to mtREV, because
a non-universal genetic code is used in the vertebrate mitochondrial DNA.
The $\Delta$AIC is improved from $435.6$ in
the JTT-F to $426.0$ in the JTT-ML91+-0.
This indicates an advantage of the present mechanistic model to 
the empirical amino acid substitution model.

The AIC values of the JTT/WAG/LG-ML91+-0
are better for all the four matrices (JTT, WAG, cpREV, and mtREV) 
than those of the physico-chemical method EI-11;
compare \Tables{\ref{tbl: optimizations_AIC} and \ref{tbl: optimizations_wij_AIC}}.
The AIC values of the KHG-200-0 
are better for all except for JTT than those of the EI-11.  
The AIC values of all the models are drastically improved 
for all the matrices by optimizing the 11 parameters; 
see \Table{\ref{tbl: optimizations_wij_AIC}}.
It is noteworthy that all the models of the JTT-ML91+-11, 
the LG-ML91+-11, and the KHG-ML200-11
yield a better AIC value for WAG than the ML-87 model does,
rejecting the null hypothesis of no multiple nucleotide change again; 
see \Tables{\ref{tbl: optimizations_wij_JTT_WAG_LG}
and \ref{tbl: optimizations_wij_AIC}}.
Thus, the ML estimates $\hat{w}^{\script{JTT/WAG/LG-ML91+}}$ 
and $\hat{w}^{\script{KHG-ML200}} $ 
sufficiently represent selective constraints on
amino acid substitutions.

In addition, 
\Table{\ref{tbl: optimizations_wij_AIC}}
indicates which parameters
are the most effective for improving AIC.
As well as the EI models,
the JTT/WAG/LG-ML91+-7, in which the parameters $m_{\xi \eta}$ are
fixed to the ML estimates for JTT/WAG/LG 
with a certain ratio of transition to transversion exchangeability,
can improve the AIC up to
the similar degree to the AIC values of the JTT/WAG/LG-ML91+-11,
respectively.
In other words, the parameters $\{ f^{\script{mut}}_{\xi} \}$ 
are very effective to improve the AIC
in comparison with the parameters $\{ m_{\xi \eta} \}$.

The log-odds values of amino acid pairs estimated by the KHG-ML200-11 
are plotted against their empirical values for the 1-PAM amino acid 
substitution matrices of JTT, WAG, LG, and mtREV
in \Fig{\ref{fig: KHG-ML200_log-odds_1pam_JTT}}.
Similar plots are shown in 
\SFigs{\ref{fig: X-ML91+_log-odds_1pam_JTT} -- \ref{fig: LG-ML91+_log-odds_1pam_KHG}}.
The comparisons of
\Fig{\ref{fig: ML-87_log-odds_1pam_JTT}} and 
\SFig{\ref{fig: ML-87_log-odds_1pam_WAG}}
for the ML-87 model with \Fig{\ref{fig: KHG-ML200_log-odds_1pam_JTT}}
and 
\SFig{\ref{fig: X-ML91+_log-odds_1pam_JTT}}
clearly indicate the good qualities of the ML estimators 
$\hat{w}_{ab}^{\script{KHG-ML200}}$ and $\hat{w}_{ab}^{\script{JTT/WAG/LG-ML91+}}$.
Relatively large disagreements between empirical and estimated log-odds 
exist for cpREV and mtREV in comparison with those for 
JTT, WAG, LG, and the KHG-derived amino acid substitution matrix (KHGaa);
see 
\Fig{\ref{fig: KHG-ML200_log-odds_1pam_MtRevRmtx}}
and
\SFigs{\ref{fig: X-ML91+_log-odds_1pam_JTT}
-- \ref{fig: X-ML91+_log-odds_1pam_KHGasm}}.
It is unknown 
whether 
the disagreements shown in these figures represent meaningful features in 
the amino acid substitutions in the chloroplast DNA and 
the mitochondrial DNA 
or result from 
the relatively small size of sequence data used for cpREV and mtREV.
However, the large disagreements in the region of low log-odds values may be 
artifacts, because
cpREV and mtREV tend to include relatively large errors in this region, 
especially for mtREV;
the log-odds values for mtREV whose values are smaller 
than about $-47.8$ are all assumed to be $-47.8$; 
see the original paper
\CITE{AH:96}.

The ML estimates of $1/\beta$
listed in 
\Tables{\ref{tbl: optimizations_wij_JTT_WAG_LG_X-ML},
\ref{tbl: optimizations_wij_CpRev_MtRevRmtx_X-ML_KHG-ML},
and
\ref{tbl: optimizations_wij_KHGasm_KHG_X-ML}} indicate that
the strength of selective constraints on amino acids is 
strong in the order of LG, WAG, and JTT.
The strength of selective constraints is also shown by
the change of the ratio of nonsynonymous to synonymous rate per codon between the
two cases without and with selective constraints, i.e.,
the cases of $w_{ab} = 0$ and $\sigma \rightarrow 0$, and $\sigma \rightarrow 0$.
As already noted, 
the ratio of these values between the two cases
represents the strength of
selective constraints.  In the  KHG-ML200-11, these ratios are equal to
$0.293/5.23=0.056$, $0.577/5.35=0.11$, and $0.499/3.71=0.13$ 
for LG, WAG, and JTT, respectively,
meaning that the selective constraints of  LG
are strongest; it should be noted that this order agrees with the increasing order of 
$1/\hat{\beta}$.

\Tables{\ref{tbl: optimizations_wij_JTT_WAG_LG_X-ML} 
and \ref{tbl: optimizations_wij_CpRev_MtRevRmtx_X-ML_KHG-ML}}
indicate that
the selective constraints $\hat{w}^{\script{KHG-ML200}}$
estimated from the KHG codon substitution matrix 
tend to estimate 
the contribution of multiple nucleotide changes ($m_{[tc][ag]}$) to be smaller,
the ratio of transition to transversion exchangeability
($m_{tc|ag}/m_{[tc][ag]}$) to be smaller,
$m_{ta}/ m_{[tc][ag]}$ to be larger,
and
variations in substitution rates ($\sigma$) to be less
than the $\hat{w}^{\script{JTT/WAG/LG-ML91+}}$
from the amino acid substitution matrices.
\Table{\ref{tbl: optimizations_wij_KHGasm_KHG_X-ML}} shows
that the same characteristic differences will be observed 
if the JTT/WAG/LG-ML91+-11 models are fitted to
the codon substitution matrix of KHG instead of its derived amino acid substitution matrix.
\Tables{\ref{tbl: optimizations_wij_JTT_WAG_LG_X-ML},
\ref{tbl: optimizations_wij_CpRev_MtRevRmtx_X-ML_KHG-ML},
and
\ref{tbl: optimizations_wij_KHGasm_KHG_X-ML}}
also show
that 
the ratio of transition to transversion exchangeability
($m_{tc|ag}/m_{[tc][ag]}$) 
tends to be estimated to be smaller 
in the order of the LG-ML91+, the WAG-ML91+, the JTT-ML91+, and the KHG-ML200.
The $m_{tc|ag}/m_{[tc][ag]}$ is estimated by the ML-91 or the ML-200 model 
to be smaller in the order of LG, WAG, JTT, and KHG; 
see \Table{\ref{tbl: optimizations_wij_JTT_WAG_LG_KHG}}.
The present ML estimates $\{ \hat{w}_{ab} \}$ for selective constraints on amino acids 
seem to reflect
the characteristics of respective substitution matrices to which the models are fitted.
It remains to be analyzed
which estimation is better among the JTT/WAG/LG-ML91+ and the KHG-ML200
and how better it is.
Irrespective of which estimation of the selection constraints is better,
the ML estimates $\hat{m}_{tc|ag}/\hat{m}_{[tc][ag]}$ indicate that
the transition to transversion bias is not so strong as previously estimated.

One of the interesting facts is that
the ratio of the total transition to the total transversion rate per codon
will be estimated to be much larger if multiple nucleotide changes
are neglected; 
$\hat{m}_{tc|ag}/\hat{m}_{[tc][ag]}$
(and the ratio of the total transition to 
the total transversion rate for $\sigma \rightarrow 0$)
are estimated for the mtREV to be 2.15 (3.32) in the JTT-ML91+-10 but 2.01 (2.52) in the JTT-ML91+-11,
4.27 (4.13) in the WAG-ML91+-10 but 3.43 (2.73) in the WAG-ML91+-11,
4.57 (4.74) in the LG-ML91+-10 but 3.82 (3.31) in the LG-ML91+-11,
and 1.81 (2.58) in the KHG-ML200-10 but 1.64 (1.96) in the KHG-ML200-11.
The same tendency is observed for JTT, WAG, cpREV, and mtREV
irrespective of the matrices,
and for the EI, the Miyata, and the Grantham models irrespective of the models. 

In the case of mtREV,
not only the transition-transversion exchangeability bias 
($\hat{m}_{tc|ag}/\hat{m}_{[tc][ag]}$) 
but also the ratio of the total transition to the total transversion rate
per codon
is larger in the JTT/WAG/LG-ML91+-11 than in the JTT/WAG/LG-ML91+-0, and
in the KHG-ML200-11 than in the KHG-ML200-0. 
Also, the JTT/WAG/LG-ML91+-11 and the KHG-ML200-11 models 
estimate $\hat{m}_{tc|ag}/\hat{m}_{[tc][ag]}$ and
the ratio of the total transition to the total transversion rate to be
larger for mtREV than for JTT, WAG, and cpREV.
These results are
consistent with a well-known fact that
transition to transversion bias is larger in mitochondrial DNA than
in nuclear DNA.

\section*{Discussion}

Halpern and Bruno 
\CITE{HB:98}
considered a codon-substitution model
in which site-specific selection is taken into account 
in terms of residue frequencies.
If site-specific codon frequencies are explicitly taken into account 
in the present model, the substitution rate $R_{\mu\nu}$ 
will be regarded as the average of 
the site-specific rate $R^{i}_{\mu\nu}$ over sites $i$.
According to  \Eq{\ref{eq: def_substitution_rate_matrix}},
the site-specific rate is defined as the product of site-independent
mutation rate $M_{\mu\nu}$ and site-dependent fixation probability,
$(f^i_{\nu}/f^{\script{mut}}_{\nu}) e^{w^{*}_{\mu\nu}}$.
\begin{eqnarray}
R^{i}_{\mu\nu} &=& C_{\script{onst}} M_{\mu\nu} 
	\frac{f^i_{\nu}}{f^{\script{mut}}_{\nu} } e^{w^{*}_{\mu\nu}}
	 \; \mbox{ for } \mu \neq \nu
\end{eqnarray}
Here the site-dependency of the fixation probability is taken into account 
only in terms of codon frequencies.
Then, the average of the site-specific rate over sites is calculated 
as follows.
\begin{eqnarray}
R_{\mu\nu} &\equiv& C_{\script{onst}} \frac{\sum_{i} f^{i}_{\mu} R^{i}_{\mu\nu}}{\sum_{i} f^{i}_{\mu}}
	= 
	C_{\script{onst}}
	M_{\mu\nu} \frac{f_{\nu}}{f^{\script{mut}}_{\nu} } e^{w_{\mu\nu}}
	\; \mbox{ for } \mu \neq \nu
	\\
e^{w_{\mu\nu}} &\equiv& 
	e^{w^{*}_{\mu\nu}} 
	\frac{\sum_i f^i_{\mu} f^i_{\nu} \sum_j 1}
	{\sum_i f^i_{\mu} \sum_j f^j_{\nu} }
	\label{eq: w_modulated_by_site-specific_frequencies}
\end{eqnarray}
where $f_{\nu}$ is the average of $f^i_{\nu}$ over sites.
Thus, the $w_{\mu\nu}$ defined here 
includes the effects of site-specific selection in terms of
codon frequencies.  

In the model of Halpern and Bruno 
\CITE{HB:98},
the term of $e^{w^{*}_{\mu\nu}}$ was not distinguished from and merged with the
mutation rate $M_{\mu\nu}$; that is, $e^{w^{*}_{\mu\nu}} = \mbox{constant}$ 
for $\mu \neq \nu$ was assumed,
Yang and Nielsen 
\CITE{YN:08}
considered mutation-selection models
of codon substitutions and estimated selective strengths on codon usage.
In their models, 
selection pressures that deviate codon frequencies from the equilibrium 
codon frequencies at the mutational level were explicitly taken into account,
and selective constraints on amino acids are 
assumed to be constant over amino acid pairs; that is, 
$e^{w_{ab}} = \mbox{constant}$ for $a \neq b$  was assumed.
However, the site-specific selection 
was not considered; that is, $f_{\mu}^i = f_{\mu}$.
In other words,
unlike the present model, 
selection was taken into account principally in terms of
codon or residue frequencies in both the models.
Also. multiple nucleotide changes were not taken into account.
Halpern and Bruno 
\CITE{HB:98}
developed their model for
distance calculation. 
As pointed out by Yang and Nielsen 
\CITE{YN:08},
taking account of site-specific codon frequencies 
is not practical for real data analysis due to the use of too many parameters.
Instead, the use of $w_{\mu\nu}$ is more practical.
The present results show that
the ML values of the JTT/WAG/cpREV/mtREV amino acid substitution matrices 
are too small
in the No-Constraints models in which $w_{ab} = 0$ is assumed,
and they can be improved by taking account of the term of the selective 
constraints $e^{w_{\mu\nu}}$.
Also, it is indicated that selective constraints on amino acids strongly depend
on the type of amino acid.

In some previous models
\CITE{MJ:93,GY:94,YNH:98},
amino acid substitutions were assumed to proceed in a stepwise manner
by successive single nucleotide changes in a codon.
The empirical amino acid substitution matrices of
JTT, WAG, LG, cpREV, and mtREV, and the codon substitution matrix KHG
all include many substitutions between amino acid or codon pairs
requiring multiple nucleotide changes.
Significance of multiple nucleotide substitutions 
was pointed out 
\CITE{YNH:98,WG:04,ARWS:00,BKOSK:04,KHG:07}.
There are two possible mechanisms to yield substitutions between 
such multi-step amino acid pairs even for a short time interval.
One is variations in substitution rates or time intervals.
Another is multiple nucleotide changes in a codon.
Here, the assumption of multiple nucleotide changes has been
directly introduced into a codon-based substitution model 
together with the use of a $\Gamma$ distribution for variations in substitution rates
and time intervals, and the effectiveness of the assumption has been examined.

In the models using any physico-chemical evaluation of 
selective constraints, 
the significance of multiple nucleotide changes
has been indicated; see 
\Tables{\ref{tbl: optimizations_AIC}
and \ref{tbl: optimizations_wij_JTT_WAG}}.
The ML-87 models fitted to JTT and WAG, 
in which the selective constraints $\{ w_{ab} \}$
for all single-step amino acid pairs are optimized by maximizing the likelihood
with the assumptions of no multiple nucleotide change for codon substitutions and
of variations in substitution rates,
reveal that large discrepancies between the observed and the estimated
log-odds values remain for multi-step amino acid pairs;
see \Fig{\ref{fig: ML-87_log-odds_1pam_JTT}}.
When multiple nucleotide changes are taken into account in the model ML-91,
these discrepancies disappear
and the AIC values significantly decrease,
indicating the significance of multiple nucleotide changes 
in codon substitutions; see \Fig{\ref{fig: ML-91_log-odds_1pam_JTT}},
\SFig{\ref{fig: ML-91_log-odds_1pam_WAG}}, 
and
\Table{\ref{tbl: optimizations_wij_JTT_WAG}}.

Evidence for multiple nucleotide changes was found by 
Averof et al. 
\CITE{ARWS:00},
and the frequency of multiple nucleotide changes was
evaluated 
\CITE{WG:04}.
On the other hand, a possibility for successive single compensatory substitutions
was pointed out by 
Bazykin et al. 
\CITE{BKOSK:04}.
As pointed out by 
Kosiol et al. 
\CITE{KHG:07}, 
the high exchangeabilities of the double nucleotide changes, Rcgt $\leftrightarrow$ Ragg
and Rcgt $\leftrightarrow$ Raga, in KHG 
may result from successive single compensatory substitutions.
On the other hand, 
a selection on synonymous substitutions is necessary for
compensatory substitutions to cause 
the higher exchangeability of Rcga $\leftrightarrow$ Ragg
than estimated, 
because the most probable paths of single nucleotide changes 
between Rcga and Ragg are
Rcga $\leftrightarrow$ Raga $\leftrightarrow$ Ragg
and
Rcga $\leftrightarrow$ Rcgg $\leftrightarrow$ Ragg both of which
do not accompany any amino acid change;
see \Fig{\ref{fig: ML200_logodds_codon_1pam_KHG_2x}}. 
Whatever causes 
multiple nucleotide changes,
the present scheme for codon substitutions could be applied to 
phylogenetic analyses of protein-coding sequences, because
the underlying time scale in the present substitution model
is much longer than that of positive selection for successive single compensatory substitutions.

The models JTT/WAG/LG-ML91+-0 and KHG-ML200-0, in which parameters
are taken to be equal to the ML estimates for JTT/WAG/LG
in the ML-91+ model and the ML estimates for KHG in the ML-200 model,
are codon-based models corresponding to the JTT/WAG/LG/KHG-F model, 
respectively.
The model ML-91+ 
can almost perfectly reproduce JTT, WAG, and LG.
The model ML-200 for the KHG codon substitution matrix can well reproduce
the codon substitution probabilities for the codon pairs for which any nucleotide change
is accompanied by an amino acid change, although the exchangeabilities 
of the other codon pairs are over-estimated for KHG.
This means that 
the JTT/WAG/LG-ML91+-0 and the KHG-ML200-0 models can be used
as a simple substitution model 
without any loss of information instead of
the empirical substitution matrices of the JTT/WAG/LG/KHG
in maximum likelihood and Bayesian inferences of 
phylogenetic trees of amino acid and codon sequences, respectively.
Although the empirical substitution matrices represent the average
tendencies of substitutions over proteins and species and 
may lack gene-level resolution\CITE{DSGMP:10,DSBGMP:10},
the present mechanistic codon model has
adjustable parameters for nucleotide mutation and
for the strength of selective constraints,
which can be tailored to specific genes.
It is possible to optimize the selective constraints $\{w_{ab}\}$
for each gene. However, such a method\CITE{HJLR:08,DSGMP:10,DSBGMP:10}
is far more computer-intensive than the present method.
The present methods, JTT/WAG/LG-ML91+-$n$ 
using 
$\hat{w}^{\script{JTT/WAG/LG-ML91+}}$ 
and the KHG-ML200-$n$ with
the $\hat{w}^{\script{KHG-ML200}}$, 
provide alternative models for amino acid/codon substitutions 
with a small number of ML parameters 
in the probabilistic inference of phylogenetic trees.
The number of ML parameters specific to the present model is 
at most 6 exchangeabilities and 3 equilibrium frequencies for nucleotide mutations,
and 2 parameters for selective constraints.
Thus, the present model requires the same order of cpu time as 
the nucleotide substitution model (GTR) does.
In other codon models\CITE{DP:07,SK:08}, 
exchangeabilities between amino acids are 
taken to be equal to their values in empirical amino acid substitution matrices.
However, in the present codon model,
amino acid and codon exchangeabilities
vary according to nucleotide mutation rates 
and the strength of selective constraints.

The parameters $m_{\xi\eta}$, $f_{\xi}$, and $\sigma$ 
are differently estimated by
the KHG-ML200-$n$
and the JTT/WAG/LG-ML91+-$n$
using different $\hat{w}$; 
see \Tables{\ref{tbl: optimizations_wij_JTT_WAG_LG_X-ML_KHG-ML},
\ref{tbl: optimizations_wij_CpRev_MtRevRmtx_X-ML_KHG-ML}},
and
{\ref{tbl: optimizations_wij_KHGasm_KHG_X-ML}}.
The $\hat{w}^{\script{KHG-ML200}}$ 
yields a smaller rate of multiple nucleotide changes,
a smaller $\sigma$, 
a smaller ratio of transition to transversion exchangeability,
and 
a smaller ratio of nonsynonymous to synonymous rate per codon than
the $\hat{w}^{\script{JTT/WAG/LG-ML91+}}$ does. 
Whichever estimation is better, 
the present ML estimators $\hat{m}_{tc|ag} / \hat{m}_{[tc][ag]}$ for
transition-transversion bias strongly indicate that
the transition-transversion bias is not so large as previously estimated.
An excess of transitional over transversional substitutions
was shown in the DNA sequences of metazoa, and has been assumed to be universal.
However, Keller et al. 
\CITE{KBN:07}
found a counter example to 
the transition-transversion bias from grasshopper pseudogenes.
The present ML estimate of the ratio of transition to transversion exchangeability for
the KHG codon substitution matrix is rather less than 1.0, i.e., 
$\hat{m}_{tc|ag}/\hat{m}_{[tc][ag]} = 0.843$ in the ML-200 model, 
which corresponds to the overall rate bias of transitions over transversions, $0.427$.
Even for the amino acid substitution matrices JTT, WAG, and LG, 
the ML-91 model estimates 
$m_{tc|ag}/m_{[tc][ag]}$ to be less than $1.9$,
making the overall rate bias of transitions over transversions less than $1.0$;
see \Table{\ref{tbl: optimizations_wij_JTT_WAG_LG_KHG}}.
It should be noted that the ratio of transition to transversion exchangeability
tends to be overestimated if no multiple nucleotide change is allowed;
see 
\STables{\ref{tbl: optimizations_EI_WAG_CpRev_MtRevRmtx}
and
\ref{tbl: optimizations_selection_JTT_WAG}}.
Thus, the present results indicate that 
transition-transversion bias is not a solid assumption.
On the other hand, the present results indicate that
transition-transversion bias is stronger in mitochondrial DNA than in nuclear DNA
in accordance with previous understanding; 
see \Tables{\ref{tbl: optimizations_wij_JTT_WAG_LG_X-ML_KHG-ML}
and 
\ref{tbl: optimizations_wij_CpRev_MtRevRmtx_X-ML_KHG-ML}}.

The ML estimates 
$\{ \hat{w}^{\script{JTT/WAG/LG-ML91+}}_{ab} \}$
and
$\{ \hat{w}^{\script{KHG-ML200}}_{ab} \}$
significantly correlate with each other 
and also with the mean energy increments due to an amino acid replacement.
However, the JTT/WAG/LG-ML91+-$n$ and KHG-ML200-$n$ models
fit substitution data significantly better than the EI-$n$ model;
see \Tables{\ref{tbl: optimizations_EI_AIC}
and 
\ref{tbl: optimizations_wij_AIC}}.
This fact indicates that the differences between the physico-chemical estimates 
and the ML estimates
$\{ \hat{w}_{ab} \}$ for selective pressure at the amino acid level
reflect the actual tendency of selective constraints
for respective types of amino acid pairs
in protein evolution.
\Eq{\ref{eq: w_modulated_by_site-specific_frequencies}}
indicates that the $w$ is modulated by site-specific 
codon frequencies and differentiated from 
the site-independent constraints, $w^{*}$, which
may be more similar to the physico-chemical estimates than the $w$.
The selective constraints estimated here
may be used as a base line
to detect evidence of positive selection.
Models\CITE{WG:04,YN:08} in which the dependences of
selective constraints on amino acid pairs are not taken into account
may be improved by introducing them.
On the other hand, it still remains to be examined
whether or not the JTT/WAG/LG-ML91+-$n$ and the KHG-ML200-$n$
perform comparably with cpREV 
for the maximum likelihood inferences 
of phylogenetic trees of chloroplast proteins 
and with mtREV for those of mitochondrial proteins.
Also, it should be examined which performs better.

A preliminary calculation has been pursued to examine the performance of 
the present substitution models in
the ML inference of a phylogenetic tree.
Log-likelihoods of the present models and the codon models corresponding to 
the mtREV-F, the JTT-F, the WAG-F, and the LG-F
are calculated and listed in \Table{\ref{tbl: LK_of_mt_tree}}
for a phylogenetic tree\CITE{AH:96}
of the concatenated sequences of 12 protein-coding sequences
encoded on the same strand of mitochondrial DNA
from 20 vertebrate species with 2 races from human.
The phylogenetic tree and the proteins used are those
which Adachi and Hasegawa\CITE{AH:96} used to estimate mtREV;
the Japanese mtDNA was not used because it couldn't be found 
in the GenBank database.
The coding sequences of each protein were aligned with codon score matrices
by the ClustalW2\CITE{LBBCMMVWWLTGH:07}, and then concatenated.
Their likelihoods on the phylogenetic tree were calculated 
by the Phyml\CITE{GG:03}.
Both the programs have been modified for the analysis of coding sequences.
Log-odds calculated by the KHG-ML200-11 fitted
to mtREV were used as the codon score matrices.
Positions with gaps are included for the calculation of the likelihoods.
The codon substitution matrices corresponding to mtREV, JTT, WAG, LG, 
and the KHG-derived amino acid substitution matrix (KHGaa) 
are calculated in such a way that
codon exchangeabilities for nonsynonymous codon pairs are taken to be equal to
$\exp w_0$ multiplied by
the exchangeability of the corresponding amino acid pair and 
those for synonymous codon pairs are assumed to be all equal to 
the mean amino acid exchangeability.
In all models, 
the parameter $w_0$ in \Eq{\ref{eq: estimation_of_fitness}} was optimized
even for the No-Constraints models,
and codon frequencies were taken to be equal to those in coding sequences.
The substitution matrices, JTT, WAG, LG, and KHG were estimated from
nuclear DNA, 
which use a different genetic code from vertebrate mtDNA.
On the other hand, 
mtREV was estimated by a maximum likelihood method from 
the almost same set of the protein sequences encoded in mtDNA.
Thus, it is expected that
the log-likelihood values of the mtDNA phylogenetic tree
for the models, KHGaa-1-F, LG-1-F, WAG-1F, and JTT-1-F are
worse than that for the mtREV-1-F.
An important thing is that
the codon models 
with the selective constraints
estimated from nuclear DNA or by the physico-chemical method
yield a much smaller value of AIC than the mtREV-1-F.
One of the effective parameters is $w_0$ that directly 
controls the ratio of nonsynonymous to synonymous substitution rate.
It also improves the likelihood to explicitly take account of 
rate variations over sites.
The discrete approximation\CITE{Y:94} 
of the $\Gamma$ distribution
with 4 categories 
was used to represent rate variations over sites in the models named with the suffix "dG4";
the shape parameter $\alpha$ is a ML parameter.
An interesting and reasonable fact is that 
averaging substitution matrices over rate becomes unnecessary, i.e., 
$\hat{\sigma} = 0.0$, in the case that
rate variations over sites are explicitly taken into account; 
in the Yang's model\CITE{Y:93,Y:94},   
the likelihood of a phylogenetic tree of each site is averaged over rate. 
Also, all the present codon-based models estimate 
$\hat{m}_{[tc][ag]} > 0.1$, which indicates the significance of 
multiple nucleotide changes.
The present results strongly indicate that 
the tendencies of nucleotide mutations and codon usage are characteristic
of a genetic system specific to each species and oranelle, but
the amino acid dependences of selective constraints
are more specifc to each type of amino acid than each species, 
organelle, and protein family.
Full evaluation will be provided in a succeeding paper.

One may question
whether the whole evolutionary process
of protein-coding sequences can be approximated
by a reversible Markov process or not.
Kinjo and Nishikawa
\CITE{KN:04}
reported that the log-odds matrices
constructed for 18 different levels of sequence identities
from structure-based protein alignments have a characteristic
dependence on time in the principal components of their eigenspectra.
Although they did not explicitly mention,
this type of temporal process peculiar to the log-odd matrix in protein evolution  
is fully encoded in the transition matrices of JTT, WAG, LG, and KHG.
In \SFig{\ref{fig: log-odds-ev_ML91+}},
it is shown
that this characteristic dependence of log-odds on time
can be reproduced by the transition matrix
based on the present reversible Markov model fitted to JTT;
see Supporting Information, Text S1, for details.
This fact supports the appropriateness of the present Markov model for codon substitutions.
The present codon-based model can be used to generate
log-odds for codon substitutions as well as amino acid substitutions. 
Such a log-odds matrix of codon substitutions would be useful
to allow us to align nucleotide sequences at the codon level rather than 
the amino acid level, increasing 
the quality of sequence alignments.

As a result, the present model would enable us to obtain more biologically meaningful
information at both nucleotide and amino acid levels 
from codon sequences and even from protein sequences, 
because this is a codon-based model.

\section*{Acknowledgments}
The author would like to thank Prof. Masami Hasegawa and Prof. Hiroyuki Toh
for their valuable advice.
I also thank reviewers for constructive suggestions 
on the manuscript.

\bibliography{plos_article_for_arxiv}

\clearpage

\renewcommand{\FigureInLegends}[1]{#1}
\newcommand{\TextFig}[1]{#1}
\newcommand{\SupFig}[1]{}

\newcommand{\FigurePanel}[1]{#1}
\newcommand{\FigureEach}[1]{}

\TextFig{

\FigureEach{

\begin{figure*}[ht]
\FigureInLegends{
\centerline{
\includegraphics*[width=130mm,angle=0]{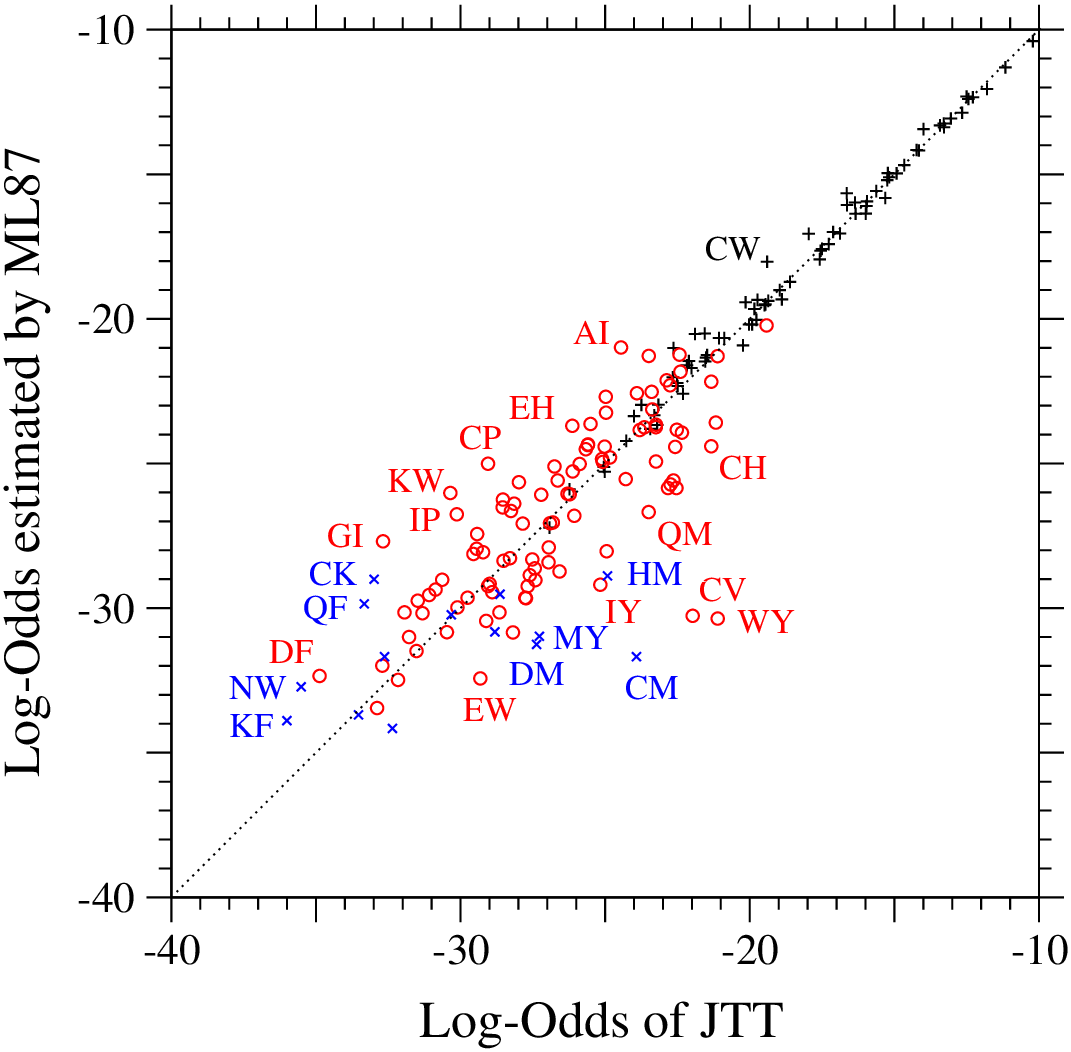}
}
} 
\caption{
\label{fig: ML-87_log-odds_1pam_JTT}
Each element 
log-$O(\langle S \rangle(\hat{\tau},\hat{\sigma}))_{ab}$
of the log-odds matrix 
of the ML-87 model fitted to the 1-PAM JTT matrix
is plotted against the log-odds log-$O(S^{\script{JTT}}(\mbox{1 PAM}))_{ab}$
calculated from JTT.
Plus, circle, and cross marks show the log-odds values for
the types of substitutions requiring single, double and triple nucleotide changes,
respectively.  
The dotted line shows the line 
of equal values between the ordinate and the abscissa.
}
\end{figure*}

\FigureInLegends{\newpage}

\begin{figure*}[ht]
\FigureInLegends{
\centerline{
\includegraphics*[width=130mm,angle=0]{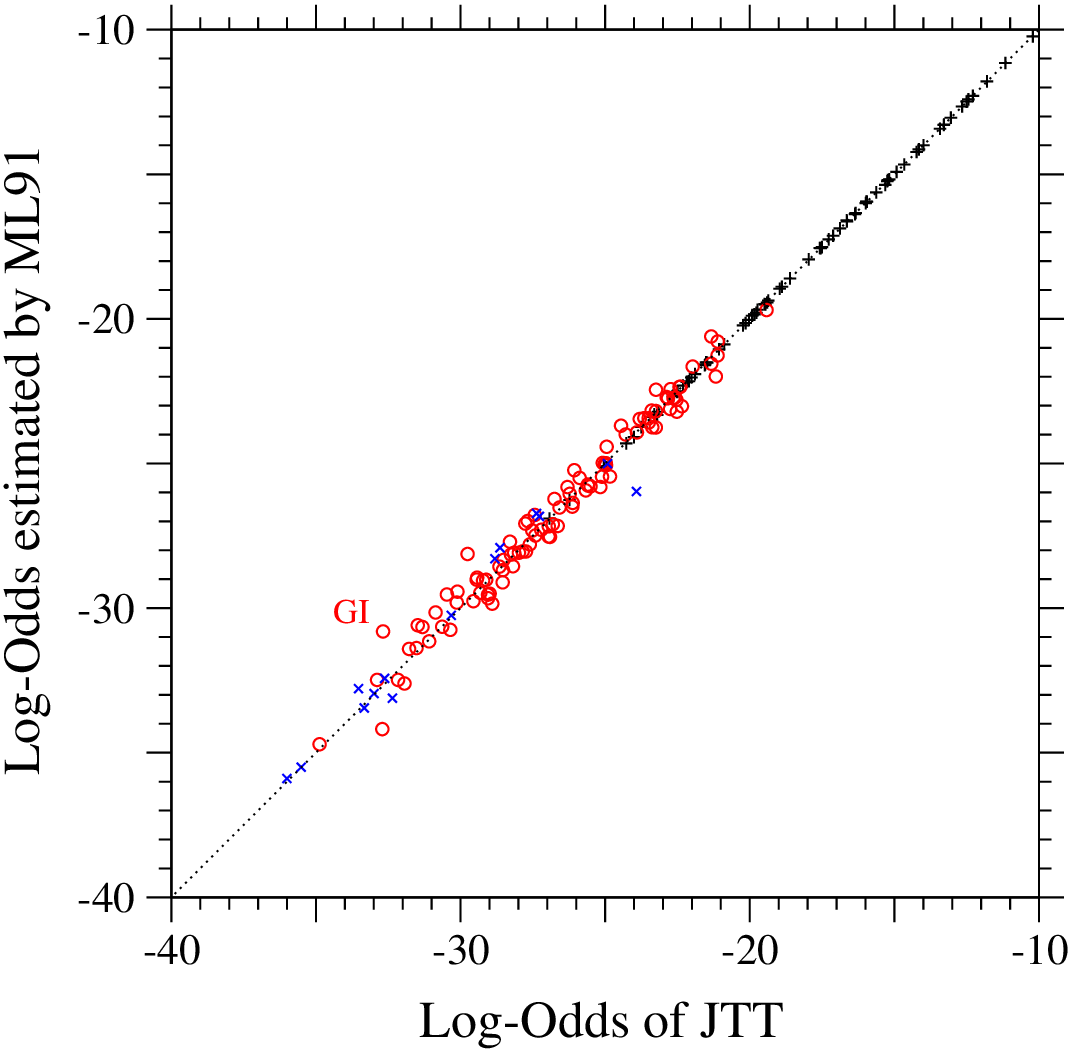}
}
} 
\caption{
\label{fig: ML-91_log-odds_1pam_JTT}
Each element 
log-$O(\langle S \rangle(\hat{\tau},\hat{\sigma}))_{ab}$
of the log-odds matrix 
of the ML-91 model fitted to the 1-PAM JTT matrix
is plotted against the log-odds log-$O(S^{\script{JTT}}(\mbox{1 PAM}))_{ab}$
calculated from JTT.
Plus, circle, and cross marks show the log-odds values for
one-, two-, and three-step amino acid pairs,
respectively.  
The dotted line shows the line 
of equal values between the ordinate and the abscissa.
}
\end{figure*}

} 

\FigurePanel{

\begin{figure}[!ht]
\FigureInLegends{
\textbf{A} \hspace*{47em} \textbf{B}

\centerline{
\includegraphics*[width=80mm,angle=0]{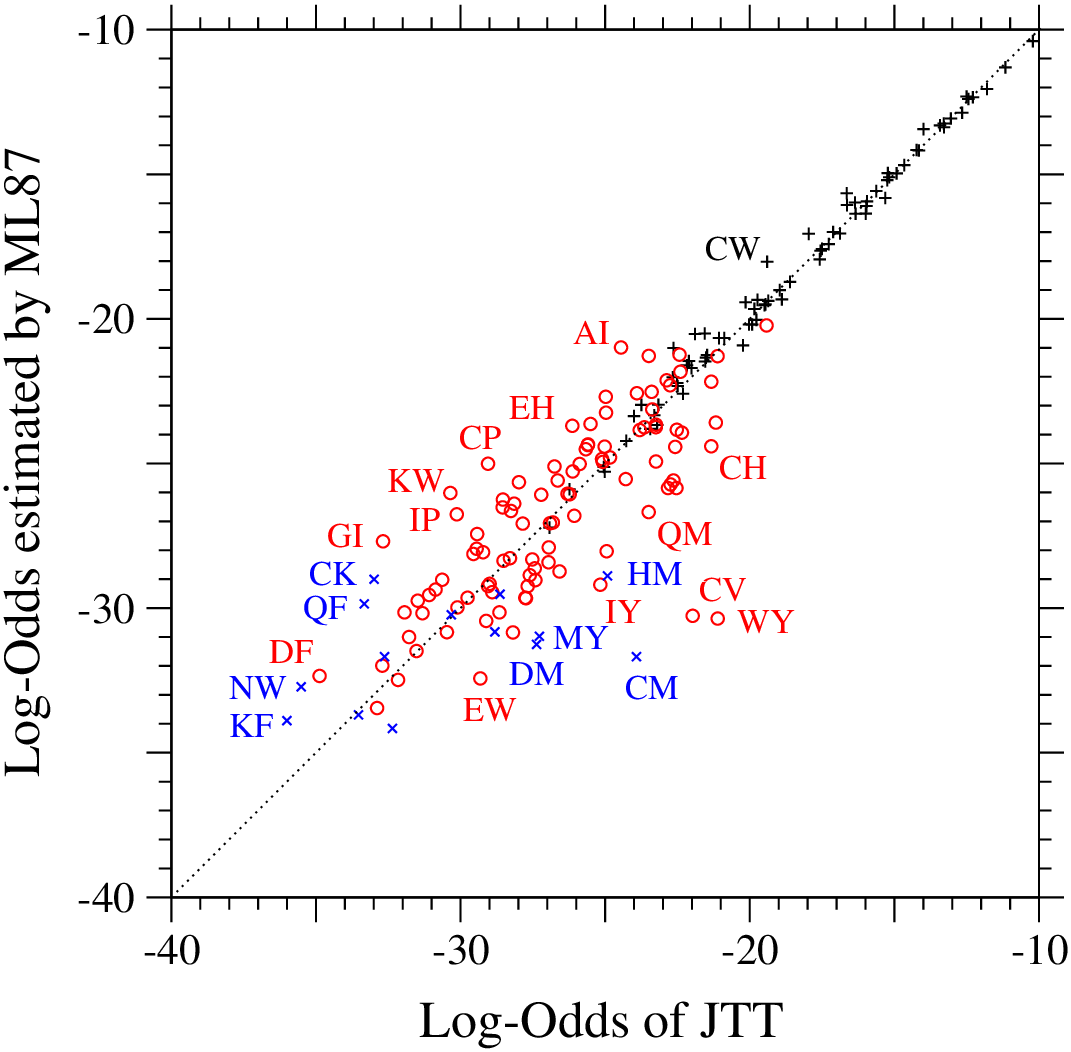}
\hspace*{5mm}
\includegraphics*[width=80mm,angle=0]{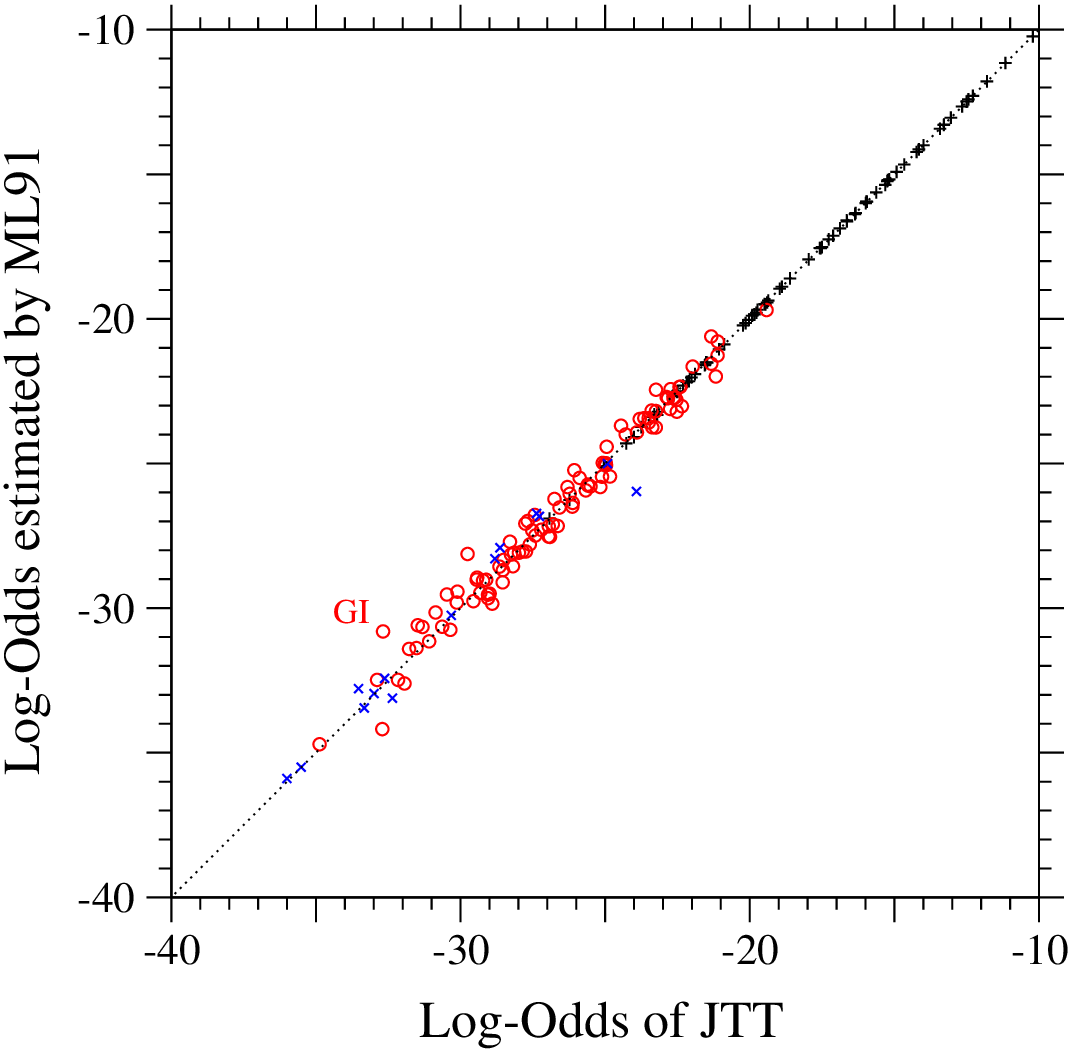}
}
} 
\caption{
\label{fig: ML-87_log-odds_1pam_JTT}
\label{fig: ML-91_log-odds_1pam_JTT}
\BF{
The ML-87 and the ML-91 models fitted to JTT.
} 
Each element 
log-$O(\langle S \rangle(\hat{\tau},\hat{\sigma}))_{ab}$
of the log-odds matrices 
of (A) the ML-87 and (B) the ML-91 models 
fitted to the 1-PAM JTT matrix
is plotted against the log-odds log-$O(S^{\script{JTT}}(\mbox{1 PAM}))_{ab}$
calculated from JTT.
Plus, circle, and cross marks show the log-odds values for
the types of substitutions requiring single, double and triple nucleotide changes,
respectively.  
The dotted line in each figure shows the line 
of equal values between the ordinate and the abscissa.
}
\end{figure}

} 

} 

\SupFig{

\FigureEach{

\begin{figure*}[ht]
\FigureInLegends{
\centerline{
\includegraphics*[width=130mm,angle=0]{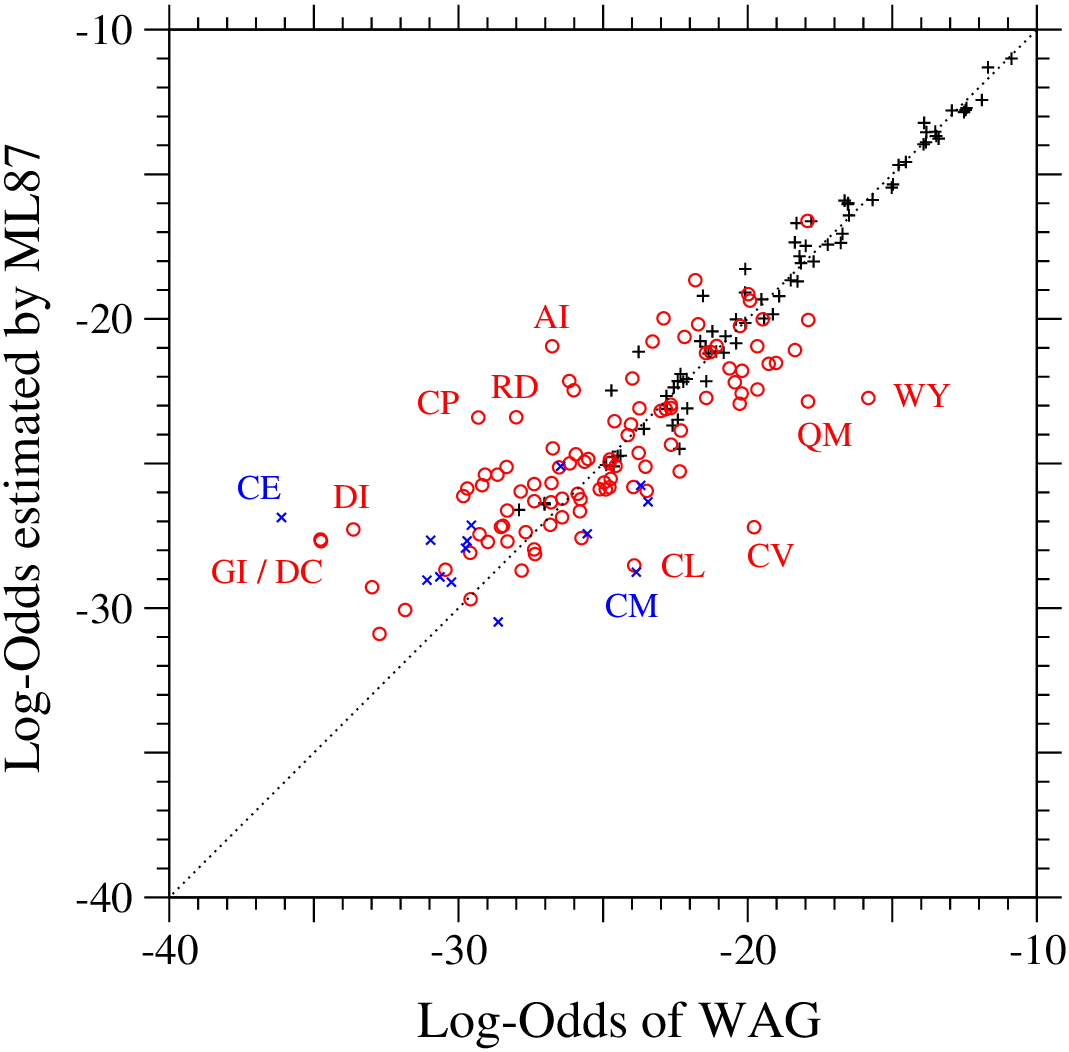}
}
} 
\caption{\label{fig: ML-87_log-odds_1pam_WAG}
Each element 
log-$O(\langle S \rangle(\hat{\tau},\hat{\sigma}))_{ab}$
of the log-odds matrix 
of the ML-87 model fitted to the 1-PAM WAG matrix
is plotted against the log-odds log-$O(S^{\script{WAG}}(\mbox{1 PAM}))_{ab}$
calculated from WAG.
Plus, circle, and cross marks show the log-odds values for
one-, two-, and three-step amino acid pairs,
respectively.  
The dotted line shows the line 
of equal values between the ordinate and the abscissa.
}
\end{figure*}

\begin{figure*}[ht]
\FigureInLegends{
\centerline{
\includegraphics*[width=130mm,angle=0]{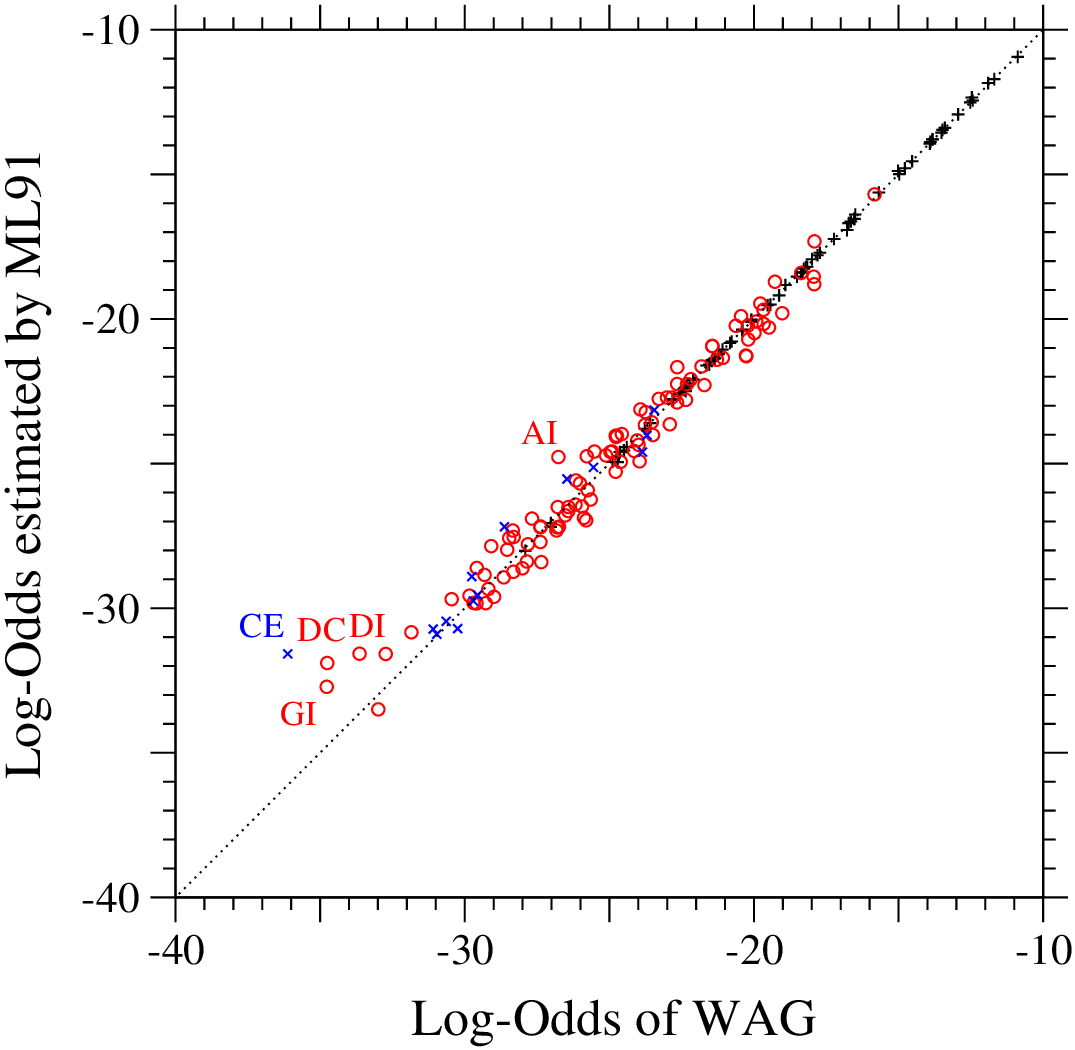}
}
} 
\caption{\label{fig: ML-91_log-odds_1pam_WAG}
Each element 
log-$O(\langle S \rangle(\hat{\tau},\hat{\sigma}))_{ab}$
of the log-odds matrix 
of the ML-91 model fitted to the 1-PAM WAG matrix
is plotted against the log-odds log-$O(S^{\script{WAG}}(\mbox{1 PAM}))_{ab}$
calculated from WAG.
Plus, circle, and cross marks show the log-odds values for
one-, two-, and three-step amino acid pairs,
respectively.  
The dotted line shows the line 
of equal values between the ordinate and the abscissa.
}
\end{figure*}

} 

\FigurePanel{

\FigureInLegends{\newpage}

\begin{figure}[t]
\FigureInLegends{
\textbf{A} \hspace*{47em} \textbf{B}

\centerline{
\includegraphics*[width=80mm,angle=0]{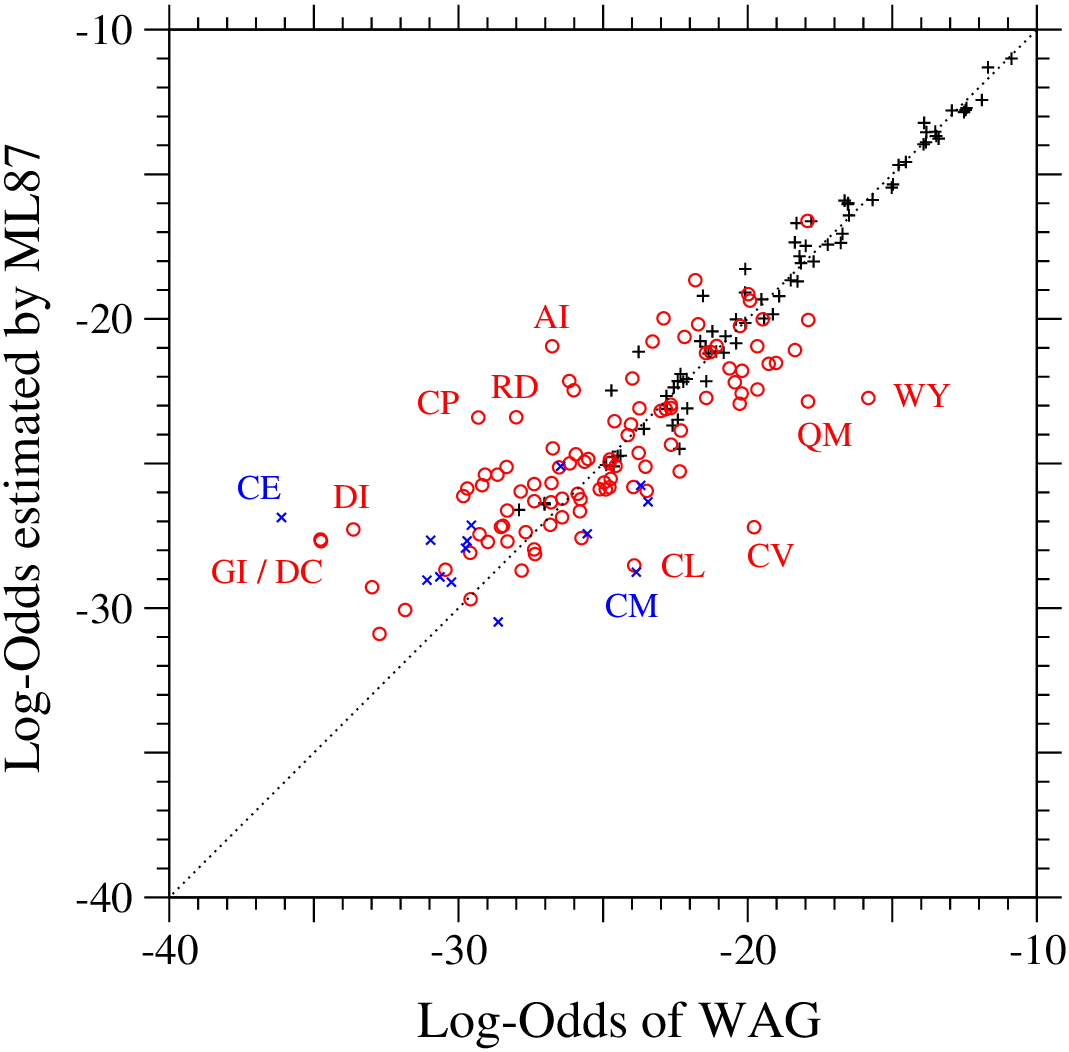}
\hspace*{5mm}
\includegraphics*[width=80mm,angle=0]{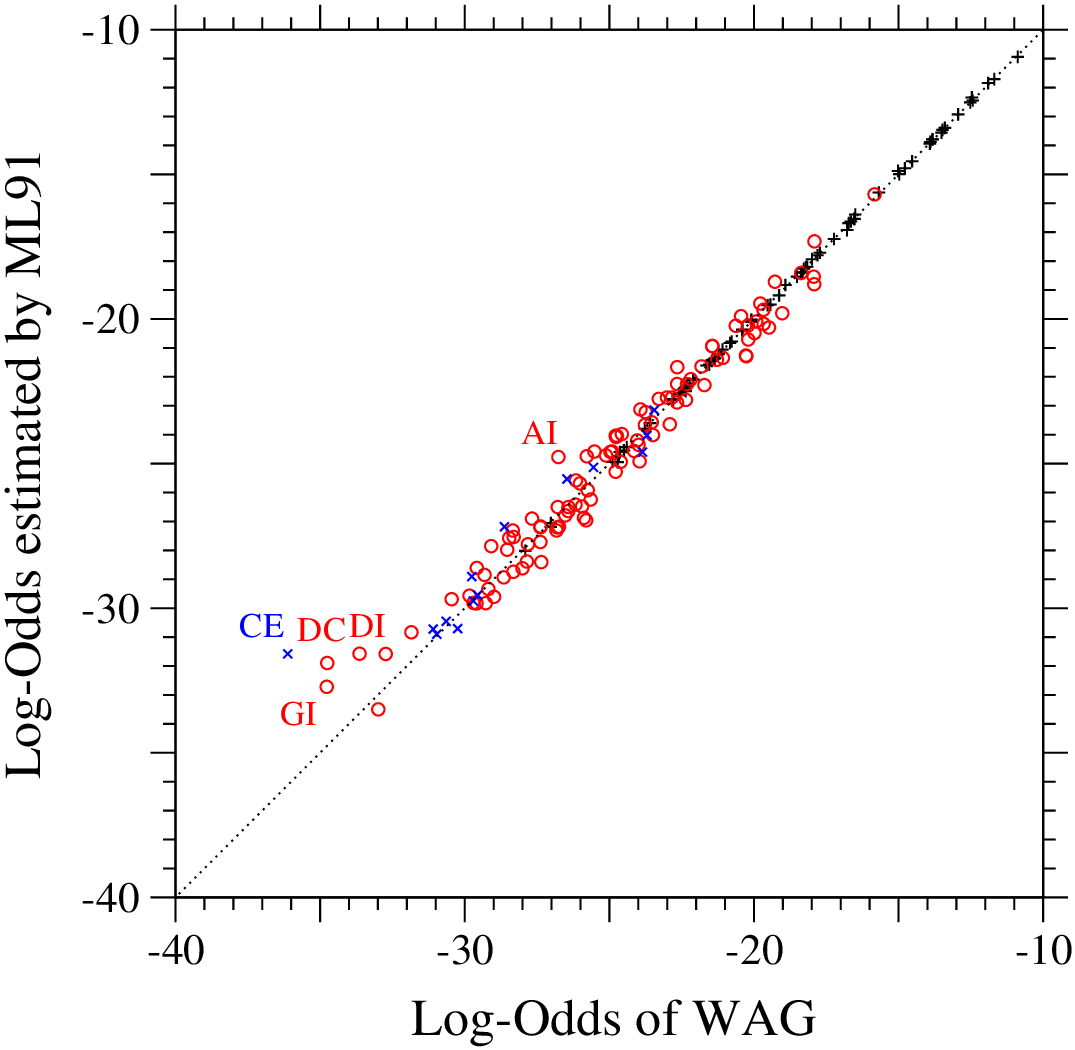}
}
} 
\caption{
\label{fig: ML-87_log-odds_1pam_WAG}
\label{fig: ML-91_log-odds_1pam_WAG}
\BF{
The ML-87 and the ML-91 models fitted to WAG.
} 
Each element 
log-$O(\langle S \rangle(\hat{\tau},\hat{\sigma}))_{ab}$
of the log-odds matrices
of (A) the ML-87 and (B) the ML-91 models fitted to the 1-PAM WAG matrix
is plotted against the log-odds log-$O(S^{\script{WAG}}(\mbox{1 PAM}))_{ab}$
calculated from WAG.
Plus, circle, and cross marks show the log-odds values for
one-, two-, and three-step amino acid pairs,
respectively.  
The dotted line in each figure shows the line 
of equal values between the ordinate and the abscissa.
}
\end{figure}

} 

} 

\TextFig{

\FigureEach{

\FigureInLegends{\newpage}

\begin{figure*}[ht]
\FigureInLegends{
\centerline{
\includegraphics*[width=130mm,angle=0]{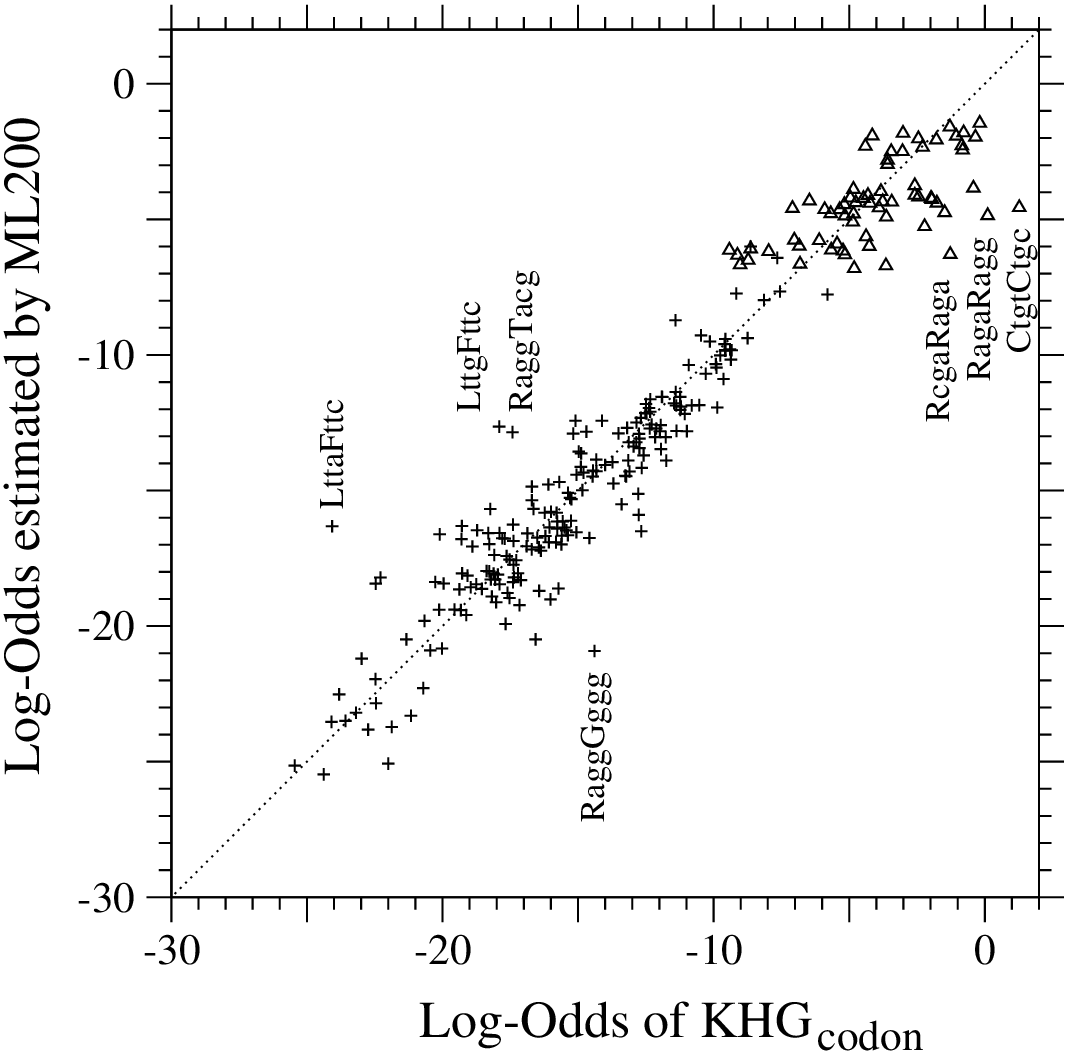}
}
} 
\caption{\label{fig: ML200_logodds_codon_1pam_KHG_1x}
Each element 
log-$O(\langle S \rangle(\hat{\tau},\hat{\sigma}))_{\mu \nu}$
of the log-odds matrix 
corresponding to single nucleotide changes 
in the ML-200 model fitted to the 1-PAM KHG codon substitution matrix
is plotted against the log-odds log-$O(S^{\script{KHG}}(\mbox{1 PAM}))_{\mu \nu}$
calculated from KHG.
Upper triangle and plus marks show the log-odds values for
synonymous pairs and single-step amino acid pirs,
respectively. 
The dotted line shows the line 
of equal values between the ordinate and the abscissa.
}
\end{figure*}

\FigureInLegends{\newpage}

\begin{figure*}[ht]
\FigureInLegends{
\centerline{
\includegraphics*[width=130mm,angle=0]{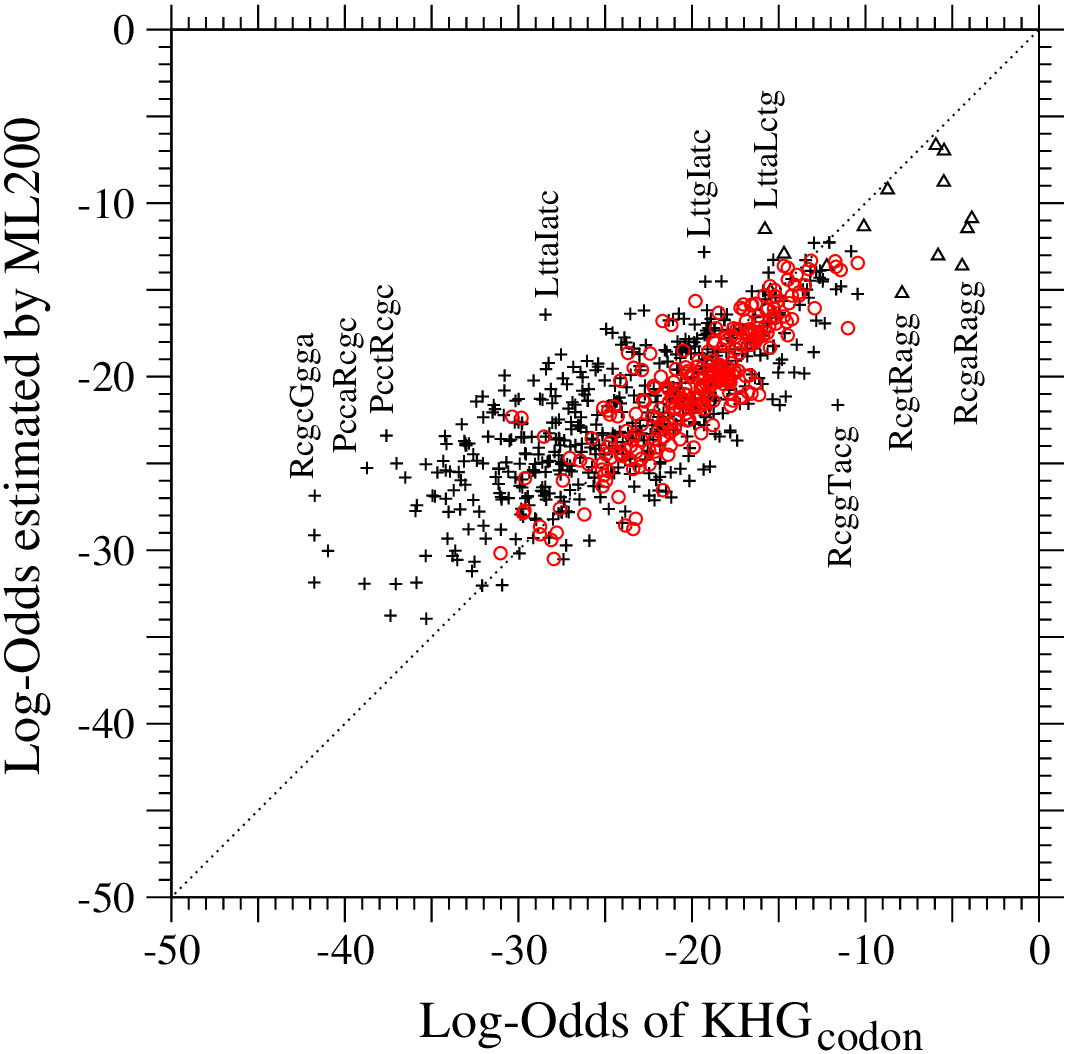}
}
} 
\caption{\label{fig: ML200_logodds_codon_1pam_KHG_2x}
Each element 
log-$O(\langle S \rangle(\hat{\tau},\hat{\sigma}))_{\mu \nu}$
of the log-odds matrix 
corresponding to double nucleotide changes 
in the ML-200 model fitted to the 1-PAM KHG codon substitution matrix
is plotted against the log-odds log-$O(S^{\script{KHG}}(\mbox{1 PAM}))_{\mu \nu}$
calculated from KHG.
Upper triangle, plus, and circle marks show the log-odds values for
synonymous pairs and one-, and two-step amino acid pairs,
respectively. 
The dotted line shows the line 
of equal values between the ordinate and the abscissa.
}
\end{figure*}

\FigureInLegends{\newpage}

\begin{figure*}[ht]
\FigureInLegends{
\centerline{
\includegraphics*[width=130mm,angle=0]{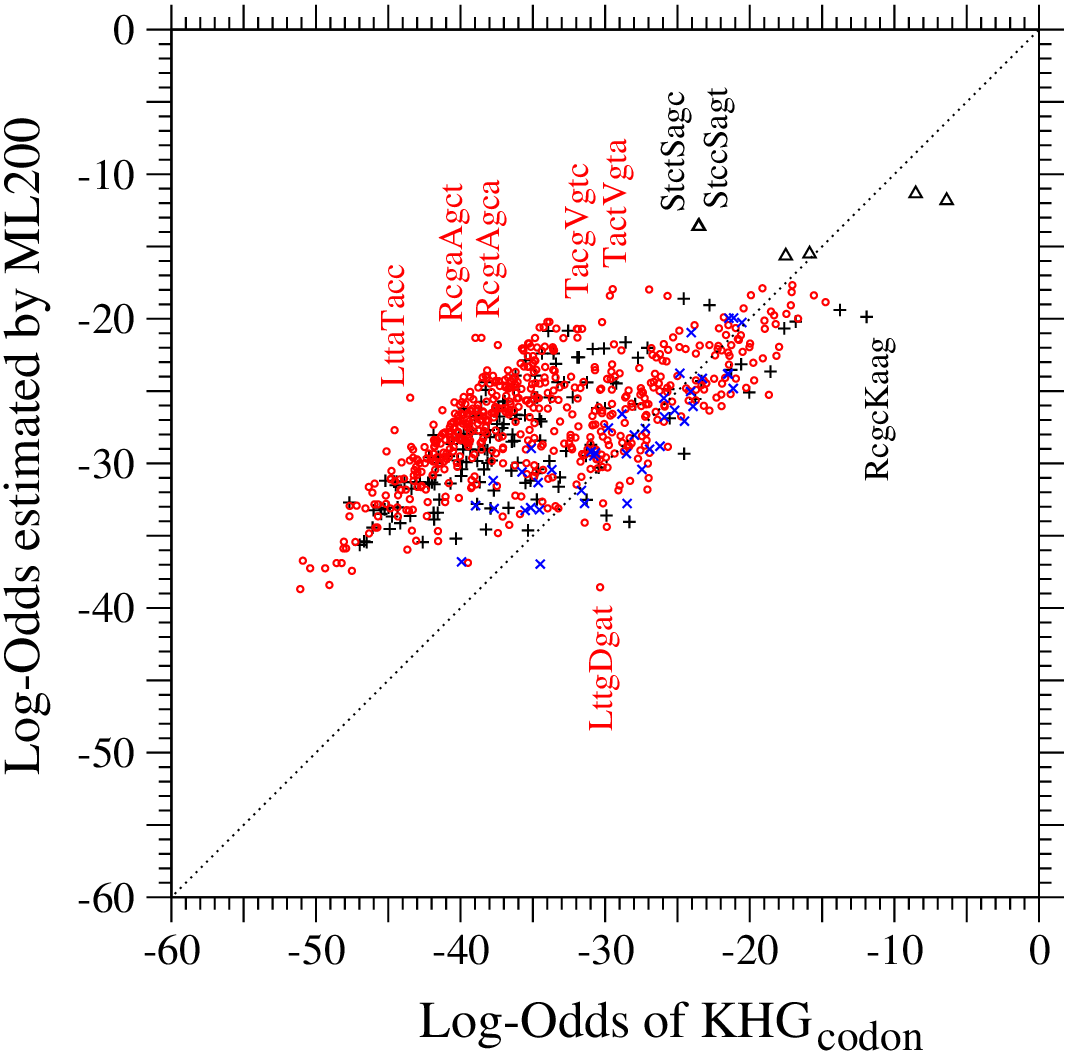}
}
} 
\caption{\label{fig: ML200_logodds_codon_1pam_KHG_3x}
Each element 
log-$O(\langle S \rangle(\hat{\tau},\hat{\sigma}))_{\mu \nu}$
of the log-odds matrix 
corresponding to triple nucleotide changes 
in the ML-200 model fitted to the 1-PAM codon KHG substitution matrix
is plotted against the log-odds log-$O(S^{\script{KHG}}(\mbox{1 PAM}))_{\mu \nu}$
calculated from KHG.
Upper triangle, plus, circle, and cross marks show the log-odds values for
synonymous pairs and one-, two-, and three-step amino acid pairs,
respectively. 
The dotted line shows the line 
of equal values between the ordinate and the abscissa.
}
\end{figure*}

\FigureInLegends{\newpage}

\begin{figure*}[ht]
\FigureInLegends{
\centerline{
\includegraphics*[width=130mm,angle=0]{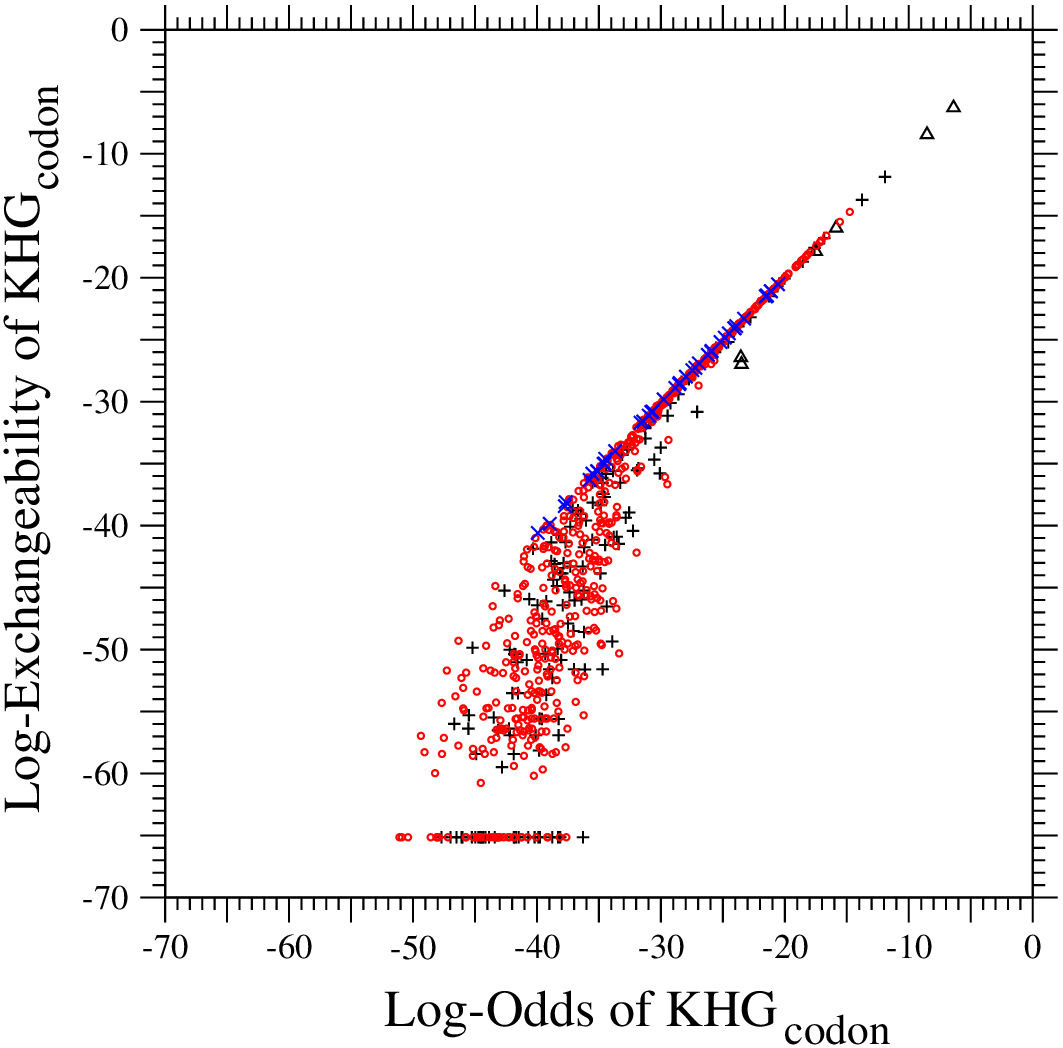}
}
} 
\caption{\label{fig: ML200_logodds_vs_exchange_scaled_1pam_observed_KHG_3x}
\label{fig: logodds_vs_exchange_scaled_1pam_observed_KHG_3x}
Codon log-exchangeabilities of codon substitutions corresponding to triple nucleotide changes in the 1-PAM KHG
are plotted against the log-odds log-$O(S^{\script{KHG}}(\mbox{1 PAM}))_{\mu \nu}$
calculated from KHG.
The log-exchangeability of the 1-PAM KHG 
is defined as $(10 / \log 10) \log [R^{\script{KHG}}_{\mu\nu} \cdot t_{\script{1-PAM}} / f_{\nu} ]$.
Upper triangle, plus, circle, and cross marks show the log-odds values for
synonymous pairs and one-, two-, and three-step amino acid pairs,
respectively. 
Log-exchangeabilities for the codon pairs whose instantaneous rate is estimated to be $0$ in KHG are
shown to be about $-65$ in this figure.
The dotted line shows the line 
of equal values between the ordinate and the abscissa.
}
\end{figure*}

} 

\FigurePanel{

\FigureInLegends{\newpage}

\begin{figure}[ht]
\FigureInLegends{
\textbf{A} \hspace*{47em} \textbf{B}

\centerline{
\includegraphics*[width=80mm,angle=0]{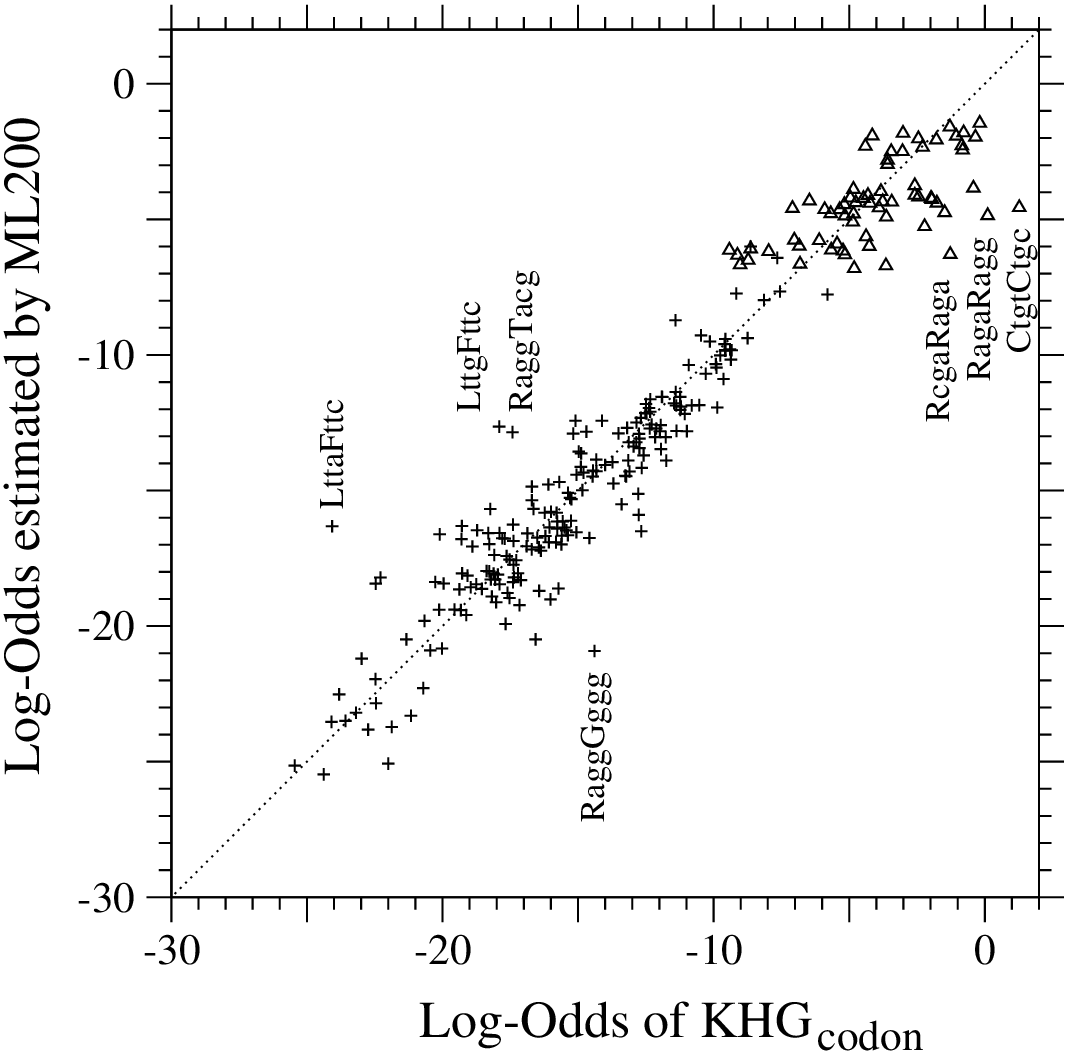}
\hspace*{5mm}
\includegraphics*[width=80mm,angle=0]{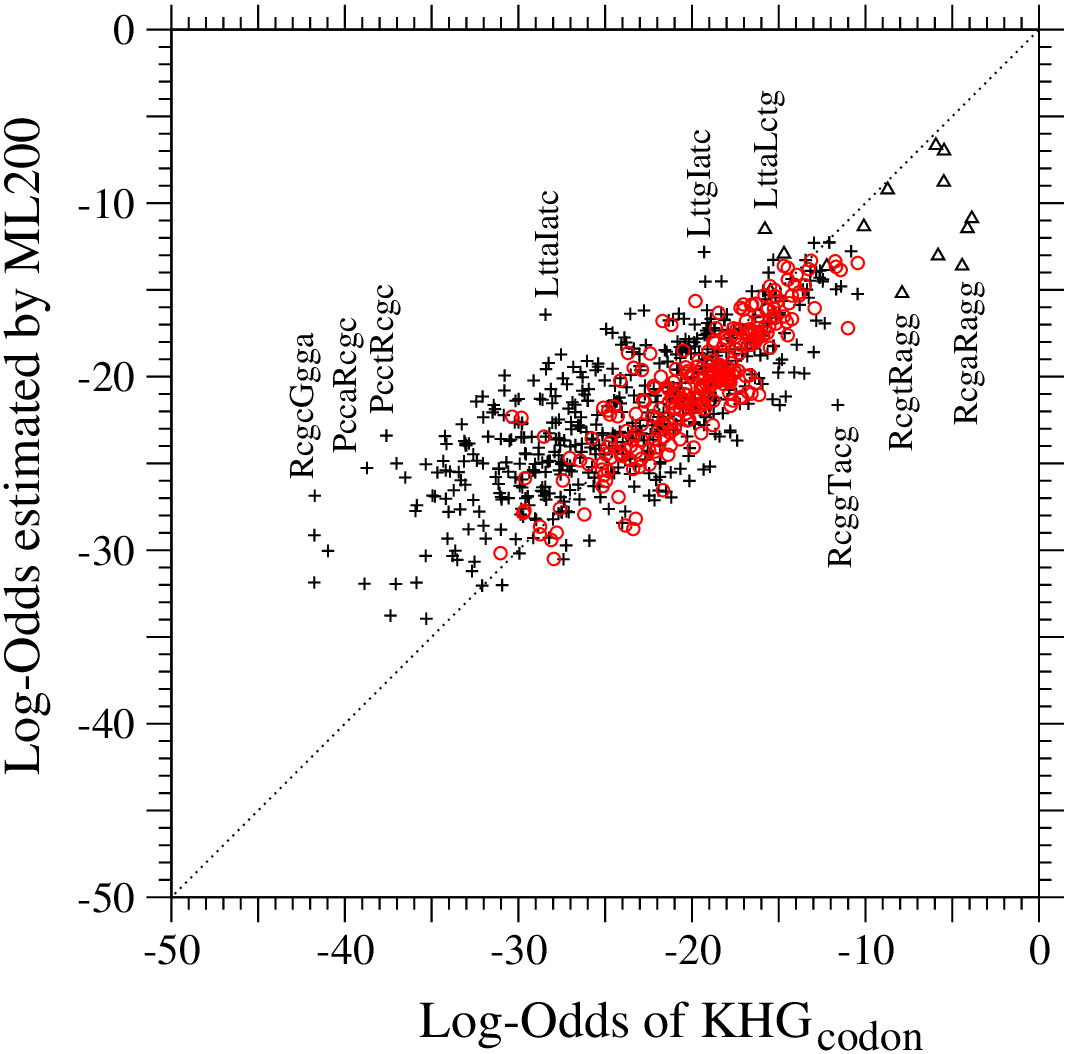}
}
\vspace*{1em}
\centerline{
\includegraphics*[width=80mm,angle=0]{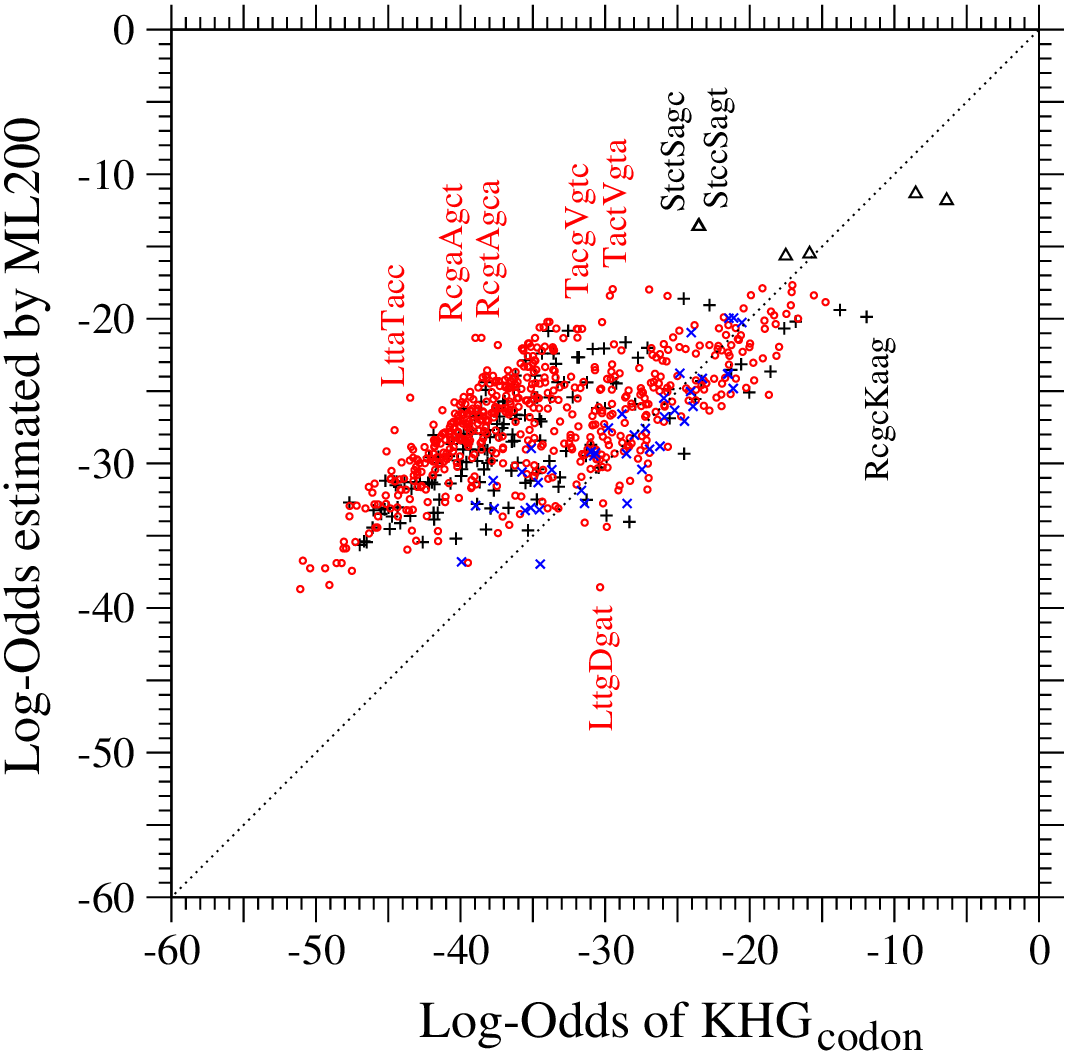}
\hspace*{5mm}
\includegraphics*[width=80mm,angle=0]{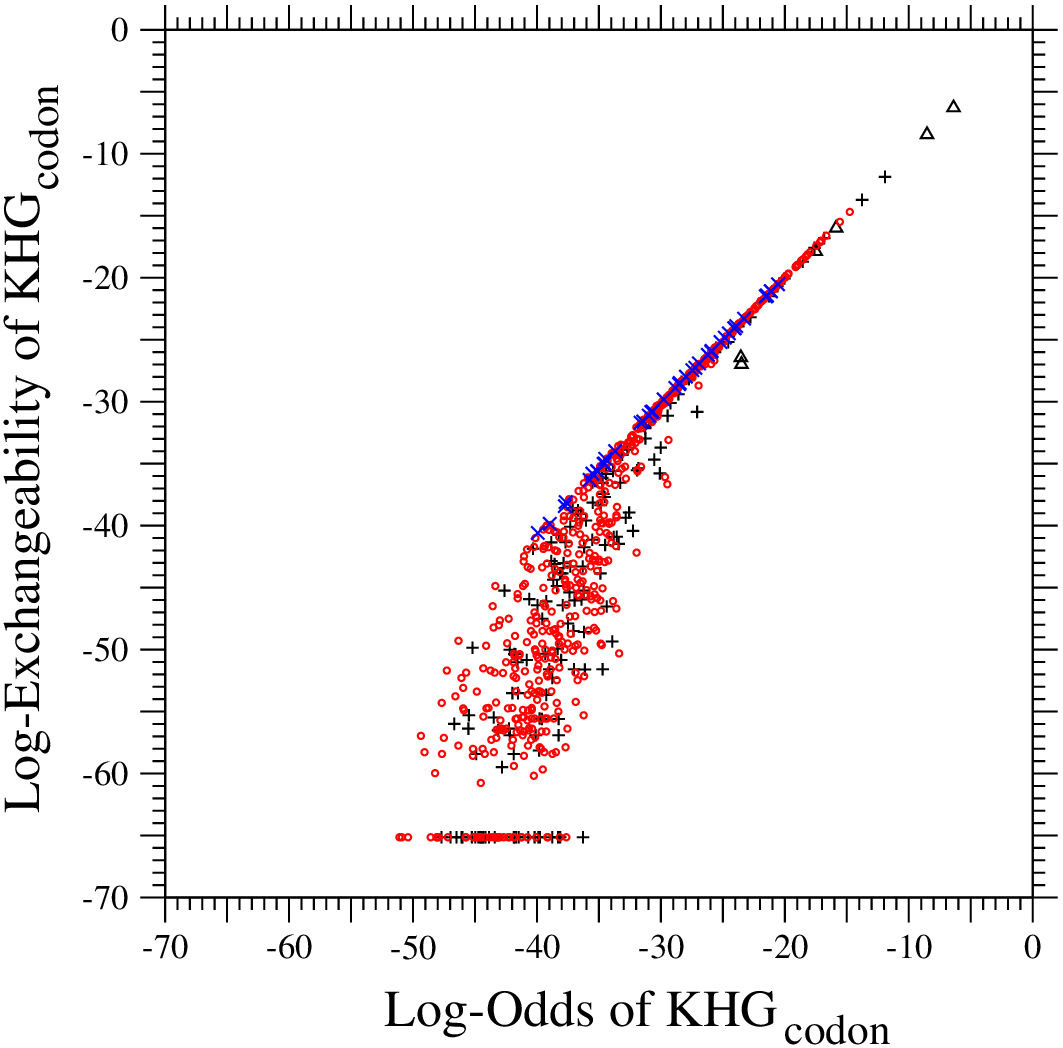}
}

\textbf{C} \hspace*{47em} \textbf{D}
} 
\caption{
\label{fig: ML200_logodds_codon_1pam_KHG}
\label{fig: ML200_logodds_codon_1pam_KHG_1x}
\label{fig: ML200_logodds_codon_1pam_KHG_2x}
\label{fig: ML200_logodds_codon_1pam_KHG_3x}
\label{fig: ML200_logodds_vs_exchange_scaled_1pam_observed_KHG_3x}
\label{fig: logodds_vs_exchange_scaled_1pam_observed_KHG_3x}
\BF{
The ML-200 model fitted to KHG.
} 
Each element 
log-$O(\langle S \rangle(\hat{\tau},\hat{\sigma}))_{\mu \nu}$
of the log-odds matrix 
corresponding to (A) single, (B) double, and (C) triple nucleotide changes 
in the ML-200 model fitted to the 1-PAM KHG codon substitution matrix
is plotted against the log-odds log-$O(S^{\script{KHG}}(\mbox{1 PAM}))_{\mu \nu}$
calculated from KHG.
In (D),
codon log-exchangeabilities of the 1-PAM KHG codon substitution matrix corresponding to triple nucleotide changes 
are plotted against the log-odds log-$O(S^{\script{KHG}}(\mbox{1 PAM}))_{\mu \nu}$
calculated from KHG.
The log-exchangeability of the 1-PAM KHG 
is defined as $(10 / \log 10) \log [R^{\script{KHG}}_{\mu\nu} \cdot t_{\script{1-PAM}} / f_{\nu} ]$.
Upper triangle, plus, circle, and cross marks show the log-odds values for
synonymous pairs and one-, two-, and three-step amino acid pairs,
respectively. 
Log-exchangeabilities for the codon pairs whose instantaneous rates are estimated to be $0$ in KHG are
shown to be about $-65$ in this figure.
The dotted line in each figure shows the line 
of equal values between the ordinate and the abscissa.
}
\end{figure}

} 

} 

\TextFig{

\FigureEach{
\FigureInLegends{\newpage}

\begin{figure*}[ht]
\FigureInLegends{
\centerline{
\includegraphics*[width=130mm,angle=0]{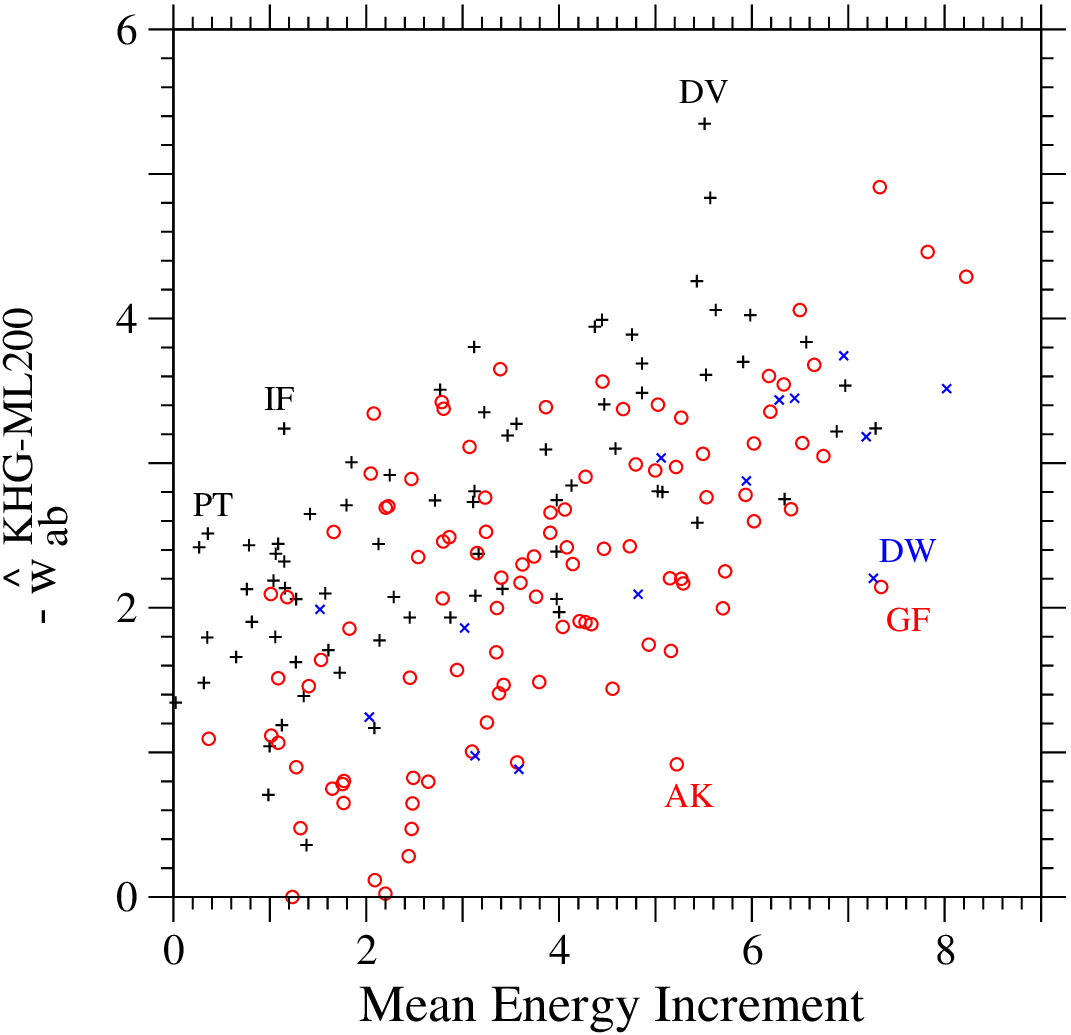}
}
} 
\caption{\label{fig: wij_KHG-ML199_vs_EI}
\label{fig: wij_KHG-ML200_vs_EI}
The ML estimate, $\hat{w}^{\script{KHG-ML200}}_{ab}$, 
of selective constraints on substitutions of the single step amino acid pair
in the ML-200 model fitted to the 1-PAM KHG matrix
is plotted against the mean energy increment,
($\Delta \hat{\varepsilon}^{\script{c}}_{ab} + \Delta \hat{\varepsilon}^{\script{v}}_{ab}$) 
defined in Supporting Information, Text S1.
due to an amino acid substitution.
Plus, circle, and cross marks show the values for
one-, two-, and three-step amino acid pairs,
respectively. 
The correlation coefficient between the axes is
equal to 0.71 for single-step amino acid pairs,
0.65 for multi-step amino acid pairs,
and 0.60 for all amino acid pairs.
}
\end{figure*}
} 

\FigurePanel{
\FigureInLegends{\newpage}

\begin{figure}[ht]
\FigureInLegends{
\textbf{A} \hspace*{47em} \textbf{B}

\centerline{
\includegraphics*[width=80mm,angle=0]{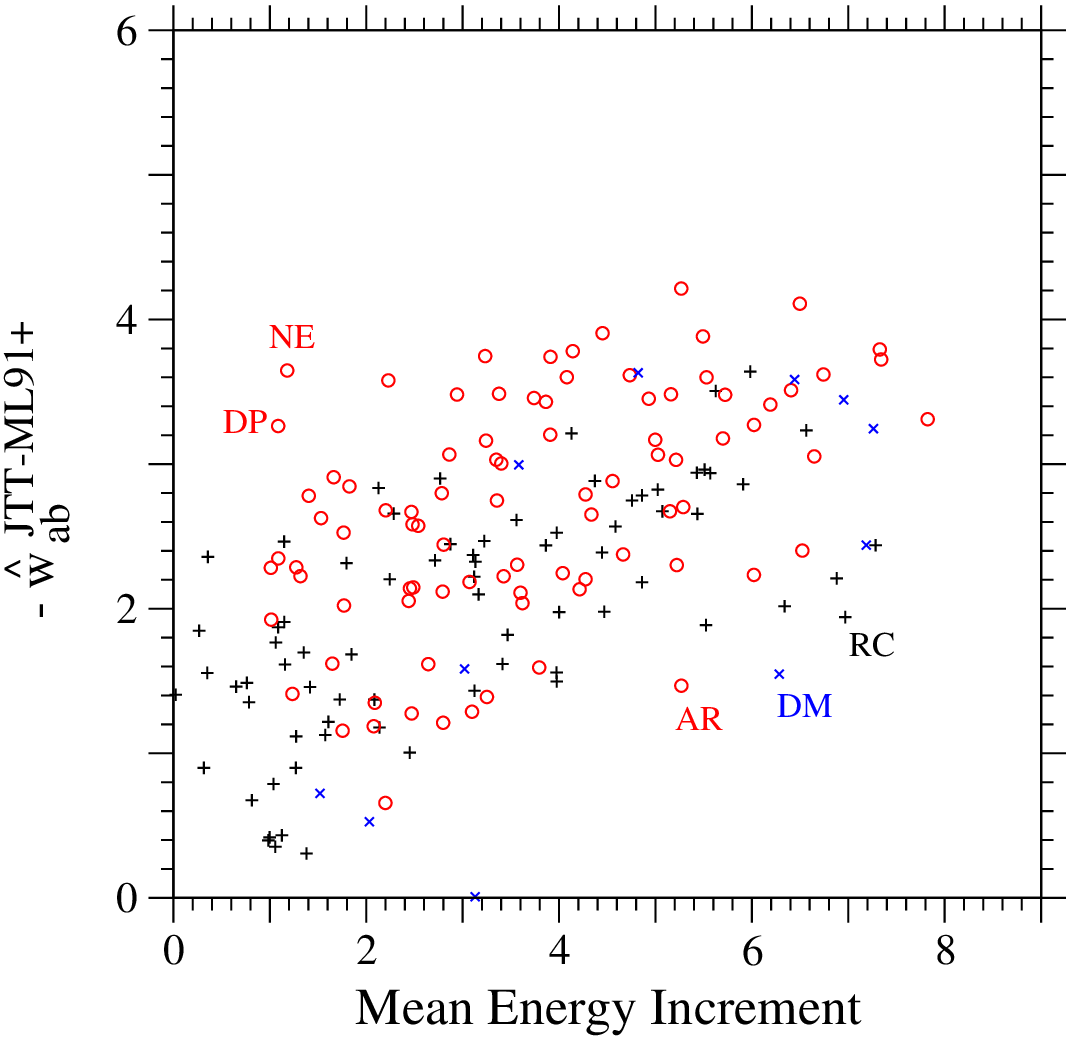}
\hspace*{5mm}
\includegraphics*[width=80mm,angle=0]{FIGS/wij_EI_vs_KHG-ML199.eps}
}
} 
\caption{
\label{fig: wij_vs_energy_increments}
\label{fig: wij_JTT-ML91+_vs_EI}
\label{fig: wij_KHG-ML199_vs_EI}
\label{fig: wij_KHG-ML200_vs_EI}
\BF{
Selective constraint for each amino acid pair estimated from JTT and from KHG.
} 
The ML estimate, (A) $- \hat{w}^{\script{JTT-ML91+}}_{ab}$ 
in the ML-91+ model fitted to the 1-PAM JTT amino acid substitution matrix
and (B) $- \hat{w}^{\script{KHG-ML200}}_{ab}$
in the ML-200 model fitted to the 1-PAM KHG codon substitution matrix,
for each amino acid pair is plotted against the mean energy increment 
due to an amino acid substitution,
($\Delta \hat{\varepsilon}^{\script{c}}_{ab} + \Delta \hat{\varepsilon}^{\script{v}}_{ab}$) 
defined 
by \SEqs{4, S1-5, and S1-6}.
In (A),
the estimates $\hat{w}_{ab}$ for
the least exchangeable class of multi-step amino acid pairs
are not shown.
Plus, circle, and cross marks show the values for
one-, two-, and three-step amino acid pairs,
respectively. 
}
\end{figure}
} 

} 

\SupFig{

\FigureEach{

\FigureInLegends{\newpage}

\begin{figure*}[ht]
\FigureInLegends{
\centerline{
\includegraphics*[width=130mm,angle=0]{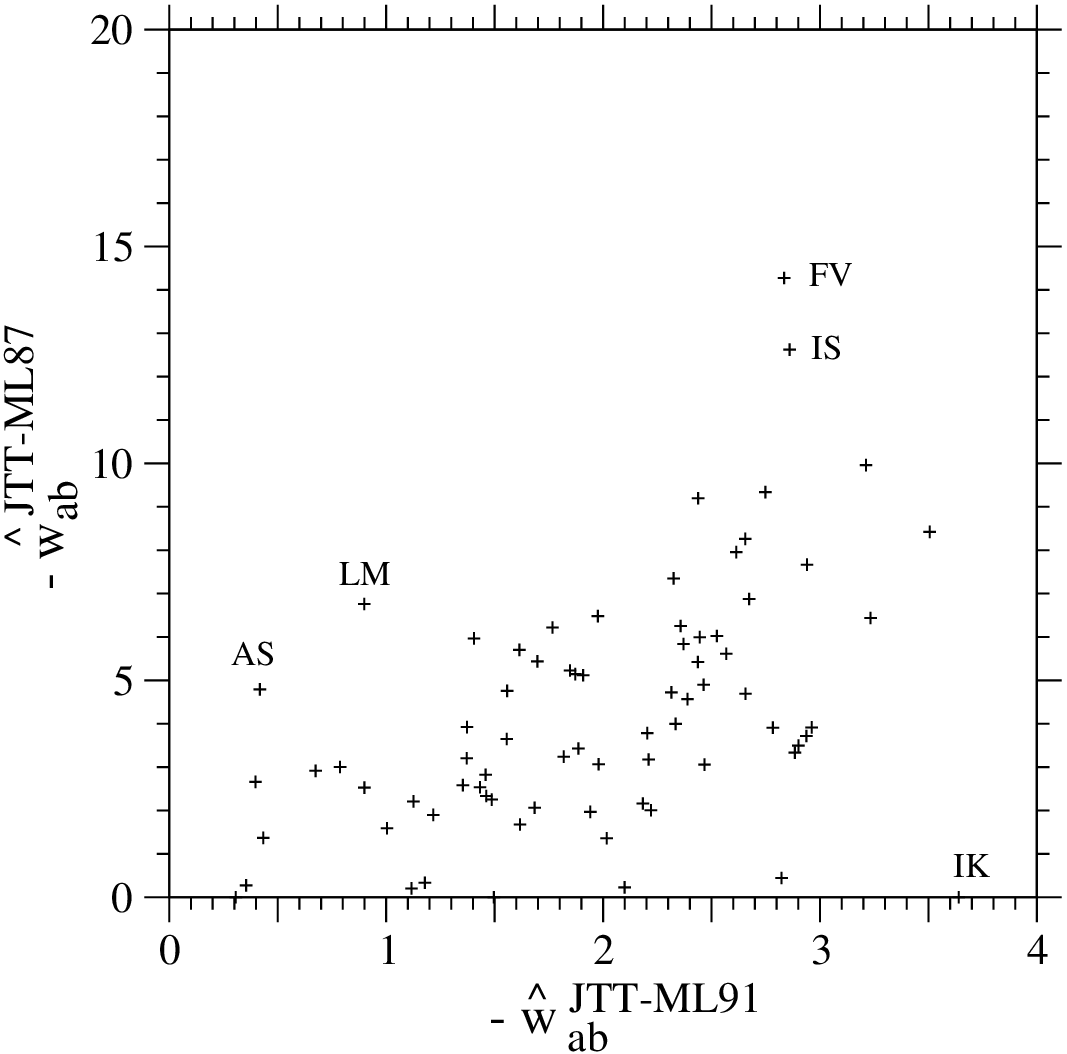}
}
} 
\caption{\label{fig: wij_JTT-ML87_vs_JTT-ML91}
\label{fig: wij1_JTT-ML87_vs_JTT-ML91}
\label{fig: wij1_ML87_vs_ML91}
The ML estimator, $\hat{w}_{ab}$, 
of selective constraints on
substitutions of the single step amino acid pair
in the ML-87 model fitted to the 1-PAM JTT matrix
is plotted against that in the ML-91 model.
}
\end{figure*}

} 

\FigureEach{

\FigureInLegends{\newpage}

\begin{figure*}[ht]
\FigureInLegends{
\centerline{
\includegraphics*[width=130mm,angle=0]{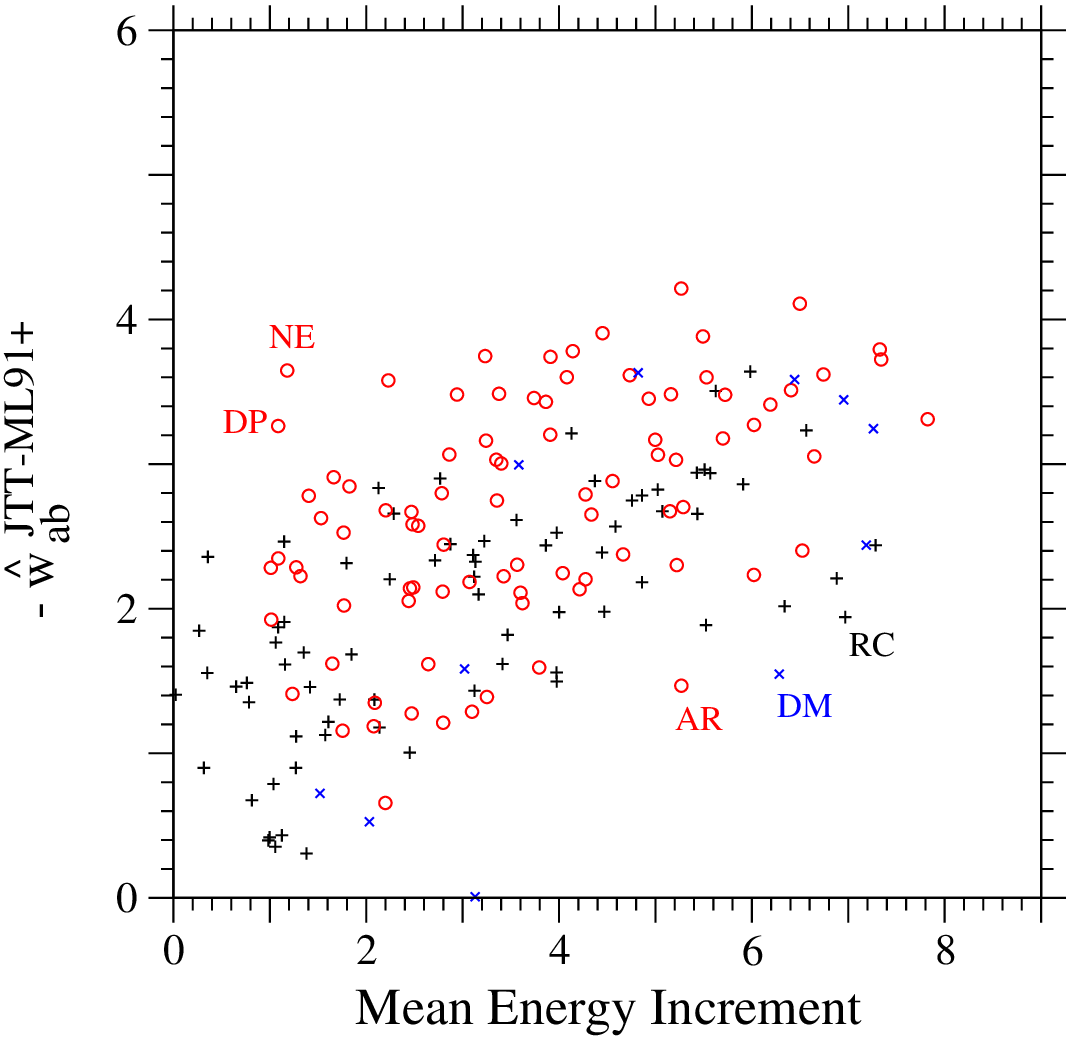}
}
} 
\caption{
\label{fig: wij_vs_energy_increments}
\label{fig: wij_JTT-ML91+_vs_EI}
The ML estimate, $\hat{w}^{\script{JTT-ML91+}}_{ab}$, for
selective constraints on substitutions of each amino acid pair
in the ML-91+ model fitted to the 1-PAM JTT matrix
is plotted against the mean energy increment
due to an amino acid substitution,
($\Delta \hat{\varepsilon}^{\script{c}}_{ab} + \Delta \hat{\varepsilon}^{\script{v}}_{ab}$) 
defined in Supporting Information, Text S1.
The estimates $\hat{w}_{ab}$ for
the least exchangeable class of multi-step amino acid pairs
are not shown.
Plus, circle, and cross marks show the values for
one-, two-, and three-step amino acid pairs,
respectively. 
The correlation coefficient for single step amino acid pairs is equal to.
0.66.
}
\end{figure*}

} 

\FigurePanel{

\FigureInLegends{\clearpage}

\FigureInLegends{\small}

\begin{figure}[ht]
\FigureInLegends{
\textbf{A} \hspace*{45em} \textbf{B}
\centerline{
\includegraphics*[width=75mm,angle=0]{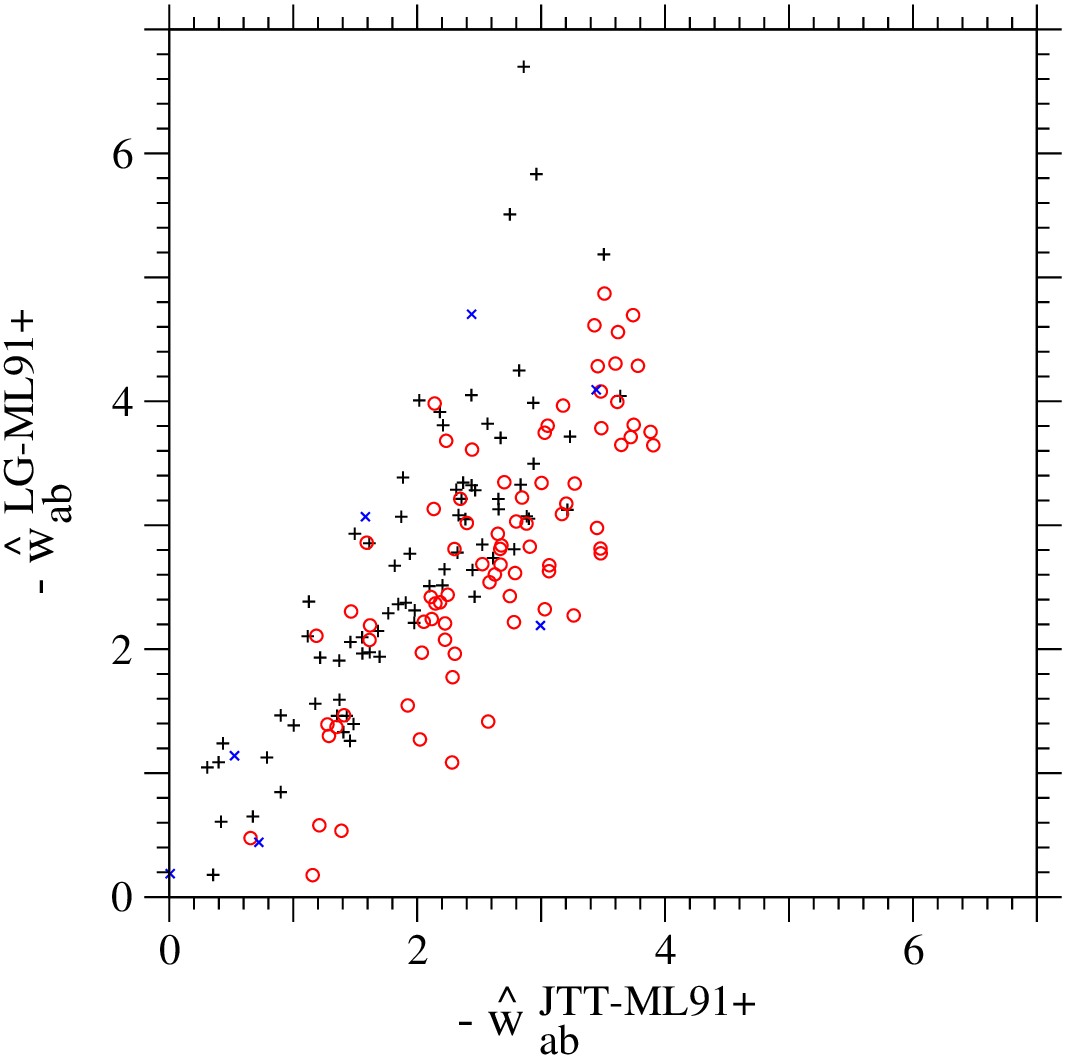}
\hspace*{5mm}
\includegraphics*[width=75mm,angle=0]{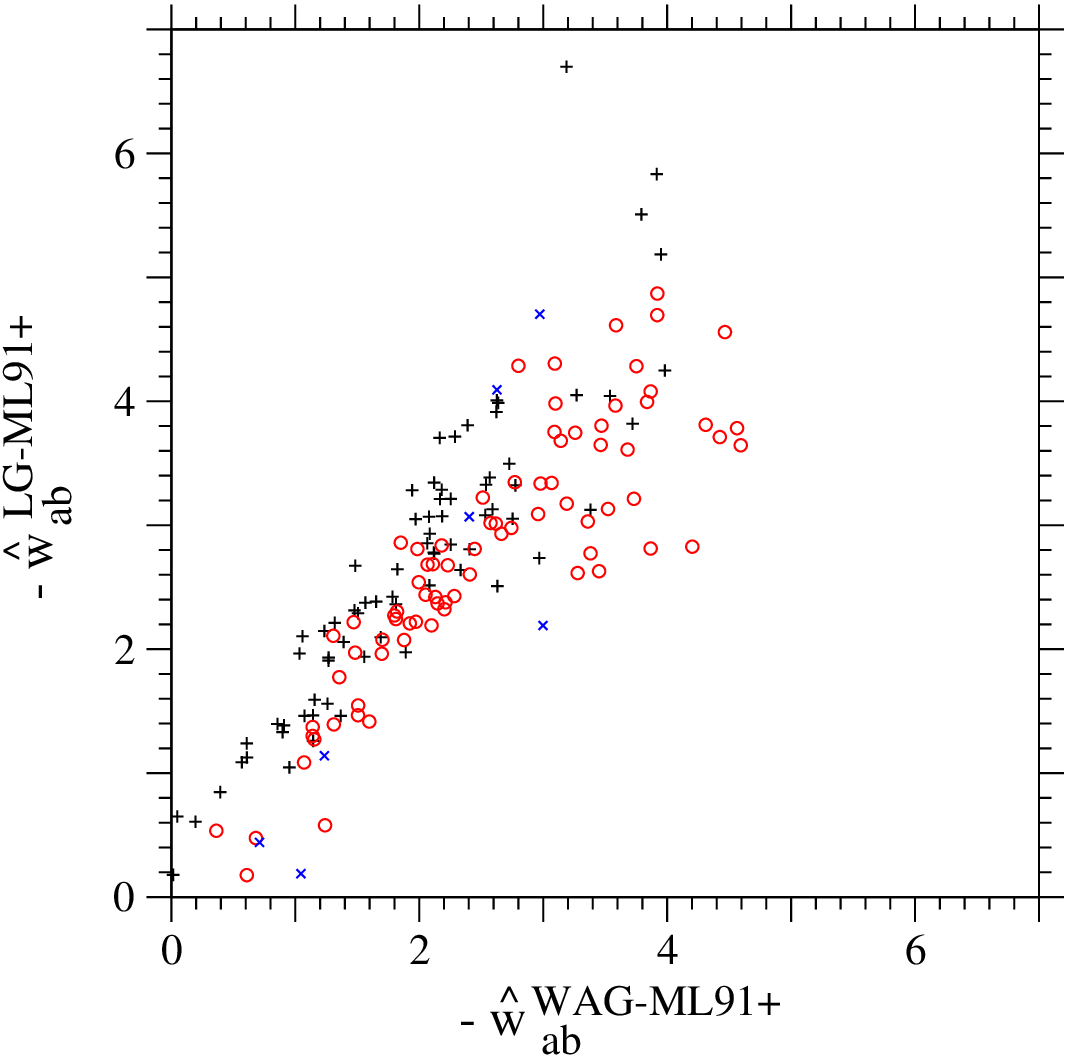}
}
\textbf{C} \hspace*{45em} \textbf{D}
\centerline{
\includegraphics*[width=75mm,angle=0]{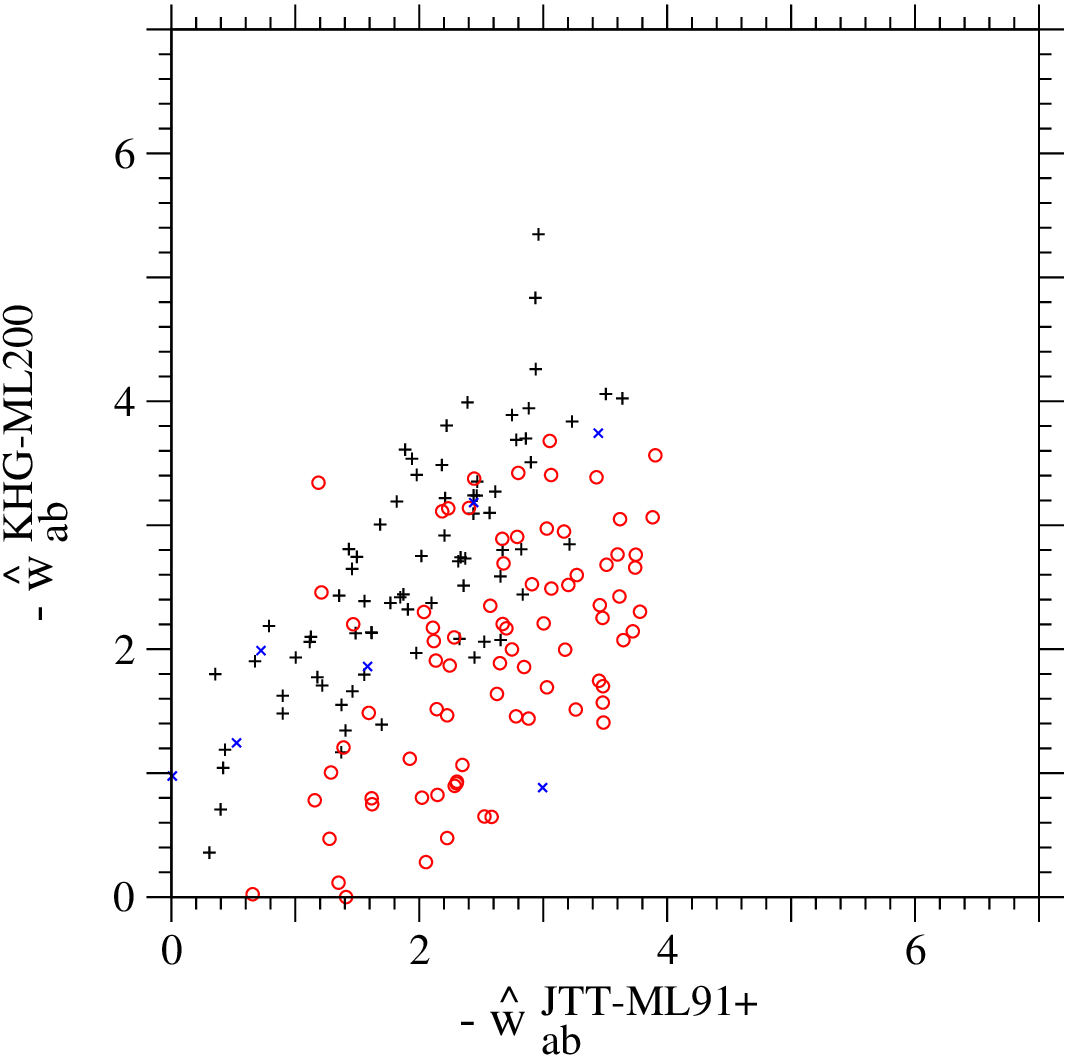}
\hspace*{5mm}
\includegraphics*[width=75mm,angle=0]{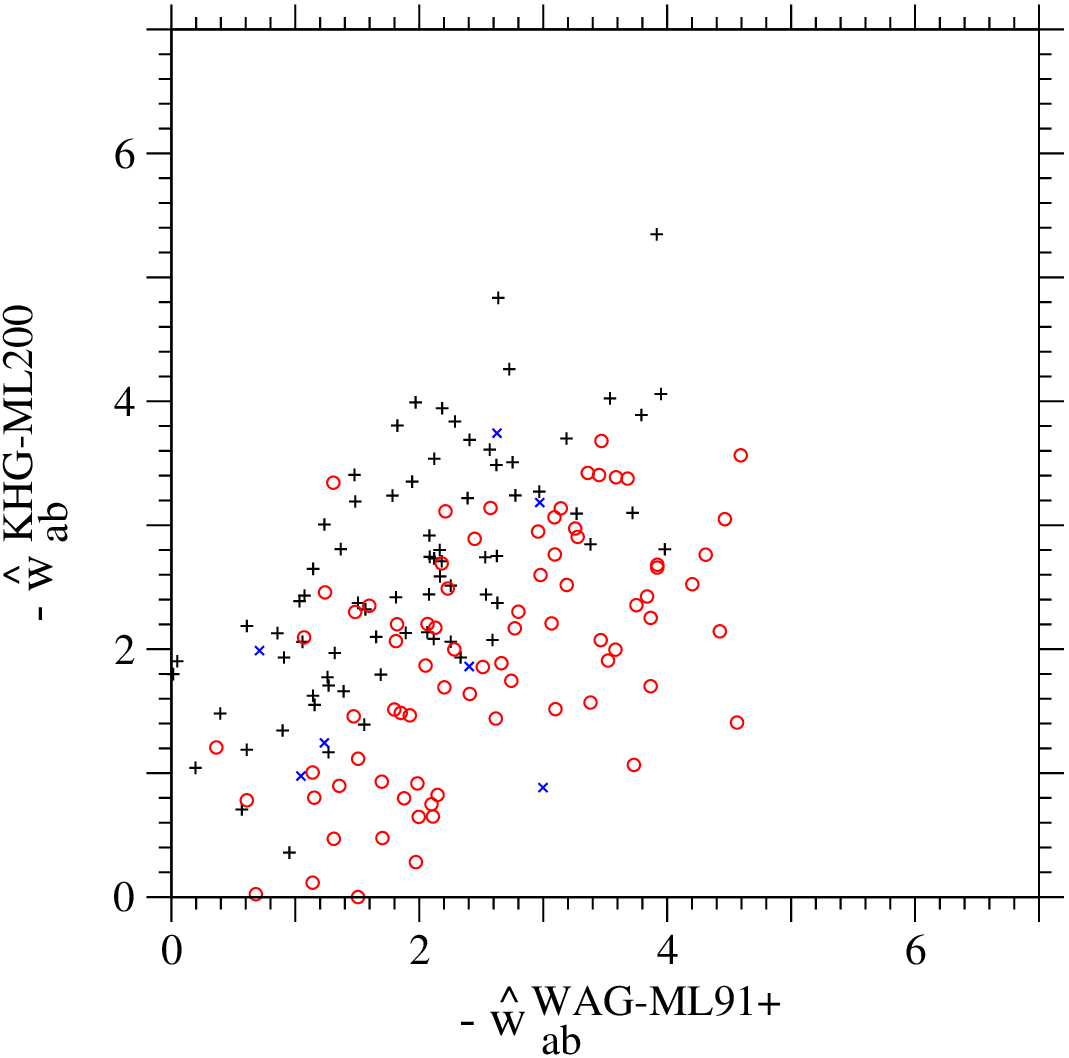}
}
\textbf{E} \hspace*{45em} \textbf{F}
\centerline{
\includegraphics*[width=75mm,angle=0]{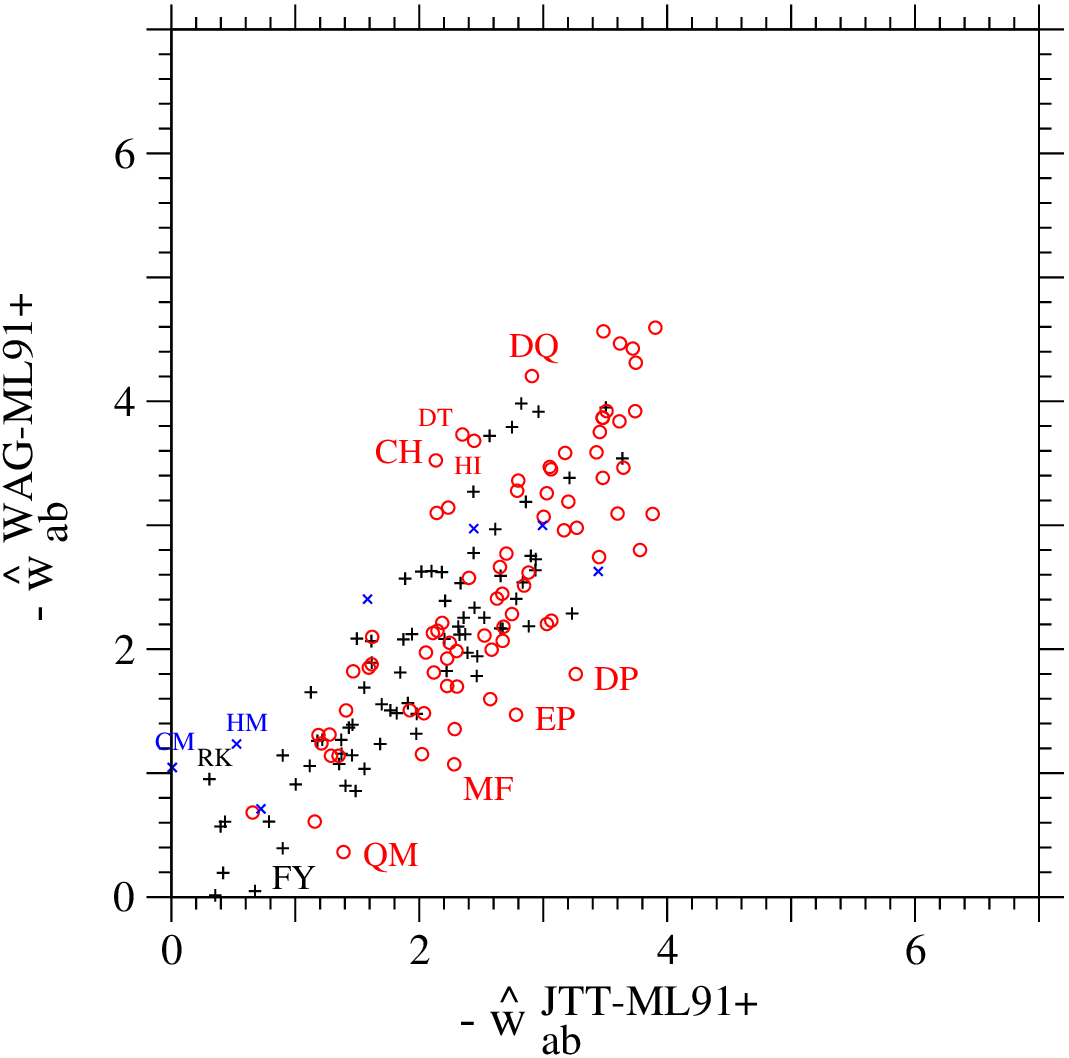}
\hspace*{5mm}
\includegraphics*[width=75mm,angle=0]{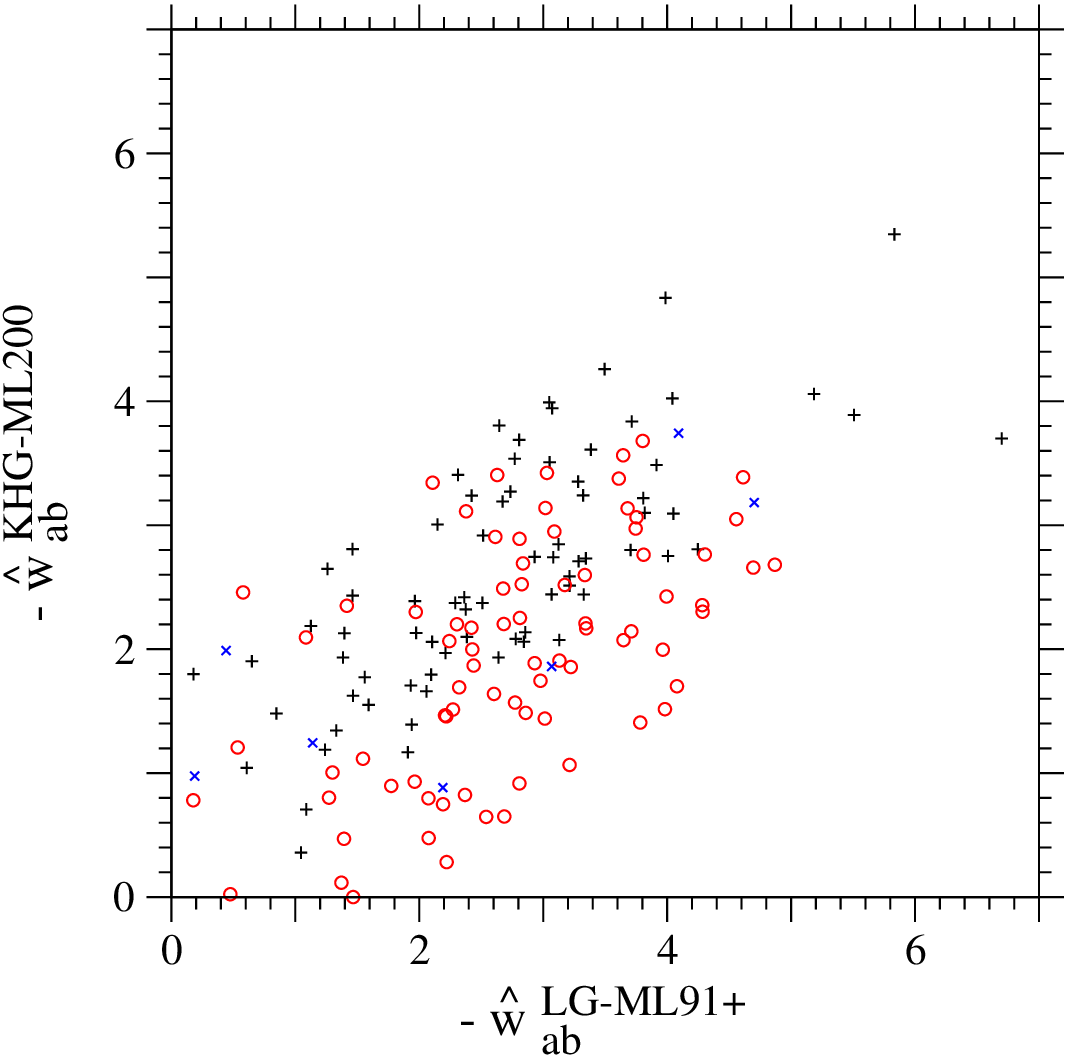}
}
} 
\caption{
\label{fig: wij_JTT_vs_WAG-ML91+}
\label{fig: wij_LG_vs_WAG-ML91+}
\label{fig: wij_JTT-ML91+_vs_KHG-ML200}
\label{fig: wij_LG-ML91+_vs_KHG-ML200}
\FigureInLegends{\small}
\BF{
Comparison between various estimates of
selective constraint for each amino acid pair 
} 
The ML estimates
of selective constraint on substitutions of each amino acid pair
are compared between the models fitted to various empirical substitution matrices.
The estimates $\hat{w}_{ab}$ for multi-step amino acid pairs
that belong to the least exchangeable class at least in one of the models
are not shown.
Plus, circle, and cross marks show the values for
one-, two-, and three-step amino acid pairs,
respectively. 
\FigureInLegends{\normalsize}
}
\end{figure}

\FigureInLegends{\normalsize}

\FigureInLegends{\clearpage}

\begin{figure}[ht]
\FigureInLegends{
\textbf{A} \hspace*{47em} \textbf{B}

\centerline{
\includegraphics*[width=80mm,angle=0]{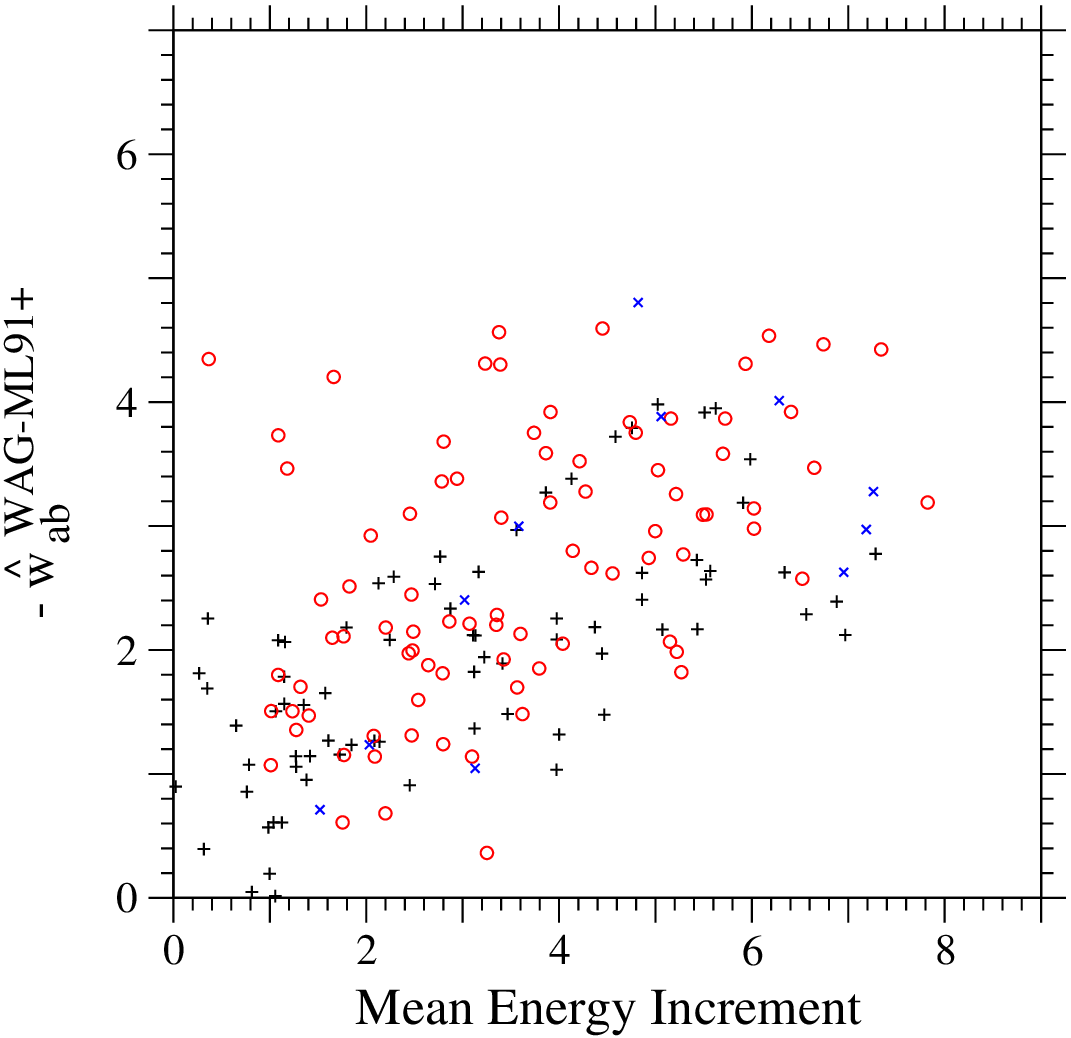}
\hspace*{5mm}
\includegraphics*[width=80mm,angle=0]{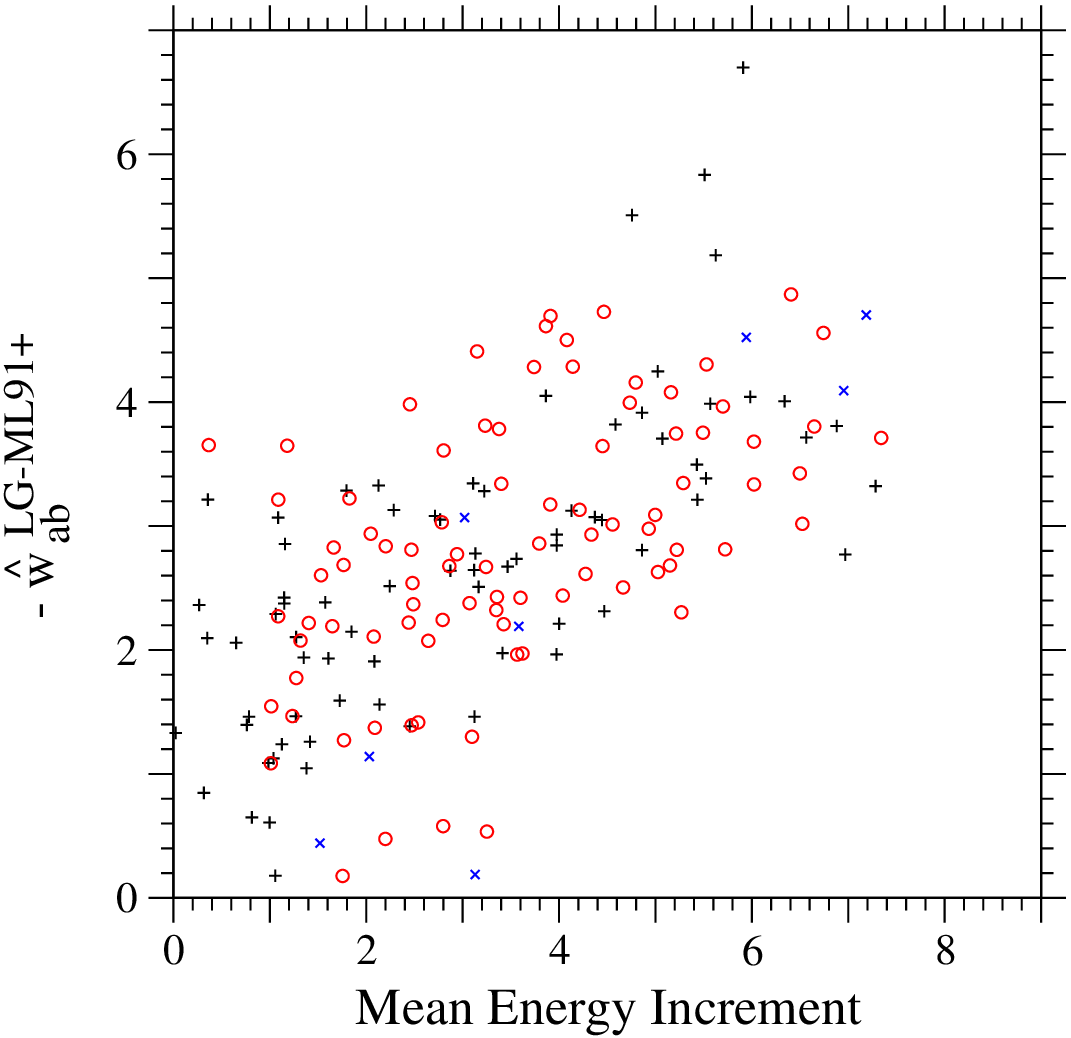}
}
} 
\caption{
\label{fig: wij_EI_vs_WAG-ML91+}
\label{fig: wij_EI_vs_LG-ML91+}
\BF{
Selective constraint for each amino acid pair 
estimated from WAG and from LG.
} 
The ML estimate, 
$- \hat{w}^{\script{WAG-ML91+}}_{ab}$ in (A) and
$- \hat{w}^{\script{LG-ML91+}}_{ab}$ in (B), 
of selective constraint on substitutions of each amino acid pair
in the ML-91+ models fitted to the 1-PAM matrices of WAG and LG
is plotted against the mean energy increment
due to an amino acid substitution,
($\Delta \hat{\varepsilon}^{\script{c}}_{ab} + \Delta \hat{\varepsilon}^{\script{v}}_{ab}$) 
defined 
by \SEqs{4, S1-5, and S1-6}.
The estimates $\hat{w}_{ab}$ for the least exchangeable class 
of multi-step amino acid pairs
are not shown.
Plus, circle, and cross marks show the values for
one-, two-, and three-step amino acid pairs,
respectively. 
}
\end{figure}

\FigureInLegends{\clearpage}

\begin{figure}[ht]
\FigureInLegends{
\textbf{A} \hspace*{47em} \textbf{B}

\centerline{
\includegraphics*[width=80mm,angle=0]{FIGS/wij_JTT-ML87_vs_JTT-ML91.eps}
\hspace*{5mm}
\includegraphics*[width=80mm,angle=0]{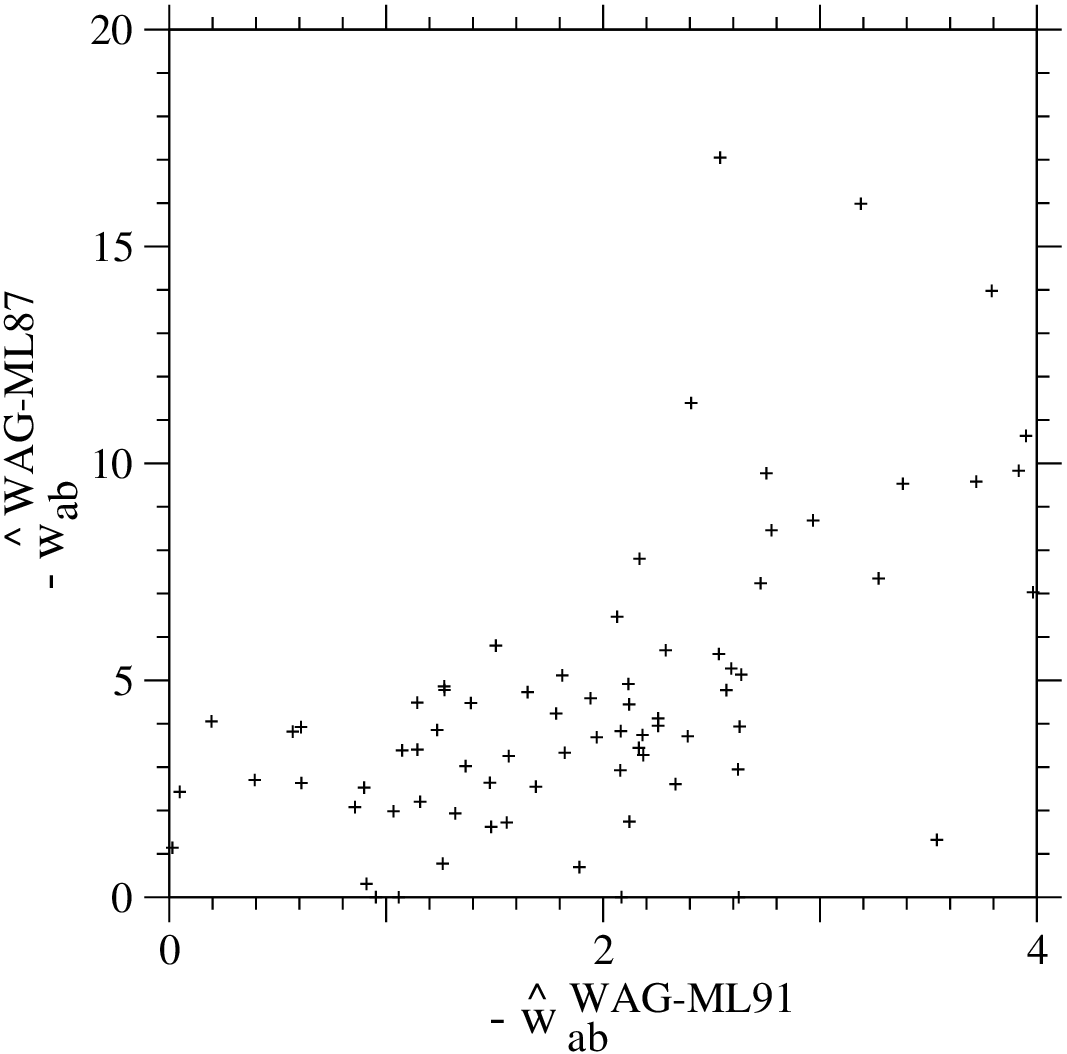}
}
} 
\caption{
\label{fig: wij_JTT-ML87_vs_JTT-ML91}
\label{fig: wij1_JTT-ML87_vs_JTT-ML91}
\label{fig: wij1_ML87_vs_ML91}
\label{sfig: wij_vs_energy_increments}
\label{fig: wij_WAG-ML87_vs_WAG-ML91}
\label{fig: wij1_WAG-ML87_vs_WAG-ML91}
\BF{
Comparison of the ML estimates of selective constraint for each amino acid pair 
between the ML-87 and the ML-91 models.
} 
The ML estimate
of selective constraint for each single step amino acid pair
in the ML-87 model fitted to 
(A) the 1-PAM JTT matrix or (B) the 1-PAM WAG matrix
is plotted against that in the ML-91 model.
}
\end{figure}

} 

} 

\SupFig{

\FigureEach{
\FigureInLegends{\newpage}

\begin{figure*}[ht]
\FigureInLegends{
\centerline{
\includegraphics*[width=130mm,angle=0]{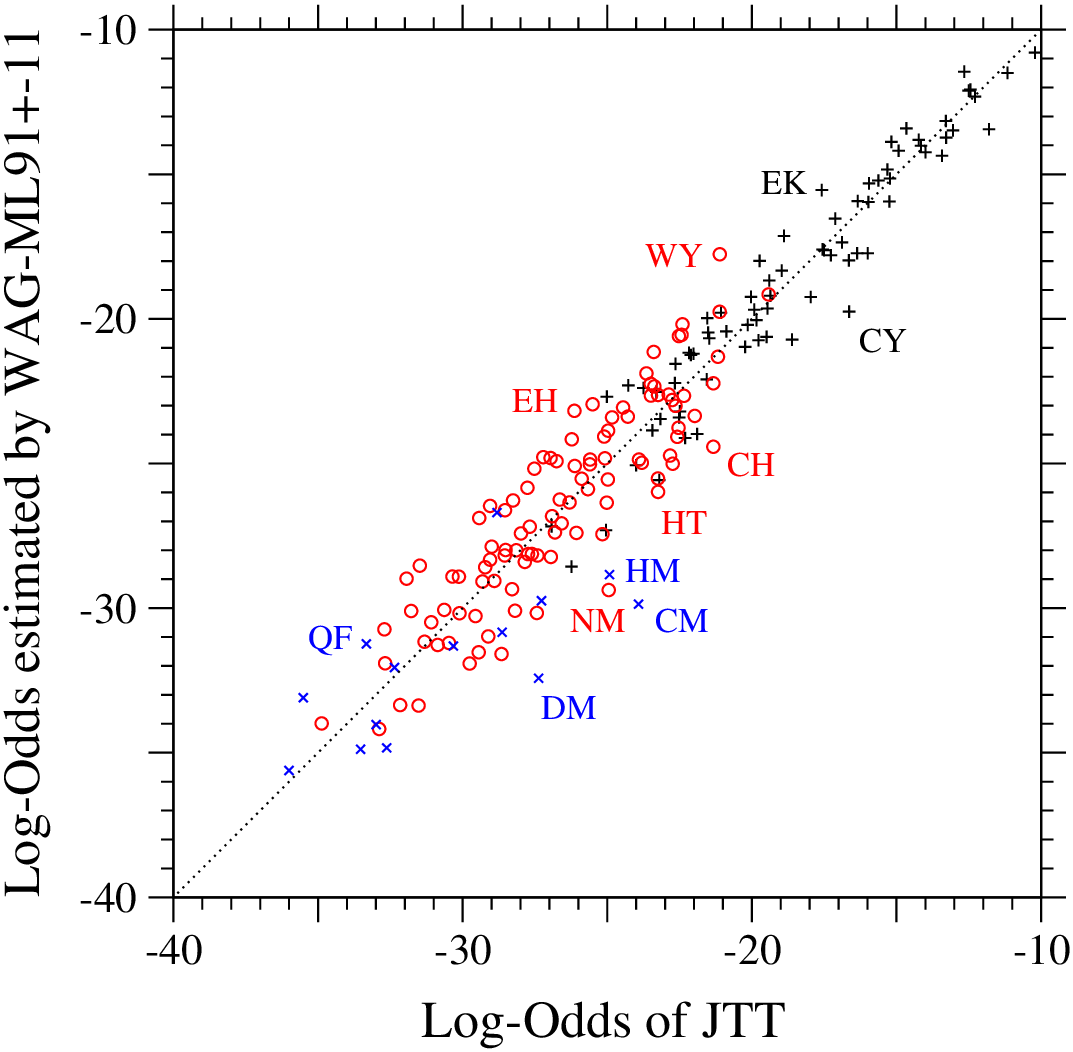}
}
} 
\caption{
\label{fig: WAG-ML-91+_log-odds_1pam_JTT}
\label{fig: WAG-ML91+_log-odds_1pam_JTT}
Each element 
log-$O(\langle S \rangle(\hat{\tau},\hat{\sigma}))_{ab}$
of the log-odds matrix 
of the WAG-ML91+-11 model fitted to the 1-PAM JTT matrix
is plotted against the log-odds log-$O(S^{\script{JTT}}(\mbox{1 PAM}))_{ab}$
calculated from the JTT.
Plus, circle, and cross marks show the log-odds values for
one-, two-, and three-step amino acid pairs,
respectively.  
The dotted line shows the line 
of equal values between the ordinate and the abscissa.
}
\end{figure*}

} 

} 

\TextFig{

\FigureEach{

\FigureInLegends{\newpage}

\begin{figure*}[ht]
\FigureInLegends{
\centerline{
\includegraphics*[width=130mm,angle=0]{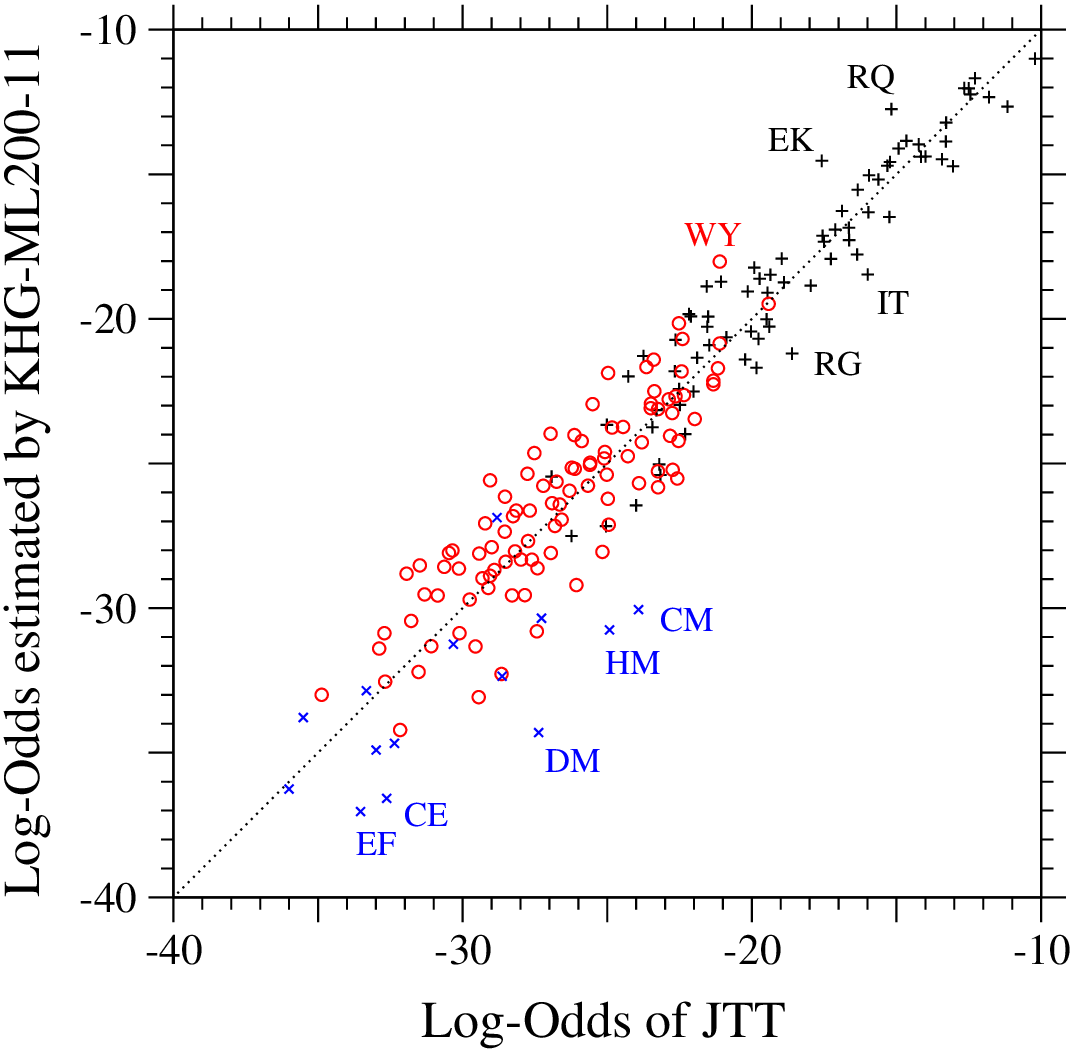}
}
} 
\caption{\label{fig: KHG-ML200_log-odds_1pam_JTT}
\label{fig: KHG-ML199_log-odds_1pam_JTT}
Each element 
log-$O(\langle S \rangle(\hat{\tau},\hat{\sigma}))_{ab}$
of the log-odds matrix 
of the KHG-ML200-11 model fitted to the 1-PAM JTT matrix
is plotted against the log-odds log-$O(S^{\script{JTT}}(\mbox{1 PAM}))_{ab}$
calculated from the JTT.
Plus, circle, and cross marks show the log-odds values for
one-, two-, and three-step amino acid pairs,
respectively.  
The dotted line shows the line 
of equal values between the ordinate and the abscissa.
}
\end{figure*}

} 

} 

\SupFig{

\FigureEach{

\FigureInLegends{\newpage}

\begin{figure*}[ht]
\FigureInLegends{
\centerline{
\includegraphics*[width=130mm,angle=0]{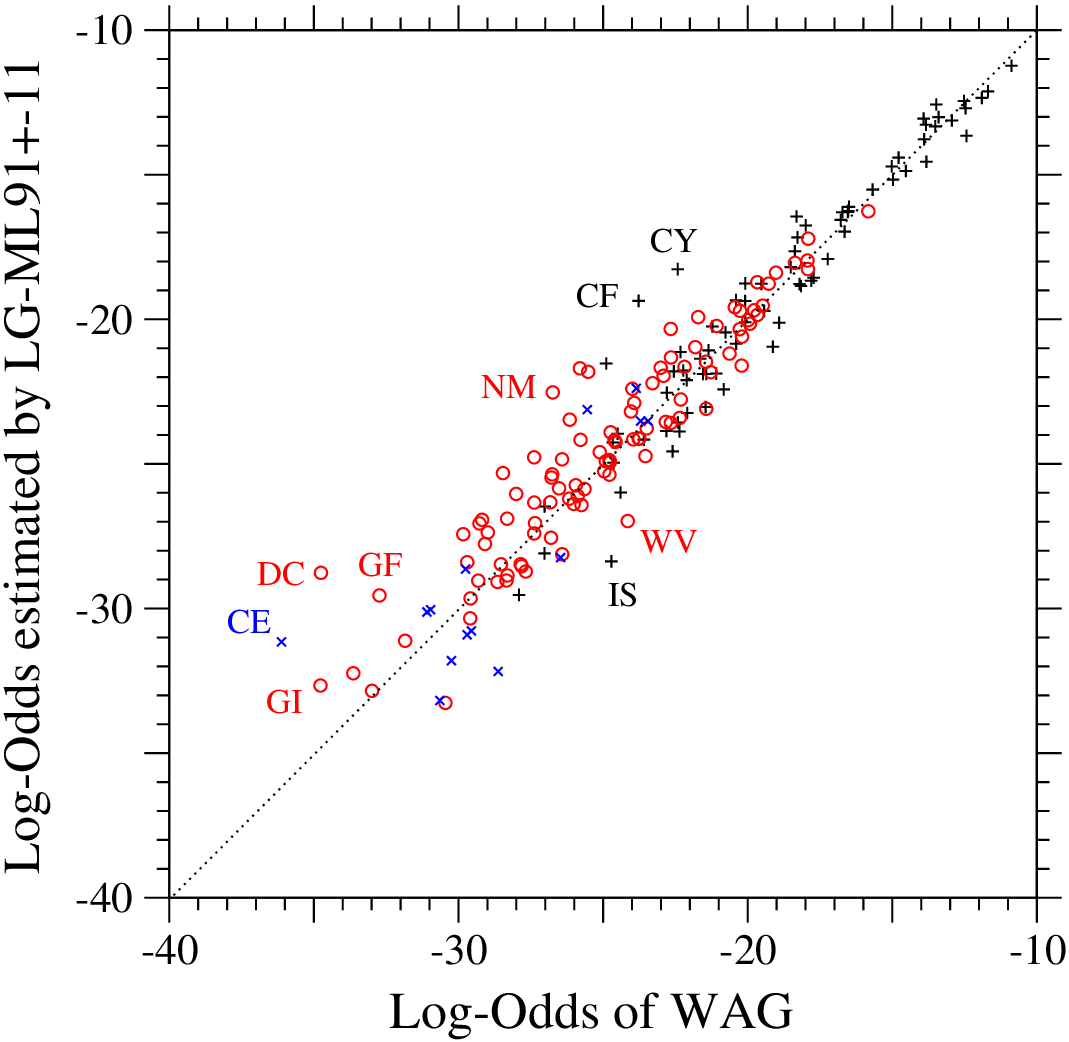}
}
} 
\caption{\label{fig: LG-ML-91+_log-odds_1pam_WAG}
\label{fig: LG-ML-91_log-odds_1pam_WAG}
Each element 
log-$O(\langle S \rangle(\hat{\tau},\hat{\sigma}))_{ab}$
of the log-odds matrix 
of the LG-ML91+-11 model fitted to the 1-PAM WAG matrix
is plotted against the log-odds log-$O(S^{\script{WAG}}(\mbox{1 PAM}))_{ab}$
calculated from WAG.
Plus, circle, and cross marks show the log-odds values for
one-, two-, and three-step amino acid pairs,
respectively.  
The dotted line shows the line 
of equal values between the ordinate and the abscissa.
}
\end{figure*}

} 

} 

\TextFig{

\FigureEach{

\FigureInLegends{\newpage}

\begin{figure*}[ht]
\FigureInLegends{
\centerline{
\includegraphics*[width=130mm,angle=0]{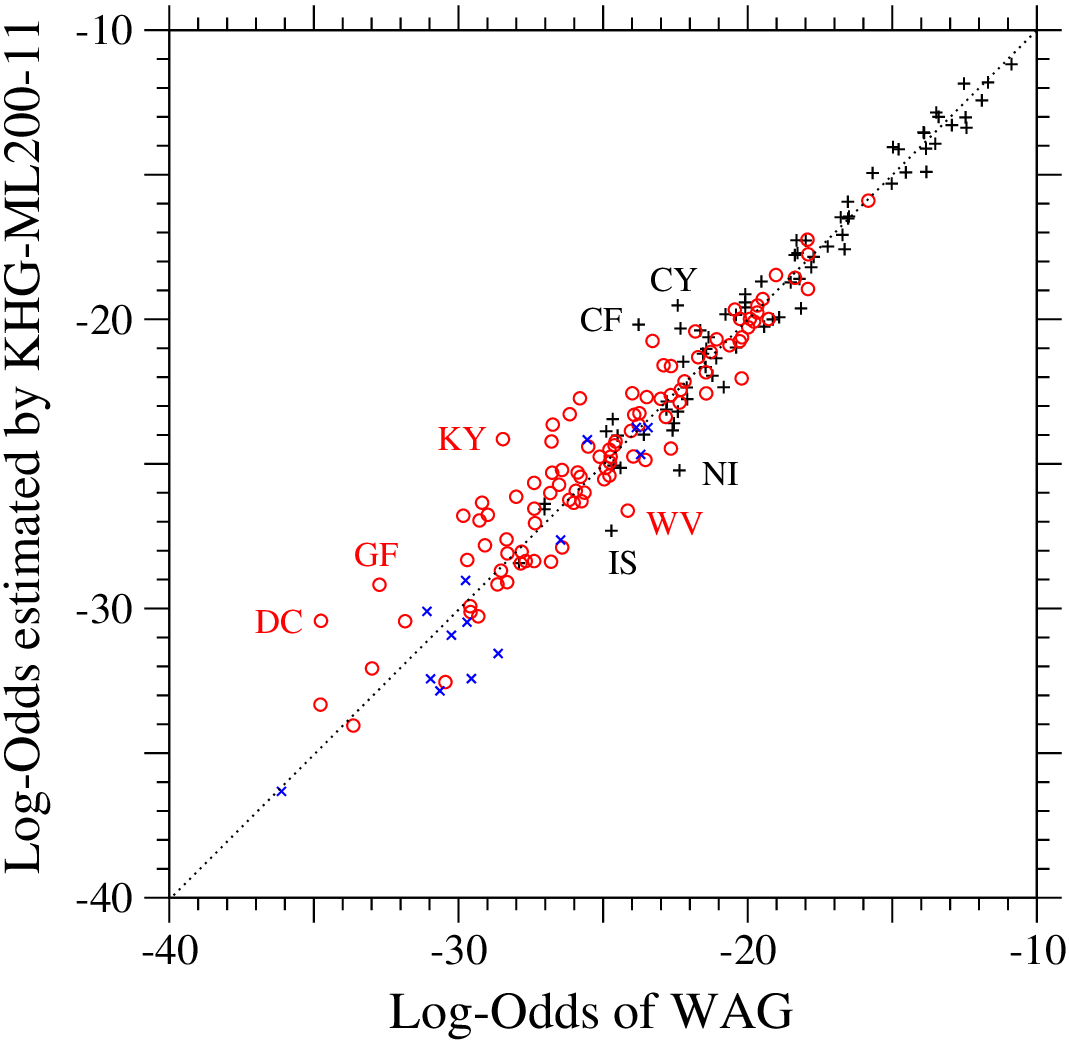}
}
} 
\caption{\label{fig: KHG-ML200_log-odds_1pam_WAG}
\label{fig: KHG-ML199_log-odds_1pam_WAG}
Each element 
log-$O(\langle S \rangle(\hat{\tau},\hat{\sigma}))_{ab}$
of the log-odds matrix 
of the KHG-ML200-11 model fitted to the 1-PAM WAG matrix
is plotted against the log-odds log-$O(S^{\script{WAG}}(\mbox{1 PAM}))_{ab}$
calculated from WAG.
Plus, circle, and cross marks show the log-odds values for
one-, two-, and three-step amino acid pairs,
respectively.  
The dotted line shows the line 
of equal values between the ordinate and the abscissa.
}
\end{figure*}

} 

} 

\SupFig{

\FigureEach{

\FigureInLegends{\newpage}

\begin{figure*}[ht]
\FigureInLegends{
\centerline{
\includegraphics*[width=130mm,angle=0]{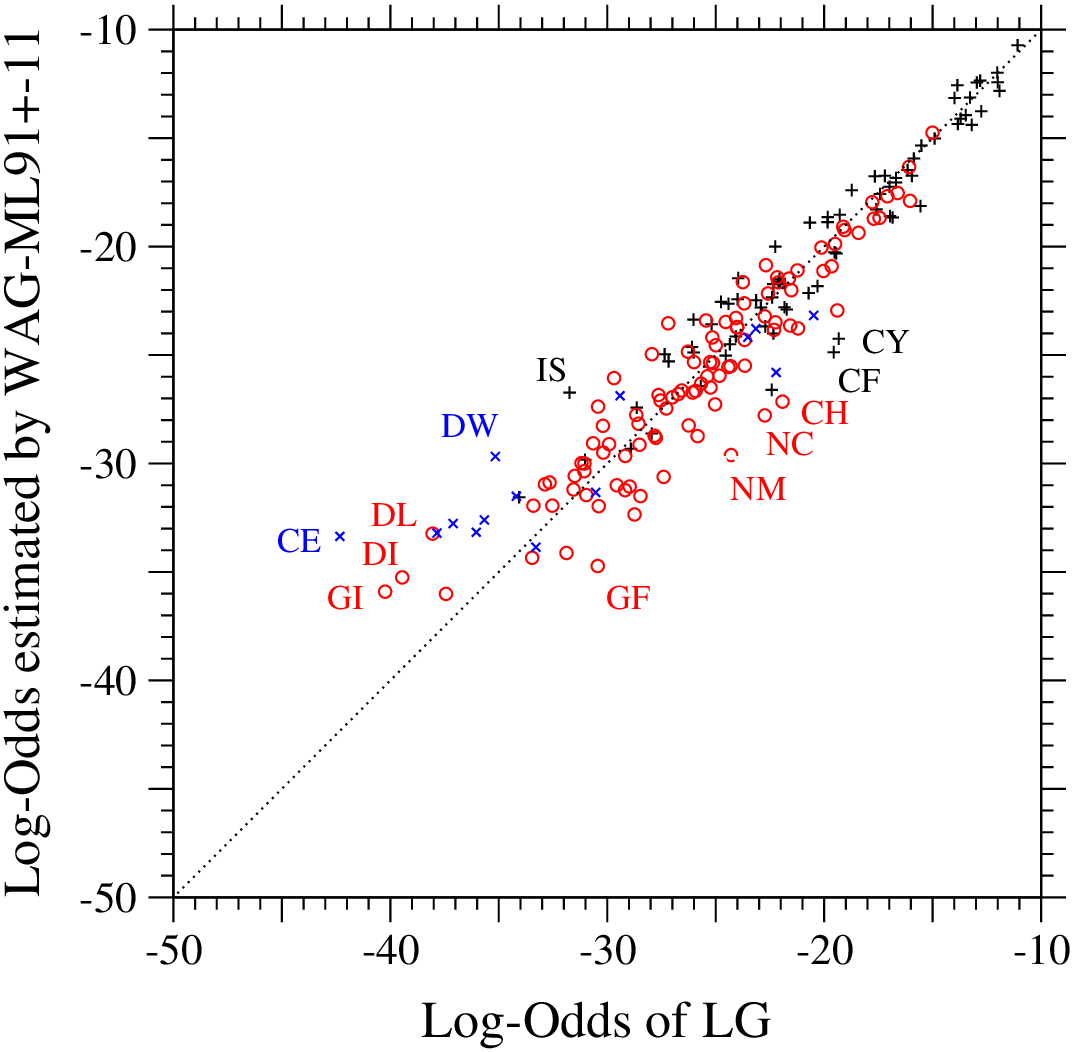}
}
} 
\caption{\label{fig: WAG-ML91+_log-odds_1pam_LG}
Each element 
log-$O(\langle S \rangle(\hat{\tau},\hat{\sigma}))_{ab}$
of the log-odds matrix 
of the WAG-ML91+-11 model fitted to the 1-PAM LG matrix
is plotted against the log-odds log-$O(S^{\script{LG}}(\mbox{1 PAM}))_{ab}$
calculated from LG.
Plus, circle, and cross marks show the log-odds values for
one-, two-, and three-step amino acid pairs,
respectively.  
The dotted line shows the line 
of equal values between the ordinate and the abscissa.
}
\end{figure*}

} 

} 

\TextFig{

\FigureEach{

\FigureInLegends{\newpage}

\begin{figure*}[ht]
\FigureInLegends{
\centerline{
\includegraphics*[width=130mm,angle=0]{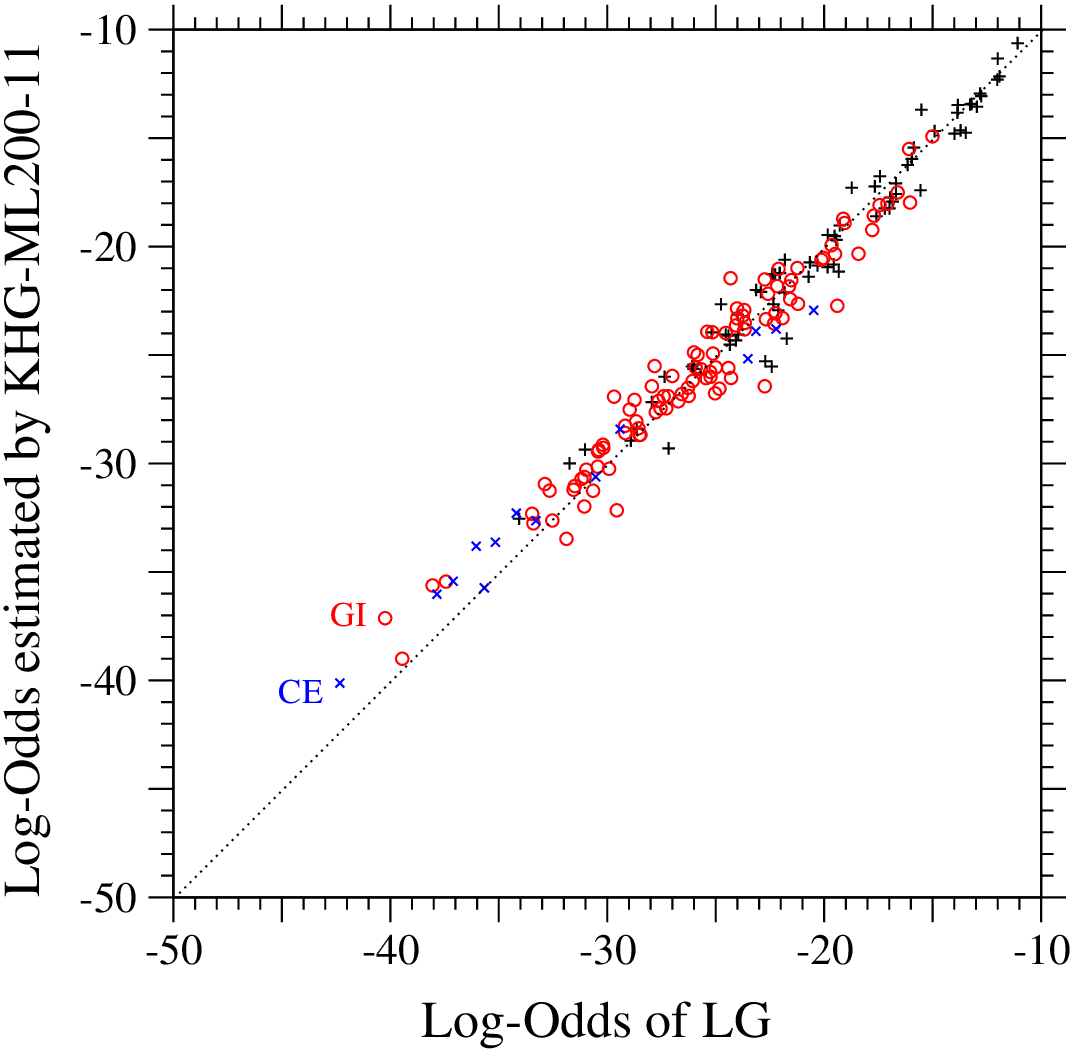}
}
} 
\caption{\label{fig: KHG-ML200_log-odds_1pam_LG}
\label{fig: KHG-ML199_log-odds_1pam_LG}
Each element 
log-$O(\langle S \rangle(\hat{\tau},\hat{\sigma}))_{ab}$
of the log-odds matrix 
of the KHG-ML200-11 model fitted to the 1-PAM LG matrix
is plotted against the log-odds log-$O(S^{\script{LG}}(\mbox{1 PAM}))_{ab}$
calculated from LG.
Plus, circle, and cross marks show the log-odds values for
one-, two-, and three-step amino acid pairs,
respectively.  
The dotted line shows the line 
of equal values between the ordinate and the abscissa.
}
\end{figure*}

} 

\FigurePanel{

\FigureInLegends{\newpage}

\begin{figure}[ht]
\FigureInLegends{
\textbf{A} \hspace*{47em} \textbf{B}

\centerline{
\includegraphics*[width=80mm,angle=0]{FIGS/KHG-ML199_logodds_JTT_KHG+kT+multi+5r+3f+sigma.eps}
\hspace*{5mm}
\includegraphics*[width=80mm,angle=0]{FIGS/KHG-ML199_logodds_WAG_KHG+kT+multi+5r+3f+sigma=0.eps}
}
\vspace*{1em}
\centerline{
\includegraphics*[width=80mm,angle=0]{FIGS/KHG-ML199_logodds_LG_KHG+kT+e0=0+multi+5r+3f+sigma=0.eps}
\hspace*{5mm}
\includegraphics*[width=80mm,angle=0]{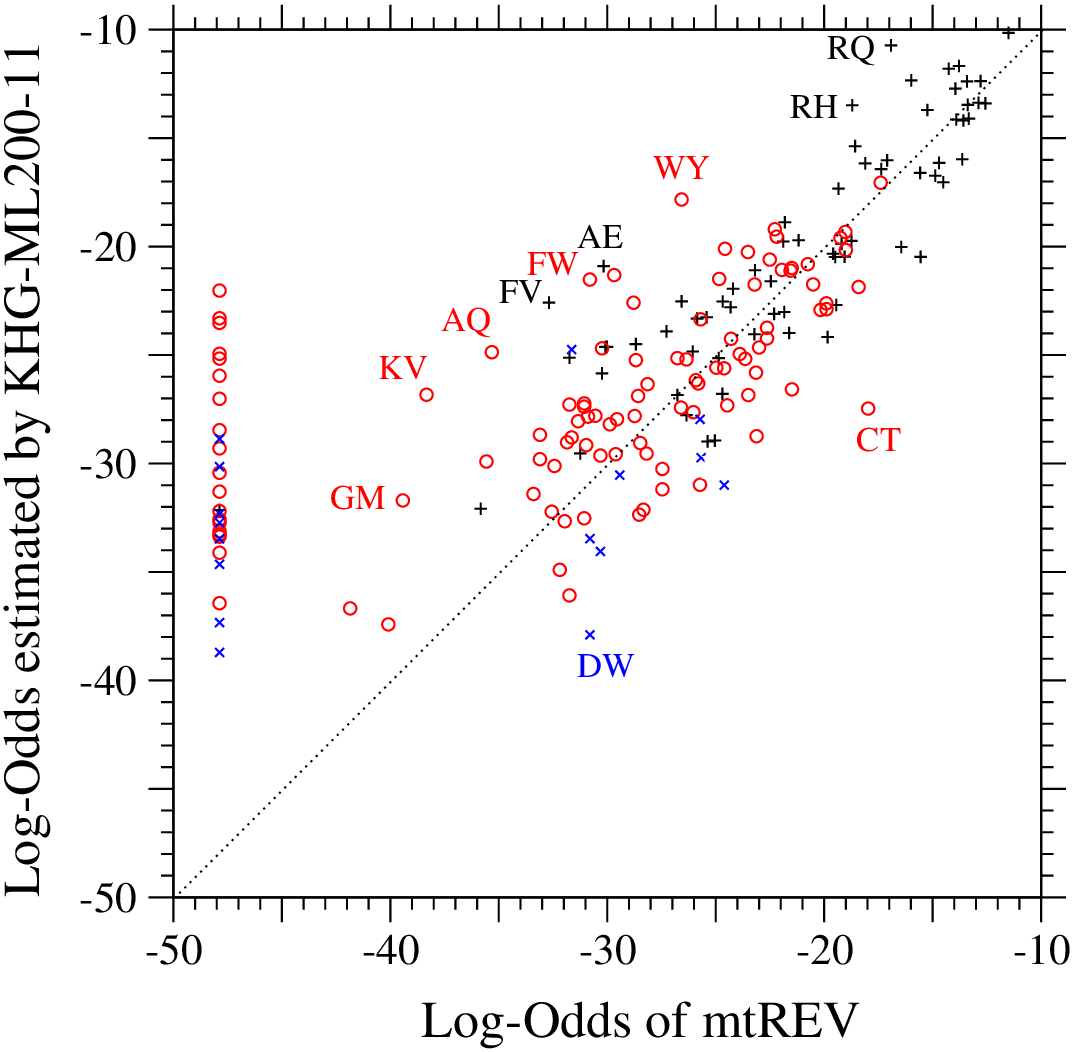}
}

\textbf{C} \hspace*{47em} \textbf{D}
} 
\caption{
\label{fig: KHG-ML200_log-odds_1pam_JTT}
\label{fig: KHG-ML199_log-odds_1pam_JTT}
\label{fig: KHG-ML200_log-odds_1pam_WAG}
\label{fig: KHG-ML199_log-odds_1pam_WAG}
\label{fig: KHG-ML200_log-odds_1pam_LG}
\label{fig: KHG-ML199_log-odds_1pam_LG}
\label{fig: KHG-ML200_log-odds_1pam_MtRevRmtx}
\label{fig: KHG-ML199_log-odds_1pam_MtRevRmtx}
\BF{
The KHG-ML200-11 model fitted to each of JTT, WAG, LG, and mtREV.
} 
Each element 
log-$O(\langle S \rangle(\hat{\tau},\hat{\sigma}))_{ab}$
of the log-odds matrices
of the KHG-ML200-11 model 
fitted to the 1-PAM matrices of 
(A) JTT, (B) WAG, (C) LG, and (D) mtREV
is plotted against the log-odds log-$O(S^{\script{LG}}(\mbox{1 PAM}))_{ab}$
calculated from the corresponding empirical substitution matrices.
Plus, circle, and cross marks show the log-odds values for
one-, two-, and three-step amino acid pairs,
respectively.  
The dotted line in each figure shows the line 
of equal values between the ordinate and the abscissa.
The log-odds elements of mtREV whose values are smaller than about $-47.8$ 
are all assumed to be $-47.8$; see the original paper\CITE{AH:96}.
}
\end{figure}

} 

} 

\SupFig{

\FigureEach{

\FigureInLegends{\newpage}

\begin{figure*}[ht]
\FigureInLegends{
\centerline{
\includegraphics*[width=130mm,angle=0]{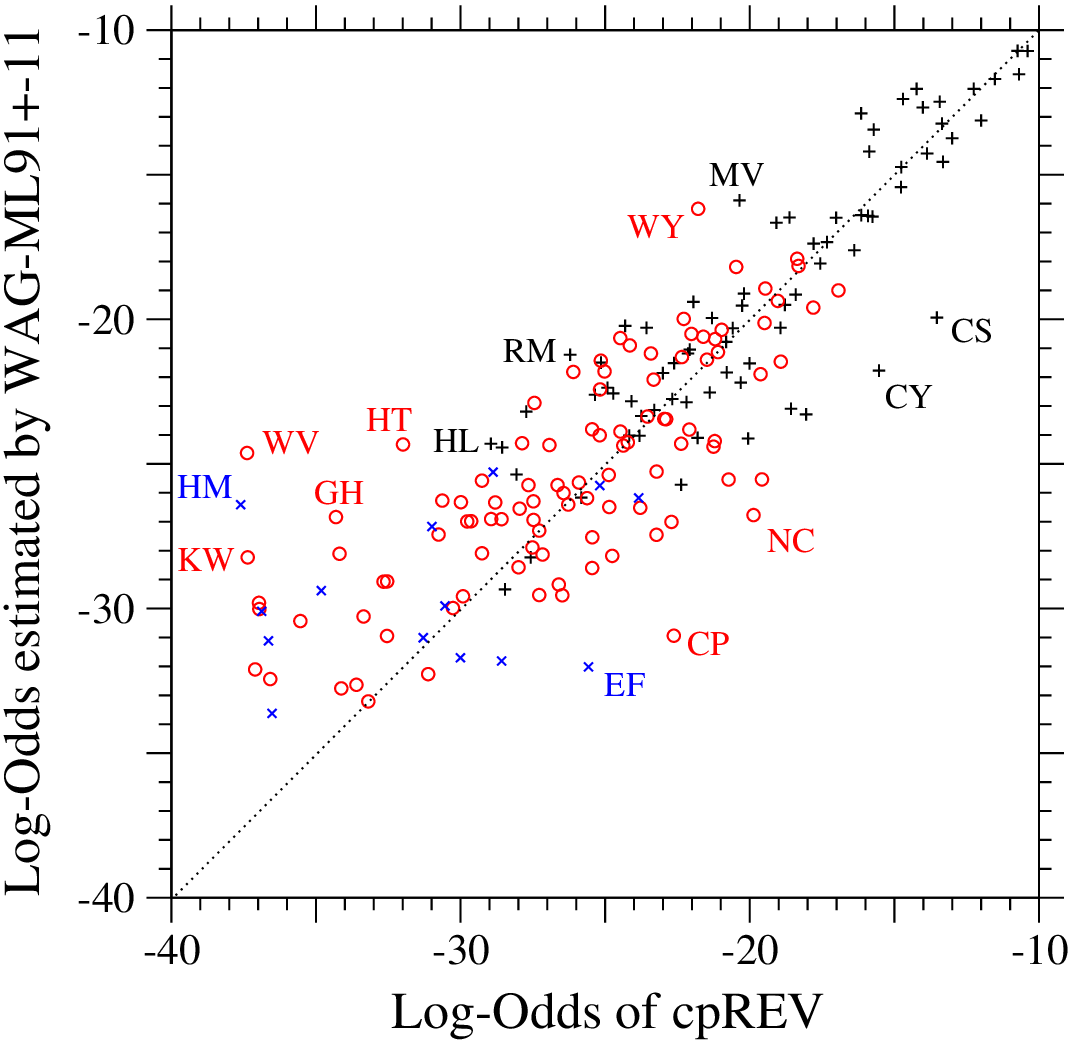}
}
} 
\caption{
\label{fig: WAG-ML91+_log-odds_1pam_CpRev}
\label{fig: WAG-ML-91+_log-odds_1pam_CpRev}
Each element 
log-$O(\langle S \rangle(\hat{\tau},\hat{\sigma}))_{ab}$
of the log-odds matrix 
of the WAG-ML91+-11 model fitted to the 1-PAM cpREV matrix
is plotted against the log-odds log-$O(S^{\script{cpREV}}(\mbox{1 PAM}))_{ab}$
calculated from cpREV.
Plus, circle, and cross marks show the log-odds values for
one-, two-, and three-step amino acid pairs,
respectively.  
The dotted line shows the line 
of equal values between the ordinate and the abscissa.
}
\end{figure*}

} 

\FigureEach{
\FigureInLegends{\newpage}

\begin{figure*}[ht]
\FigureInLegends{
\centerline{
\includegraphics*[width=130mm,angle=0]{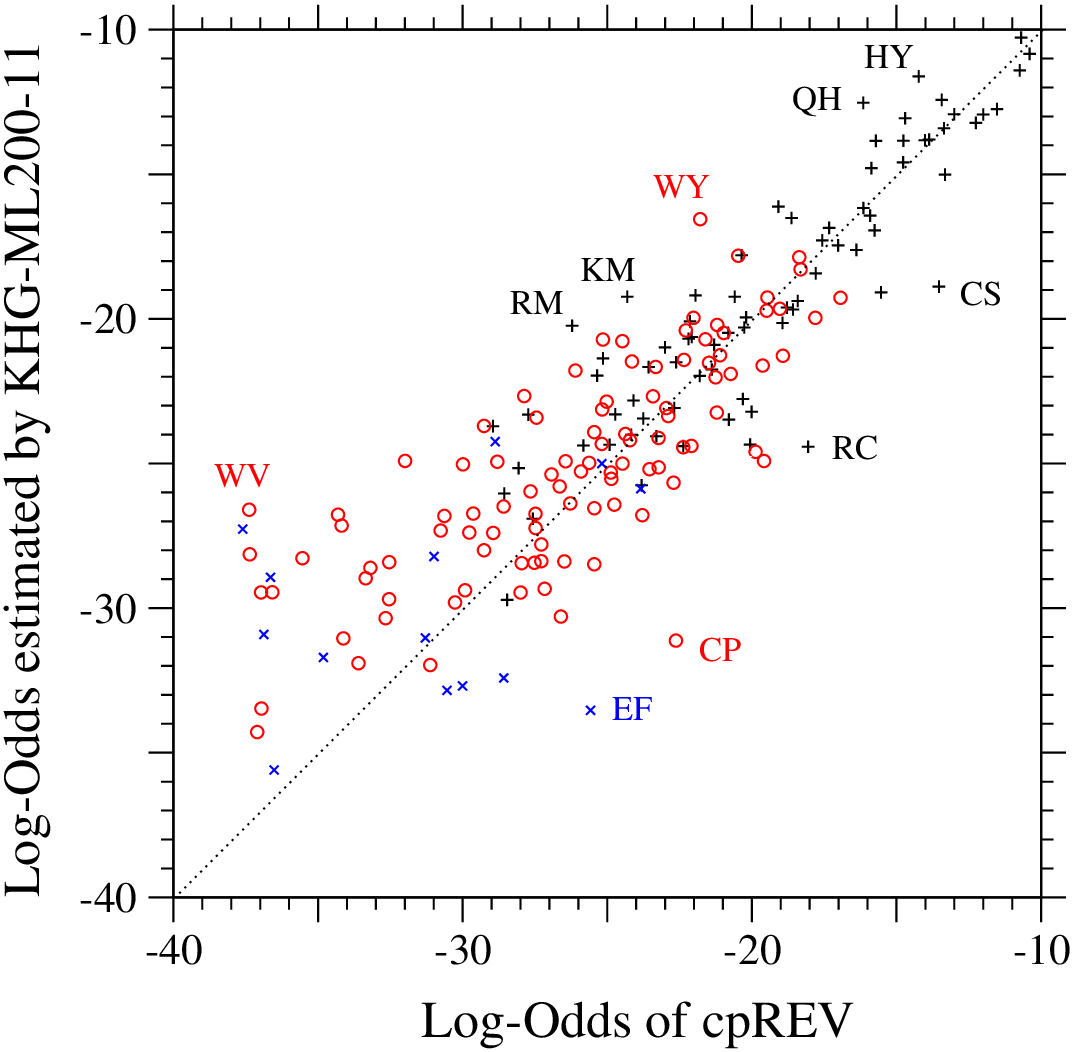}
}
} 
\caption{\label{fig: KHG-ML200_log-odds_1pam_CpRev}
\label{fig: KHG-ML199_log-odds_1pam_CpRev}
Each element 
log-$O(\langle S \rangle(\hat{\tau},\hat{\sigma}))_{ab}$
of the log-odds matrix 
of the KHG-ML200-11 model fitted to the 1-PAM cpREV matrix
is plotted against the log-odds log-$O(S^{\script{cpREV}}(\mbox{1 PAM}))_{ab}$
calculated from cpREV.
Plus, circle, and cross marks show the log-odds values for
one-, two-, and three-step amino acid pairs,
respectively.  
The dotted line shows the line 
of equal values between the ordinate and the abscissa.
}
\end{figure*}

} 

\FigureEach{
\FigureInLegends{\newpage}

\begin{figure*}[ht]
\FigureInLegends{
\centerline{
\includegraphics*[width=130mm,angle=0]{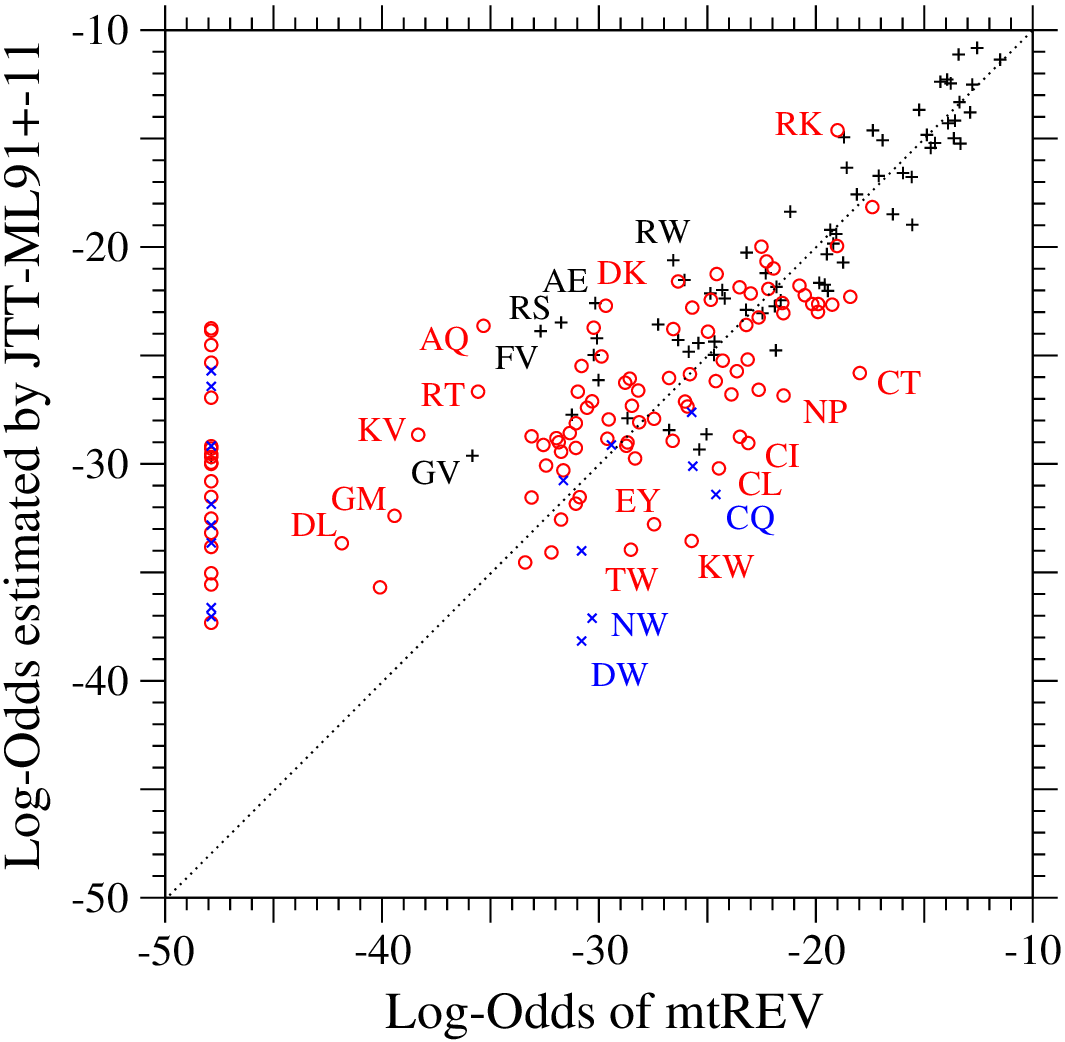}
}
} 
\caption{
\label{fig: JTT-ML-91+_log-odds_1pam_MtRevRmtx}
\label{fig: JTT-ML91+_log-odds_1pam_MtRevRmtx}
\label{fig: JTT-ML-91_log-odds_1pam_MtRevRmtx}
Each element 
log-$O(\langle S \rangle(\hat{\tau},\hat{\sigma}))_{ab}$
of the log-odds matrix 
of the JTT-ML91+-11 model fitted to the 1-PAM mtREV matrix
is plotted against the log-odds log-$O(S^{\script{mtREV}}(\mbox{1 PAM}))_{ab}$
calculated from mtREV.
Plus, circle, and cross marks show the log-odds values for
one-, two-, and three-step amino acid pairs,
respectively.  
The log-odds elements of mtREV whose values are smaller than about $-47.8$ 
are all assumed to be $-47.8$; see the original paper\CITE{AH:96}
The dotted line shows the line 
of equal values between the ordinate and the abscissa.
}
\end{figure*}

} 

\FigureEach{

\FigureInLegends{\newpage}

\begin{figure*}[ht]
\FigureInLegends{
\centerline{
\includegraphics*[width=130mm,angle=0]{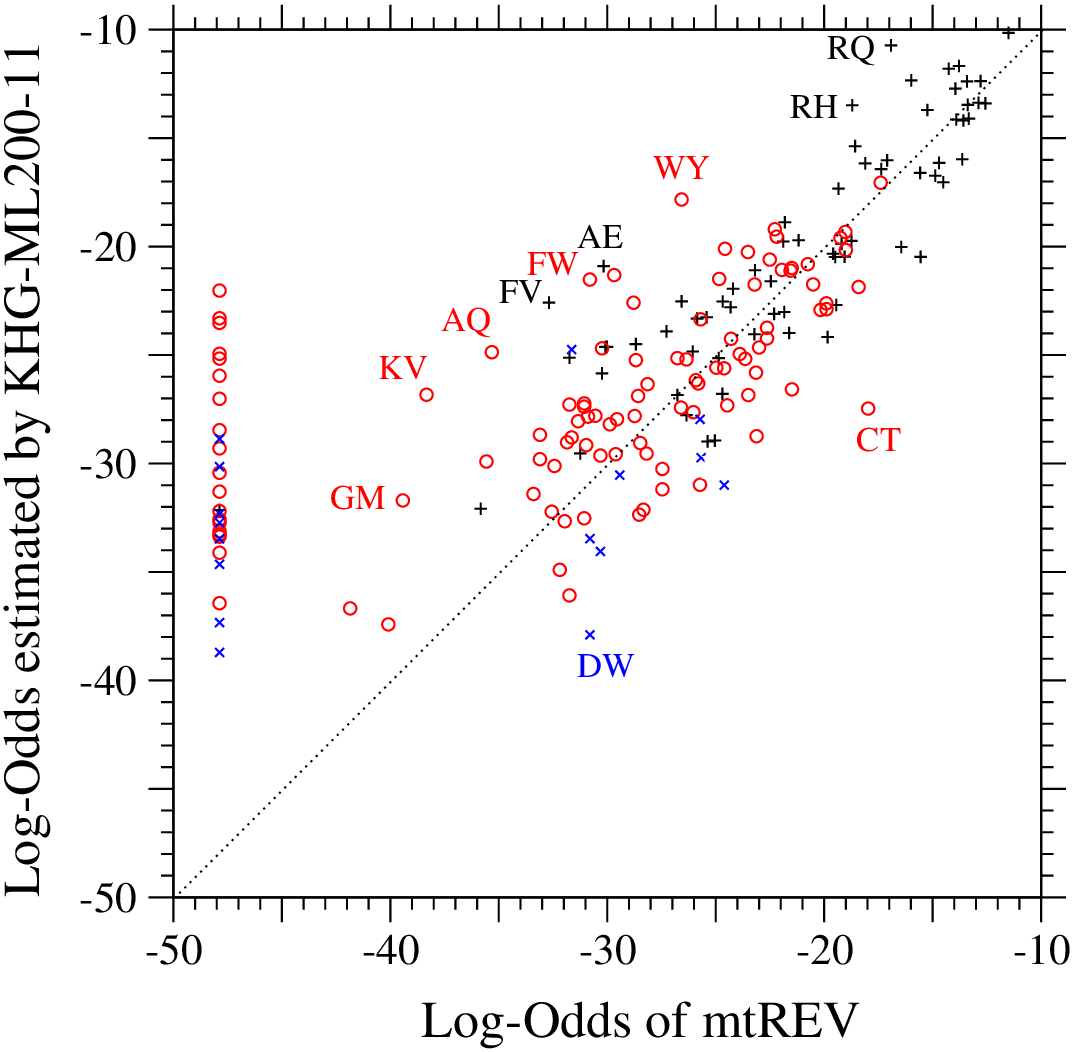}
}
} 
\caption{\label{fig: KHG-ML200_log-odds_1pam_MtRevRmtx}
\label{fig: KHG-ML199_log-odds_1pam_MtRevRmtx}
Each element 
log-$O(\langle S \rangle(\hat{\tau},\hat{\sigma}))_{ab}$
of the log-odds matrix 
of the KHG-ML200-11 model fitted to the 1-PAM mtREV matrix
is plotted against the log-odds log-$O(S^{\script{mtREV}}(\mbox{1 PAM}))_{ab}$
calculated from mtREV.
Plus, circle, and cross marks show the log-odds values for
one-, two-, and three-step amino acid pairs,
respectively.  
The log-odds elements of mtREV whose values are smaller than about $-47.8$ 
are all assumed to be $-47.8$; see the original paper\CITE{AH:96}
The dotted line shows the line 
of equal values between the ordinate and the abscissa.
}
\end{figure*}

} 

\FigurePanel{

\FigureInLegends{\newpage}
\FigureInLegends{\small}

\begin{figure}[ht]
\FigureInLegends{
\textbf{A} \hspace*{45em} \textbf{B}
\centerline{
\includegraphics*[width=75mm,angle=0]{FIGS/WAG-ML91+mod_logodds_1pam_JTT_WAG+kT+multi+5r+3f+sigma.eps}
\hspace*{5mm}
\includegraphics*[width=75mm,angle=0]{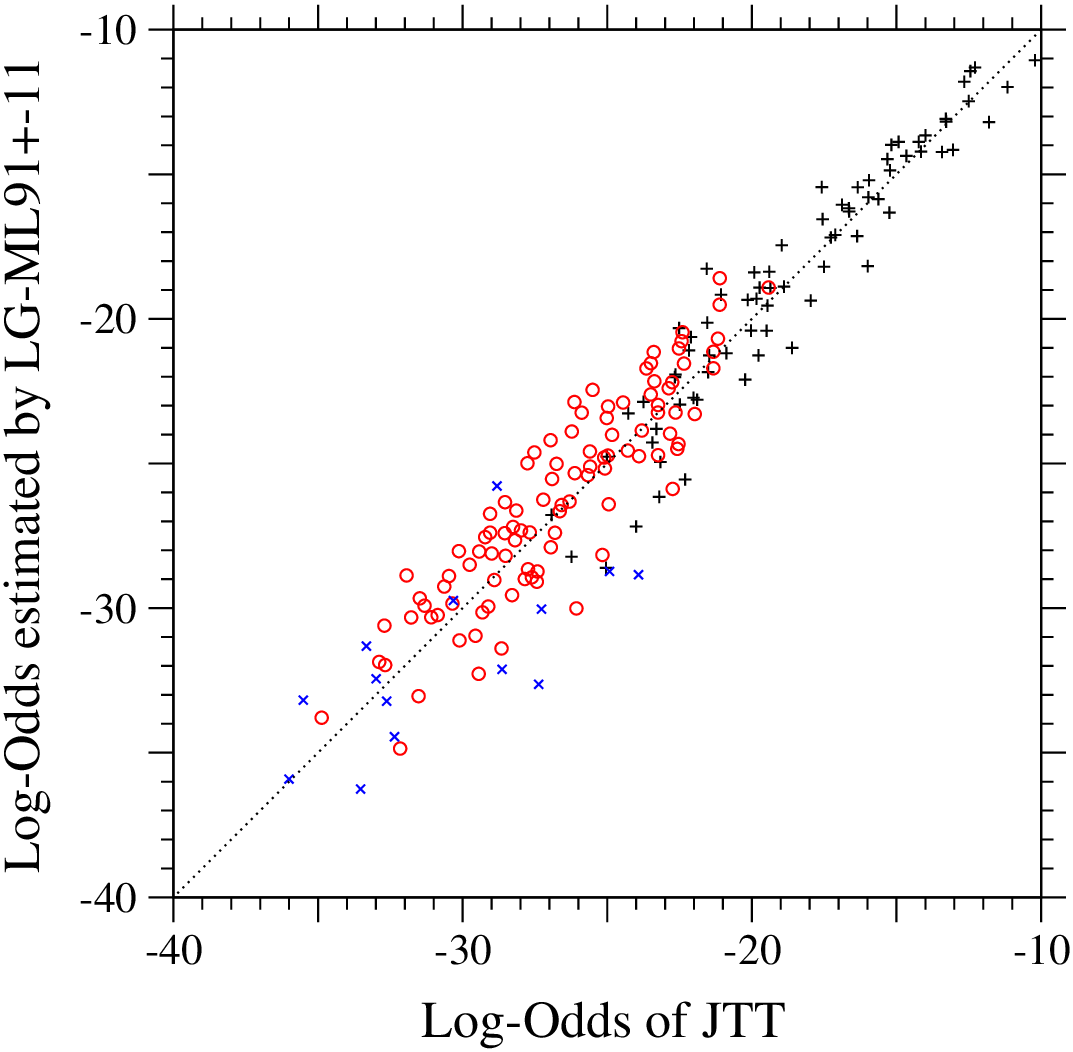}
}
\textbf{C} \hspace*{45em} \textbf{D}
\centerline{
\includegraphics*[width=75mm,angle=0]{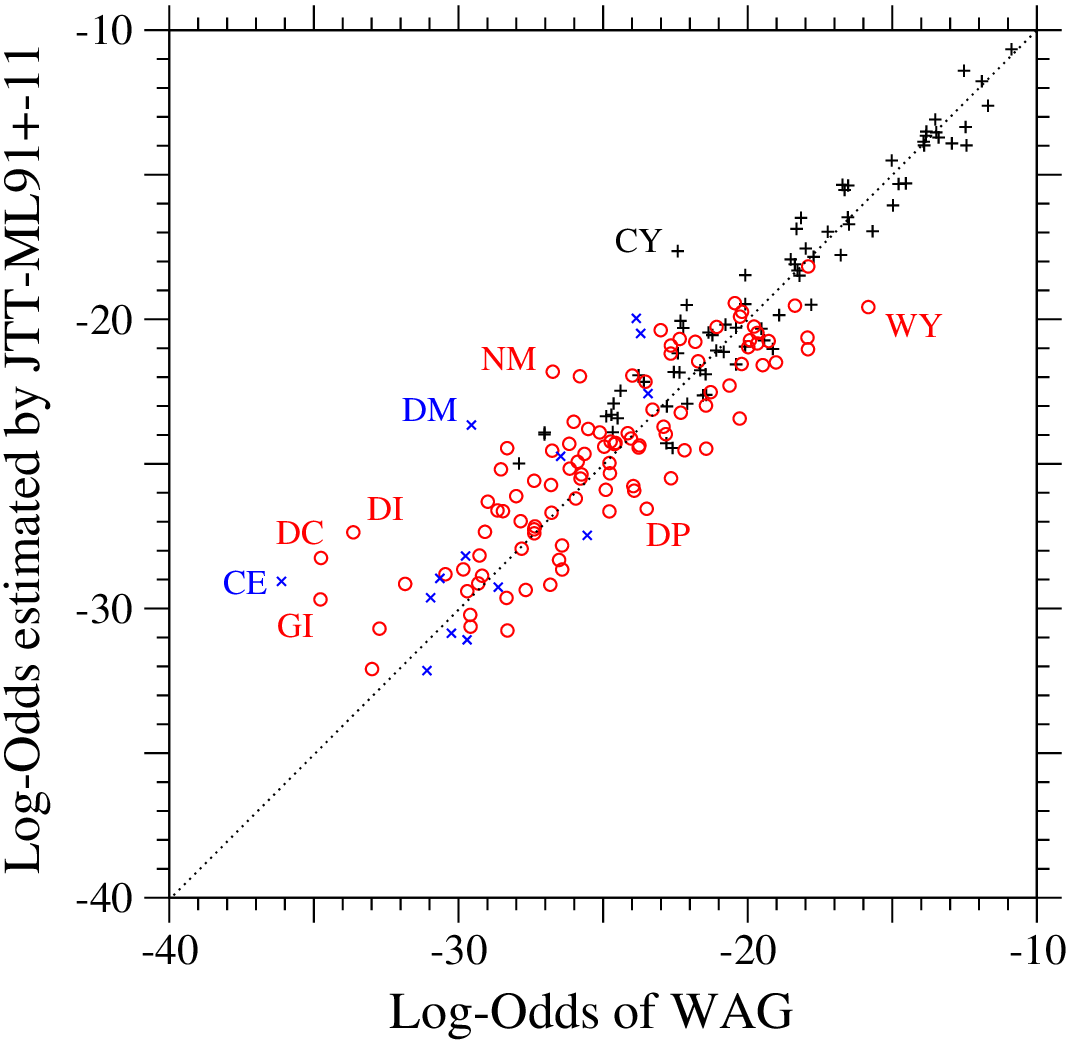}
\hspace*{5mm}
\includegraphics*[width=75mm,angle=0]{FIGS/LG-ML91+mod_logodds_1pam_WAG_LG+kT+multi+5r+3f+sigma.eps}
}
\textbf{E} \hspace*{45em} \textbf{F}
\centerline{
\includegraphics*[width=75mm,angle=0]{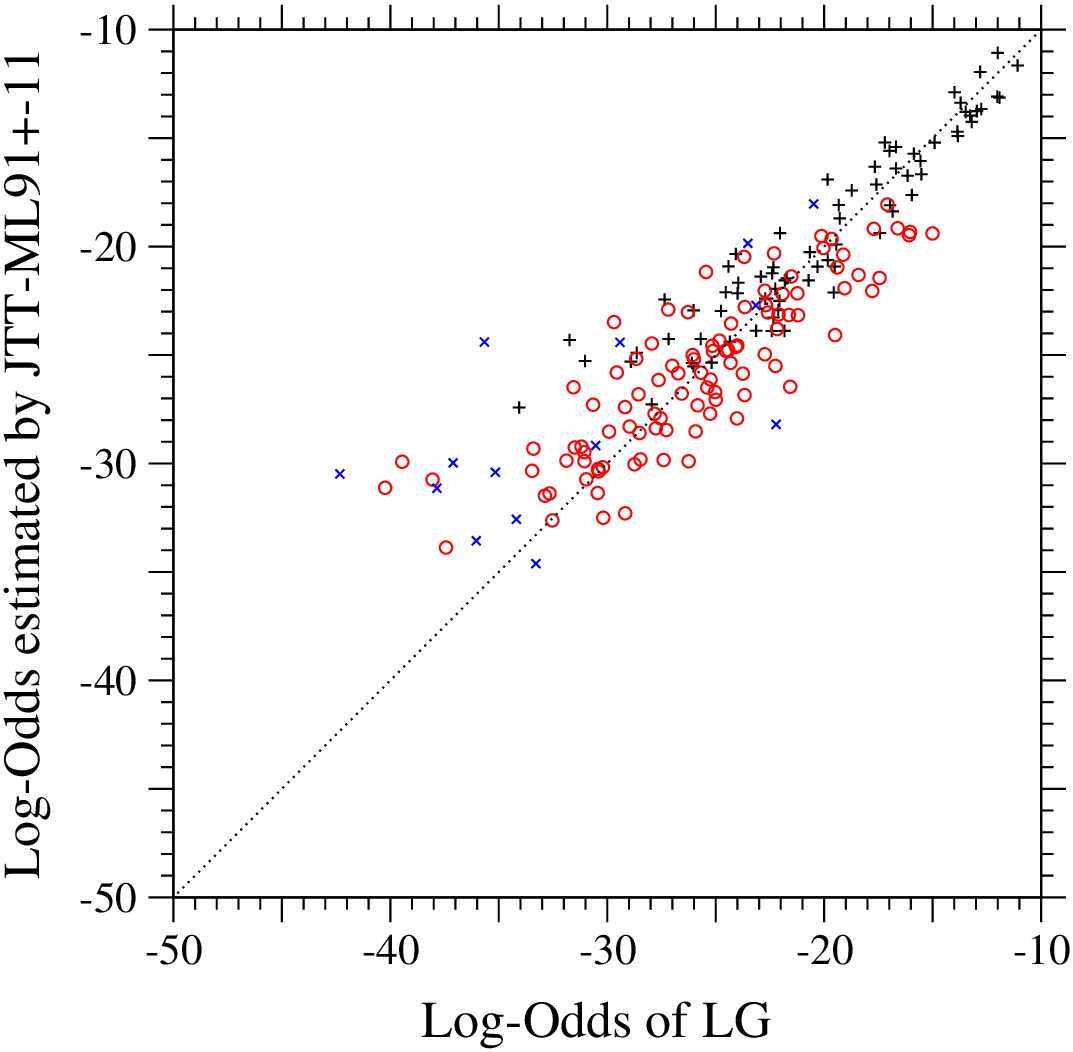}
\hspace*{5mm}
\includegraphics*[width=75mm,angle=0]{FIGS/WAG-ML91+mod_logodds_1pam_LG_WAG+kT+multi+5r+3f+sigma.eps}
}

} 
\caption{
\label{fig: X-ML91+_log-odds_1pam_JTT}
\label{fig: X-ML91+_log-odds_1pam_WAG}
\label{fig: X-ML91+_log-odds_1pam_LG}
\FigureInLegends{\small}
\BF{
Models fitted to each of JTT, WAG, and LG.
} 
Each element 
log-$O(\langle S \rangle(\hat{\tau},\hat{\sigma}))_{ab}$
of the log-odds matrix 
of the model fitted to each empirical substitution matrix
is plotted against the log-odds log-$O(S^{\script{obs}}(\mbox{1 PAM}))_{ab}$
calculated from the corresponding empirical substitution matrix.
Plus, circle, and cross marks show the log-odds values for
one-, two-, and three-step amino acid pairs,
respectively.  
The dotted line in each figure shows the line 
of equal values between the ordinate and the abscissa.
\FigureInLegends{\normalsize}
}
\end{figure}

\FigureInLegends{\normalsize}

} 

\FigurePanel{

\FigureInLegends{\newpage}

\FigureInLegends{\small}

\begin{figure}[ht]
\FigureInLegends{
\textbf{A} \hspace*{45em} \textbf{B}
\centerline{
\includegraphics*[width=75mm,angle=0]{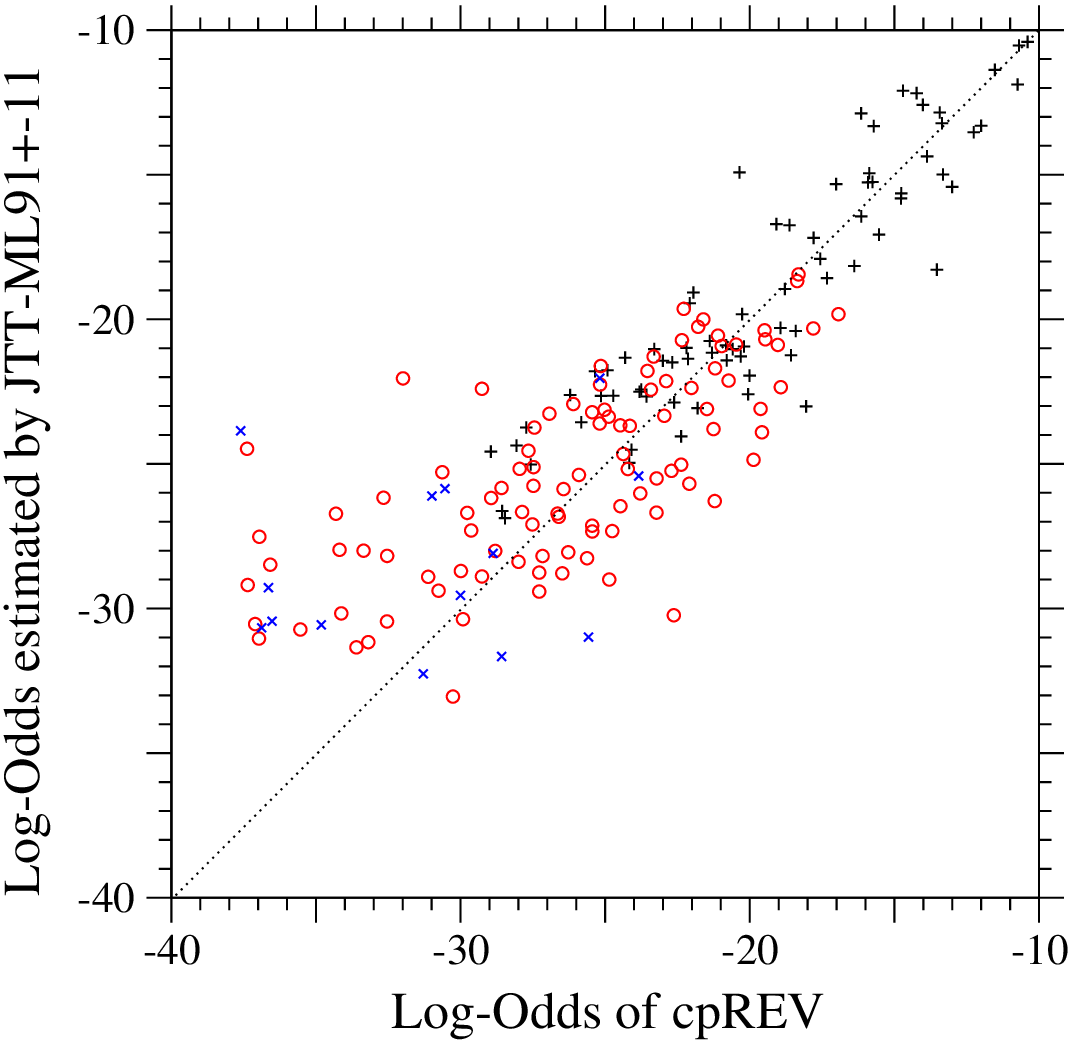}
\hspace*{5mm}
\includegraphics*[width=75mm,angle=0]{FIGS/WAG-ML91+mod_logodds_1pam_CpRev_WAG+kT+multi+5r+3f+sigma.eps}
}
\textbf{C} \hspace*{45em} \textbf{D}
\centerline{
\includegraphics*[width=75mm,angle=0]{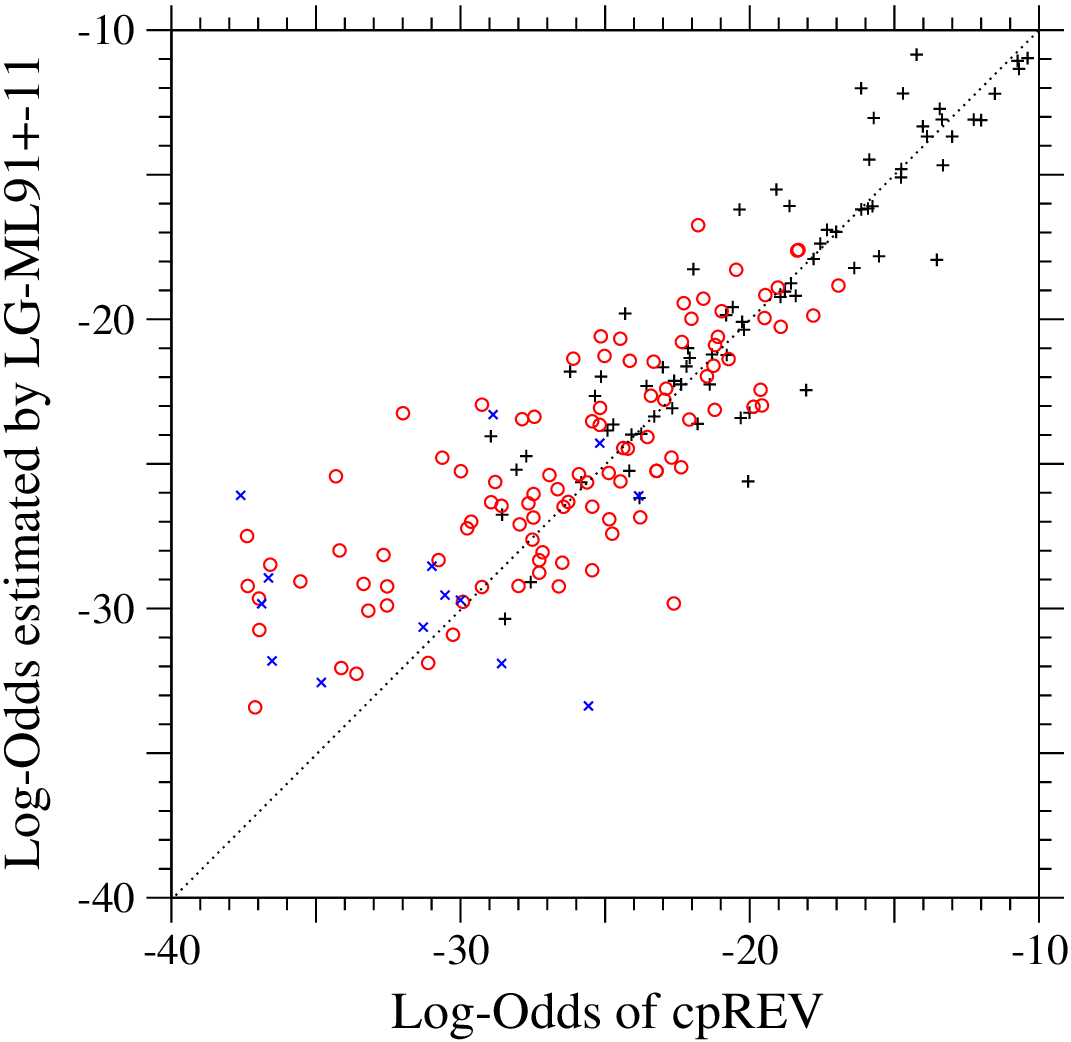}
\hspace*{5mm}
\includegraphics*[width=75mm,angle=0]{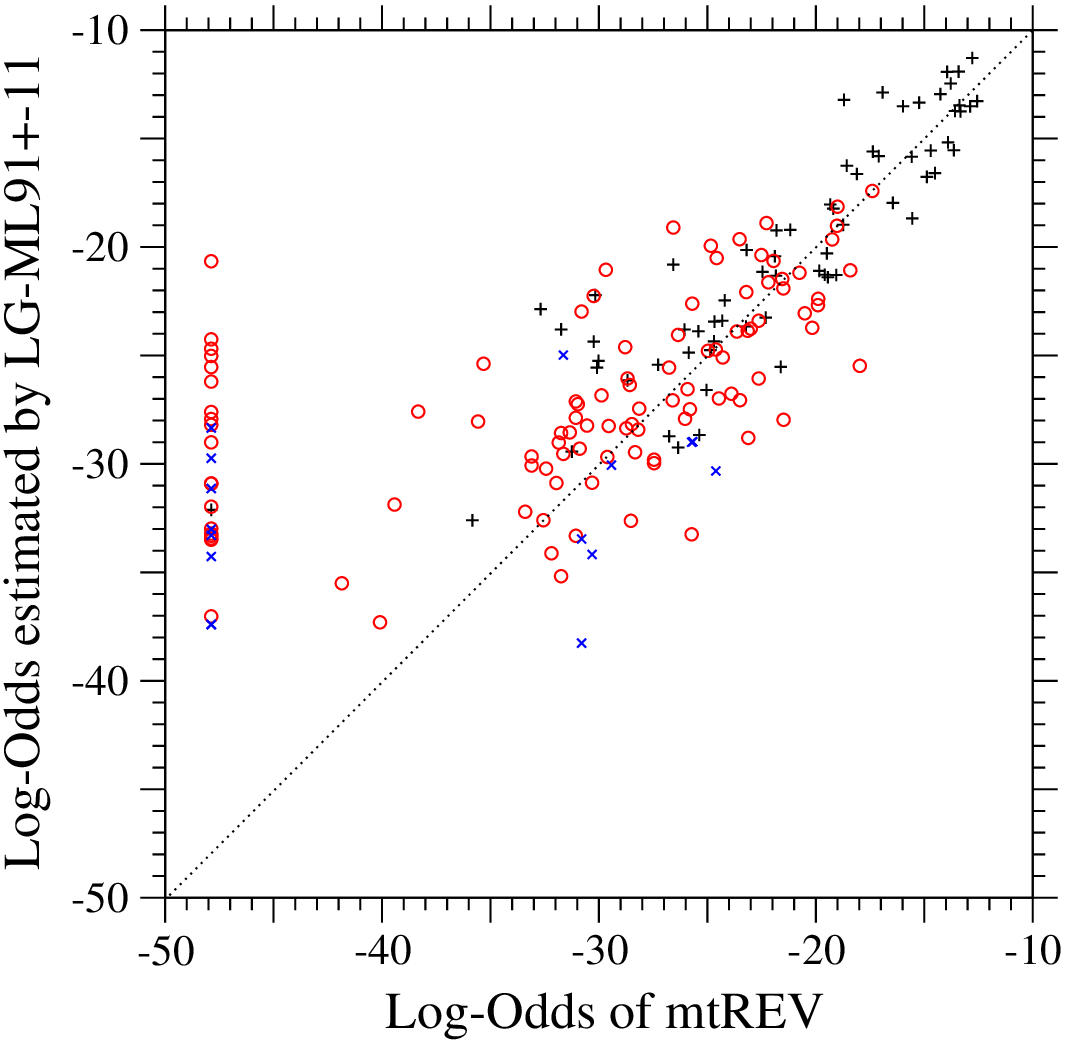}
}
\textbf{E} \hspace*{45em} \textbf{F}
\centerline{
\includegraphics*[width=75mm,angle=0]{FIGS/JTT-ML91+mod_logodds_1pam_MtRevRmtx_JTT+kT+multi+5r+3f+sigma.eps}
\hspace*{5mm}
\includegraphics*[width=75mm,angle=0]{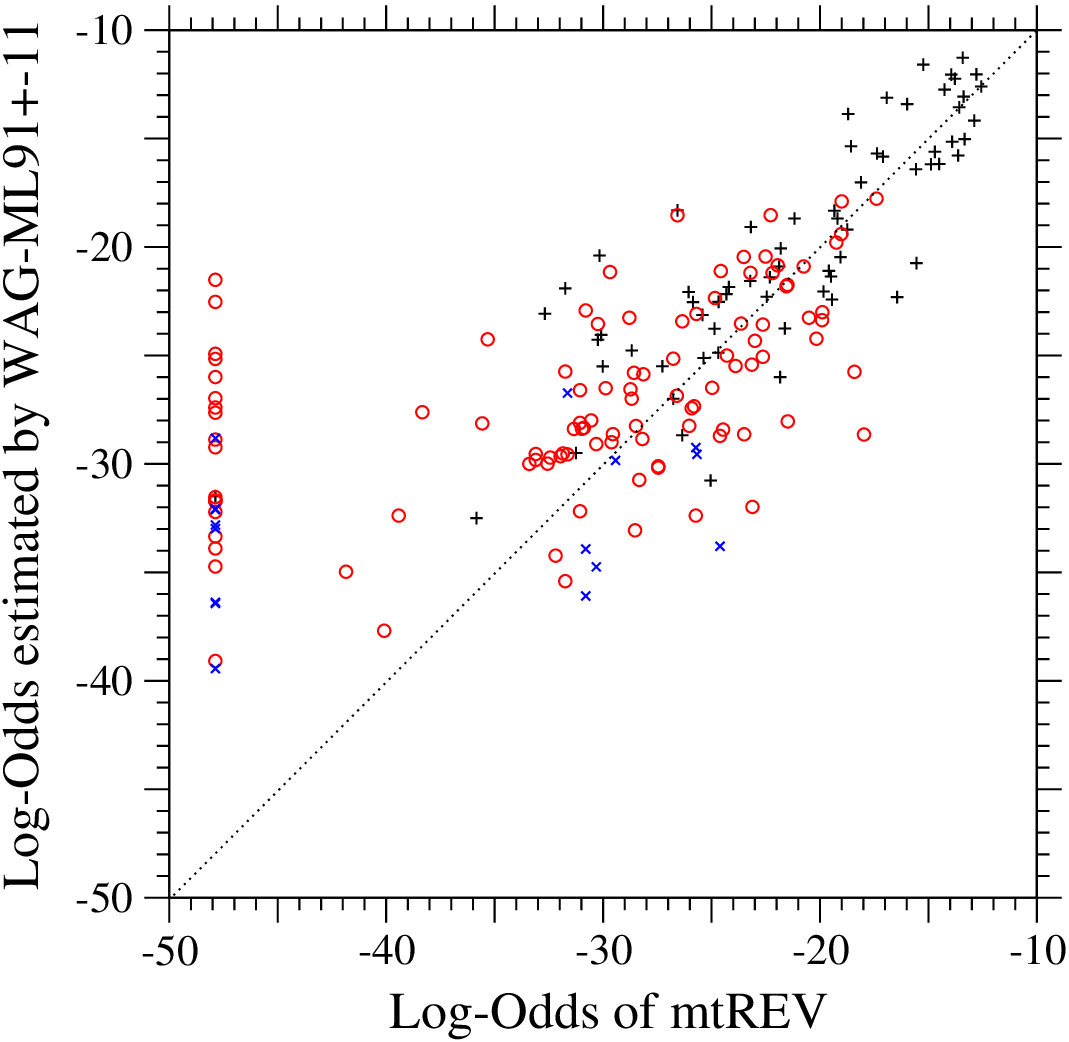}
}

} 
\caption{
\label{fig: X-ML91+_log-odds_1pam_CpRev}
\label{fig: X-ML91+_log-odds_1pam_MtRevRmtx}
\FigureInLegends{\small}
\BF{
Models fitted to each of cpREV and mtREV.
} 
Each element 
log-$O(\langle S \rangle(\hat{\tau},\hat{\sigma}))_{ab}$
of the log-odds matrix 
of the model fitted to each empirical substitution matrix
is plotted against the log-odds log-$O(S^{\script{obs}}(\mbox{1 PAM}))_{ab}$
calculated from the corresponding empirical substitution matrix.
Plus, circle, and cross marks show the log-odds values for
one-, two-, and three-step amino acid pairs,
respectively.  
The dotted line in each figure shows the line 
of equal values between the ordinate and the abscissa.
\FigureInLegends{\normalsize}
}
\end{figure}

\FigureInLegends{\normalsize}

} 

\FigurePanel{

\FigureInLegends{\newpage}

\begin{figure}[ht]
\FigureInLegends{
\textbf{A} \hspace*{47em} \textbf{B}

\centerline{
\includegraphics*[width=80mm,angle=0]{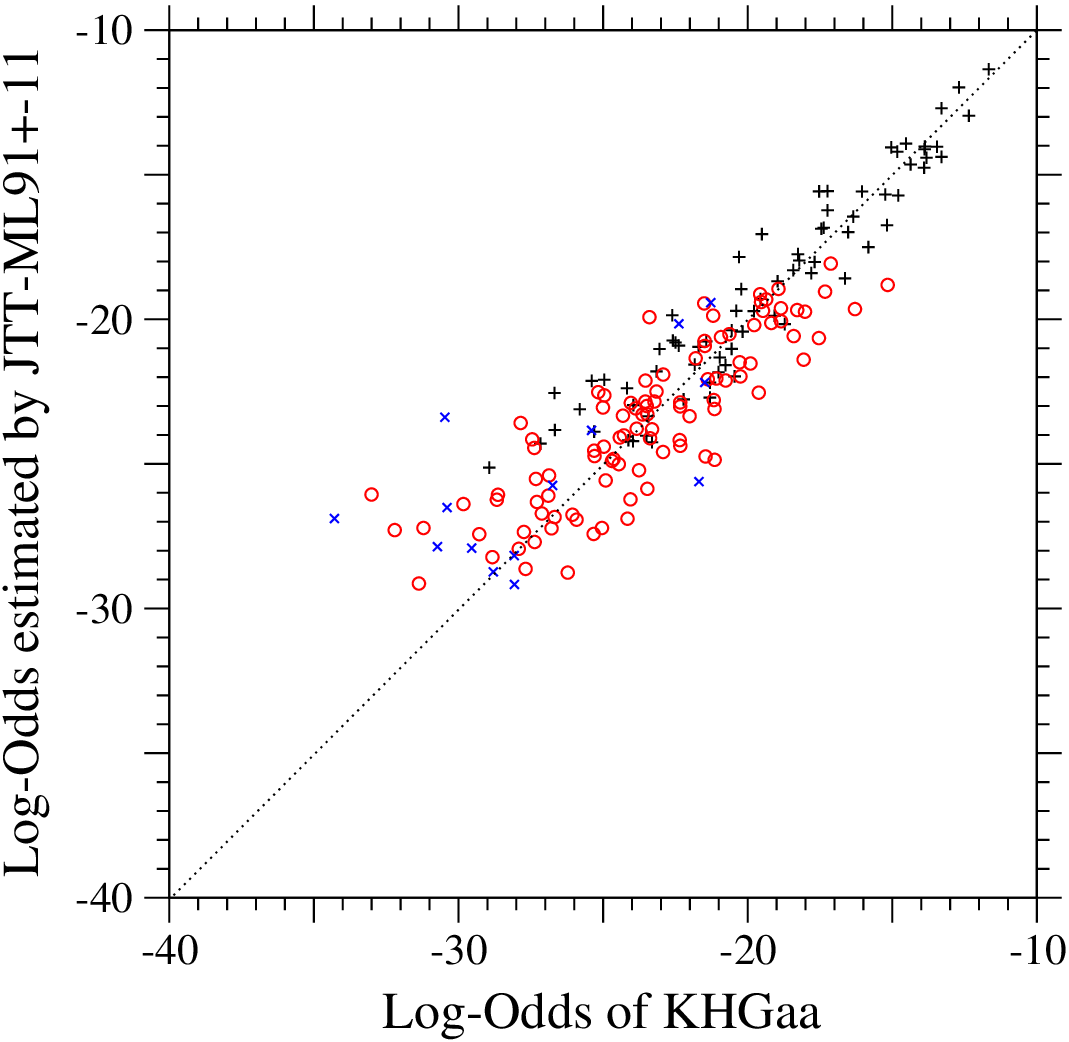}
\hspace*{5mm}
\includegraphics*[width=80mm,angle=0]{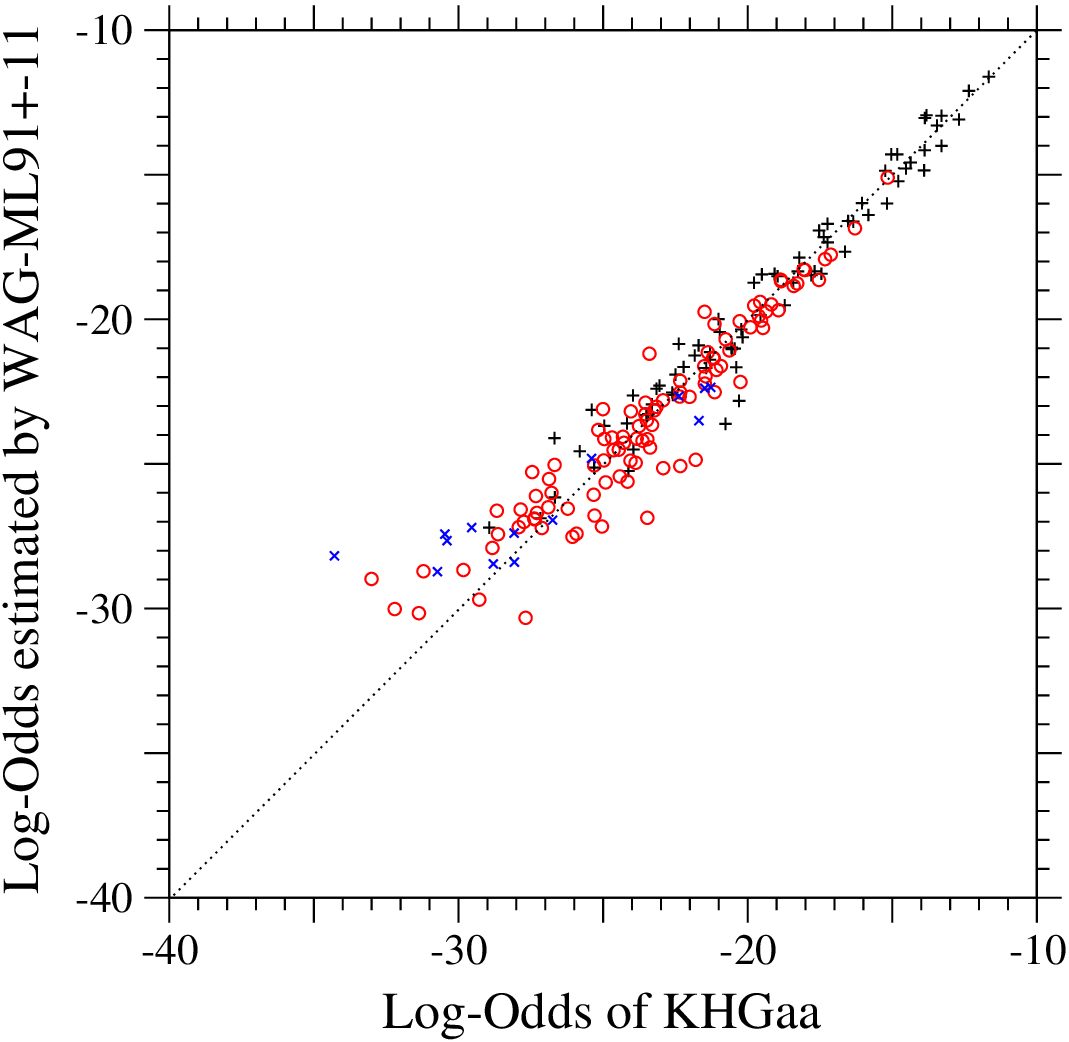}
}
\vspace*{1em}
\centerline{
\includegraphics*[width=80mm,angle=0]{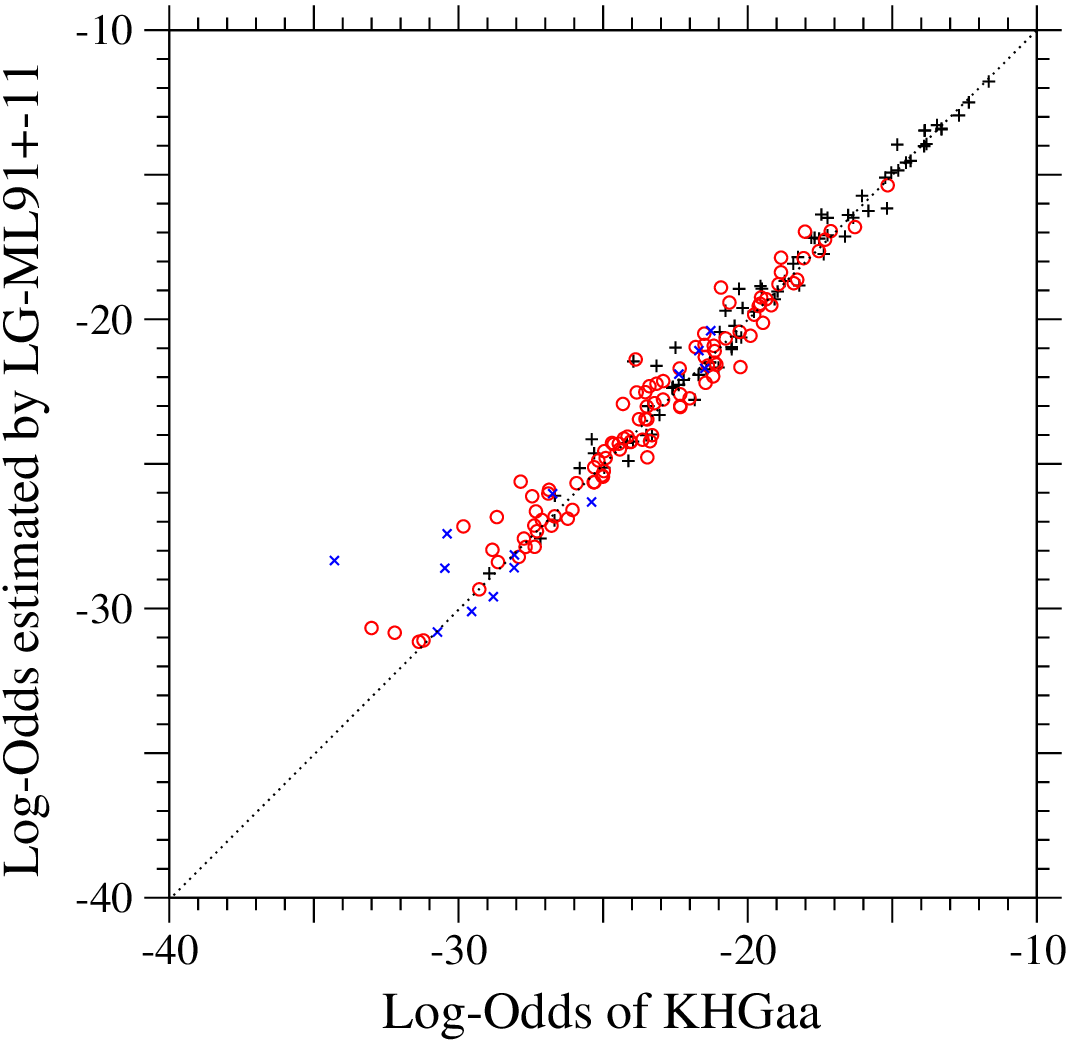}
\hspace*{5mm}
\hspace*{80mm}
}

\textbf{C} 
} 
\caption{
\label{fig: X-ML91+_log-odds_1pam_KHGasm}
\BF{
Models fitted to the KHG-derived amino acid substitution matrix.
} 
Each element 
log-$O(\langle S \rangle(\hat{\tau},\hat{\sigma}))_{ab}$
of the log-odds matrix 
of the model 
fitted to the 1-PAM KHG-derived amino acid substitution matrix (KHGaa)
is plotted against the log-odds log-$O(S^{\script{obs}}(\mbox{1 PAM}))_{ab}$
calculated from KHGaa.
Plus, circle, and cross marks show the log-odds values for
one-, two-, and three-step amino acid pairs,
respectively.  
The dotted line in each figure shows the line 
of equal values between the ordinate and the abscissa.
}
\end{figure}

} 

\FigurePanel{

\FigureInLegends{\newpage}

\begin{figure}[ht]
\FigureInLegends{
\textbf{A} \hspace*{47em} \textbf{B}

\centerline{
\includegraphics*[width=80mm,angle=0]{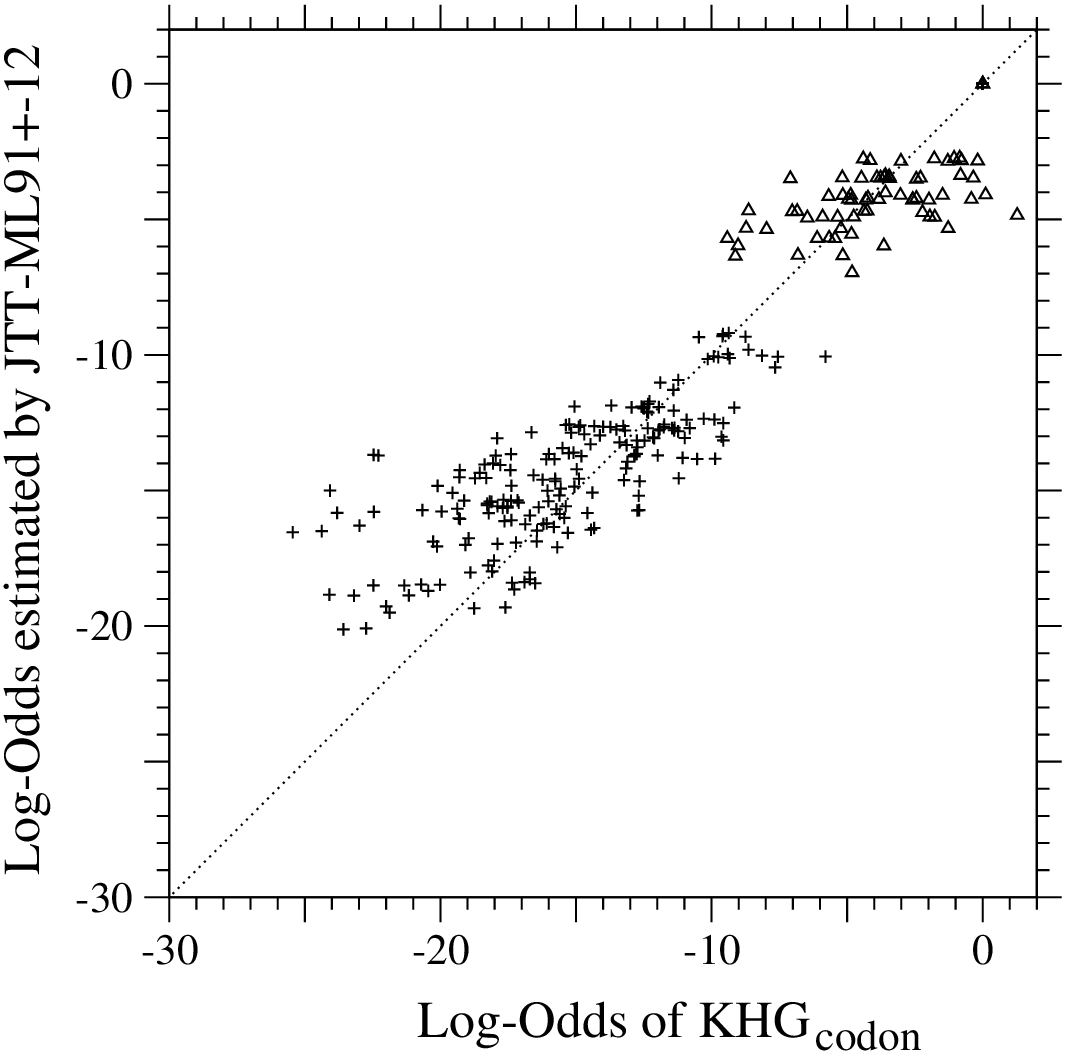}
\hspace*{5mm}
\includegraphics*[width=80mm,angle=0]{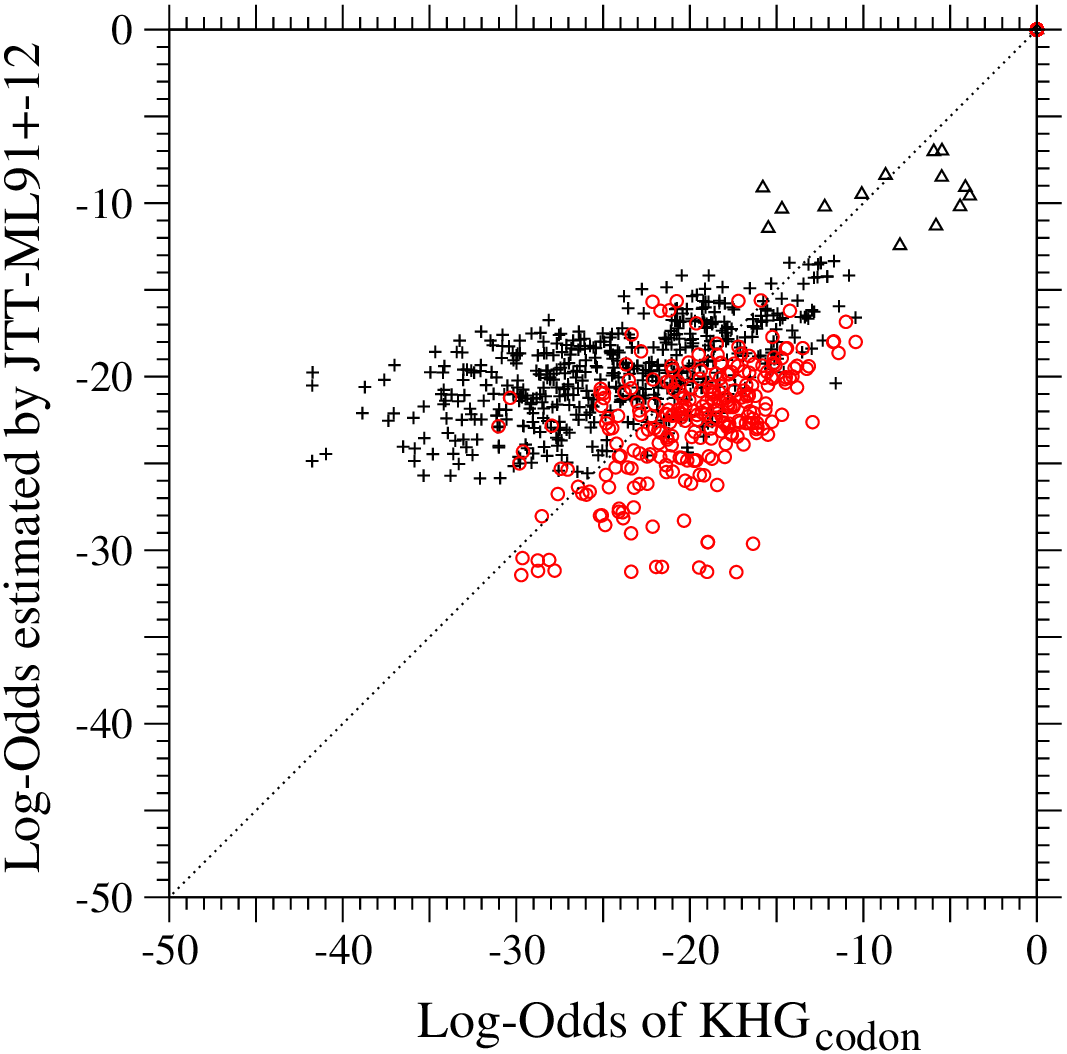}
}
\vspace*{1em}
\centerline{
\includegraphics*[width=80mm,angle=0]{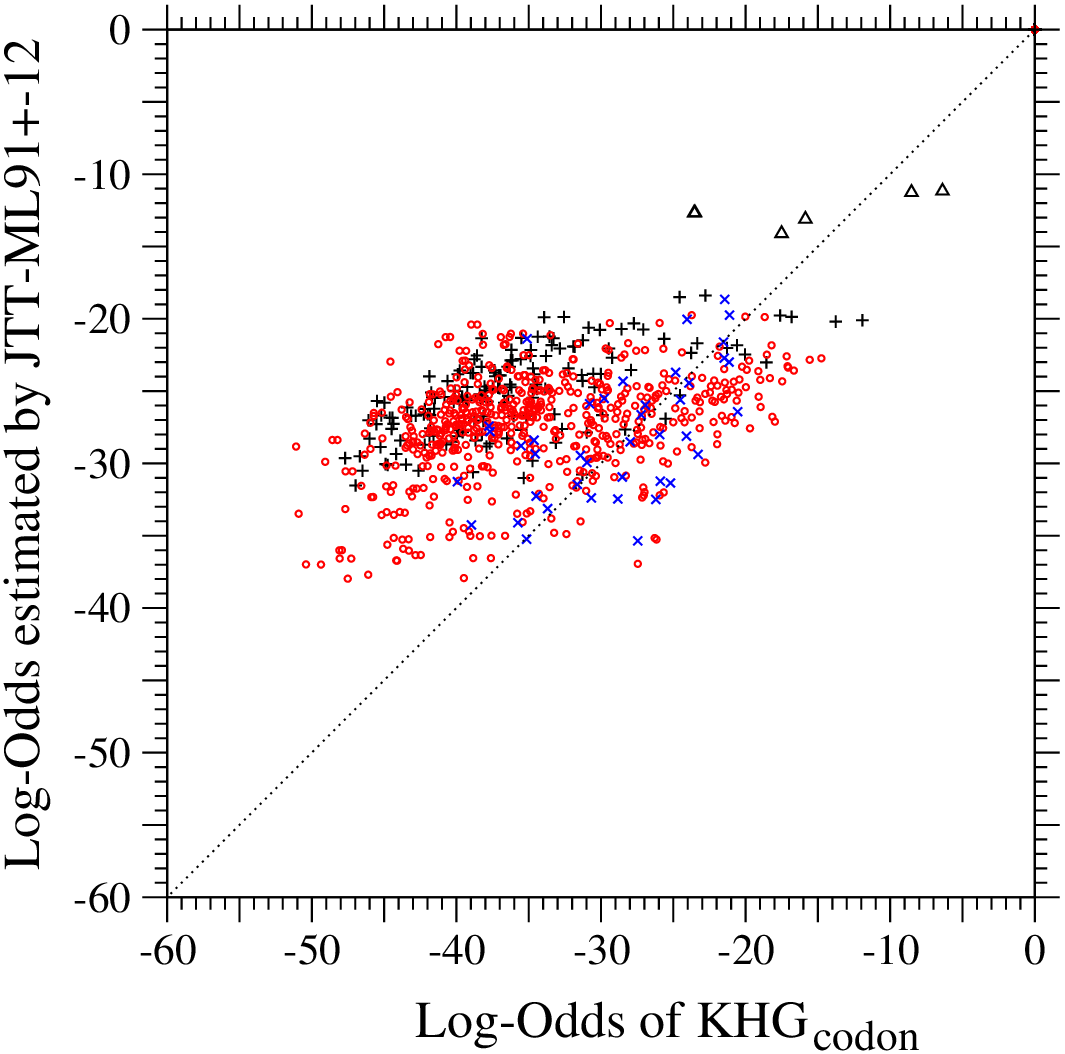}
\hspace*{5mm}
\hspace*{80mm}
}

\textbf{C} 
} 
\caption{
\label{fig: JTT-ML91+_log-odds_1pam_KHG}
\BF{
The JTT-ML91+-12 model fitted to the 1-PAM KHG codon substitution matrix.
} 
Each element 
log-$O(\langle S \rangle(\hat{\tau},\hat{\sigma}))_{\mu \nu}$
of the log-odds matrix 
corresponding to (A) single, (B) double, and (C) triple nucleotide changes 
in the JTT-ML91+-12 model fitted to the 1-PAM KHG codon substitution matrix
is plotted against the log-odds log-$O(S^{\script{KHG}}(\mbox{1 PAM}))_{\mu \nu}$
calculated from KHG.
Upper triangle, plus, circle, and cross marks show the log-odds values for
synonymous pairs and one-, two-, and three-step amino acid pairs,
respectively. 
The dotted line in each figure shows the line 
of equal values between the ordinate and the abscissa.
}
\end{figure}

} 

\FigurePanel{

\FigureInLegends{\newpage}

\begin{figure}[ht]
\FigureInLegends{
\textbf{A} \hspace*{47em} \textbf{B}

\centerline{
\includegraphics*[width=80mm,angle=0]{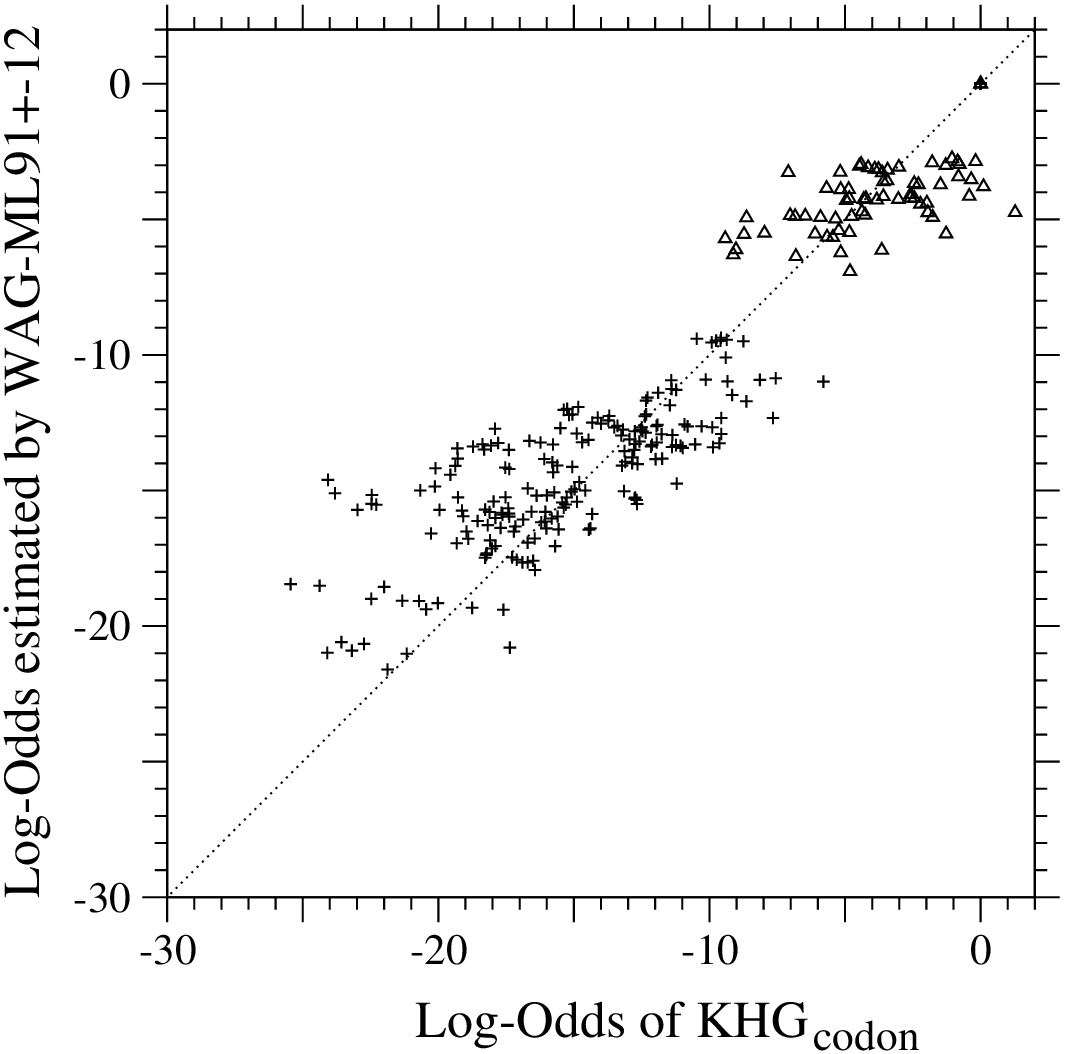}
\hspace*{5mm}
\includegraphics*[width=80mm,angle=0]{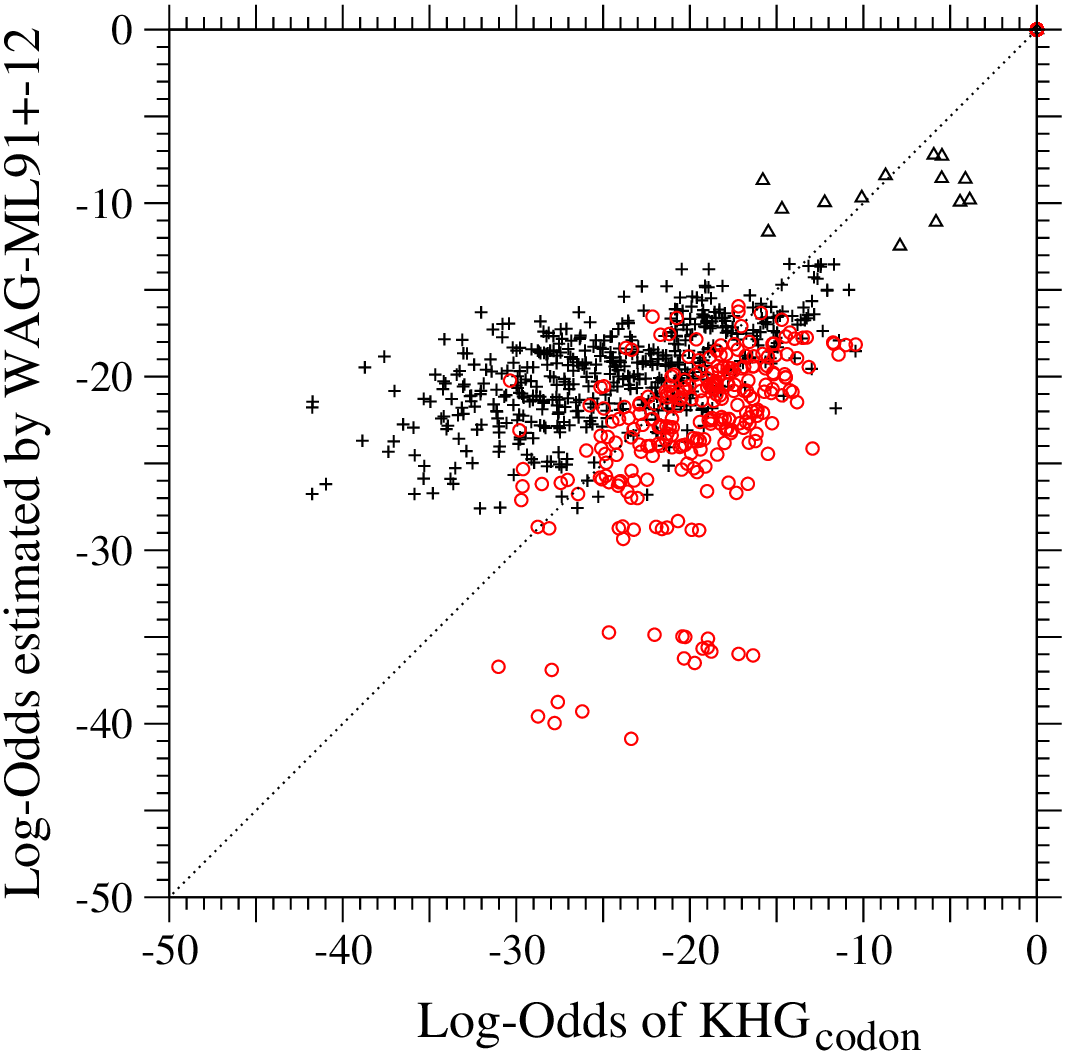}
}
\vspace*{1em}
\centerline{
\includegraphics*[width=80mm,angle=0]{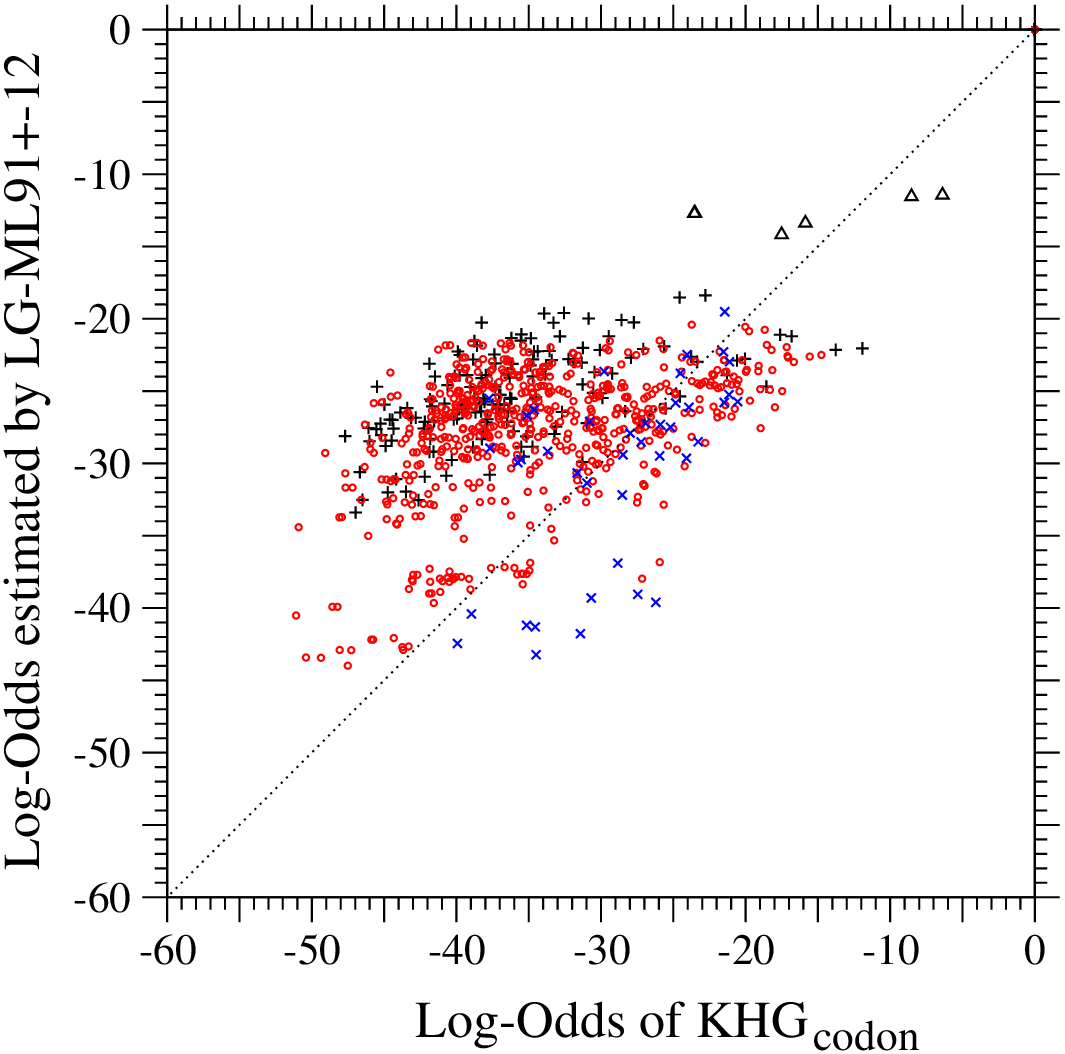}
\hspace*{5mm}
\hspace*{80mm}
}

\textbf{C} 
} 
\caption{
\label{fig: WAG-ML91+_log-odds_1pam_KHG}
\BF{
The WAG-ML91+-12 model fitted to the 1-PAM KHG codon substitution matrix.
} 
Each element 
log-$O(\langle S \rangle(\hat{\tau},\hat{\sigma}))_{\mu \nu}$
of the log-odds matrix 
corresponding to (A) single, (B) double, and (C) triple nucleotide changes 
in the WAG-ML91+-12 model fitted to the 1-PAM KHG codon substitution matrix
is plotted against the log-odds log-$O(S^{\script{KHG}}(\mbox{1 PAM}))_{\mu \nu}$
calculated from KHG.
Upper triangle, plus, circle, and cross marks show the log-odds values for
synonymous pairs and one-, two-, and three-step amino acid pairs,
respectively. 
The dotted line in each figure shows the line 
of equal values between the ordinate and the abscissa.
}
\end{figure}

} 

\FigurePanel{

\FigureInLegends{\newpage}

\begin{figure}[ht]
\FigureInLegends{
\textbf{A} \hspace*{47em} \textbf{B}

\centerline{
\includegraphics*[width=80mm,angle=0]{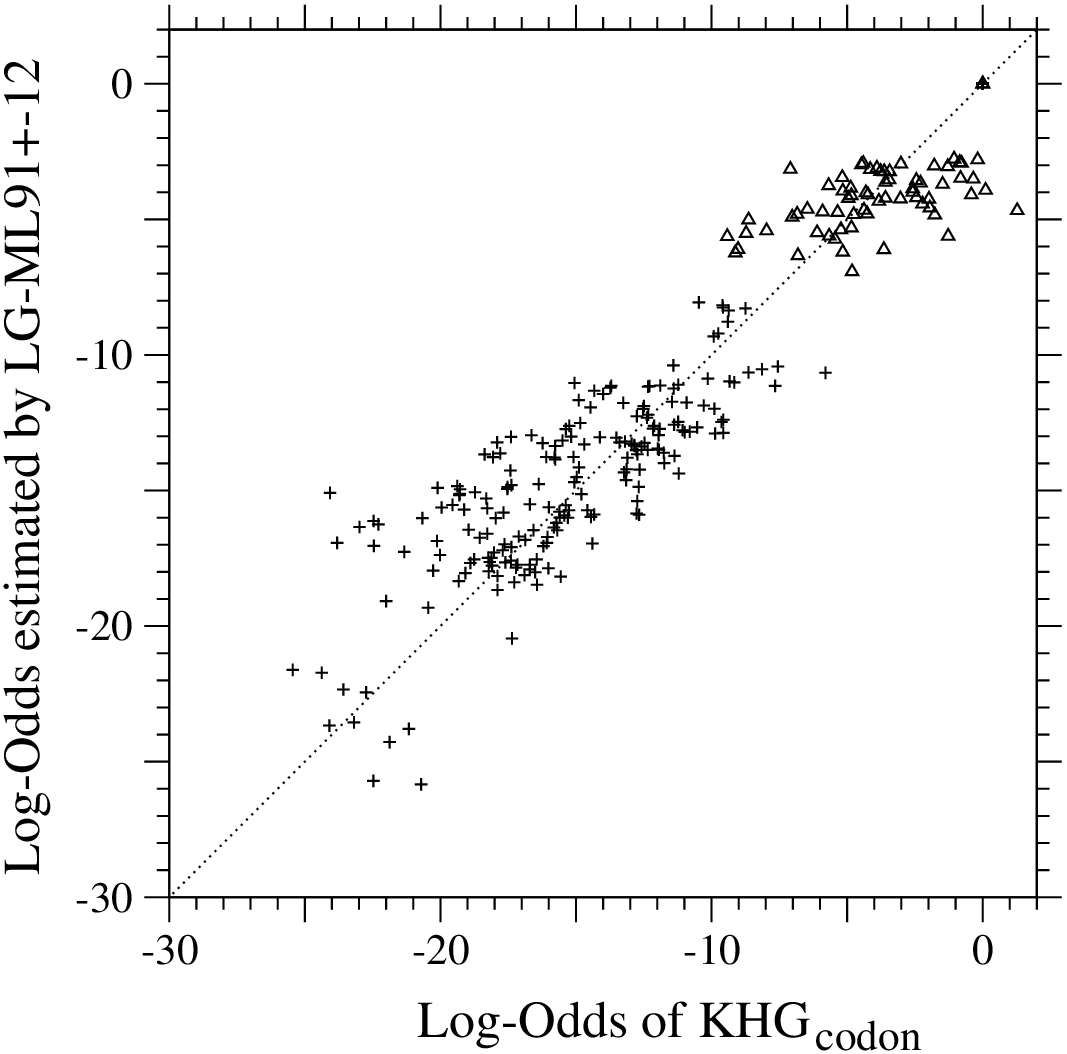}
\hspace*{5mm}
\includegraphics*[width=80mm,angle=0]{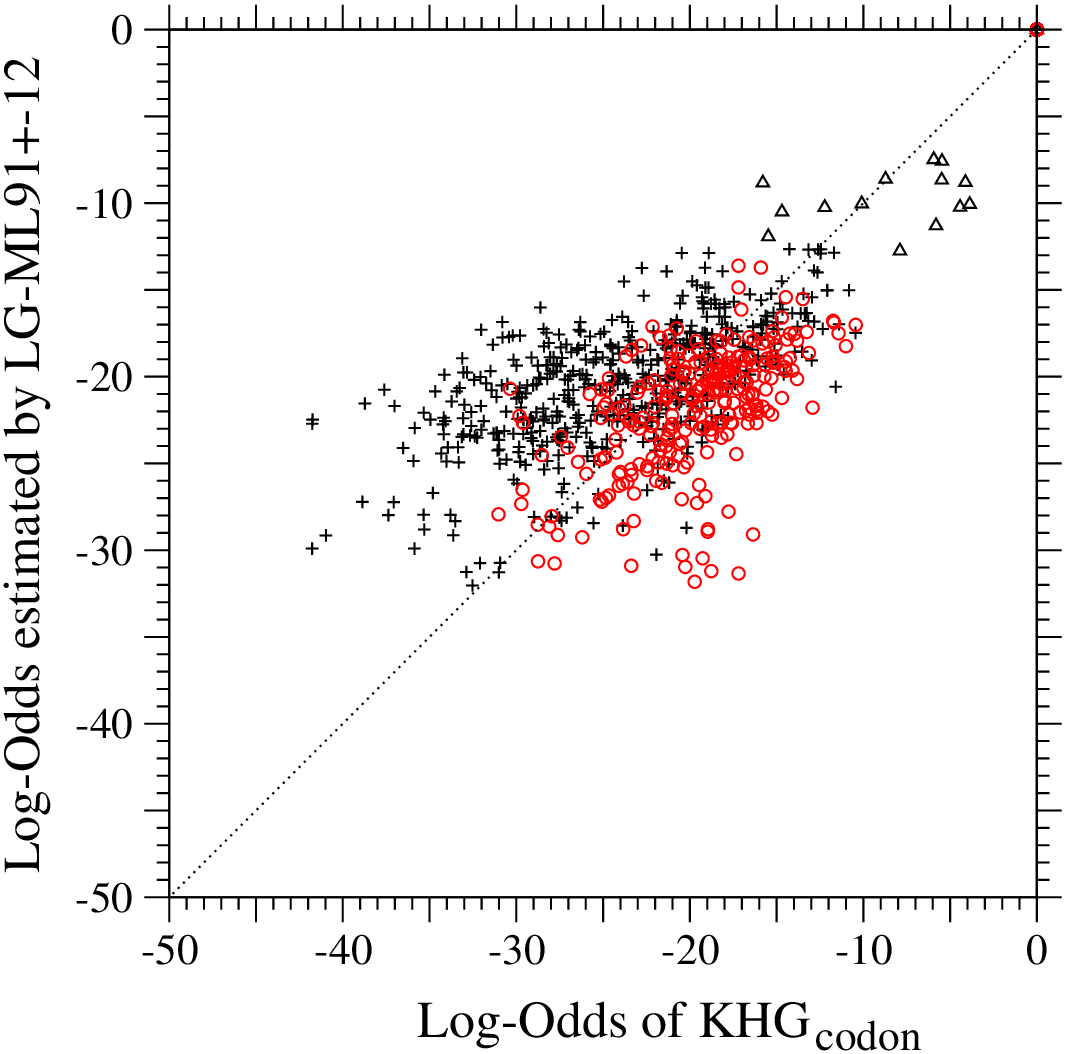}
}
\vspace*{1em}
\centerline{
\includegraphics*[width=80mm,angle=0]{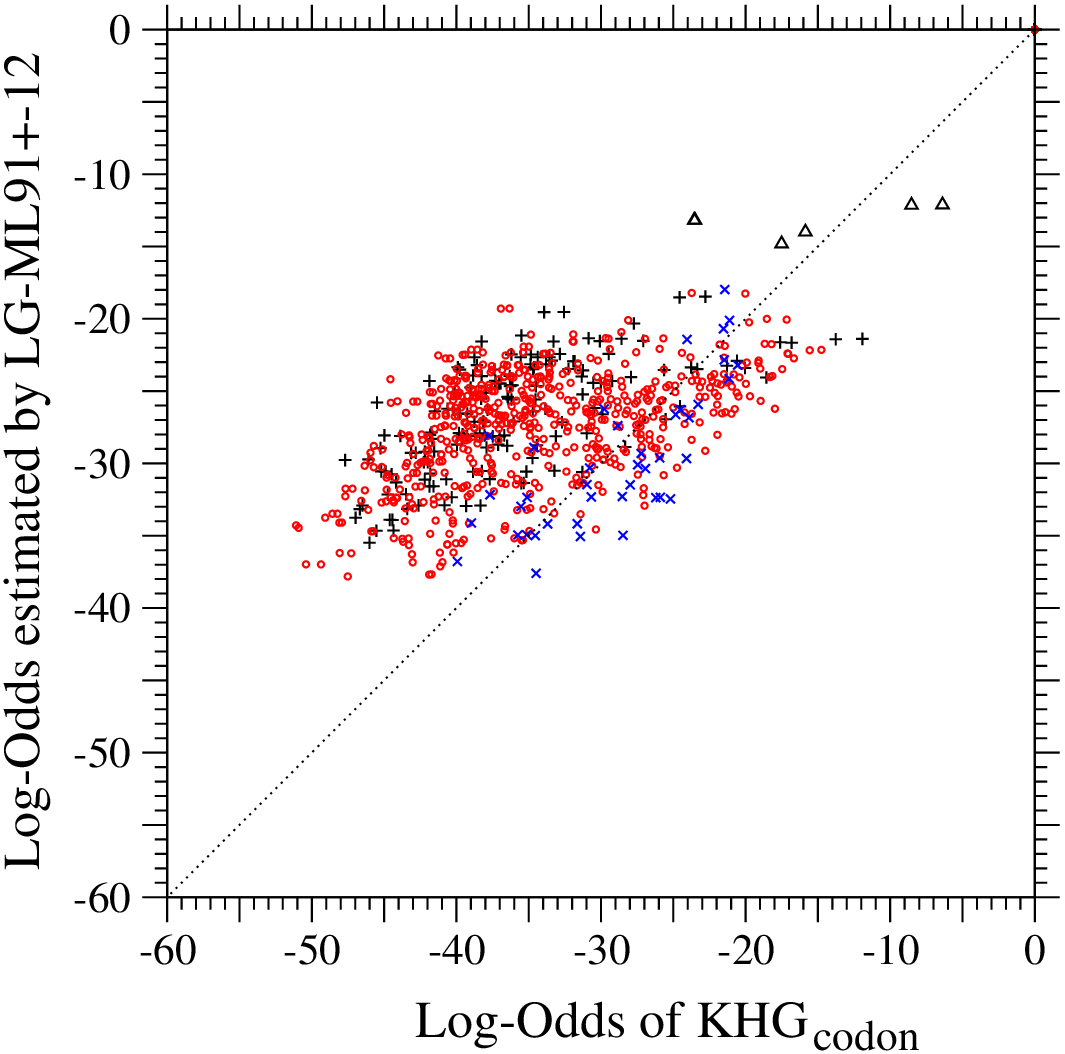}
\hspace*{5mm}
\hspace*{80mm}
}

\textbf{C} 
} 
\caption{
\label{fig: LG-ML91+_log-odds_1pam_KHG}
\BF{
The LG-ML91+-12 model fitted to the 1-PAM KHG codon substitution matrix.
} 
Each element 
log-$O(\langle S \rangle(\hat{\tau},\hat{\sigma}))_{\mu \nu}$
of the log-odds matrix 
corresponding to (A) single, (B) double, and (C) triple nucleotide changes 
in the LG-ML91+-12 model fitted to the 1-PAM KHG codon substitution matrix
is plotted against the log-odds log-$O(S^{\script{KHG}}(\mbox{1 PAM}))_{\mu \nu}$
calculated from KHG.
Upper triangle, plus, circle, and cross marks show the log-odds values for
synonymous pairs and one-, two-, and three-step amino acid pairs,
respectively. 
The dotted line in each figure shows the line 
of equal values between the ordinate and the abscissa.
}
\end{figure}

} 

} 

\SupFig{

\FigureInLegends{\newpage}

\FigureInLegends{\small}

\begin{figure*}[ht]
\FigureInLegends{
\centerline{
\textbf{A}
\includegraphics*[width=133mm,angle=0]{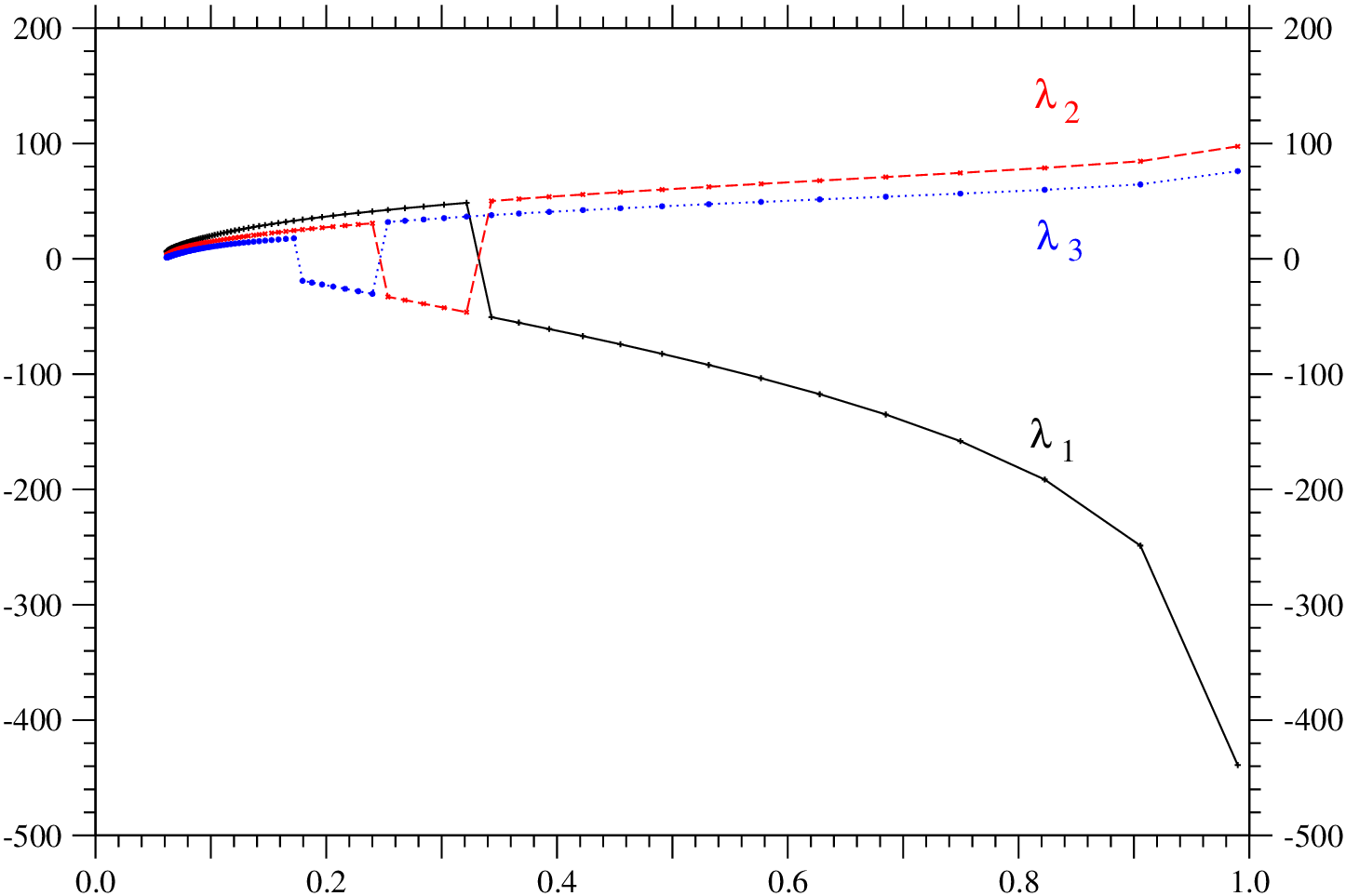}
}
\centerline{
\textbf{B}
\includegraphics*[width=130mm,angle=0]{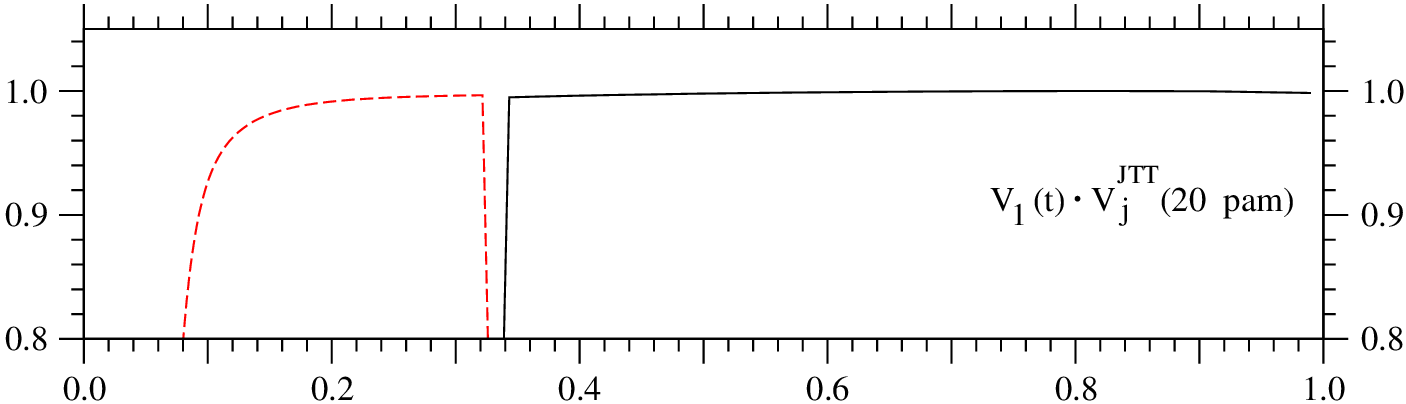}
}
\centerline{
\textbf{C}
\includegraphics*[width=130mm,angle=0]{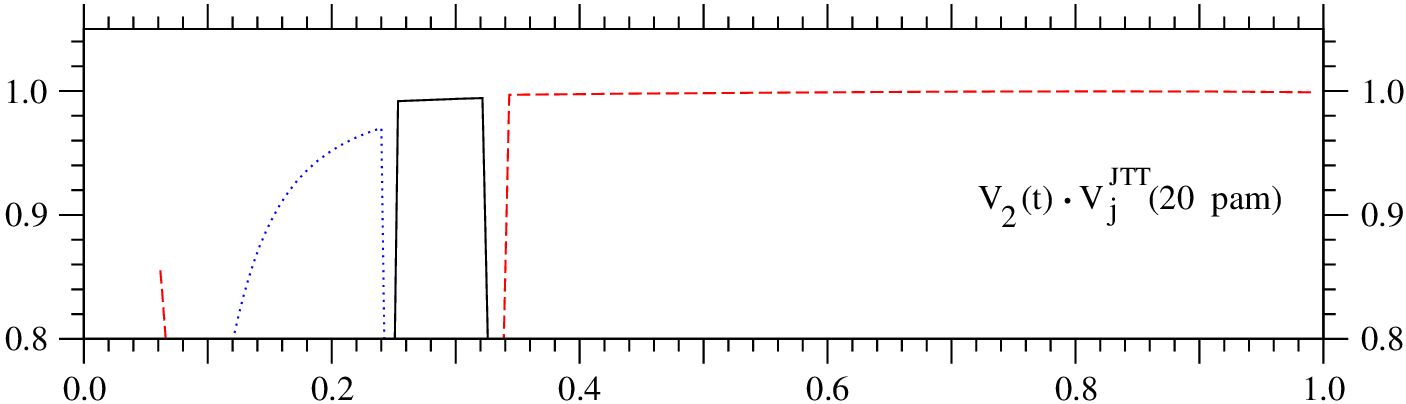}
}
\centerline{
\textbf{D}
\includegraphics*[width=130mm,angle=0]{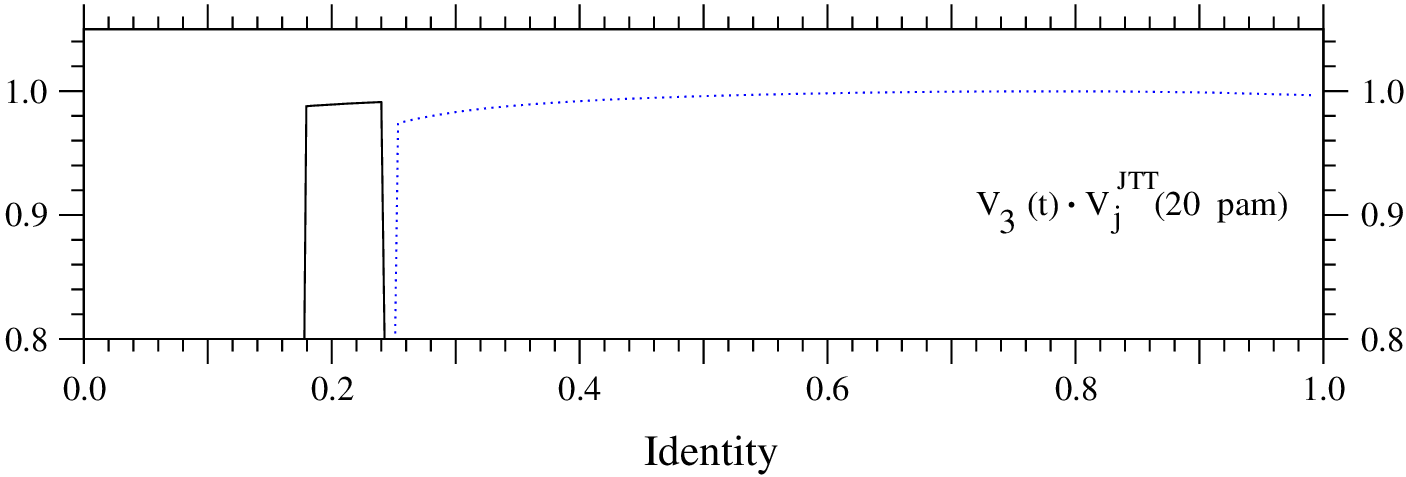}
}
} 
\caption{\label{fig: log-odds-ev_ML91+}
\label{fig: log-odds-ev_ML91}
\FigureInLegends{\small}
\BF{
Temporal changes of the eigenvalues and the eigenvectors
of the log-odds matrix 
log-$O(\langle S \rangle(t))$
calculated by the ML-91+ model fitted to JTT
as a function of sequence identity.
} 
In (A), 
the solid, the broken, and the dotted lines show 
the temporal changes of the first ($\lambda_1$),
the second ($\lambda_2$), and the third ($\lambda_3$) principal eigenvalues, 
respectively.
The inner products of the eigenvectors with the eigenvectors of the JTT 20-PAM log-odds matrix, 
$\VEC{V}_i(t) \cdot \VEC{V}_j^{\script{JTT}}(\mbox{20-PAM})$,
are shown 
in (B) for the first principal eigenvector ($i = 1$),
in (C) for the second principal eigenvector ($i = 2$), 
and in (D) for the third principal eigenvector ($i = 3$),
by solid lines for $j = 1$, by broken lines for $j = 2$, and 
by dotted lines for $j = 3$.
\FigureInLegends{\normalsize}
}
\end{figure*}
\FigureInLegends{\normalsize}

} 
\clearpage

\renewcommand{\TableInLegends}[1]{#1}
\newcommand{\TextTable}[1]{#1}
\newcommand{\SupTable}[1]{}

\renewcommand{\Red}[1]{#1}

\newcommand{\FullTable}[1]{#1}

\TextTable{ 

\begin{table}[!ht]
\caption{\label{tbl: model_names}
\BF{
Brief description of models.
} 
}
\vspace*{2em}

\begin{tabular}{p{3cm}|p{12cm}}
\hline
Model name	
		&	Description	\\
\hline
No-Constraints-$n$	& No amino acid dependences of selective constraints; $\beta = 0$.
			 The suffix $n$ means the number of ML parameters.
		\\
\hline
EI-$n$		
		& $\hat{w}_{ab}^{\script{estimate}} \equiv \Delta \hat{\varepsilon}^{\script{c}}_{ab} + \Delta \hat{\varepsilon}^{\script{v}}_{ab}$
		based on the Energy-Increment-based (EI) method, which is described in
		Supporting Information, Text S1, is used 
		to estimate $w_{ab}$ in 
		\Eq{\ref{eq: estimation_of_fitness}}.
		The suffix $n$ means the number of ML parameters.
		\\
\hline
Miyata-$n$	& The amino acid pair distance $d_{ab}$ estimated by 
		Miyata et al.\CITE{MMY:79} 
		is used as $w_{ab}^{\script{estimate}} = - d_{ab}$ 
		to estimate $w_{ab}$ in \Eq{\ref{eq: estimation_of_fitness}}.
		The suffix $n$ means the number of ML parameters.
		\\
\hline
Grantham-$n$	& The amino acid distance $d_{ab}$ estimated by Grantham\CITE{G:74}
		is used as $w_{ab}^{\script{estimate}} = - d_{ab}$ 
		to estimate $w_{ab}$ in \Eq{\ref{eq: estimation_of_fitness}}.
		The suffix $n$ means the number of ML parameters.
		\\
\hline
ML-$n$		& Selective constraints $\{ w_{ab} \}$ are estimated 
		by maximizing the likelihood of JTT\CITE{JTT:92}, 
	WAG\CITE{WG:01}, or LG\CITE{LG:08}, 
	and called $\{ w_{ab}^{\script{JTT/WAG/LG-ML}n} \}$.
	The suffix $n$ means the number of ML parameters.
	In the ML-87, multiple nucleotide changes are 
	disallowed, and $\{ w_{ab} \}$ for all 75 single-step amino acid pairs
	are estimated.
	In the ML-91 and the ML-94, multiple nucleotide changes are allowed,
	and $\{ w_{ab} \}$ for all 75 single-step amino acid pairs  and for 
	6 groups of multiple-step amino acid pairs are estimated.
	In the ML-91, equal codon usage is assumed.
	In the ML-200 for codon substitution matrices,
	$\{ w_{ab} \}$ for all 190 amino acid pairs are estimated.
		\\
\hline
ML-$n+$		
	&
	First, the ML-$n$ is used to estimate parameters, and then
	$\{ w_{ab} \}$ for all multiple-step amino acid pairs are estimated
	by maximizing the likelihood with fixing all other 
	parameters to the values estimated by the ML-$n$.
		\\
\hline
JTT-ML91-$n$,	
WAG-ML91-$n$,
LG-ML91-$n$
		& Selective constraints 
	$\{ w_{ab}^{\script{JTT/WAG/LG-ML91}} \}$ estimated 
		by maximizing the likelihood of 
	JTT/WAG/LG\CITE{JTT:92,WG:01,LG:08}
		in the ML-91 model are used 
		as $\{ w_{ab}^{\script{estimate}} \}$
                in \Eq{\ref{eq: estimation_of_fitness}}.
	The suffix $n$ means the number of ML parameters.
		\\
\hline
JTT-ML91+-$n$,	
WAG-ML91+-$n$,
LG-ML91+-$n$
		& Selective constraints 
	$\{ w_{ab}^{\script{JTT/WAG/LG-ML91+}} \}$ estimated 
		by maximizing the likelihood of 
	JTT/WAG/LG\CITE{JTT:92,WG:01,LG:08}
		in the ML-91+ model are used 
		as $\{ w_{ab}^{\script{estimate}} \}$
                in \Eq{\ref{eq: estimation_of_fitness}}.
	The suffix $n$ means the number of ML parameters.
	The JTT/WAG/ LG-ML91+-0 models correspond to the JTT/WAG/LG-F models, respectively.
		\\
\hline
KHG-ML200-$n$
		& Selective constraints $\{ w_{ab}^{\script{KHG-ML200}} \}$ estimated 
	by maximizing the likelihood of 
	the KHG codon substitution matrix\CITE{KHG:07} 
		in the ML-200 model are used 
		as $\{ w_{ab}^{\script{estimate}} \}$
                in \Eq{\ref{eq: estimation_of_fitness}}.
	The suffix $n$ means the number of ML parameters.
	The KHG-ML200-0 models correspond to the KHG-F model.
		\\
\hline
\end{tabular}

\end{table}

} 

\SupTable{

\TableInLegends{
\newpage
} 
\begin{table}[ht]
\caption{\label{tbl: optimizations_no-selection_JTT_WAG_CpRev_MtRevRmtx}
\label{tbl: optimizations_no-selection_JTT_CpRev_MtRevRmtx}
\label{tbl: optimizations_no-selection}
ML estimates 
of the present models without 
selective constraints on amino acids for the 1-PAM substitution matrices of 
JTT, WAG, cpREV, and mtREV.
}
\vspace*{2em}
\TableInLegends{

\footnotesize
\begin{tabular}{ll|rr|rr|rr|rr}
\hline
	&	& \multicolumn{2}{c|}{JTT}	&\multicolumn{2}{c|}{WAG}	& \multicolumn{2}{c|}{cpREV}	& \multicolumn{2}{c}{mtREV}	\\
		\cline{3-10}
	&	& \multicolumn{2}{c|}{No-Constraints- $^a$}	& \multicolumn{2}{c|}{No-Constraints- $^a$}	& \multicolumn{2}{c|}{No-Constraints- $^a$}	& \multicolumn{2}{c}{No-Constraints- $^a$} \\
id no.	&parameter
		&1	
							&10	
		&1					&10	
		&1	
							&10	
		&1	
							&10	\\
\hline
0&
$-\hat{w}_0$		&(0.0)	& (0.0) & (0.0)	& (0.0)	& (0.0) & (0.0)	&(0.0)	& (0.0)	 	\\
1&
$1/\hat{\beta}$	&($\infty$) &($\infty$) &($\infty$) &($\infty$) & ($\infty$) & ($\infty$) &($\infty$) & ($\infty$) \\
2&
$\hat{m}_{[tc][ag]}$	&($\rightarrow 0$)	
											&$\rightarrow 0$	
			&($\rightarrow 0$)	&0.279	
			&($\rightarrow 0$)	
											&0.0455	
			&($\rightarrow 0$)	
											&0.0405	\\
3&
$\hat{m}_{tc|ag}/\hat{m}_{[tc][ag]}$ 	&2.16	
											&2.20	
					&1.61	&1.54	

					&2.17	
											&2.62	
					&2.32	
											&3.24	\\
4&
$\hat{m}_{ag}/\hat{m}_{tc|ag}$		&(1.0)	
											&1.28	
					&(1.0)	&1.36	
					&(1.0)	
											&1.50	
					&(1.0)	
											&1.47	\\
5&
$\hat{m}_{ta}/\hat{m}_{[tc][ag]}$	&(1.0)	
											&0.629	
					&(1.0)	&0.687	
					&(1.0)	
											&0.480	
					&(1.0)	
											&0.595	\\
6&
$\hat{m}_{tg}/\hat{m}_{[tc][ag]}$	&(1.0)	
											&0.708	
					&(1.0)	&0.622	
					&(1.0)	
										 	&0.775	
					&(1.0)	
										 	&0.373	\\
7&
$\hat{m}_{ca}/\hat{m}_{[tc][ag]}$	&(1.0)	
											&1.28	
					&(1.0)	&1.45	
					&(1.0)	
											&1.64	
					&(1.0)	
											&1.96	\\
8&
$\hat{f}^{\script{mut}}_{t+a}$		&(0.5)	
											&0.495	
					&(0.5)	&0.401	
					&(0.5)	
										 	&0.279	
					&(0.5)	
										 	&0.226	\\
9&
$\hat{f}^{\script{mut}}_t/\hat{f}^{\script{mut}}_{t+a}$	
					&(0.5)	
											&0.486	
					&(0.5)	&0.503	
					&(0.5)	
											&0.563	
					&(0.5)	
											&0.583	\\
10&
$\hat{f}^{\script{mut}}_c/\hat{f}^{\script{mut}}_{c+g}$	
					&(0.5)	
											&0.335	
					&(0.5)	&0.354	
					&(0.5)	
										 	&0.306	
					&(0.5)	
										 	&0.223	\\
14&
$\hat{\sigma}$				&($\rightarrow 0$)	
											&1.76	
					&($\rightarrow 0$)	&1.58	
					&($\rightarrow 0$)	
										 	&2.96	
					&($\rightarrow 0$)	
										 	&2.46	\\
\hline
&
$\hat{\tau} \hat{\sigma}$		&0.0137	
											&0.0228	
					&0.0136	&0.0206	
					&0.0139	
											&0.0296	
					&0.0149	
											&0.0296	\\
&
\#parameters				&21	
											&30	
					&21	&30	
					&21	
										 	&30	
					&21	
										 	&30	\\
&
$\hat{I}_{KL}(\hat{\VEC{\theta}}) \times 10^8 \ ^b$
					&$729533$	
																	&207260	
					&$1156393$	&233841	
					&$1014962$	
																	&249448	
					&$945289$	
																	&305500	\\

&
$\Delta \mbox{AIC} \ ^c$		&86428.1	&24595.5	
					&37917.6	&7719.1	
					&3478.0		&904.5	
					&2644.1		&901.0	\\
\FullTable{
\hline
&
Ratio of substitution rates 			&	&	&	&	&	&	&	&	\\
& \ \  per codon					&	&	&	&	&	&	&	&	\\
&
\hspace*{1em} the total base/codon 		&1.0	
												&1.30	
						&1.0	&1.47	
						&1.0	
												&1.40	
						&1.0	
												&1.35	\\
&
\hspace*{1em} transition/transversion           &1.13	
												&1.00	
						&0.848	&0.752	
						&1.11	
												&1.02	
						&1.24	
												&1.10	\\
&
\hspace*{1em} nonsynonymous/synonymous$^d$          &2.75	
												&4.15	
						&2.84	&5.77	
						&2.60	
												&4.91	
						&2.09	
												&3.30	\\
\hline
&
Ratio of substitution rates 	 		&	&	&	&	&	&	&	&	\\
& \ \  per codon for $\sigma \rightarrow 0$	&	&	&	&	&	&	&	&	\\
&
\hspace*{1em} the total base/codon    		&1.0	
												&1.0	
						&1.0	&1.21	
						&1.0	
												&1.04	
						&1.0	
												&1.02	\\
&
\hspace*{1em} transition/transversion           &1.13	
												&1.20	
						&0.848	&0.853	
						&1.11	
												&1.43	
						&1.24	
												&1.45	\\
&
\hspace*{1em} nonsynonymous/synonymous$^d$          &2.75	
												&2.83	
						&2.84	&4.26	
						&2.60	
												&3.19	
						&2.09	
												&2.08	\\
} 
\hline
\end{tabular}

\vspace*{1em}

\noindent
$^a$ In all models,
equal codon usage 
($\hat{f}^{\script{usage}}_{t} = \hat{f}^{\script{usage}}_{a} = \hat{f}^{\script{usage}}_{c} = \hat{f}^{\script{usage}}_{g} = 0.25$)
is assumed. 
If the value of a parameter is parenthesized, the parameter is not variable but fixed to the value specified.

\noindent
$^b$ $\hat{I}_{KL}(\hat{\VEC{\theta}}) = -(\ell(\hat{\VEC{\theta}})/N + 2.98607330)$ for JTT, 
						$- (\ell(\hat{\VEC{\theta}})/N + 2.97444860)$ for WAG,
						$- (\ell(\hat{\VEC{\theta}})/N + 2.95801048)$ for cpREV,
					and  $- (\ell(\hat{\VEC{\theta}})/N + 2.85313622)$ for mtREV; see text for details.

\noindent
$^c$ $\Delta \mbox{AIC} \equiv 2 N \hat{I}_{KL}(\hat{\VEC{\theta}}) + 2 \times $ \#parameters 
with $N \simeq 5919000$ for JTT, 
$N \approx 1637663$ for WAG,
$N \approx 169269$ for cpREV 
and $N \approx 137637$ for mtREV; see text for details.

$^d$  Note that these ratios are not the ratios of the rates per site but per codon; see text for details.

\normalsize
} 
\end{table}

} 

\TextTable{ 

\begin{table}[ht]
\caption{\label{tbl: optimizations_EI_AIC}
\label{tbl: optimizations_AIC}
\BF{
$\Delta$AIC values
of the present models 
without and with
the selective constraints on amino acids, 
} 
which are based on 
mean energy increments due to an amino acid substitution (EI), 
the Miyata's and the Grantham's physico-chemical distances,
for the 1-PAM amino acid substitution matrices of 
JTT, WAG, cpREV, and mtREV.
}
\vspace*{2em}

\footnotesize

\begin{tabular}{l|l|r|r|r|r}
\hline
	&	& \multicolumn{4}{c}{$\Delta \mbox{AIC} \ ^a$}  \\
	\hline
Model &\#parameters	& JTT    &WAG    &cpREV  &mtREV
							\\
	&(id no. $^b$)	&	&	& 	&	\\
\hline
No-Constraints-	&		&		&	&	&	\\
\ \ 1	&21($\beta=0$, 3)	&86428.1	&37917.6 &3478.0 &2644.1 \\
\ \ 10	&30($\beta=0$, 2-10,14)	&24595.6	&7719.1	&904.5	&901.0	\\
\ \ 13	&33($\beta=0$, 2-14)	&22913.6	&7141.5	& 874.9	& 798.8	\\
\hline
EI-	&		&		&	&	&	\\
\ \ 2	&22(1,3)	&77337.9	&35058.8&3186.0	&2396.6	\\
\ \ 2G	&22(1,14)	&24197.7	&5571.6	&974.0	&1066.8	\\
\ \ 3	&23(1,3,14)	&16463.7	&4995.0	&761.5	&776.4	\\
\ \ 4	&24(1-3,14)	&15808.7	&4443.6	&743.0	&753.9	\\
\ \ 8	&28(1-7,14)	&15715.0	&4327.8	&722.0	&728.2	\\
\ \ 7	&27(1-3,8-10,14)&15081.0	&4312.6	&650.7	&688.7	\\
\ \ 10	&30(1,3-10,14)	&15435.7	&4801.8	&670.7	&702.8	\\
\ \ 10M	&30(1-10)	&15270.7	&4250.4	&645.3	&674.3	\\
\ \ 11	&31(1-10,14)	&14999.0	&4202.5	&636.0	&674.3	\\
\ \ 10MU&30(1-3,8-14)	&13464.3	&3959.7	&578.9	&662.4	\\
\ \ 12	&32(1,3-13)	&72316.3	&33908.4&2939.7	&2215.0	\\
\ \ 13	&33(1,3-14)	&13819.7	&4554.2	&623.6 	&655.5	\\
\ \ 13M	&33(1-13)	&13436.2	&3822.4	&551.1	&623.3	\\
\ \ 14	&34(1-14)	&13151.9	&3748.0	&541.9	&614.8	\\
\hline
Miyata-	&		&		&	&	&	\\
\ \ 4	&24(1-3,14)	&16090.1	&4938.1	&750.3	&783.0	\\
\ \ 7	&27(1-3,8-10,14)&15767.2	&4715.4	&654.5	&701.6	\\
\ \ 10	&30(1,3-10,14)  &16446.1	&5124.9	&679.2	&708.5	\\
\ \ 11	&31(1-10,14)    &15536.8	&4429.5	&628.4	&658.4	\\
\ \ 13	&33(1,3-14)     &15058.2	&4943.1	&656.5	&682.3	\\
\ \ 14	&34(1-14)       &14338.5	&4254.0	&603.7	&613.6	\\
\hline
Grantham-	&		&		&	&	&	\\
\ \ 4	&24(1-3,14)	&20505.1	&5953.7	&916.4	&887.1	\\
\ \ 7	&27(1-3,8-10,14)&18898.2	&5814.0	&840.6	&832.9	\\
\ \ 10	&30(1,3-10,14)  &18744.5	&5749.0	&805.4	&799.8	\\
\ \ 11	&31(1-10,14)    &18680.9	&5579.7	&803.2	&796.5	\\
\ \ 13	&33(1,3-14)     &16784.9	&5512.9	&765.0	&741.0	\\
\ \ 14	&34(1-14)       &16729.7	&5477.1	&755.0	&739.5	\\
\hline
\end{tabular}

\vspace*{1em}

\noindent
$^a$ $\Delta \mbox{AIC} \equiv 2 N \hat{I}_{KL}(\hat{\VEC{\theta}}) + 2 \times $ \#parameters
with 
$N \simeq 5919000$ for JTT,
$N \approx 1637663$ for WAG,
$N \approx 169269$ for cpREV,
and $N \approx 137637$ for mtREV; see text for details.

$^b$
ML parameters in each model are specified by
the parameter id numbers in the parenthesis,
and other parameters are fixed at
$\mbox{id}_{0} = 0$, $\mbox{id}_{1} = \infty$, $\mbox{id}_{2} \rightarrow 0$,
$\mbox{id}_{3-7} = 1.0$, $\mbox{id}_{8-13} = 0.5$, and $\mbox{id}_{14} \rightarrow 0$.
Each id number corresponds to the parameter id number listed in 
\Table{\ref{tbl: optimizations_wij_JTT_WAG_LG_KHG}}.

\end{table}

} 

\SupTable{

\TableInLegends{
\newpage
} 
\begin{table}[ht]
\caption{\label{tbl: optimizations_EI_JTT_WAG_CpRev_MtRevRmtx}
\label{tbl: optimizations_EI-14_JTT}
\label{tbl: optimizations_EI-14_WAG}
\label{tbl: optimizations_EI-14_CpRev}
\label{tbl: optimizations_EI-14_MtRevRmtx}
\label{tbl: optimizations_EI-14_WAG_CpRev_MtRevRmtx}
\label{tbl: optimizations_EI_WAG_CpRev_MtRevRmtx}
ML estimates 
of the present models with 
the selective constraints based on 
mean energy increments due to an amino acid substitution (EI) 
for the 1-PAM substitution matrices of JTT, WAG, cpREV, and mtREV.
}
\vspace*{2em}
\TableInLegends{

\footnotesize

\begin{tabular}{l|rr|rr|rr|rr}
\hline
				&\multicolumn{2}{c|}{JTT}	&\multicolumn{2}{c|}{WAG}	&\multicolumn{2}{c|}{cpREV} 	&\multicolumn{2}{c}{mtREV}  \\
				\cline{2-9}
				&EI-10 $^a$	&EI-11 $^a$	&EI-10 $^a$	&EI-11 $^a$	&EI-10 $^a$	&EI-11 $^a$	&EI-10 $^a$	&EI-11 $^a$	\\
\hline
$-\hat{w}_0$				& (0.0)	& (0.0)	& (0.0)	& (0.0)	& (0.0)	& (0.0)	& (0.0)	& (0.0)	\\
$1/\hat{\beta}$			& 2.50	& 2.60	&1.78	&2.14	&2.15	&2.26	&2.14	&2.29	\\			
$\hat{m}_{[tc][ag]}$		&($\rightarrow 0$) &0.308	&($\rightarrow 0$) &0.916	&($\rightarrow 0$) &0.684	&($\rightarrow 0$) &0.737	\\
$\hat{m}_{tc|ag}/\hat{m}_{[tc][ag]}$
				& 2.51	& 2.22	&1.82	&1.58	&2.82	&2.24	&4.21	&3.06	\\
$\hat{m}_{ag}/\hat{m}_{tc|ag}$	& 1.01	& 1.01	&1.13	&1.10	&1.19	&1.14	&1.05	&1.01	\\
$\hat{m}_{ta}/\hat{m}_{[tc][ag]}$ &1.02	&1.07	&1.26	&1.22	&0.992	&1.14	&1.48	&1.44	\\
$\hat{m}_{tg}/\hat{m}_{[tc][ag]}$ &1.06	& 1.09	&0.985	&1.01	&1.34	&1.23	&0.792	&0.797	\\
$\hat{m}_{ca}/\hat{m}_{[tc][ag]}$ &0.937 & 0.891 &1.04	&0.949	&0.974	&0.925	&1.17	&1.08	\\
$\hat{f}^{\script{mut}}_{t+a}$	& 0.582 & 0.565 &0.516	&0.486	&0.376	&0.405	&0.359	&0.403	\\
$\hat{f}^{\script{mut}}_t/\hat{f}^{\script{mut}}_{t+a}$	
				& 0.522 & 0.525	&0.603	&0.575	&0.647	&0.642	&0.671	&0.646	\\
$\hat{f}^{\script{mut}}_c/\hat{f}^{\script{mut}}_{c+g}$	
				& 0.432	& 0.450	&0.495	&0.511	&0.450	&0.462	&0.388	&0.404	\\
$\hat{\sigma}$ 			&3.20	&0.918	&11.7	&0.998	&7.26	&0.969	&5.25	&0.339	\\
\hline
$\hat{\tau} \hat{\sigma}$	&0.0358	&0.0217	&0.0709	&0.0204	&0.0558	&0.0211	&0.0531	&0.0185	\\
\#parameters			&30	&31	& 30 	& 31	&30	&31	&30	&31	\\
$\hat{I}_{KL}(\hat{\VEC{\theta}}) \times 10^8 \ ^b$ 
				&$129885$	&$126178$	&144772	&126415			&180379	&169548			&233525	&222441	\\
$\Delta \mbox{AIC} \ ^c$	&15435.7	&14999.0	&4801.8	&4202.5			&670.7	&636.0			&702.8	&674.3	\\
\FullTable{
\hline
Ratio of substitution rates     &	&	&	&			&	&			&	&       \\
\ \ per codon 			&	&	&	&			&	&			&	&       \\
\hspace*{1em} the total base/codon
				&1.36	&1.35	&1.53	&1.54	&1.45	&1.48	&1.38	&1.44	\\
\hspace*{1em} transition/transversion           
				&1.09	&1.11	&0.803	&0.834	&1.08	&1.13	&1.34	&1.41	\\
\hspace*{1em} nonsynonymous/synonymous$^d$          
				&2.09	&2.13	&2.48	&2.82	&2.45	&2.65	&1.75	&1.92	\\
\hline
Ratio of substitution rates per codon & &	&	&	&	&			&	&\\
\ \ for $\sigma \rightarrow 0$			&	&	&	&			&	&			&	&       \\
\hspace*{1em} total base/codon  &1,0	&1.18	&1.0	&1.38	&1.0	&1.31	&1.0	&1.37	\\
\hspace*{1em} transition/transversion           
				&1.49	&1.28	&1.25	&0.944	&1.93	&1.36	&2.35	&1.56	\\
\hspace*{1em} nonsynonymous/synonymous$^d$          
				&1.12	&1.59	&0.945	&2.13	&1.15	&1.99	&0.767	&1.64	\\
\hline
Ratio of substitution rates per codon 	 	& 	&	&	&	&	&			&	&\\
\ \ for $w_{ab}=0$ and $\sigma \rightarrow 0$	&	&	&	&			&	&			&	&       \\
\hspace*{1em} total base/codon  &1.0	&1.28	&1.0	&1.59	&1.0	&1.48	&1.0	&1.59	\\
\hspace*{1em} transition/transversion           
				&1.31	&1.15	&0.983	&0.830	&1.51	&1.50	&2.15	&1.57	\\
\hspace*{1em} nonsynonymous/synonymous$^d$          
				&2.57	&3.83	&2.82	&6.53	&2.74	&1.16	&1.84	&4.51	\\
} 
\hline
\end{tabular}

\vspace*{1em}

\noindent
$^a$ In all models, 
equal codon usage (
$\hat{f}^{\script{usage}}_{t} = \hat{f}^{\script{usage}}_{a} = \hat{f}^{\script{usage}}_{c} = \hat{f}^{\script{usage}}_{g} = 0.25$
) is assumed.            
If the value of a parameter is parenthesized, the parameter is not variable but fixed to the value specified.

\noindent
$^b$ $\hat{I}_{KL}(\hat{\VEC{\theta}}) = $
				$-(\ell(\hat{\VEC{\theta}})/N + 2.98607330)$ for JTT, 
				$- (\ell(\hat{\VEC{\theta}})/N + 2.97444860)$ for WAG,
				$- (\ell(\hat{\VEC{\theta}})/N + 2.95801048)$ for cpREV,
				and  $- (\ell(\hat{\VEC{\theta}})/N + 2.85313622)$ for mtREV; see text for details.

\noindent
$^c$ $\Delta \mbox{AIC} \equiv 2 N \hat{I}_{KL}(\hat{\VEC{\theta}}) + 2 \times $ \#parameters
with 
$N \simeq 5919000$ for JTT,
$N \approx 1637663$ for WAG,
$N \approx 169269$ for cpREV,
and $N \approx 137637$ for mtREV; see text for details.

$^d$  Note that these ratios are not the ratios of the rates per site but per codon; see text for details.

} 
\end{table}

\TableInLegends{
\newpage
} 
\begin{table}[ht]
\caption{\label{tbl: optimizations_selection_JTT_WAG}
\label{tbl: optimizations_selection_JTT}
\label{tbl: optimizations_selection_WAG}
ML estimates 
of the present models with 
the selective constraints based on the Grantham's and the Miyata's amino acid distances 
for the 1-PAM substitution matrices of JTT and WAG.
}
\vspace*{2em}
\TableInLegends{

\footnotesize
\begin{tabular}{l|rr|rr|rr|rr}
\hline
		& \multicolumn{4}{c|}{JTT}	&\multicolumn{4}{c}{WAG}	\\
		\cline{2-9}
		& \multicolumn{2}{c|}{Grantham- $^a$} & \multicolumn{2}{c|}{Miyata- $^a$}
		& \multicolumn{2}{c|}{Grantham- $^a$} & \multicolumn{2}{c}{Miyata- $^a$}	\\
					&10 	
							&11	&10 	& 11  		
										&10	&11	&10	&11	\\
\hline
$-\hat{w}_0$					& (0.0)	
							&(0.0)	& (0.0)	& (0.0)	
										&(0.0)	&(0.0)	&(0.0)	&(0.0) 	\\
$1/\hat{\beta}$				& 82.0	
							&81.9	& 1.71	& 1.82	
										&58.9	&65.1	&1.28	&1.59 	\\
$\hat{m}_{[tc][ag]}$			&($\rightarrow 0$) 
							&0.0392	& ($\rightarrow 0$) & 0.617 
										&($\rightarrow 0$)	&0.353	&($\rightarrow 0$)	&1.33 	\\
$\hat{m}_{tc|ag}/\hat{m}_{[tc][ag]}$	& 2.12	
							&2.09	& 2.32	& 1.92	
										&1.49	&1.44	&1.64	&1.40 	\\
$\hat{m}_{ag}/\hat{m}_{tc|ag}$		& 1.08	
							&1.08	& 1.05	& 1.05	
										&1.18	&1.17	&1.15	&1.11 	\\
$\hat{m}_{ta}/\hat{m}_{[tc][ag]}$	& 0.864	
							&0.863	& 0.925	& 0.983	
										&0.987	&0.938	&1.02	&1.02 	\\
$\hat{m}_{tg}/\hat{m}_{[tc][ag]}$	& 0.961	
							&0.983	& 0.922	& 0.985	
										&0.816	&0.907	&0.813	&0.912 	\\
$\hat{m}_{ca}/\hat{m}_{[tc][ag]}$	& 1.16	
							&1.16	& 1.26	& 1.12	
										&1.39	&1.32	&1.55	&1.23 	\\
$\hat{f}^{\script{mut}}_{t+a}$		& 0.582	
							&0.581	&0.574	& 0.543	
										&0.528	&0.517	&0.499	&0.466 	\\
$\hat{f}^{\script{mut}}_t/\hat{f}^{\script{mut}}_{t+a}$	
					& 0.512	
							&0.513	&0.513	& 0.505	
										&0.573	&0.562	&0.575	&0.531 	\\
$\hat{f}^{\script{mut}}_c/\hat{f}^{\script{mut}}_{c+g}$	
					& 0.384	
							&0.385	&0.448	& 0.479	
										&0.412	&0.420	&0.513	&0.541 	\\
$\hat{\sigma}$				&2.80	
							&2.37	&2.98	& $0.00938$	
										&9.00	&2.97	&9.87	& $0.00118$ \\
\hline
$\hat{\tau} \hat{\sigma}$		&0.0330	
							&0.0306 & 0.0342 &0.0147	
										&0.0596	&0.0317	&0.0632	&0.0135 	\\
\#parameters				&30	
							&31	& 30	& 31	
										&30	& 31	& 30	&31	\\
$\hat{I}_{KL}(\hat{\VEC{\theta}}) \times 10^8 \ ^b$	
					&$157835$	
								&157281 &$138419$	&$130721$	
													&173694	&168463	&154639	&133347 \\
$\Delta \mbox{AIC} \ ^c$		& 18744.5	
								&18680.9	& 16446.1	& 15536.8	
															&5749.0	&5579.7	&5124.9	&4429.5 \\ 
\FullTable{
\hline
Ratio of substitution rates per codon          	&	
								&	&	&	
											&	&	&	&	\\
\hspace*{1em} the total base/the total codon	&1.35	&1.35	&1.35	&1.34	&1.51	&1.50	&1.51	&1.53	\\
\hspace*{1em} transition/transversion		&1.04	&1.04	&1.07	&1.10	&0.768	&0.779	& 0.791	& 0.812	\\
\hspace*{1em} nonsynonymous/synonymous$^d$	&2.21	&2.20	&2.14	&2.18	&2.54	&2.65	&2.53	&2.93	\\
\hline
Ratio of substitution rates per codon 		&	
								&	&	&      
											&	&	&	&	\\
\ \  for $\sigma \rightarrow 0$			&	&	&	&	&	&	&	&	\\
\hspace*{1em} the total base/the total codon	&1.0	&1.02	&1.0	&1.33	&1.0	&1.16	&1.0	&1.53	\\
\hspace*{1em} transition/transversion		&1.33	&1.31	&1.42	&1.10	&1.06	&0.951	&1.17	&0.813	\\
\hspace*{1em} nonsynonymous/synonymous$^d$	&1.22	&1.28	&1.17	&2.17	&1.04	&1.52	&1.02	&2.93	\\
Ratio of substitution rates per codon 		&	
								&	&	&       
											&	&	&	&	\\
\ \  for $w_{ab}=0$ and $\sigma \rightarrow 0$	&	&	&	&	&	&	&	&	\\
\hspace*{1em} the total base/the total codon	&1.0	&1.04	&1.0	&1.48	&1.0	&1.26	&1.0	&1.74	\\
\hspace*{1em} transition/transversion		&1.12	&1.10	&1.21	&0.990	&0.803	&0.771	&0.881	&0.736	\\
\hspace*{1em} nonsynonymous/synonymous$^d$	&2.67	&2.81	&2.63	&5.24	&2.97	&4.20	&2.92	&8.49	\\
} 
\hline
\end{tabular}

\vspace*{1em}

\noindent
$^a$ In all models, 
equal codon usage
($\hat{f}^{\script{usage}}_{t} = \hat{f}^{\script{usage}}_{a} = \hat{f}^{\script{usage}}_{c} = \hat{f}^{\script{usage}}_{g} = 0.25$)
is assumed.
If the value of a parameter is parenthesized, the parameter is not variable but fixed to the value specified.

\noindent
$^b$ $\hat{I}_{KL}(\hat{\VEC{\theta}}) = -(\ell(\hat{\VEC{\theta}})/N + 2.98607330)$ 
for JTT, and 
$- (\ell(\hat{\VEC{\theta}})/N + 2.97444860)$ for WAG; see text for details.

\noindent
$^c$ $\Delta \mbox{AIC} \equiv 2 N \hat{I}_{KL}(\hat{\VEC{\theta}}) + 2 \times $ \#parameters 
with $N = 5919000$
for JTT, and 
$N \approx 1637663$ for WAG; see text for details.

$^d$  Note that these ratios are not the ratios of the rates per site but per codon; see text for details.

\normalsize
} 
\end{table}

} 

\TextTable{ 

\newpage
\begin{table}[ht]
\caption{\label{tbl: optimizations_wij_JTT_WAG_LG_KHG}
\label{tbl: optimizations_wij_JTT_WAG}
\label{tbl: optimizations_wij_JTT_WAG_LG}
\BF{
ML estimates and $\Delta$AIC values
of the present models
for the 1-PAM amino acid substitution matrices of JTT, WAG, and LG, 
and the 1-PAM codon substitution matrix of KHG.
} 
}
\vspace*{2em}

\footnotesize

\begin{tabular}{ll|rrr|rrr|rr|r}
\hline
	&
						& \multicolumn{3}{c|}{JTT} 
						& \multicolumn{3}{c|}{WAG}
						& \multicolumn{2}{c|}{LG}
						& \multicolumn{1}{c}{KHG}	\\
		&	&	&	&	&	&	&	&	&	&(codon)	\\
\hline
id	& parameter
						& ML--87$^a$ 
							& ML--91$^a$ & ML--94 
						& ML--87$^a$ 
							& ML--91$^a$ & ML--94 
							& ML--91$^a$ & ML--94 
							& ML--200	\\
no.		&	&	&	&	&	&	&	&	&	&		\\
\hline
0& $-\hat{w}_0$				& N/A 	
							& N/A	& N/A
						& N/A
							& N/A	& N/A
							& N/A	& N/A
							& N/A				\\
1& $1/\hat{\beta}$				& N/A   	
							& N/A  	& N/A  
						& N/A  
							& N/A  	& N/A  
							& N/A  	& N/A  
							& N/A  				\\
2&
$\hat{m}_{[tc][ag]}$			
						&($\rightarrow 0$)	
									&0.637	& 0.662 
						&($\rightarrow 0$)
									&1.28	& 1.29	
									&1.08	& 1.19	
									&0.939	\\
3&
$\hat{m}_{tc|ag}/\hat{m}_{[tc][ag]}$	
						&0.0919	
							&1.57	&1.59	
						&0.746	
							&1.70	& 1.69 
							&1.85	& 1.81 
							&0.843	\\
4&
$\hat{m}_{ag}/\hat{m}_{tc|ag}$		
						&1.77	
							&1.14	& 1.15	
						&1.98	
							&1.32	&1.31	
							&1.23	&1.21	
							&0.945	\\
5&
$\hat{m}_{ta}/\hat{m}_{[tc][ag]}$	
						&0.0293	
							&0.729	&0.730	
						&0.0477	
							&0.791	&0.784	
							&0.676	&0.682	
							&1.52	\\
6&
$\hat{m}_{tg}/\hat{m}_{[tc][ag]}$	
						&3.21	
							&0.940	&0.950	
						&3.64	
							&1.04	&1.01	
							&1.07	&1.07	
							&0.554	\\
7&
$\hat{m}_{ca}/\hat{m}_{[tc][ag]}$	
						&0.719	
							&1.19	&1.18	
						&0.110	
							&1.23	&1.23	
							&1.28	&1.25	
							&0.573	\\
8&
$\hat{f}^{\script{mut}}_{t+a}$		
						&0.408	
							&0.459	&0.446 	
						&0.372	
							&0.367	&0.392 	
							&0.388	&0.403 	
							&0.497	\\
9&
$\hat{f}^{\script{mut}}_t/\hat{f}^{\script{mut}}_{t+a}$	
					
						&0.113	
							&0.501	&0.522	
					
						&0.234	
							&0.587	&0.513	
							&0.450	&0.439	
							&0.513	\\
10&
$\hat{f}^{\script{mut}}_c/\hat{f}^{\script{mut}}_{c+g}$	
					
						&0.698	
							&0.429	&0.436 	
					
						&0.425	
							&0.479	&0.471 	
							&0.427	&0.383 	
							&0.470	\\
11&
$\hat{f}^{\script{usage}}_{t+a}$	
						&0.0682	
							&(0.5)	&0.483	
						&0.0669	
							&(0.5)	&0.221	
							&(0.5)	&0.447	
							&NA	\\
12&
$\hat{f}^{\script{usage}}_t/\hat{f}^{\script{usage}}_{t+a}$
					
						&0.461	
							&(0.5)	&0.491	
					
						&0.330	
							&(0.5)	&0.429	
							&(0.5)	&0.555	
							&NA	\\
13&
$\hat{f}^{\script{usage}}_c/\hat{f}^{\script{usage}}_{c+g}$
					
						&0.386	
							&(0.5)	&0.558	
					
						&0.310	
							&(0.5)	&0.306	
							&(0.5)	&0.249	
							&NA	\\
14&
$\hat{\sigma}$				
						&27.3	
							&0.738	&0.740	
						&43.3	
							&0.905	&0.840	
							&0.415	&0.395	
							&$\rightarrow 0$	\\
\hline
\multicolumn{2}{l|}	
{\hspace*{1em} $\hat{\tau} \hat{\sigma}$}		
						&0.334	
							&0.0243	&0.0246	
						&0.317	
							&0.0223	&0.0207	 
							&0.0246	&0.0240	 
							&0.0240	\\
\multicolumn{2}{l|}	
{\hspace*{1em} \#parameters}				
						&107	
							&111	&114	
						&107	
							&111	&114	
							&111	&114	
							&261	\\
\multicolumn{2}{l|}	
{\hspace*{1em} $\hat{I}_{KL}(\hat{\VEC{\theta}}) \times 10^8 \ ^b$}
					&$15695$
							&638	& $613$	
					&$35319$
							&1903	& $1438$ 
							&2771	&2335	
							&269946	\\
\multicolumn{2}{l|}	
{\hspace*{1em} $\Delta \mbox{AIC} \ ^c$}			&2072.0
							&297.5	&300.6	
						&1370.8
							&284.3	&275.1	
								&782.5	&700.4	
							& \mbox{unknown}	\\
\hline					
\multicolumn{2}{l|}	
{Ratio of substitution rates} 	 	&	&	&	&	&	&	&	&	& \\
\multicolumn{2}{l|}	
{\hspace*{1em} per codon}		&       &       &       &       &       &       &       &       & \\
\multicolumn{2}{l|}	
{\hspace*{1em} the total base/codon}    		
							&1.28	
								&1.35	&1.35	  
							&1.38	
								&1.53	&1.52	  
								&1.38	&1.39	
								&1.29 \ \ \	\\
		&	&	&	&	&	&	&	&	&	&(1.29)$^d$	\\
\multicolumn{2}{l|}	
{\hspace*{1em} transition/transversion}           
							&0.464	
								&1.08	&1.08	  
							&0.482	
								&0.932	&0.806	  
								&1.18	&1.20	
								&0.764 \ \ \ \\
		&	&	&	&	&	&	&	&	&	&(0.765)$^d$	\\
\multicolumn{2}{l|}	
{\hspace*{1em} nonsynonymous/synonymous$^e$ }        
							&1.13
								&1.37	&1.34	  
							&1.57	
								&2.07	&2.40	  
								&1.05     &1.20	
								&0.726 \ \ \ \\
		&	&	&	&	&	&	&	&	&	&(0.723)$^d$	\\
\hline
\multicolumn{2}{l|}	
{Ratio of substitution rates }			&	&	&	&	&	&	&	&	& \\
\multicolumn{2}{l|}	
{\hspace*{1em} per codon for $\sigma \rightarrow 0$} &       &       &       &       &       &       &       &       & \\
\multicolumn{2}{l|}	
{\hspace*{1em} total base/codon}    		
							&1.0	
								&1.22	&1.22	  
							&1.0	
								&1.38	&1.40	  
								&1.31	&1.33	
								&1.29	\\
\multicolumn{2}{l|}	
{\hspace*{1em} transition/transversion}           
							&0.101	
								&1.21	&1.22	  
							&0.647	
								&1.11	&0.932	  
								&1.31	&1.35	
								&0.764	\\
\multicolumn{2}{l|}	
{\hspace*{1em} nonsynonymous/synonymous$^e$}          
							&0.0644
								&1.04	&1.02	  
							&0.138	
								&1.50	&1.79	  
								&0.853	&0.889	
								&0.726	\\
\hline
\multicolumn{2}{l|}	
{Ratio of substitution rates per}			 	&	&	&	&	&	&	&	&	& \\
\multicolumn{2}{l|}	
{\hspace*{1em} codon for $w_{ab}=0$ and $\sigma \rightarrow 0$}	&       &       &       &       &       &       &       &       & \\
\multicolumn{2}{l|}	
{\hspace*{1em} total base/codon}                  
                                                        &1.0    
                                                                &1.45	&1.46     
                                                        &1.0
                                                                &1.72	&1.74     
								&1.67	&1.71	
								&1.51	\\
\multicolumn{2}{l|}	
{\hspace*{1em} transition/transversion}           
                                                        &0.0605 
                                                                &0.829	&0.831    
                                                        &0.499
                                                                &0.933	&0.849    
								&0.992	&0.981	
								&0.427	\\
\multicolumn{2}{l|}	
{\hspace*{1em} nonsynonymous/synonymous$^e$}      
                                                        &11.3   
                                                                &5.58	&5.74     
                                                        &11.1
                                                                &8.68	&11.1     
								&7.45	&8.46	
								&6.81	\\
\hline
\end{tabular}

\vspace*{1em}

\noindent
$^a$ If the value of a parameter is parenthesized, the parameter is not variable but fixed to the value specified.

\noindent
$^b$ $\hat{I}_{KL}(\hat{\VEC{\theta}}) = -(\ell(\hat{\VEC{\theta}})/N + 2.98607330)$ for JTT,
$-(\ell(\hat{\VEC{\theta}})/N + 2.97444860)$ for WAG,
$-(\ell(\hat{\VEC{\theta}})/N + 2.96853414)$ for LG,
and
$-(\ell(\hat{\VEC{\theta}})/N + 4.19073314)$ for KHG; see text for details. 

\noindent
$^c$ $\Delta \mbox{AIC} \equiv 2 N \hat{I}_{KL}(\hat{\VEC{\theta}}) + 2 \times $ \#parameters 
with $N \simeq 5919000$ for JTT, 
$N \approx 1637663$ for WAG,
$N \approx 10114373$ for LG,
and
the value of $N$ is unknown for KHG; see text for details.

\noindent
$^d$  The value in the parenthesis corresponds to the one for the KHG codon substitution probability matrix.

\noindent
$^e$  Note that these ratios are not the ratios of the rates per site but per codon; see text for details.

\normalfont
\end{table}

\newpage
\begin{table}[ht]
\caption{\label{tbl: correlation_of_wij_between_EI_JTT_WAG_LG_KHG}
\label{tbl: correlation_of_wij}
\BF{
Correlations of $\hat{w}_{ab}$ between various estimates;
} 
the lower half shows the correlation coefficients of $\hat{w}_{ab}$ 
for 75 single-step amino acid pairs
and the upper half does those of $\hat{w}_{ab}$ 
for 86 multi-step amino acid pairs
by excluding 29 amino acid pairs of the least exchangeable category in 
the JTT-ML91, the WAG-ML91 or the LG-ML91.
}
\vspace*{2em}

\begin{tabular}{l|cccccc}
\hline
Model		& EI	& JTT-ML91+	& WAG-ML91+	& LG-ML91+	&  \multicolumn{2}{c}{KHG-ML200} 	\\
\hline
EI		&	&0.45           &0.51           &0.59           &0.55 &(0.65)$^a$	\\
JTT-ML91+	&0.66   &               &0.80           &0.80           &0.51		\\
WAG-ML91+	&0.68   &0.87           &               &0.86           &0.55		\\
LG-ML91+	&0.71   &0.82           &0.90           &               &0.58		\\
KHG-ML200	&0.71   &0.77           &0.69           &0.74		&		\\
\hline
\end{tabular}

\vspace*{1em}
$^a$ The value in the parenthesis is the correlation coefficient
for which the $\hat{w}_{ab}$ for all multi-step amino acid pairs are taken into account.
The correlation coefficient of $\hat{w}_{ab}$ for all amino acid pairs 
between the EI and the KHG-ML200 is equal to 0.60.

\end{table}

} 

\SupTable{

} 

\TextTable{ 

\newpage
\begin{table}[th]
\caption{\label{tbl: optimizations_wij_AIC}
\label{tbl: optimizations_ML91+mod_AIC}
\BF{
$\Delta$AIC values 
of the present models with 
the respective selective constraints on amino acids,
} 
$\hat{w}^{\script{JTT-ML91+}}$, $\hat{w}^{\script{WAG-ML91+}}$,
$\hat{w}^{\script{LG-ML91+}}$, and $\hat{w}^{\script{KHG-M200}}$,
for the various 1-PAM substitution matrices.
}
\vspace*{2em}

\footnotesize

\begin{tabular}{l|l|rrrrr|r|r}
\hline
	& \#parameters	& \multicolumn{5}{c}{$\Delta \mbox{AIC} \ ^b$}  & \multicolumn{2}{|c}{$\hat{I}_{KL}(\hat{\VEC{\theta}}) \times 10^8 \ ^c$}    \\
		\hline
Model name	& \#parameters	 	&JTT	&WAG	&LG	&cpREV	 &mtREV	&KHG & KHG	\\
	& (id no. $^a$)		&	&	&	&	&	&(amino acid)	&(codon)	\\
\hline
JTT-ML91+- &			&	&	&	&	&	&	&	\\
\ \ 0	&20			&	&2657.5	&20807.0 &461.7  &426.0	&	\\
\ \ 1	&21(14)			&	&2065.1	&20382.6 &433.9  &424.4	&	\\
\ \ 4	&24(1-3,14)		&	&1773.7	&16148.3 &439.2  &401.9	&	\\
\ \ 7	&27(1-3,8-10,14)	&	&1257.8	&12330.2 &303.4  &295.5	&	\\
\ \ 11	&31(1-10,14)	&	&1152.9	&12140.0 &291.5  &286.5	& 40931	\\
\ \ 12	&32(0-10,14)	&	&	&	&	&	&	& 473668	\\
\hline
WAG-ML91+- &			&	&	&	&	&	&	&	\\
\ \ 0	&20			&9095.4	&	&10537.3 &316.2  &535.1	& 	\\
\ \ 1	&21(14)			&8928.9	&	&9196.3  &317.1  &532.8	&	\\
\ \ 4	&24(1-3,14)		&6274.9	&	&6354.9	 &281.4  &414.0	&	\\
\ \ 7	&27(1-3,8-10,14)	&3658.3 &	&5294.9	 &261.6  &383.6	&	\\
\ \ 11	&31(1-10,14)	&3299.2 &	&4813.3	 &259.1  &365.1	& 12789 \\
\ \ 12	&32(0-10,14)	&	&	&	&	&	&	& 496804	\\
\hline
LG-ML91+- &			&	&	&	&	&	&	&	\\
\ \ 0	&20			&13669.8 &1806.0 &	 &487.1  &593.4	&	\\
\ \ 1	&21(14)			&12176.2 &1188.8 &	 &421.4  &558.0	&	\\
\ \ 4	&24(1-3,14)		&6325.7	 &811.6	 &	 &340.6  &391.6	&	\\
\ \ 7	&27(1-3,8-10,14)	&3983.0	 &636.0	 &	 &267.0  &329.8	&	\\
\ \ 11	&31(1-10,14)	&3878.5	 &574.7	 &	 &267.1  &314.9	& 5732 	\\
\ \ 12	&32(0-10,14)	&	&	&	&	&	&	& 436557	\\
\hline
KHG-ML200- &			&	&	&	&	&	&	&	\\
\ \ 0	&20			&15063.5 &953.4 &12568.9 &403.6 &593.6	&	&	\\
\ \ 1	&21(14)			&15078.6 &955.4 &12570.9 &405.6 &595.6	&	&	\\
\ \ 4	&24(1-3,14)		&6398.0	 &540.7	&5683.3	&297.4	&399.3	&	&	\\
\ \ 7	&27(1-3,8-10,14)	&4611.5	 &533.4	&3804.2	&259.9	&358.0	&	&	\\
\ \ 11	&31(1-10,14)	&4429.9  &518.7	&3006.1	&251.7	&334.1	&	&	\\
\hline
\end{tabular}

\vspace*{1em}

$^a$
Parameter id numbers in the parenthesis mean 
ML parameters in each model
and other parameters except for $\beta = 1$ and $w_0 = 0$ are fixed              
to the value of the corresponding parameter
listed in the column of the ML-91 or the ML-200 in \Table{\ref{tbl: optimizations_wij_JTT_WAG}};
each id number 
corresponds to the parameter id number listed in   
\Table{\ref{tbl: optimizations_wij_JTT_WAG}}. 

\noindent
$^b$ $\Delta \mbox{AIC} \equiv 2 N \hat{I}_{KL}(\hat{\VEC{\theta}}) + 2 \times $ \#parameters
with 
$N \simeq 5919000$ for JTT,
$N \approx 1637663$ for WAG,
$N \approx 10114373$ for LG,
$N \approx 169269$ for cpREV,
and
$N \approx 137637$ for mtREV; see text for details.

\noindent
$^c$ $\hat{I}_{KL}(\hat{\VEC{\theta}}) = $
$- (\ell(\hat{\VEC{\theta}})/N + 2.97009788)$ for the KHG-derived amino acid substitution probability matrix,
and $- (\ell(\hat{\VEC{\theta}})/N + 4.19073314)$ for the KHG codon substitution probability
matrix; see text for details.

\noindent

\end{table}

} 

\TextTable{

\setlength{\topmargin}{-3.0cm}

\newpage
\begin{table}[ht]
\caption{
\label{tbl: optimizations_wij_WAG_JTT_CpRev_MtRevRmtx_X-ML91+}
\label{tbl: optimizations_wij_WAG_JTT_CpRev_MtRevRmtx_X-ML94+}
\label{tbl: optimizations_wij_JTT_WAG_LG_X-ML91+}
\label{tbl: optimizations_wij_JTT_WAG_LG_X-ML}
\label{tbl: optimizations_wij_JTT_WAG_LG_X-ML_KHG-ML}
\BF{
ML estimates
of the present models with 
the respective selective constraints
for the 1-PAM amino acid substitution matrices of JTT, WAG, and LG.
} 
}
\vspace*{2em}

\small

\begin{tabular}{l|rrr|rrr|rrr}
\hline
		&\multicolumn{3}{c}{JTT} &\multicolumn{3}{|c}{WAG} 	  &\multicolumn{3}{|c}{LG}		\\
		\hline
	 	&WAG- $^a$	&LG- $^a$	&KHG- $^a$    &JTT- $^a$ 	& LG- $^a$ 	&KHG- $^a$    &JTT- $^a$ 	&WAG- $^a$  	&KHG- $^a$  \\
		\hline
	 	&\multicolumn{2}{c|}{ML91+-11} 		&{\footnotesize{ML200-11}}	&\multicolumn{2}{|c|}{ML91+-11}		&{\footnotesize{ML200-11}}	&\multicolumn{2}{|c|}{ML91+-11} &{\footnotesize{ML200-11}}	\\
\hline
$- \hat{w}_0$					&(0.0)	& (0.0)	&(0.0)	&(0.0)	&(0.0)	&(0.0)	&(0.0)	&(0.0)	& (0.0)	\\
$1/\hat{\beta}$				&1.08	&1.32	&1.07	&1.04 	&1.28	&1.01	&0.830	&0.798	&0.757	\\
$\hat{m}_{[tc][ag]}$			&0.429	&0.304	&0.257	&1.29	&0.921	&0.648	&1.45	&1.543	&0.577	\\
$\hat{m}_{tc|ag}/\hat{m}_{[tc][ag]}$	&2.36	&2.42	&1.26	&1.19	&1.71	&0.850	&1.16	&1.82	&0.783	\\
$\hat{m}_{ag}/\hat{m}_{tc|ag}$		&1.22 	&1.16	&0.915	&1.26	&1.27	&1.00	&1.20	&1.26	&0.869	\\
$\hat{m}_{ta}/\hat{m}_{[tc][ag]}$	&0.649	&0.654	&1.32	&0.814	&0.802	&1.54	&0.668	&0.634	&1.59	\\
$\hat{m}_{tg}/\hat{m}_{[tc][ag]}$	&1.13	&1.01	&0.622	&0.862	&0.947	&0.568	&0.988	&1.20	&0.524	\\
$\hat{m}_{ca}/\hat{m}_{[tc][ag]}$	&1.18	&1.31	&0.605	&1.27	&1.33	&0.597	&1.24	&1.20	&0.446	\\
$\hat{f}^{\script{mut}}_{t+a}$		&0.481	&0.507	&0.578	&0.351	&0.405	&0.512	&0.333	&0.335	&0.534	\\
$\hat{f}^{\script{mut}}_t/\hat{f}^{\script{mut}}_{t+a}$	
					&0.527	&0.488	&0.490	&0.548	&0.527	&0.519	&0.462	&0.518	&0.463	\\
$\hat{f}^{\script{mut}}_c/\hat{f}^{\script{mut}}_{c+g}$	
					&0.429	&0.390	&0.413	&0.461	&0.435	&0.463	&0.455	&0.468	&0.446	\\
$\hat{\sigma}$				&1.09	&1.28	&0.604	&0.893  &0.751	&$\rightarrow 0$	&0.886	&0.718	&$\rightarrow 0$	\\
\hline
$\hat{\tau} \hat{\sigma}$		&0.0263	&0.0310	&0.0363	&0.0220	&0.0230	&0.0275	&0.0246	&0.0231	&0.0444	\\
\#parameters				&31	&31	&31	&31	&31	&31	&31	&31	&31	\\
$\hat{I}_{KL}(\hat{\VEC{\theta}}) \times 10^8 \ ^b$	
					&$27346$&$32239$&36897	&$33306$&$15653$&13945	&$59707$&$23488$&14554	\\
$\Delta \mbox{AIC} \ ^c$		&3299.2	&3878.5	&4429.9	&1152.9  &574.7	&518.7	&12140.0&4813.3 &3006.1	\\
\hline
Ratio of substitution	 		&	&	&	&	&	&	&	&	&	\\
\ rates per codon			&	&	&	&	&	&	&	&	&	\\
\ the total base/codon    		&1.35	&1.32 	&1.19	& 1.51  &1.45	&1.19	&1.47 	&1.49	&1.12	\\
\ transition/transversion           	&1.23 	&1.25	&1.02	& 0.815 &0.959	&0.753	&0.902	&1.08	&0.789	\\
\ non-/synonymous$^d$          		&1.49 &1.17 &0.612 & 2.07  &1.59 &0.577	&1.56 &1.60 &0.293	\\
\hline
For $\sigma \rightarrow 0$		&	&	&	&	&	&	&	&	&	\\
\ the total base/codon    		&1.19	&1.13	&1.09	& 1.37  &1.33	&1.19	&1.34 	&1.39	&1.12	\\
\ transition/transversion           	&1.51	&1.57	&1.06	& 0.923 &1.10	&0.753	&1.03	&1.29	&0.789	\\
\ non-/synonymous$^d$  	        	&1.03 &0.755 &0.449 & 1.54  &1.19 &0.577 &1.14 &1.20 &0.293	\\
\hline
For $w_{ab}=0$ and $\sigma \rightarrow 0$ &	&	&	&	&	&	&	&	&	\\
\ the total base/codon    		&1.38	&1.29	&1.18	& 1.66  &1.60	&1.38	&1.68	&1.80	&1.34	\\
\ transition/transversion           	&1.27	&1.28	&0.642	& 0.645 &0.926	&0.440	&0.622 	&0.989	&0.390	\\
\ non-/synonymous$^d$  	        	&4.67 &3.99 &3.71	& 8.62  &7.02	&5.35	&8.79 &9.49	&5.23	\\
\hline
\end{tabular}

\vspace*{1em}

\footnotesize

\noindent
$^a$ In all models, 
equal codon usage 
($\hat{f}^{\script{usage}}_{t} = \hat{f}^{\script{usage}}_{a} = \hat{f}^{\script{usage}}_{c} = \hat{f}^{\script{usage}}_{g} = 0.25$)
is assumed.                          
If the value of a parameter is parenthesized, the parameter is not variable but fixed to the value specified.

\noindent
$^b$ $\hat{I}_{KL}(\hat{\VEC{\theta}}) = $
$- (\ell(\hat{\VEC{\theta}})/N + 2.98607330)$ for JTT, 
$- (\ell(\hat{\VEC{\theta}})/N + 2.97444860)$ for WAG,
and
$- (\ell(\hat{\VEC{\theta}})/N + 2.96853414)$ for LG.

\noindent
$^c$ $\Delta \mbox{AIC} \equiv 2 N \hat{I}_{KL}(\hat{\VEC{\theta}}) + 2 \times $ \#parameters
with 
$N \simeq 5919000$ for JTT,
$N \approx 1637663$ for WAG,
and
$N \approx 10114373$ for LG;
see text for details.

\noindent
$^d$  Note that these ratios are not the ratios of the rates per site but per codon; see text for details.

\normalsize
\end{table}

\newpage
\begin{table}[ht]
\caption{
\label{tbl: optimizations_wij_CpRev_MtRevRmtx_X-ML91+}
\label{tbl: optimizations_wij_CpRev_MtRevRmtx_X-ML}
\label{tbl: optimizations_wij_CpRev_MtRevRmtx_X-ML_KHG-ML}
\BF{
ML estimates
of the present models with 
the respective selective constraints
for the 1-PAM amino acid substitution matrices of cpREV and mtREV.
} 
}
\vspace*{2em}

\small

\begin{tabular}{l|rrrr|rrrr}
\hline
			&\multicolumn{4}{c}{cpREV} 	&\multicolumn{4}{|c}{mtREV}  \\
		\hline
	 	&JTT- $^a$ 	&WAG- $^a$	&LG- $^a$ 	&KHG- $^a$    & JTT- $^a$ 	&WAG- $^a$ 	&LG- $^a$  	&KHG- $^a$    \\
		\hline
	 	&\multicolumn{3}{c|}{ML91+-11} 	&ML200-11	&\multicolumn{3}{|c|}{ML91+-11}	&ML200-11	\\
\hline
$- \hat{w}_0$					&(0.0)	&(0.0)	&(0.0)	&(0.0)	&(0.0)	&(0.0)	& (0.0)	&(0.0)	\\
$1/\hat{\beta}$				&0.940	&0.977	&1.18	&1.02	&0.690	&0.845	&0.977	&0.752	\\
$\hat{m}_{[tc][ag]}$			&0.865	&0.917	&0.611	&0.521	&0.564	&0.524	&0.321	&0.228	\\
$\hat{m}_{tc|ag}/\hat{m}_{[tc][ag]}$	&1.50	&2.23	&2.353	&1.14	&2.01	&3.43	&3.82	&1.64	\\
$\hat{m}_{ag}/\hat{m}_{tc|ag}$		&1.28	&1.30	&1.24	&0.973	&1.06	&1.13	&1.08	&0.752	\\
$\hat{m}_{ta}/\hat{m}_{[tc][ag]}$	&0.746	&0.705	&0.733	&1.61	&0.681	&0.595	&0.638	&2.00	\\
$\hat{m}_{tg}/\hat{m}_{[tc][ag]}$	&1.17	&1.37	&1.25	&0.747	&0.792	&0.893	&0.839	&0.411	\\
$\hat{m}_{ca}/\hat{m}_{[tc][ag]}$	&1.23	&1.17	&1.26	&0.566	&1.65	&1.67	&1.76	&0.623	\\
$\hat{f}^{\script{mut}}_{t+a}$		&0.283	&0.306	&0.328	&0.442	&0.262	&0.270	&0.287	&0.426	\\
$\hat{f}^{\script{mut}}_t/\hat{f}^{\script{mut}}_{t+a}$	
					&0.611	&0.654	&0.609	&0.597	&0.601	&0.652	&0.598	&0.631	\\
$\hat{f}^{\script{mut}}_c/\hat{f}^{\script{mut}}_{c+g}$	
					&0.425	&0.446	&0.393	&0.425	&0.349	&0.304	&0.260	&0.332	\\
$\hat{\sigma}$				&1.93	&1.43	&1.75	&0.158	&3.48	&2.18	&3.37	&2.89	\\
\hline
$\hat{\tau} \hat{\sigma}$		&0.0325	&0.0285	&0.0339	&0.0288	&0.0603	&0.0445	&0.0653	&0.0923	\\
\#parameters				&31	&31	&31	&31	&31	&31	&31	&31	\\
$\hat{I}_{KL}(\hat{\VEC{\theta}}) \times 10^8 \ ^b$	
					&$67803$	&$58229$	&$60586$	&56032	&$81541$	&$110126$	&$91860$	&98837	\\
$\Delta \mbox{AIC} \ ^c$		&291.5	&259.1	&267.1	&251.7	&286.5	&365.1	&314.9	&334.1	\\
\hline
Ratio of substitution	 		&	&	&	&	&	&	&	&	\\
\ rates per codon			&	&	&	&	&	&	&	&	\\
\ the total base/codon    		& 1.45	& 1.46	&1.41	&1.20	& 1.36	& 1.37	&1.33	&1.23	\\
\ transition/transversion           	& 1.05	& 1.20	&1.25	&1.05	& 1.44	& 1.65	&1.74	&1.45	\\
\ non-/synonymous$^d$          		& 1.74 & 1.80 &1.38	&0.631	& 0.908 & 1.04 &0.772 &0.403	\\
\hline
For $\sigma \rightarrow 0$		&	&	&	&	&	&	&	&	\\
\ the total base/codon    		& 1.21	& 1.26	&1.20	&1.16	& 1.11	& 1.15	&1.09	&1.05	\\
\ transition/transversion           	& 1.42	& 1.66	&1.77	&1.07	& 2.52	& 2.73	&3.31	&1.96	\\
\ non-/synonymous$^d$  	        	& 1.03 & 1.10 &0.794 &0.573 & 0.387 & 0.515 &0.312 &0.163	\\
\hline
For $w_{ab}=0$ and $\sigma \rightarrow 0$ &	&	&	&	&	&	&	&	\\
\ the total base/codon    		&1.45	&1.55	&1.44	&1.33	&1.31	&1.37	&1.26	&1.16	\\
\ transition/transversion           	&0.797	&1.20	&1.25	&0.569	&1.06	&1.78	&1.98	&0.883	\\
\ non-/synonymous$^d$  	        	&6.06 &6.33	&5.14	&4.97	&3.40	&3.09	&2.58	&3.02	\\
\hline
\end{tabular}

\vspace*{1em}

\footnotesize

\noindent
$^a$ In all models, 
equal codon usage 
($\hat{f}^{\script{usage}}_{t} = \hat{f}^{\script{usage}}_{a} = \hat{f}^{\script{usage}}_{c} = \hat{f}^{\script{usage}}_{g} = 0.25$)
is assumed.                          
If the value of a parameter is parenthesized, the parameter is not variable but fixed to the value specified.

\noindent
$^b$ $\hat{I}_{KL}(\hat{\VEC{\theta}}) = $
$- (\ell(\hat{\VEC{\theta}})/N + 2.95801048)$ for cpREV, 
and
$- (\ell(\hat{\VEC{\theta}})/N + 2.85313622)$ for mtREV; see text for details.

\noindent
$^c$ $\Delta \mbox{AIC} \equiv 2 N \hat{I}_{KL}(\hat{\VEC{\theta}}) + 2 \times $ \#parameters
with 
$N \approx 169269$ for cpREV,
and $N \approx 137637$ for mtREV; see text for details.

$^d$  Note that these ratios are not the ratios of the rates per site but per codon; see text for details.

\normalsize

\end{table}

\newpage
\begin{table}[ht]
\caption{
\label{tbl: optimizations_wij_KHGasm_KHG_X-ML91+}
\label{tbl: optimizations_wij_KHGasm_KHG_X-ML}
\BF{
ML estimates
of the present models with 
the respective selective constraints
for the 1-PAM KHG-derived amino acid and KHG codon substitution matrices.
} 
}
\vspace*{2em}

\small

\begin{tabular}{l|rrr|rrr}
\hline
			&\multicolumn{3}{c}{KHG (amino acid)} 	&\multicolumn{3}{|c}{KHG (codon)}  \\
		\hline
	 	&JTT- $^a$\ \ \ \ 	&WAG- $^a$\ \ \ \	&LG- $^a$\ \ \ \ 	&JTT- $^a$\ \ \ \ 	&WAG- $^a$\ \ \ \	&LG- $^a$\ \ \ \ 	\\
		\hline
	 	&\multicolumn{3}{c}{ML91+-11} 	&\multicolumn{3}{|c}{ML91+-12}	\\
\hline
$- \hat{w}_0$					&(0.0)	&(0.0)	&(0.0)	&1.29	&1.50	&1.11	\\
$1/\hat{\beta}$				&0.952	&0.912	&1.22	&1.72	&2.02	&1.91	\\
$\hat{m}_{[tc][ag]}$			&1.545	&1.68	&1.33	&1.23	&1.21	&1.15	\\
$\hat{m}_{tc|ag}/\hat{m}_{[tc][ag]}$	&1.19	&1.73	&1.69	&0.992	&1.07	&1.09	\\
$\hat{m}_{ag}/\hat{m}_{tc|ag}$		&1.24	&1.28	&1.22	&1.09	&1.12	&1.10	\\
$\hat{m}_{ta}/\hat{m}_{[tc][ag]}$	&0.689	&0.682	&0.748	&1.26	&1.25	&1.25	\\
$\hat{m}_{tg}/\hat{m}_{[tc][ag]}$	&0.855	&1.07	&0.943	&0.646	&0.662	&0.671	\\
$\hat{m}_{ca}/\hat{m}_{[tc][ag]}$	&1.32	&1.26	&1.31	&0.815	&0.806	&0.813	\\
$\hat{f}^{\script{mut}}_{t+a}$		&0.317	&0.334	&0.377	&0.480	&0.484	&0.488	\\
$\hat{f}^{\script{mut}}_t/\hat{f}^{\script{mut}}_{t+a}$	
					&0.533	&0.579	&0.512	&0.499	&0.499	&0.493	\\
$\hat{f}^{\script{mut}}_c/\hat{f}^{\script{mut}}_{c+g}$	
					&0.460	&0.480	&0.441	&0.464	&0.459	&0.459	\\
$\hat{\sigma}$				&2.64	&2.25	&1.30	&$\rightarrow 0$	&0.0496	&$\rightarrow 0$	\\
\hline
$\hat{\tau} \hat{\sigma}$		&0.0308	&0.0286	&0.0247	&0.0240	&0.0247	&0.0240	\\
\#parameters				&31	&31	&31	&32	&32	&32	\\
$\hat{I}_{KL}(\hat{\VEC{\theta}}) \times 10^8 \ ^b$	
					&40931	&12789	&5732	&473668	&496804	&436557	\\
\hline
Ratio of substitution	 		&	&	&	&	&	&	\\
\ rates	per codon			&	&	&	&	&	&	\\
\ the total base/codon    		& 1.64	& 1.66	&1.59	& 1.29	& 1.29	&1.29	\\
\ transition/transversion           	& 0.772	& 0.859	&0.891	& 0.759	& 0.765	&0.767	\\
\ non-/synonymous$^c$          		& 2.56 & 2.61	&2.03	& 0.728	& 0.727	&0.724	\\
\hline
For $\sigma \rightarrow 0$		&	&	&	&	&	&	\\
\ the total base/codon    		& 1.39	& 1.45	&1.43	& 1.29	& 1.28	&1.29	\\
\ transition/transversion           	& 0.977	& 1.15	&1.08	& 0.759	& 0.770	&0.767	\\
\ non-/synonymous$^c$  	        	& 1.48 & 1.54 	&1.36	& 0.728	& 0.704	&0.724	\\
\hline
For $w_{ab}=0$ and $\sigma \rightarrow 0$		&	&	&	&	\\
\ the total base/codon    		&1.71	&1.83	&1.75	&1.65	&1.65	&1.64	\\
\ transition/transversion           	&0.637	&0.926	&0.892	&0.51	&0.552	&0.561	\\
\ non-/synonymous$^c$  	        	&9.41 	&10.3 	&8.86	&8.16	&8.07	&7.77	\\
\hline
\end{tabular}

\vspace*{1em}

\footnotesize

\noindent
$^a$ In all models, codon frequencies are taken to be equal to the observed ones.
If the value of a parameter is parenthesized, the parameter is not variable but fixed to the value specified.

\noindent
$^b$ $\hat{I}_{KL}(\hat{\VEC{\theta}}) = $
$- (\ell(\hat{\VEC{\theta}})/N + 2.97009788)$ for the KHG-derived amino acid substitution probability matrix,
and $- (\ell(\hat{\VEC{\theta}})/N + 4.19073314)$ for the KHG codon substitution probability 
matrix; see text for details.

$^d$  Note that these ratios are not the ratios of the rates per site but per codon; see text for details.

\normalsize
\end{table}

} 

\TextTable{

} 

\TextTable{

\newpage
\begin{table}[ht]
\caption{\label{tbl: LK_of_mt_tree}
\BF{
Log-likelihoods of a phylogenetic tree\CITE{AH:96} 
} 
of the concatenated sequences of 12 protein-coding sequences 
encoded on the same strand of mitochondrial DNA 
from 20 vertebrate species with 2 races from human.
}
\vspace*{2em}

\begin{tabular}{l|rrrrrr}
\hline
Codon Substitution 
	&\#p$^b$ & $\ell +$ \hspace*{3em} &$\mbox{AIC} -$ \hspace*{2em}	&$\hat{\sigma}$ & $\hat{m}_{[tc][ag]}$
	& $\hat{m}_{tc|ag}/\hat{m}_{[tc][ag]}$
		\\
Model$^a$	&	&$\Red{116898.6}$	&$\Red{233917.3}$	&		&
	&
		\\
\hline
 LG-1-F$^c$	&60      
					&$\Red{-1293.8}$	&$\Red{2587.6}$		\\
 KHGaa-1-F$^{cd}$	&60      
					&$\Red{-1293.0}$	&$\Red{2586.1}$		\\
 WAG-1-F$^c$	&60      
					&$\Red{-1108.1}$	&$\Red{2216.1}$		\\
 JTT-1-F$^c$    &60      
					&$\Red{-836.4}$       &$\Red{1672.8}$		\\
 mtREV-1-F$^c$  &60      
					&$0.0$		&$0.0$			\\
		\\
 No-Constraints-1-F$^e$	&60     
					&$\Red{-1731.0}$	&$\Red{3462.1}$
					&$(2.46)$     	&$(0.040)$	&$(3.24)$ 	\\
 WAG-ML91+-1-F$^e$	&60	
					&$\Red{1021.4}$	&$\Red{-2042.7}$	
					&$(2.18)$	&$(0.524)$	&$(3.43)$	\\
 JTT-ML91+-1-F$^e$	&60	
					&$\Red{1237.7}$	&$\Red{-2475.5}$	
					&$(3.48)$	&$(0.564)$	&$(2.01)$	\\
 LG-ML91+-1-F$^e$	&60	
					&$\Red{1382.2}$	&$\Red{-2764.4}$
					&$(3.37)$	&$(0.321)$	&$(3.82)$	\\
 EI-1-F$^e$		&60	
					&$\Red{1395.8}$	&$\Red{-2791.6}$
					&$(0.339)$	&$(0.737)$	&$(3.06)$	\\
 KHG-ML200-1-F$^e$	&60	
					&$\Red{1676.9}$	&$\Red{-3353.9}$
					&$(2.89)$	&$(0.228)$	&$(1.64)$	\\
		\\
 No-Constraints-11-F	&70      
					&$\Red{772.2}$	&$\Red{-1524.4}$	
					&$0.906$	&$0.273$	&$3.37$	\\
 EI-12-F	        &71      
					&$\Red{1966.6}$	&$\Red{-3911.2}$
					&$0.326$	&$0.549$	&$3.60$	\\
 WAG-ML91+-12-F         &71      
					&$\Red{2268.3}$	&$\Red{-4514.5}$	
					&$1.84$     	&$0.471$ 	&$4.16 $	\\
 JTT-ML91+-12-F         &71      
					&$\Red{2275.1}$	&$\Red{-4528.1}$
					&$3.57 $	&$0.506 $	&$2.91 $	\\
 KHG-ML200-12-F         &71      
					&$\Red{2355.7}$	&$\Red{-4689.4}$	
					&$0.469 $	&$0.226 $	&$2.50 $	\\
 LG-ML91+-12-F          &71      
					&$\Red{2510.0}$	&$\Red{-4997.9}$	
					&$1.26 $ 	&$0.357 $	&$4.32 $	\\
			\\
 No-Constraints-11-F-dG4  &71      
					&$\Red{2495.4}$	&$\Red{-4968.9}$
					&$0.000$	&$0.182 $	&$3.62 $	\\
 EI-12-F-dG4            &72      
					&$\Red{3742.4}$	&$\Red{-7460.7}$
					&$0.000$	&$0.392 $	&$3.95 $	\\
 JTT-ML91+-12-F-dG4     &72      
					&$\Red{4156.9}$	&$\Red{-8289.8}$
					&$0.064 $	&$0.385 $	&$3.11 $	\\
 KHG-ML200-12-F-dG4     &72      
					&$\Red{4190.0}$	&$\Red{-8356.0}$
					&$0.000$	&$0.147 $	&$2.60 $	\\
 WAG-ML91+-12-F-dG4     &72      
					&$\Red{4196.4}$	&$\Red{-8368.7}$
					&$0.042 $	&$0.342 $	&$4.61 $	\\
 LG-ML91+-12-F-dG4      &72      
					&$\Red{4412.6}$	&$\Red{-8801.1}$
					&$0.029 $	&$0.253 $	&$4.83 $	\\
\hline
\end{tabular}

\vspace*{1em}

$^a$ In all models named with a suffix "F", 
codon frequencies are taken to be equal to those in coding sequences.
A suffix "dG4" means the discrete approximation of the $\Gamma$ distribution
with 4 categories\CITE{Y:94} for rate variation.
The parameter $w_0$ in \Eq{\ref{eq: estimation_of_fitness}} is optimized in all models.

$^b$ The number of parameters; the value for the mtREV-1-F
is not quite correct, because mtREV was estimated 
from the almost same set of protein sequences\CITE{AH:96}.

$^c$ The exchangeabilties of nonsynonymous and synonymous codon pairs are equal to 
$\exp w_0$ multiplied by 
those of the corresponding amino acid pairs 
and all equal to the mean amino acid exchangeability
in the empirical amino acid substitution matrix specified, respectively.

$^d$ KHGaa means the amino acid substitution matrix derived from KHG.

$^e$ All parameters except $w_0$ and codon frequencies are fixed to those ML estimates
of each model fitted to mtREV.

\end{table}

} 
\clearpage

\section*{Supporting Information Legends}

\noindent
\begin{description}
\item Text S1. 
Supporting information consisting of the following sections.
\begin{enumerate}
\item A method for the physico-chemical evaluation of selective constraints on amino acid replacement.
\item Models with no amino acid dependences of selective constraints.
\item A physico-chemical evaluation of selective constraints on amino acids.
\item Other physico-chemical evaluations of selective constraints on amino acids.
\item Evolutionary process of amino acid substitutions in terms of log-odds.
\end{enumerate}

\item Data S1.
A computer-readable dataset of the ML estimates of parameters in
the ML-200 for KHG, and the ML-91 and the ML-91+ for LG, WAG, and JTT as well as the EI.
\end{description}
\clearpage

\clearpage

\setcounter{page}{1}
\pagestyle{myheadings}
\renewcommand{\thepage}{S1-\arabic{page}}

\vspace*{2em}

\begin{center}
{\large
Supporting Information
	\\
for
	\\
Selective Constraints on Amino Acids
	\\
Estimated by a Mechanistic Codon Substitution Model 
	\\
with Multiple Nucleotide Changes
}
\end{center}
\vspace{4ex}

\vspace{3ex}
\begin{center}
Sanzo Miyazawa
\end{center}
\begin{center}
\ 			\\
Graduate School of Engineering	\\
Gunma University	\\
Kiryu, Gunma 376-8515, Japan	\\
Phone: +81-277-30-1940  \\
E-Mail: miyazawa@smlab.sci.gunma-u.ac.jp \\
	sanzo.miyazawa@gmail.com \\

\ 		\\
\ 		\\
(\today)
\end{center}

\renewcommand{\SEq}[1]{Eq. S1-{#1}}
\renewcommand{\SEqs}[1]{Eqs. S1-{#1}}

\newcommand{\Text}[1]{}
\newcommand{\Supp}[1]{#1}

\setcounter{equation}{0}
\renewcommand{\theequation}{S1-\arabic{equation}}
\newpage

\section*{Supplementary Methods}

\noindent
\subsection*{A method for the physico-chemical evaluation of selective constraints on amino acid replacements}
\vspace*{1em}

Physico-chemical evaluations of $\{ w_{ab} \}$ are not meaningless, even though
selective constraints $\{ w_{ab} \}$ in \Eq{\ref{eq: ratio_of_neutral_mutations}}
in the text
can be optimized for observed data. 
Their performance in reproducing observed substitution data
indicates how extensively 
selective constraints on amino acid substitutions
can be explained by physico-chemical requirements on
amino acid substitutions to preserve protein structures 
and functions.
In this section, a new physico-chemical method for the evaluation of
the selective constraints is introduced.

The rate of acceptance
in amino acid replacements is
assumed here to be proportional 
to the mean relative stability of the native conformation $\mathcal{C}$
of the mutant type of sequence $\mathcal{S'}$ to 
that of the wild type of sequence $\mathcal{S}$.
The probability $P(\mathcal{C}|\mathcal{S})$ of a conformation $\mathcal{C}$
that a sequence $\mathcal{S}$ takes
is equal to the Boltzmann factor of $\mathcal{C}$ divided by 
the conformational partition function of $\mathcal{S}$.
The conformational partition function of a protein may be crudely 
approximated in the high temperature expansion.
\begin{eqnarray}
\lefteqn{
\log P(\mathcal{C} | \mathcal{S})  
}
	\nonumber \\
	&\simeq& - \frac{1}{kT} \mathcal{E}(\mathcal{C}, \mathcal{S})
	\nonumber \\
	& & \vspace*{1em}
			- [ \log (\sum_{\mathcal{C} \in \{\script{compact}\}} 1) 
		- \frac{1}{kT} \langle \mathcal{E}(\mathcal{C}, \mathcal{S}) \rangle_{\script{compact}, T \rightarrow \infty} ]
	\label{eq: approx_of_Z_in_high_T}
\end{eqnarray}
where $k$ is the Boltzmann constant, $T$ is temperature,
and $\mathcal{E}(\mathcal{C}, \mathcal{S})$ is the conformational free energy of
the conformation $\mathcal{C}$ taken by the sequence $\mathcal{S}$.
First, the sum over conformations $\mathcal{C}$ are approximated by the sum over compact/nativelike
conformations whose energies are significantly lower than those of extended conformations.
Then, the logarithm of the partition function is approximated 
by the sum of the first and the second terms in the high temperature expansion.
Thus,
the relative stability of the native conformation of sequence $\mathcal{S'}$ 
to that of sequence $\mathcal{S}$ is estimated by
\begin{eqnarray}
\lefteqn{
\log [P( \mathcal{C} | \mathcal{S}') / P( \mathcal{C} | \mathcal{S}) ] 
}
	\nonumber \\
	&\simeq& - \frac{1}{kT} (\mathcal{E}(\mathcal{C}, \mathcal{S}') - \mathcal{E}(\mathcal{C}, \mathcal{S}))
	 \mbox{ , }
	\nonumber \\
	& &
	\mbox{ if the amino acid composition does not change.}
	\label{eq: relative_stability}
\end{eqnarray}
The mean energy of compact conformations 
does not depend on the details of the amino acid order 
in protein sequences but primarily on the amino acid composition.
Therefore, if the amino acid composition keeps constant during amino acid substitutions,
as indicated  by the present assumption of the stationary state for amino acid
substitutions, 
the relative stability can be approximated by the difference of the native conformational energies
of the two sequences.

As a result, the parameter $w_{ab}$, whose exponent is 
the acceptance rate of substitutions between
amino acids of type $a$ and type $b$,
is evaluated here to be proportional 
to the mean free energy increment caused by a substitution between amino acids
of type $a$ and type $b$.  Then, the mean free energy increment is 
approximated by the sum of two terms one of which
results from the increment of contact energy 
between amino acids in a protein structure
and the other from the change of side-chain volume.
\begin{eqnarray}
        w_{ab} 
	&=& - \beta [ \Delta \hat{\varepsilon}^{\script{c}}_{ab} + \Delta \hat{\varepsilon}^{\script{v}}_{ab} ]
		+ w_0 (1- \delta_{ab})
        \label{eq: def_fitness}
\end{eqnarray}
where $\beta$ is a parameter, 
$\Delta \hat{\varepsilon}^c_{ab}$ is 
the mean increment of contact energy between amino acids 
due to an amino acid exchange between amino acids of type $a$ and type $b$ 
in a protein structure, and $\Delta \hat{\varepsilon}^v_{ab}$ is 
the mean increment of free energy caused by the change of side-chain volume
between amino acids of type $a$ and type $b$.
The exponent of the constant term, $e^{w_0}$, may represent
the ratio of replaceable amino acid sites
in a protein sequence, and then the first term represents the ratio of neutral
substitutions at such mutable sites; the ratio of nonsynonymous to synonymous
mutations is primarily determined by $w_0$.
However, $w_0$ may be positive, meaning positive selection.

\vspace*{2em}
\noindent
\textit{Mean energy increment for each type of amino acid substitutions} 

To consider the mean contact energy increment due to
an amino acid replacement between amino acids of type $a$ and type $b$,
we must note that the evolutionary process 
of amino acid substitutions in proteins 
is assumed here to be in the stationary process, which means
that the amino acid composition of proteins must be kept constant 
in the whole process of amino acid substitutions.
To keep the amino acid composition constant,  an exchange of amino acids
in a protein may be considered as the process of substitutions.
A mean contact energy increment, $2 \Delta \varepsilon^c_{ab}$, 
due to an exchange between amino acids of type $a$ and type $b$
in a protein can be estimated 
\CITE{MJ:93}
by averaging the difference of interaction energies over 
surrounding residues as
\begin{eqnarray}
\Delta \hat{\varepsilon}^{\script{c}}_{ab} = \Delta \hat{\varepsilon}^{\script{c}}_{ba}
	&=& \sum_c (e_{bc} - e_{ac}) ( \frac{N_{ac}}{N_{a}} - \frac{N_{bc}}{N_{b}} ) 
	\geq 0
	\label{eq: def_contact_energy_increment}
\end{eqnarray}
where
$e_{ac}(=e_{ca})$ is the contact energy between amino acids of type $a$ and $c$,
and
$N_{ac}(=N_{ca})$ is a half of the observed number of contacts between amino acids of 
type $a$ and type $c$, and $N_a$ is the number of amino acids of type $a$ in
protein structures.  The contact energies $e_{ab}$ 
and the number of contacts $N_{ab}$
are the ones evaluated from 
the numbers of contacts between amino acids
observed in representative protein structures
\CITE{MJ:03}.
The mean energy increment due to an amino acid exchange
is non-negative for any pair of amino acids
\CITE{MJ:93},
because the contact energies are derived by assuming
that the native conformations of proteins
are at the minimum of the total contact energy.
This means that no favorable substitutions
occur in protein evolution in which amino acid substitutions 
are in the stationary state.
Thus, the assumption of the stationary state for
amino acid substitutions 
is consistent\CITE{MJ:03}
with the neutral theory\CITE{KO:74}
of molecular evolution.

A contact potential used is
a statistical estimate
\CITE{MJ:03} 
of contact energies 
with a correction
\CITE{MJ:99} 
for the Bethe approximation
\CITE{MJ:96,MJ:85}.
The contact energy between amino acids of type $a$ and type $b$ was estimated as
\begin{eqnarray}
e_{ab} &=& e_{rr} + \alpha' [ \Delta e^{\script{Bethe}}_{ar} + \Delta e^{\script{Bethe}}_{rb} 
	+ \frac{\beta'}{\alpha'} \delta e^{\script{Bethe}}_{ab}]
	\label{eq: contact_potential}
\end{eqnarray}
$e_{rr}$ is part of contact energies irrespective of residue types and is called a collapse energy, which 
is essential for a protein to fold by cancelling out the large
conformational entropy of extended conformations but
cannot be estimated explicitly from contact frequencies 
between amino acids in protein structures.
$\Delta e^{\script{Bethe}}_{ar}$ and $\delta e^{\script{Bethe}}_{ab}$ are
the values of $\Delta e_{ar}$ and $\delta e_{ab}$
evaluated by the Bethe approximation 
from the observed numbers of contacts between amino acids.
$\Delta e_{ar} + e_{rr}$ is a partition energy or hydrophobic energy for
a residue of type $a$.  $\delta e_{ab}$ is an intrinsic contact energy
for a contact between residues of type $a$ and type $b$;
refer to
\CITE{MJ:99,MJ:85} 
for their exact definitions. 
The proportional constants for correction
were estimated as $\beta'/ \alpha' = 2.2$ and 
$\alpha' \leq 1$
\CITE{MJ:99}.
Here, energy is measured in $k$T units.
The scaling constant $\beta$ in \Eq{\ref{eq: def_fitness}} 
in the text
is given for $\alpha' = 1$.

The energy increment $\Delta \hat{\varepsilon}^v_{ab}$,
which results from a replacement between amino acids of different sizes,
is assumed here to be proportional to the volume difference 
between amino acids of the type $a$ and type $b$:
\begin{eqnarray}
\Delta \hat{\varepsilon}^{\script{v}}_{ab}
	&=& \upsilon \; [\frac{\sum_{a,b} \Delta \hat{\varepsilon}^{\script{c}}_{ab}}{ \sum_{a,b} | V_{a}  -  V_{b} | } ] | V_{a}  -  V_{b} |
	\label{eq: def-energy-change-due-to-volume-change}
\end{eqnarray}
where $V_{a}$ is the volume of amino acid $a$,
and $\upsilon$ is a proportional constant.
The value of $\upsilon$ is taken to be equal to one,
otherwise specified; 
that is, the contact energy increment and
the volume change are assumed to contribute
to the total free energy increment and the acceptance rate 
with an equal weight.
The amino acid volumes used here are 
the mean volume occupied by each type of amino acid in protein structures, 
and taken from the set named BL+ in Table 6 of 
Tsai et al. 
\CITE{TTCG:99};
the volume of a half cystine (labeled as "cys" in the table) 
is used here for a cysteine.

The values of $[ \Delta \hat{\varepsilon}^{\script{c}}_{ab} + \Delta \hat{\varepsilon}^{\script{v}}_{ab} ]$
for all amino acid pairs are provided in Supporting Information, Data S1.

\newpage
\section*{Supplementary Results}

\vspace*{2em}

\noindent
\subsection*{Models with no amino acid dependences of selective constraints}
\vspace*{1em}

Before examining the effects of selective constraints ($w_{ab}$) on likelihood,
ML values 
for the models 
with no amino acid dependences of selective constraints, i.e., $\beta = 0$
in \Eq{\ref{eq: estimation_of_fitness}},
were calculated for JTT, WAG, cpREV, and mtREV.
The $\Delta \mbox{AIC}$ value and the ML estimators of
$m_{\xi \eta}$, $f_{\xi}^{\script{mut}}$, $f_{\xi}^{\script{usage}}$, and $\sigma$ 
for each model are listed in
\Table{\ref{tbl: optimizations_AIC}} and
\STable{\ref{tbl: optimizations_no-selection}}, 
respectively.
Please note that
$w_0$ is fixed here to $0$, 
and so there is completely no   
selection pressure on nonsynonymous replacements;
the likelihoods of amino acid substitution matrices 
do not strongly depend on $w_0$ and 
codon substitution data are required to
reliably estimate the value of $w_0$.
ML parameters in each model are specified by the parameter id numbers
written in the parenthesis in the second column; 
each id number corresponds to the parameter id number listed 
in \Table{\ref{tbl: optimizations_wij_JTT_WAG_LG}}. 
Each model is called the No-Constraints model with a suffix meaning the number of ML
parameters; see \Table{\ref{tbl: model_names}}.
Although No-Constraints models corresponding to 
the Kimura's two-parameter model
\CITE{K:80},
the model of Hasegawa et al.
\CITE{HKY:85},
the Tamura-Nei model
\CITE{TN:93} 
and the general reversible model
\CITE{LPSS:84} 
were examined, only three models
for each matrix are shown in 
\Table{\ref{tbl: optimizations_AIC}}. 

The bias toward transition has been often pointed out
\CITE{F:04}.
In the present results for the No-Constraints models,
t
the ratio of transition to transversion exchangeability
$m_{tc|ag}/m_{[tc][ag]}$ is evaluated to be 
between 1.5 and 3.3
for all four matrices of JTT, WAG, cpREV, and mtREV, although
that for mtREV is larger than those for the others.
For the No-Constraints-1 of mtREV, 
its parameter is evaluated to be
$\hat{m}_{tc|ag}/\hat{m}_{[tc][ag]} = 2.32$ 
and the ratio of the total transition to the total transversion rate
is equal to $1.24$.
This estimate of transition to transversion exchangeability bias
for mitochondrial proteins is significantly smaller
than the previous estimate by a maximum likelihood method for phylogeny.
Yang et al.
\CITE{YNH:98} 
estimated
$\hat{m}_{tc|ag}/\hat{m}_{[tc][ag]} = 9.157$
for the model corresponding to the No-Constraints-1 in the analyses of
the most likely phylogeny of mitochondrial DNA encoding proteins.

Although the significance of each parameter is indicated by the AIC values
of the No-Constraints models with the various sets of parameters,
its discussion is postponed until the next section 
where results for models with selective constraints 
are presented, because
no selective constraints on amino acids is a completely wrong assumption.

\vspace*{2em}
\noindent
\subsection*{A physico-chemical evaluation of selective constraints on amino acids}
\vspace*{1em}

Let us examine how the likelihood of
JTT is improved
by using the present formula for selective constraints,
\Eq{\ref{eq: estimation_of_fitness}}.
The first evaluation of selective constraints on amino acids
is based on the mean energy increments
due to an amino acid replacement 
that result from the changes of pairwise contact energies
\CITE{MJ:93,MJ:96,MJ:99,MJ:03}
and the volume change 
\CITE{TTCG:99}
of an amino acid side chain by an amino acid replacement.
\Text{
The detailed description for the mean energy increments is
provided in Supplementary Methods of Supporting Information, Text S1.
} 
This model in which selective constraints on amino acids
are evaluated from mean energy increments due to an amino acid
replacement is called here an Energy-Increment-based (EI) model
with a suffix meaning the number of ML parameters; 
see \Table{\ref{tbl: model_names}}.
The ML values for the EI models with various sets of parameters
are listed in \Table{\ref{tbl: optimizations_EI_AIC}}, and
the ML estimates for the EI-10 and the EI-11 
are listed in 
\STable{\ref{tbl: optimizations_EI_JTT_WAG_CpRev_MtRevRmtx}}.

The No-Constraints-1, the No-Constraints-10, and the No-Constraints-13 models
correspond to a special case of $\beta = 0$
in the EI-2, the EI-11, and the EI-14 models, respectively.
As a matter of course,
the selective constraints on amino acids
that represent conservative selection against amino acid substitutions
significantly improve the $\Delta \mbox{AIC}$ values 
for all substitution matrices.

The significance of multiple nucleotide changes in a codon
is indicated by
the improvements of the $\Delta \mbox{AIC}$ 
between the EI-3 and the EI-4, between the EI-12 and the EI-13M,
between the EI-10 and the EI-11,
and between the EI-13 and the EI-14 models, in the latter of which 
the parameter $\hat{m}_{[tc][ag]}$ for multiple
nucleotide changes is optimized as a free variable.
Also, the $\Delta \mbox{AIC}$ is improved
by the inclusion of the scale parameter $\sigma$; 
compare the $\Delta \mbox{AIC}$ values
between the EI-2 and the EI-3,
between the EI-10M and the EI-11,
between the EI-12 and the EI-13,
and between the EI-13M and the EI-14.
Thus,
taking account of both multiple nucleotide changes in a codon 
and variations in substitution rates
is essential to obtain the reasonably large ML values.

The most effective one of the remaining parameters on likelihood
is the parameter for transition-transversion bias, $m_{tc|ag}/m_{[tc][ag]}$. 
The next effective parameters are $f^{\script{mut}}_{\xi}$ and $f^{\script{usage}}_{\xi}$,
and finally the remaining rate parameters.
The $\Delta \mbox{AIC}$ values of the models EI-2G, EI-3, EI-7, EI-11, EI-10MU,
and EI-14 indicate that
all parameters are effective to significantly improve the likelihood of
each of the observed matrices.
The ML estimates of the parameters $f^{\script{mut}}_{\xi}$
and $f^{\script{usage}}_{\xi}$
show the similar tendencies between the models, although
this tendency differs among the substitution matrices, 
JTT, WAG, cpREV, and mtREV.
The comparison of the $\Delta \mbox{AIC}$ values
between the EI-10MU and the EI-14 models
indicates that the parameters for exchangeabilities except for 
transition-transversion bias,
are statistically significant but are not so effective 
as $f^{\script{mut}}_{\xi}$ and $f^{\script{usage}}_{\xi}$
on the improvement of the likelihood.

The relative weight $\upsilon$ of 
the effects of volume change due to an amino acid replacement 
on selective constraints in 
\SEq{\ref{eq: def-energy-change-due-to-volume-change}}
is assumed to be equal to one but may be varied.
Optimizing $\upsilon$ as a free variable
can improve the value of $\Delta \mbox{AIC}$ from 13151.9 to 12932.1 for JTT.
This model may be justified because the effects of volume change
due to an amino acid replacement on protein structures
may be different among the types of protein structures, 
i.e., between membrane and soluble proteins, 
and between $\alpha$ and $\beta$ proteins.

\Table{\ref{tbl: optimizations_EI_AIC}} shows that
the parameters $\{ f^{\script{usage}}_{\xi} \}$ for codon usage are
significant to improve likelihood, however,
the ML estimator of $f^{\script{usage}}_{\xi}$ often takes
extremely small or large values.
Thus, it may be better to assume equal codon usage by fixing
$f^{\script{usage}}_{\xi} = 0.25$ if codon frequencies are unknown.
In the following, equal codon usage is assumed in most cases
of unknown codon frequencies.

\vspace*{2em}
\noindent
\subsection*{Other physico-chemical evaluations of selective constraints on amino acids}
\vspace*{1em}

Grantham
\CITE{G:74} 
and Miyata et al.
\CITE{MMY:79} 
introduced physico-chemical distances between amino acids
in attempts to model selective restraints against amino acid substitutions.
Their physico-chemical distances were also used by 
Goldman and Yang 
\CITE{GY:94} 
and Yang et al.
\CITE{YNH:98},
in which
the acceptance ratio ($\exp w_{ab}$) was represented
by using a linear formula of Miyata et al.
\CITE{MMY:79} 
($\exp w_{ab} = \alpha (1 - \beta d_{ab})$) 
or a geometric formula ($\exp w_{ab} = \alpha \exp( - \beta d_{ab})$) of
physico-chemical distance $d_{ab}$ between amino acids of type $a$ and $b$;
where $\alpha$ and $\beta$ are parameters.
In their models, stepwise substitutions through single nucleotide changes
were assumed,
and codon substitutions due to multiple nucleotide changes 
were completely neglected;
in other words,
$m_{[tc][ag]} \rightarrow 0$ 
with $m_{\xi\eta}/m_{[tc][ag]} = \mbox{constant}$ 
in  \Eq{\ref{eq: def_mutation_rate_matrix}} 
was assumed.
Yang et al.
\CITE{YNH:98} 
reported that
the use of the Miyata's distance
\CITE{MMY:79}
for the acceptance ratio
in their codon-based model lead to a better fit 
to the small data of mitochondrial protein sequences
than the JTT-F and the mtREV24-F models,
in which the rate matrix of JTT or mtREV24
with an adjustment 
for the equilibrium frequencies of amino acids is used;
their codon-based models correspond to the present model with  
$m_{[tc][ag]} \rightarrow 0$, $m_{tc} = m_{ag}$, and
$m_{ta} = m_{tg} = m_{ca} = m_{cg}$, i.e.,
the two parameter model for nucleotide mutations with the adjustment for 
amino acid frequencies.

\Table{\ref{tbl: optimizations_EI_AIC}} 
and
\STable{\ref{tbl: optimizations_selection_JTT_WAG}} 
list
the ML values and the ML estimates 
for JTT and WAG in the present models
in which either the Grantham's distance or the Miyata's distance 
($d_{ab}$ for an amino acid pair $a$ and $b$) is used
as $w_{ab}^{\script{estimate}} = - d_{ab}$ 
to evaluate the selective constraints $w_{ab}$ 
in \Eq{\ref{eq: estimation_of_fitness}};
\begin{eqnarray}
w_{ab} &\equiv& - \beta d_{ab} + w_0 (1 - \delta_{ab})
	\label{eq: estimation_of_fitness_from_aa_distance}
\end{eqnarray}
where $w_0$ is always fixed to the value $0$,
because the likelihoods of amino acid substitution matrices 
do not significantly depend on $w_0$.
These models are called here Grantham and Miyata with
a suffix meaning the number of ML parameters; see \Table{\ref{tbl: model_names}}.
Both the selective constraints based on the Grantham's and 
on the Miyata's distances significantly improve the $\Delta \mbox{AIC}$.

Miyata et al.\CITE{MMY:79} 
claimed that their new scale can explain
the tendencies of amino acid replacements
better than the Grantham's distance scale.
\Table{\ref{tbl: optimizations_EI_AIC}}
shows that 
the Miyata's physico-chemical distance performs better
in all parameter sets than the Grantham's distance.
This result is 
consistent with
that of Yang et al.\CITE{YNH:98}
for mitochondrial proteins.
The present physico-chemical evaluation of selective constraints 
(EI model)
fits JTT and WAG even better than the 
Miyata's distance scale, although
the performances of both the methods are almost same for cpREV and mtREV.

One of the important facts in these results is that
allowing multiple nucleotide changes in a codon
significantly improve the AIC irrespective of
the estimations of selective constraints; compare 
the $\Delta$ AIC values between the Grantham-10 and the Grantham-11,
and between the Miyata-10 and the Miyata-11. 
In other words, the improvement of the AIC value
is not an artifact due to the present physico-chemical 
estimation of selective constraints.

\vspace*{2em}
\noindent
\subsection*{Evolutionary process of amino acid substitutions in terms of log-odds}
\vspace*{1em}

Kinjo and Nishikawa
\CITE{KN:04} 
reported that
the most principal component of log-odds matrices exhibits
a sharp transition at the sequence identity of 30-35\%, which almost coincides with
the twilight zone in homology search.
This interesting feature of log-odds matrices 
was found by analyzing the eigenspectra
of the log-odds matrices for 18 different levels of sequence identities, which
were constructed from the structure-based alignments of protein sequences 
in the Homstrad database 
\CITE{MDBO:98} 
with the procedure of 
the BLOSUM substitution matrices
\CITE{HH:92}.
Although they did not mention, this feature is also
encoded in an amino acid or codon substitution probability matrix for a short time interval
such as JTT, WAG, LG, and KHG.
Here, we show that this feature is encoded in the transition matrix estimated 
by the ML-91+ model
that precisely reproduces JTT.

\Figure{\ref{fig: log-odds-ev_ML91+}A} shows the first, 
the second and the third principal eigenvalues 
of the log-odds matrix  
$( \mbox{log-}O(\langle S \rangle(t))_{ab} )$ of the ML-91+
are drawn on amino acid identity by solid, broken and dotted lines, 
respectively.
The dependences of these eigenvalues on the amino acid identity
are almost exactly the same as those shown 
in the Fig. 1A of their paper
\CITE{KN:04};
i.e., the first principal eigenvalue changes its sign from negative to positive 
at about 35 \% identity,
and the second principal eigenvalue takes the place of a negative eigenvalue
by changing its sign from positive to negative.
A similar event of exchanging 
the second and the third principal eigenvalues in the order 
occurs between 15 and 20 \% identity in their case
and at about 25 \% identity in the present JTT-ML91+ matrix;
note that the value of sequence identity $x$ \% on the abscissa in their Fig. 1A
\CITE{KN:04}
represents a log-odds matrix compiled from alignments
with sequence identity $\geq x$ \% and $< (x + 10)$ \%.

From \Fig{\ref{fig: log-odds-ev_ML91}A}, one infers that the vector
corresponding to the first principal eigenvector at about 80 \% identity
becomes the second principal eigenvector at about 35 \% identity and 
the third principal eigenvector at about 25 \% identity.
Likewise one infers that
the vector corresponding to the second principal eigenvector 
at about 80 \% identity
becomes the first principal eigenvector below about 35 \% identity,
and
the vector being equal to the third principal eigenvector
at about 80 \% identity 
becomes the second principal eigenvector below 25 \% identity.
This inference is exactly correct, as shown in 
\Figs{\ref{fig: log-odds-ev_ML91}B,
\ref{fig: log-odds-ev_ML91}C,
and
\ref{fig: log-odds-ev_ML91}D}
and in Fig. 1B of Kinjo and Nishikawa
\CITE{KN:04}.
In \Figs{\ref{fig: log-odds-ev_ML91}B,
\ref{fig: log-odds-ev_ML91}C,
and
\ref{fig: log-odds-ev_ML91}D},
the inner product $\VEC{V}_i(t) \cdot \VEC{V}^{\script{JTT}}_j(20\mbox{PAM})$
of the $i$th principal eigenvectors $\VEC{V}_i(t)$ of the JTT-ML91+ log-odds matrix 
at time $t$
and the $j$th principal eigenvectors $\VEC{V}^{\script{JTT}}_j(20\mbox{PAM})$ of 
the JTT log-odds matrix at 20 PAM
is plotted against sequence identity at time $t$.
\Fig{\ref{fig: log-odds-ev_ML91}} indicates that
the eigenvalues change but the eigenvectors remain
almost the same until sequence identity attains about 20 \%.
The sharp exchange between the first and the second principal eigenvalues
is not peculiar to the present substitution matrices but
can occur in any transition matrix in which diagonal elements 
differ from each other; transition matrices generated with
$R_{ab} = \mbox{const} \cdot f_b$ 
have such a characteristic feature.
A critical point is what the principal eigenvectors are
as well as those eigenvalues.

The first principal eigenvalues of the log-odds matrices are
large negative in $t > 40$ \% identity, 
contributing negative values to
the diagonal elements of the log-odds matrices.
Thus, the first principal eigenvector 
with a large negative eigenvalue is a
primary contribution to the mutability of each amino acid, 
as pointed out in Kinjo and Nishikawa
\CITE{KN:04}.
On the other hand, the second and the third principal eigenvalues are
positive, so that the product of $i$th and $j$th elements
of their eigenvectors represents how often the $i$th and the $j$th types of amino acids
can be replaced to each other.  Kinjo and Nishikawa
\CITE{KN:04} 
showed that
the second principal eigenvector is well correlated 
with a hydrophobicity scale of amino acids.

Thus, the sharp transition in the order of the eigenvalues 
contributing to the mutabilities of amino acids and 
to the replaceabilities of amino acid pairs
at about 35 \% identity means that 
the memory of ancestral sequences disappear 
and amino acids in the sequences are replaced with
similar physico-chemical types of amino acids
at about 35 \% identity.
This explains why it becomes hard to identify 
homologous relationships between sequences whose
similarities are less than 35 \% identity
\CITE{KN:04}.
Barriers for identifying sequence homologies
may also exist at about 25 \% and 15 \%, where
the second and the third sharp transitions in 
the order of the eigenvalues occur.
Because conservative substitutions in respect to
physico-chemical properties of amino acids
are required for proteins to fold into their native structures,
the second barrier at about 25 \% corresponds to a threshold
for being able to detect structural homology between proteins.
The similar characteristic features are observed in the mtREV
and the cpREV matrices, too.
Thus, the characteristic features becoming manifest
after a long evolutionary history of proteins
are completely encoded in the transition matrices based 
on the reversible Markov model.
This fact supports in some extent the appropriateness of
the present Markov model to describe the evolutionary process
of codon substitutions.

\clearpage

\pagestyle{empty}

\renewcommand{\FigureInLegends}[1]{#1}

\setcounter{figure}{0}

\renewcommand{\thefigure}{S\arabic{figure}}
\newpage

\renewcommand{\TextFig}[1]{}
\renewcommand{\SupFig}[1]{#1}

\renewcommand{\FigurePanel}[1]{#1}
\renewcommand{\FigureEach}[1]{}

\TextFig{

\FigureEach{

\begin{figure*}[ht]
\FigureInLegends{
\centerline{
\includegraphics*[width=130mm,angle=0]{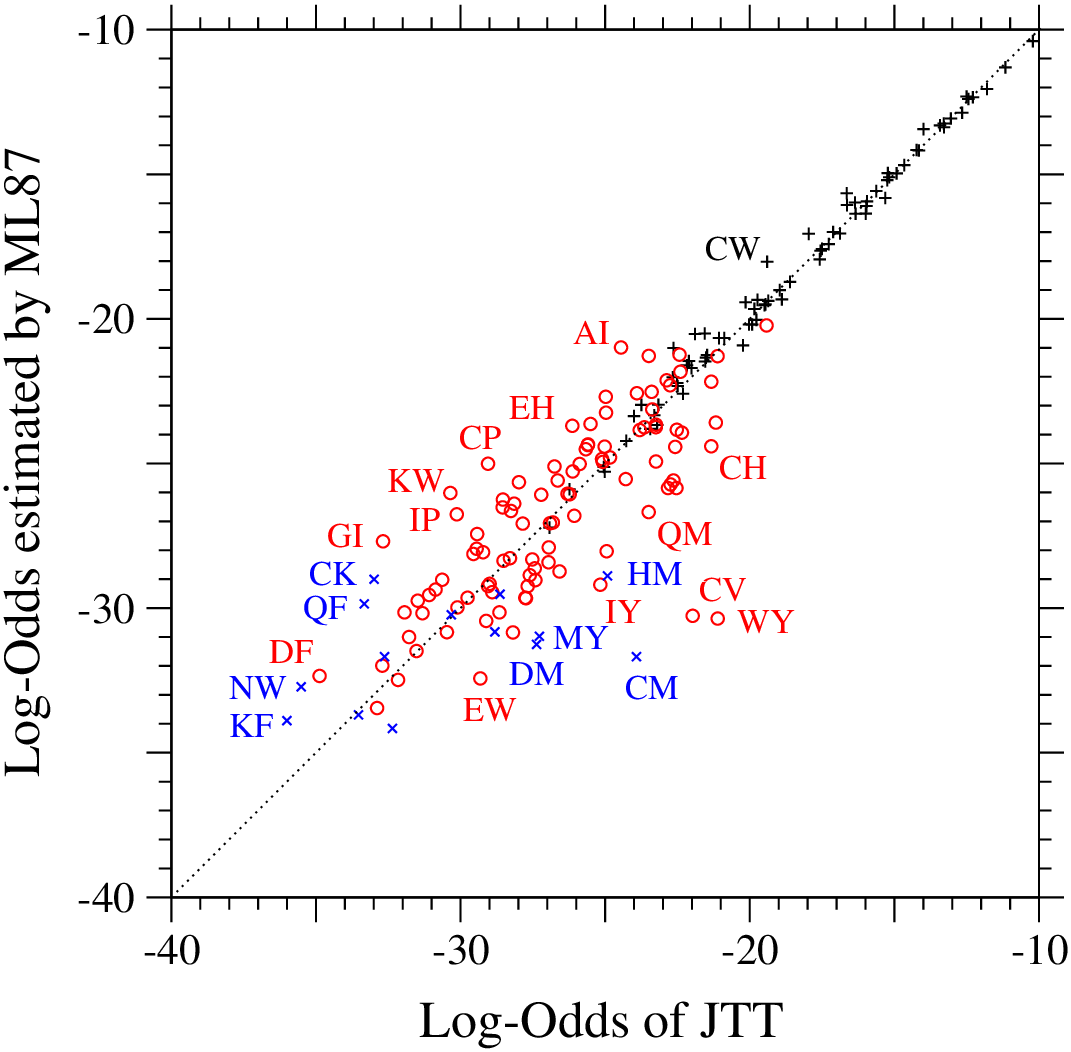}
}
} 
\caption{
\label{fig: ML-87_log-odds_1pam_JTT}
Each element 
log-$O(\langle S \rangle(\hat{\tau},\hat{\sigma}))_{ab}$
of the log-odds matrix 
of the ML-87 model fitted to the 1-PAM JTT matrix
is plotted against the log-odds log-$O(S^{\script{JTT}}(\mbox{1 PAM}))_{ab}$
calculated from JTT.
Plus, circle, and cross marks show the log-odds values for
the types of substitutions requiring single, double and triple nucleotide changes,
respectively.  
The dotted line shows the line 
of equal values between the ordinate and the abscissa.
}
\end{figure*}

\FigureInLegends{\newpage}

\begin{figure*}[ht]
\FigureInLegends{
\centerline{
\includegraphics*[width=130mm,angle=0]{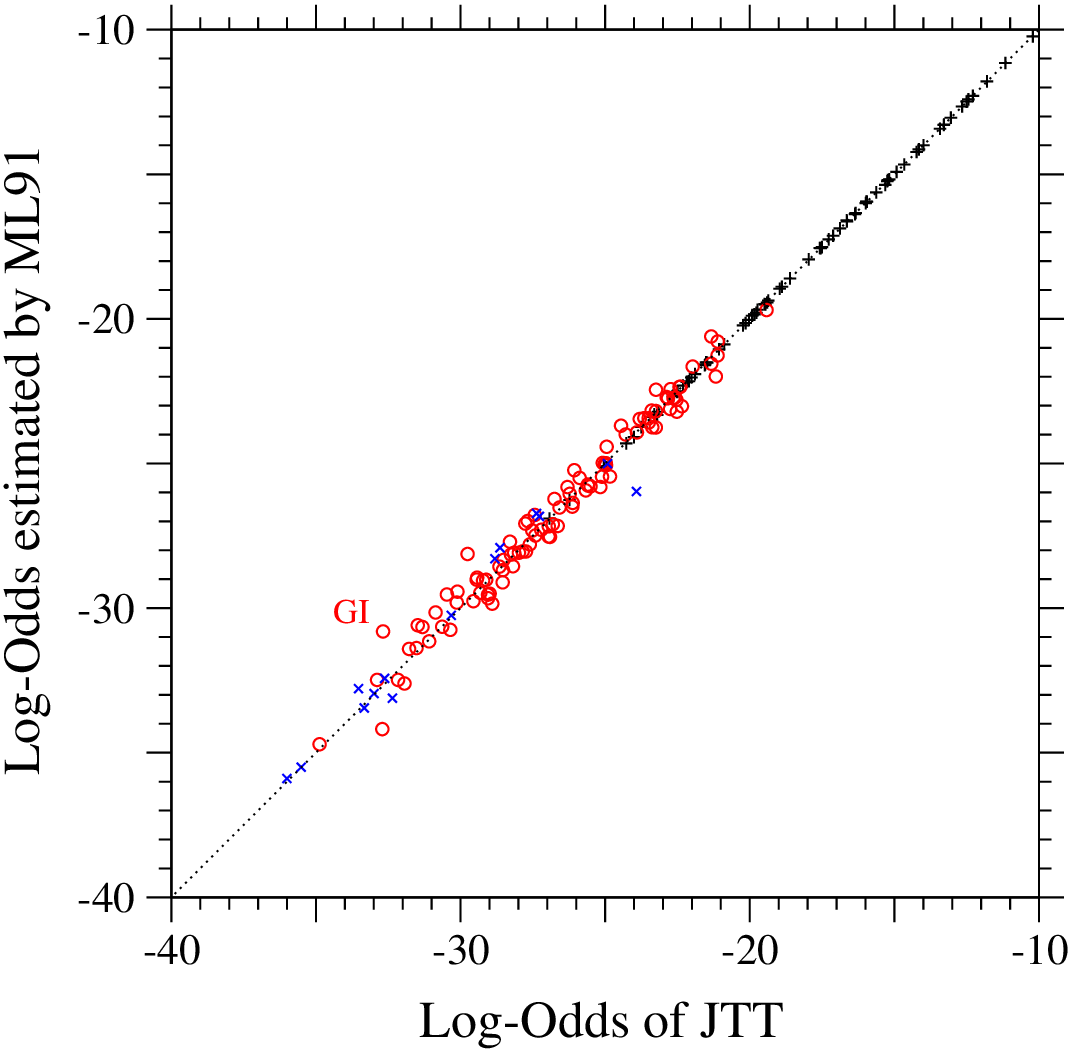}
}
} 
\caption{
\label{fig: ML-91_log-odds_1pam_JTT}
Each element 
log-$O(\langle S \rangle(\hat{\tau},\hat{\sigma}))_{ab}$
of the log-odds matrix 
of the ML-91 model fitted to the 1-PAM JTT matrix
is plotted against the log-odds log-$O(S^{\script{JTT}}(\mbox{1 PAM}))_{ab}$
calculated from JTT.
Plus, circle, and cross marks show the log-odds values for
one-, two-, and three-step amino acid pairs,
respectively.  
The dotted line shows the line 
of equal values between the ordinate and the abscissa.
}
\end{figure*}

} 

\FigurePanel{

\begin{figure}[!ht]
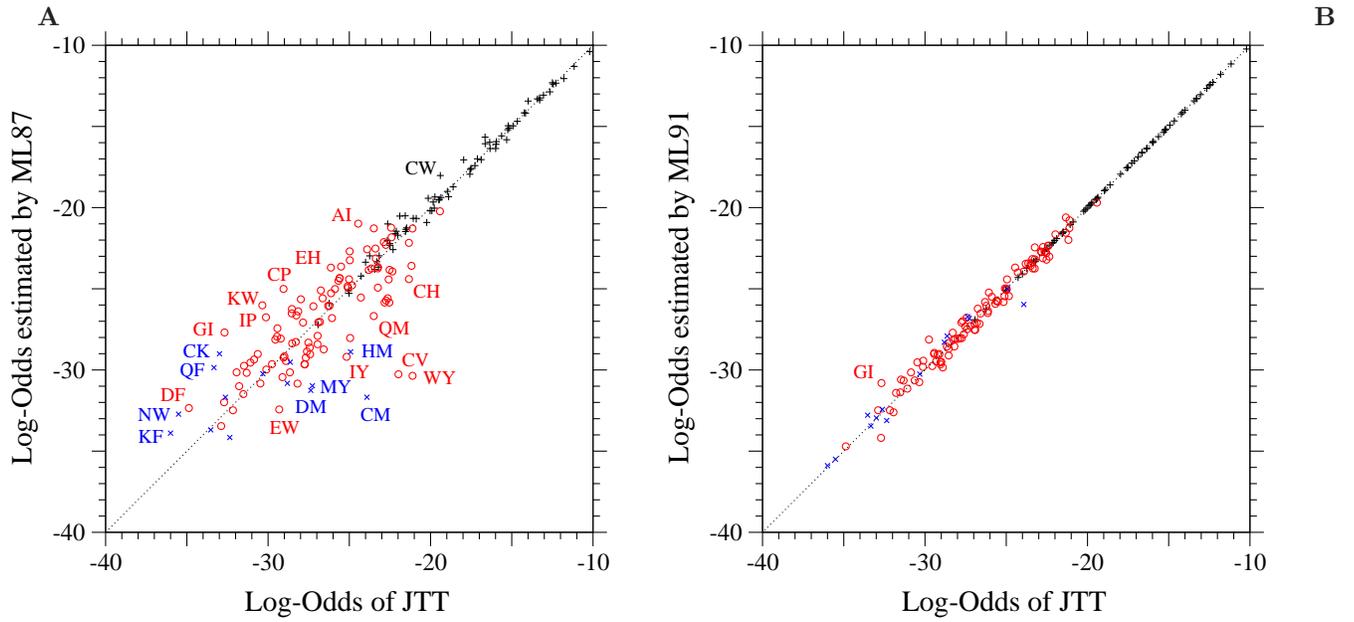

\FigureInLegends{
\textbf{A} \hspace*{47em} \textbf{B}

\centerline{
\includegraphics*[width=80mm,angle=0]{FIGS/ML-87_logodds_1pam.eps}
\hspace*{5mm}
\includegraphics*[width=80mm,angle=0]{FIGS/ML-91_logodds_1pam_JTT.eps}
}
} 
\caption{
\label{fig: ML-87_log-odds_1pam_JTT}
\label{fig: ML-91_log-odds_1pam_JTT}
\BF{
The ML-87 and the ML-91 models fitted to JTT.
} 
Each element 
log-$O(\langle S \rangle(\hat{\tau},\hat{\sigma}))_{ab}$
of the log-odds matrices 
of (A) the ML-87 and (B) the ML-91 models 
fitted to the 1-PAM JTT matrix
is plotted against the log-odds log-$O(S^{\script{JTT}}(\mbox{1 PAM}))_{ab}$
calculated from JTT.
Plus, circle, and cross marks show the log-odds values for
the types of substitutions requiring single, double and triple nucleotide changes,
respectively.  
The dotted line in each figure shows the line 
of equal values between the ordinate and the abscissa.
}
\end{figure}

} 

} 

\SupFig{

\FigureEach{

\begin{figure*}[ht]
\FigureInLegends{
\centerline{
\includegraphics*[width=130mm,angle=0]{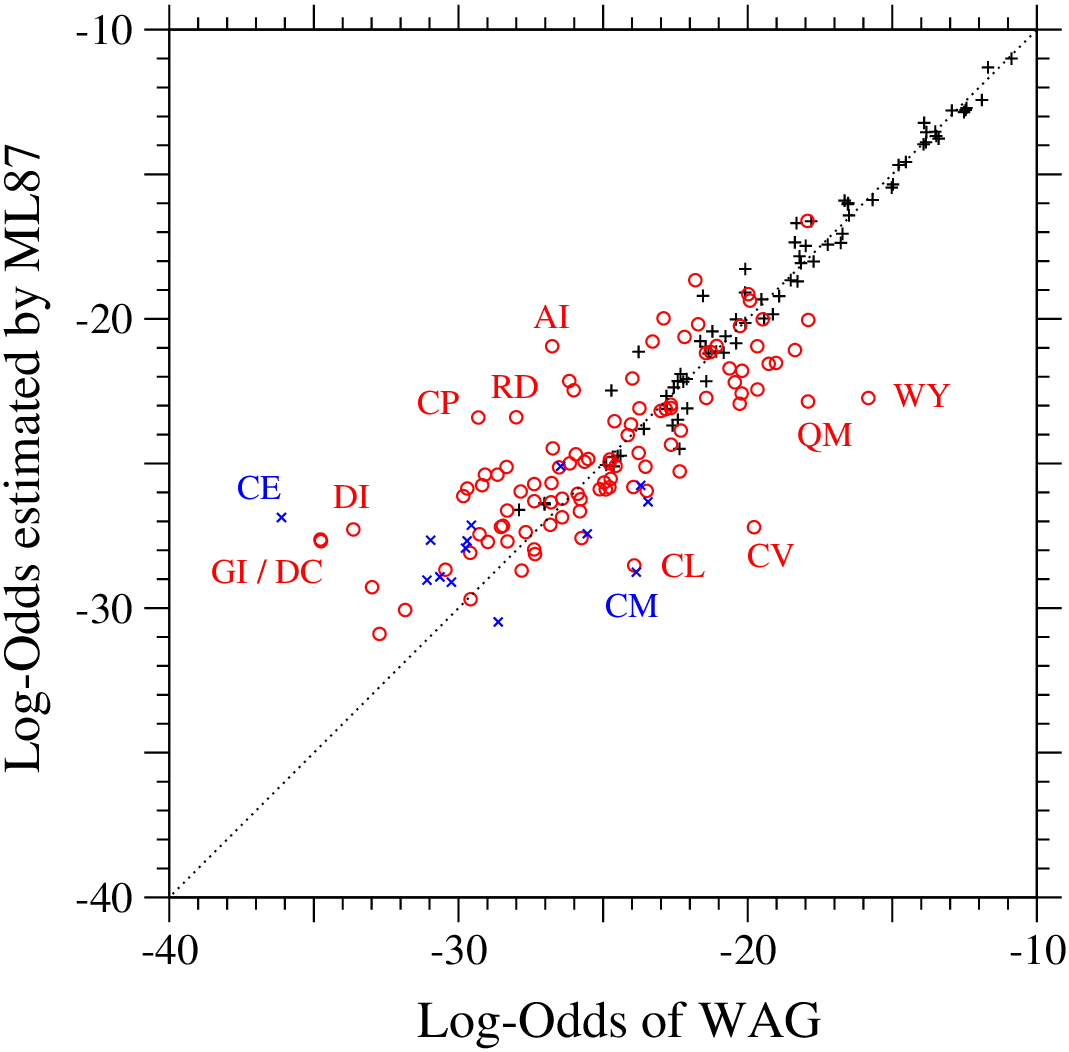}
}
} 
\caption{\label{fig: ML-87_log-odds_1pam_WAG}
Each element 
log-$O(\langle S \rangle(\hat{\tau},\hat{\sigma}))_{ab}$
of the log-odds matrix 
of the ML-87 model fitted to the 1-PAM WAG matrix
is plotted against the log-odds log-$O(S^{\script{WAG}}(\mbox{1 PAM}))_{ab}$
calculated from WAG.
Plus, circle, and cross marks show the log-odds values for
one-, two-, and three-step amino acid pairs,
respectively.  
The dotted line shows the line 
of equal values between the ordinate and the abscissa.
}
\end{figure*}

\begin{figure*}[ht]
\FigureInLegends{
\centerline{
\includegraphics*[width=130mm,angle=0]{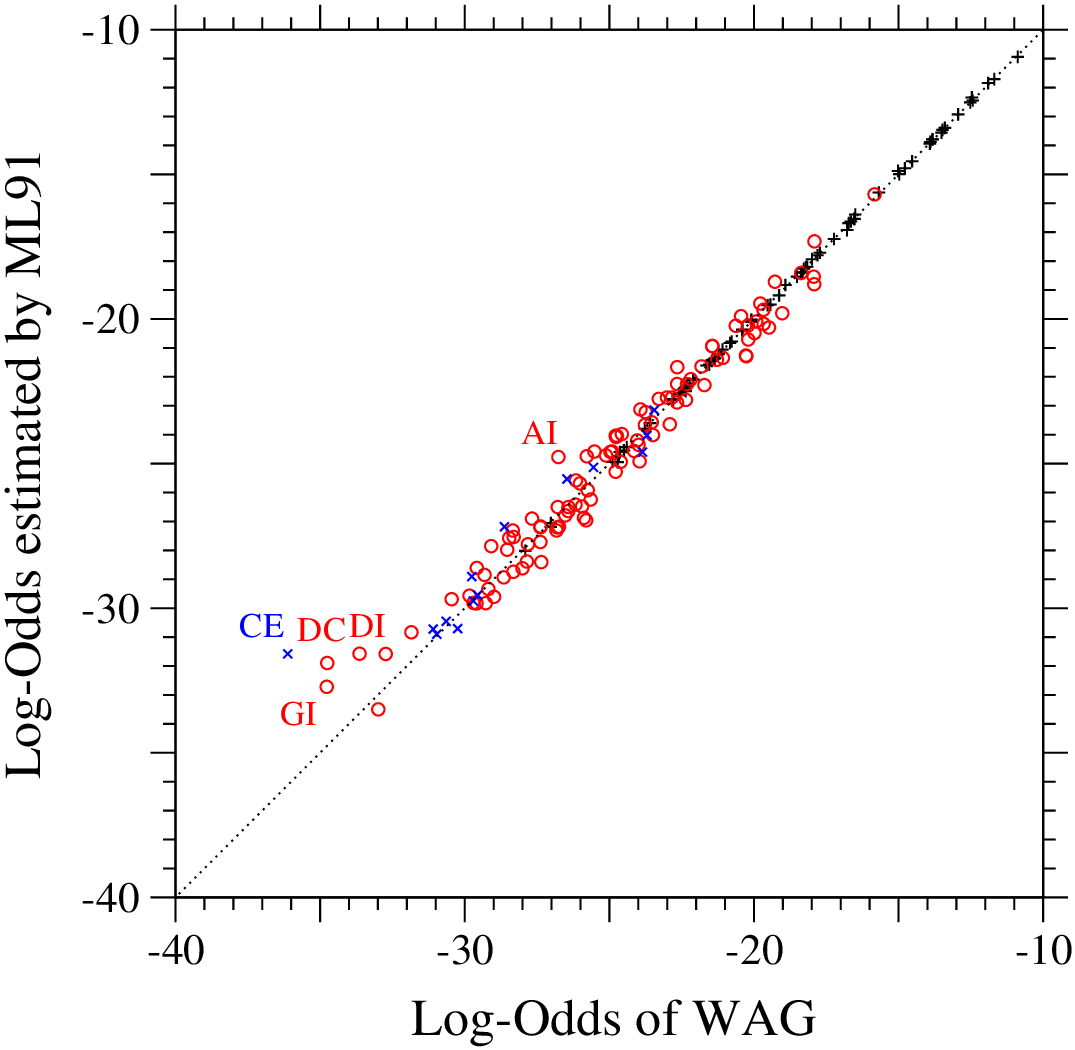}
}
} 
\caption{\label{fig: ML-91_log-odds_1pam_WAG}
Each element 
log-$O(\langle S \rangle(\hat{\tau},\hat{\sigma}))_{ab}$
of the log-odds matrix 
of the ML-91 model fitted to the 1-PAM WAG matrix
is plotted against the log-odds log-$O(S^{\script{WAG}}(\mbox{1 PAM}))_{ab}$
calculated from WAG.
Plus, circle, and cross marks show the log-odds values for
one-, two-, and three-step amino acid pairs,
respectively.  
The dotted line shows the line 
of equal values between the ordinate and the abscissa.
}
\end{figure*}

} 

\FigurePanel{

\FigureInLegends{\newpage}

\begin{figure}[t]
\FigureInLegends{
\textbf{A} \hspace*{47em} \textbf{B}

\centerline{
\includegraphics*[width=80mm,angle=0]{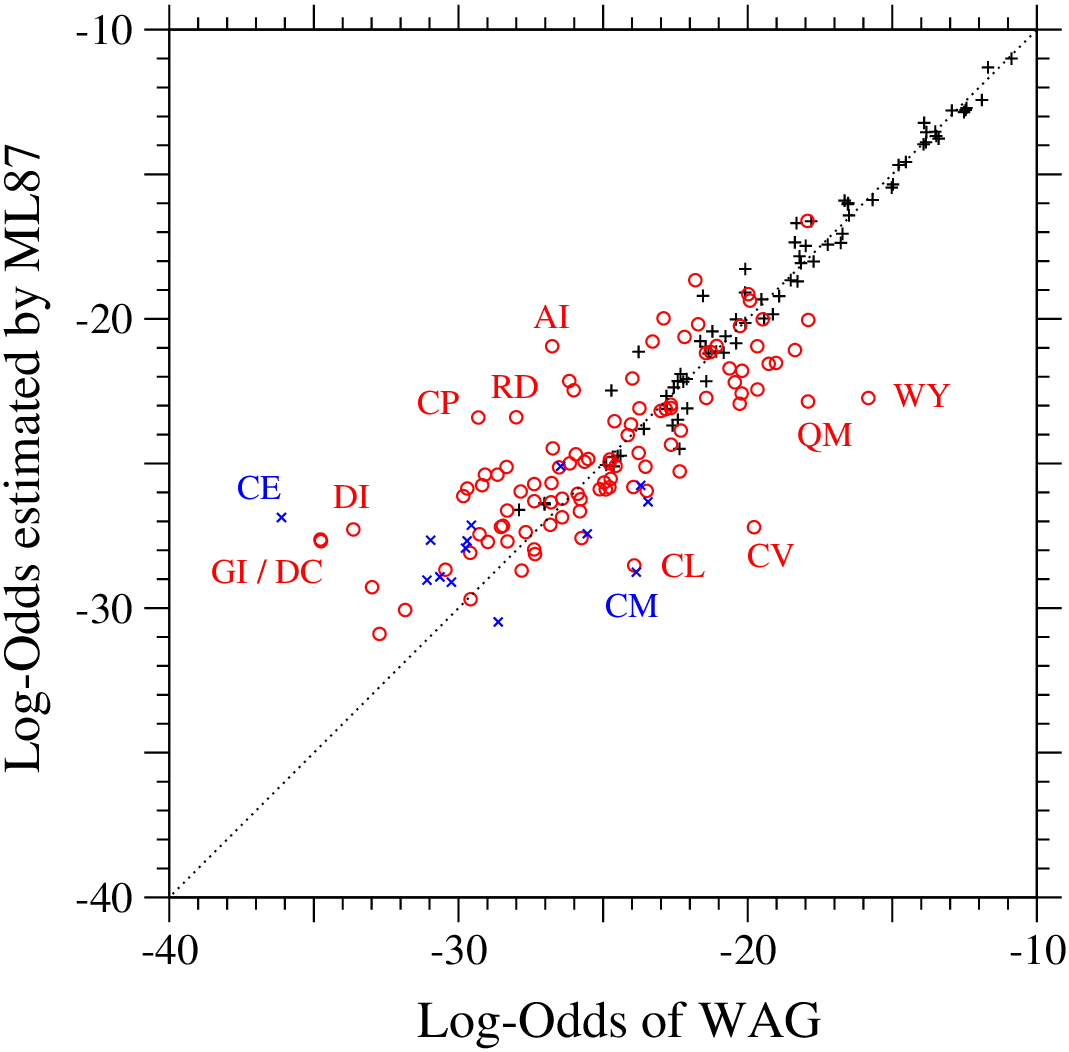}
\hspace*{5mm}
\includegraphics*[width=80mm,angle=0]{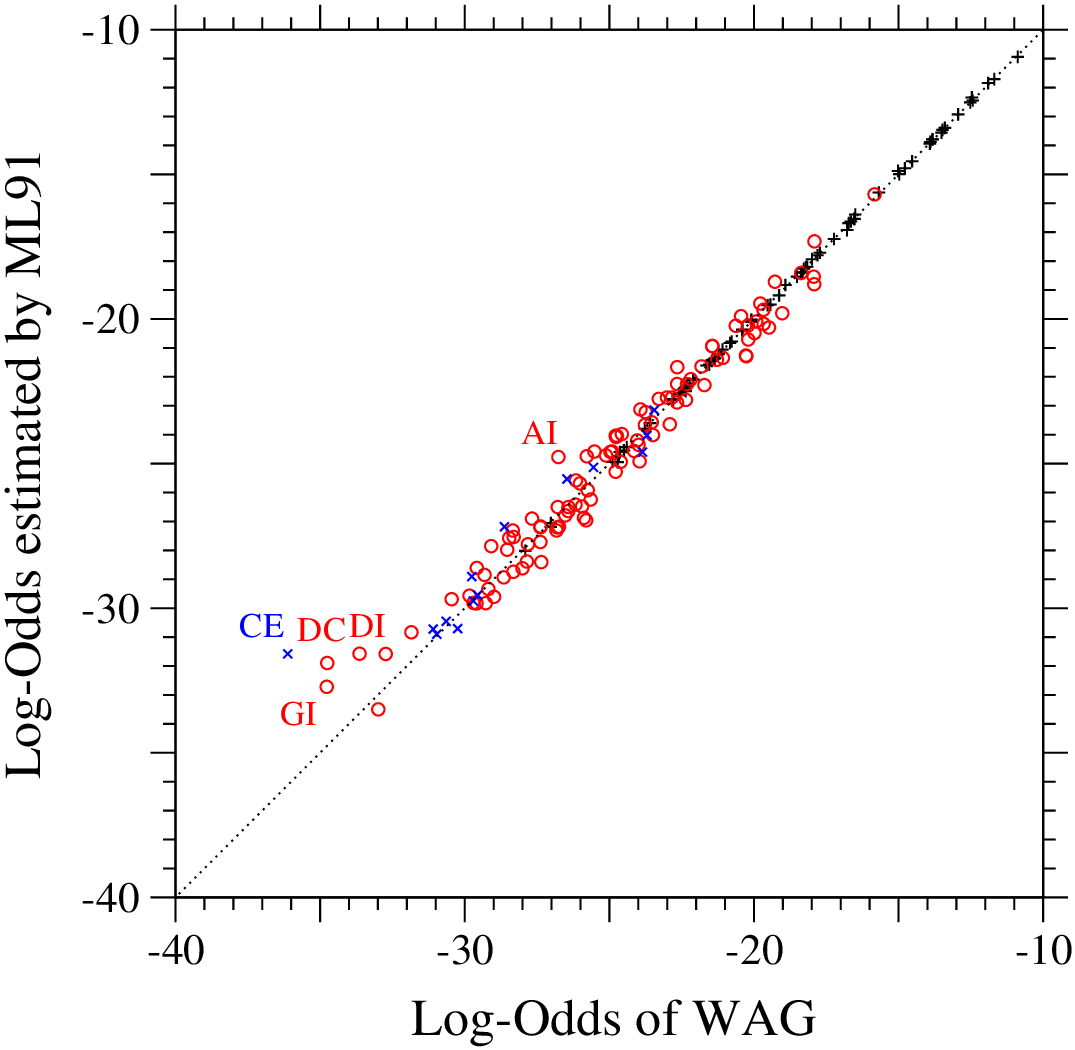}
}
} 
\caption{
\label{fig: ML-87_log-odds_1pam_WAG}
\label{fig: ML-91_log-odds_1pam_WAG}
\BF{
The ML-87 and the ML-91 models fitted to WAG.
} 
Each element 
log-$O(\langle S \rangle(\hat{\tau},\hat{\sigma}))_{ab}$
of the log-odds matrices
of (A) the ML-87 and (B) the ML-91 models fitted to the 1-PAM WAG matrix
is plotted against the log-odds log-$O(S^{\script{WAG}}(\mbox{1 PAM}))_{ab}$
calculated from WAG.
Plus, circle, and cross marks show the log-odds values for
one-, two-, and three-step amino acid pairs,
respectively.  
The dotted line in each figure shows the line 
of equal values between the ordinate and the abscissa.
}
\end{figure}

} 

} 

\TextFig{

\FigureEach{

\FigureInLegends{\newpage}

\begin{figure*}[ht]
\FigureInLegends{
\centerline{
\includegraphics*[width=130mm,angle=0]{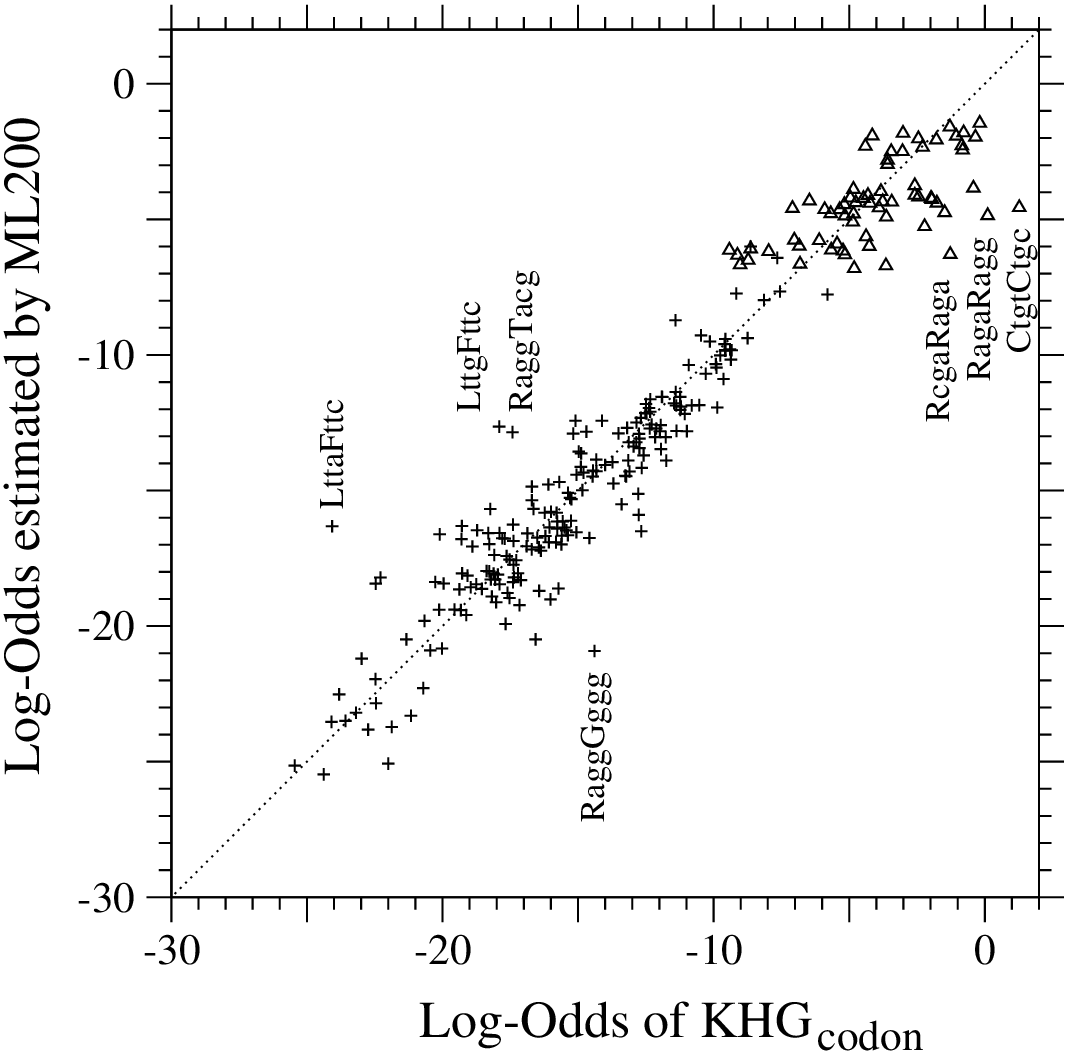}
}
} 
\caption{\label{fig: ML200_logodds_codon_1pam_KHG_1x}
Each element 
log-$O(\langle S \rangle(\hat{\tau},\hat{\sigma}))_{\mu \nu}$
of the log-odds matrix 
corresponding to single nucleotide changes 
in the ML-200 model fitted to the 1-PAM KHG codon substitution matrix
is plotted against the log-odds log-$O(S^{\script{KHG}}(\mbox{1 PAM}))_{\mu \nu}$
calculated from KHG.
Upper triangle and plus marks show the log-odds values for
synonymous pairs and single-step amino acid pirs,
respectively. 
The dotted line shows the line 
of equal values between the ordinate and the abscissa.
}
\end{figure*}

\FigureInLegends{\newpage}

\begin{figure*}[ht]
\FigureInLegends{
\centerline{
\includegraphics*[width=130mm,angle=0]{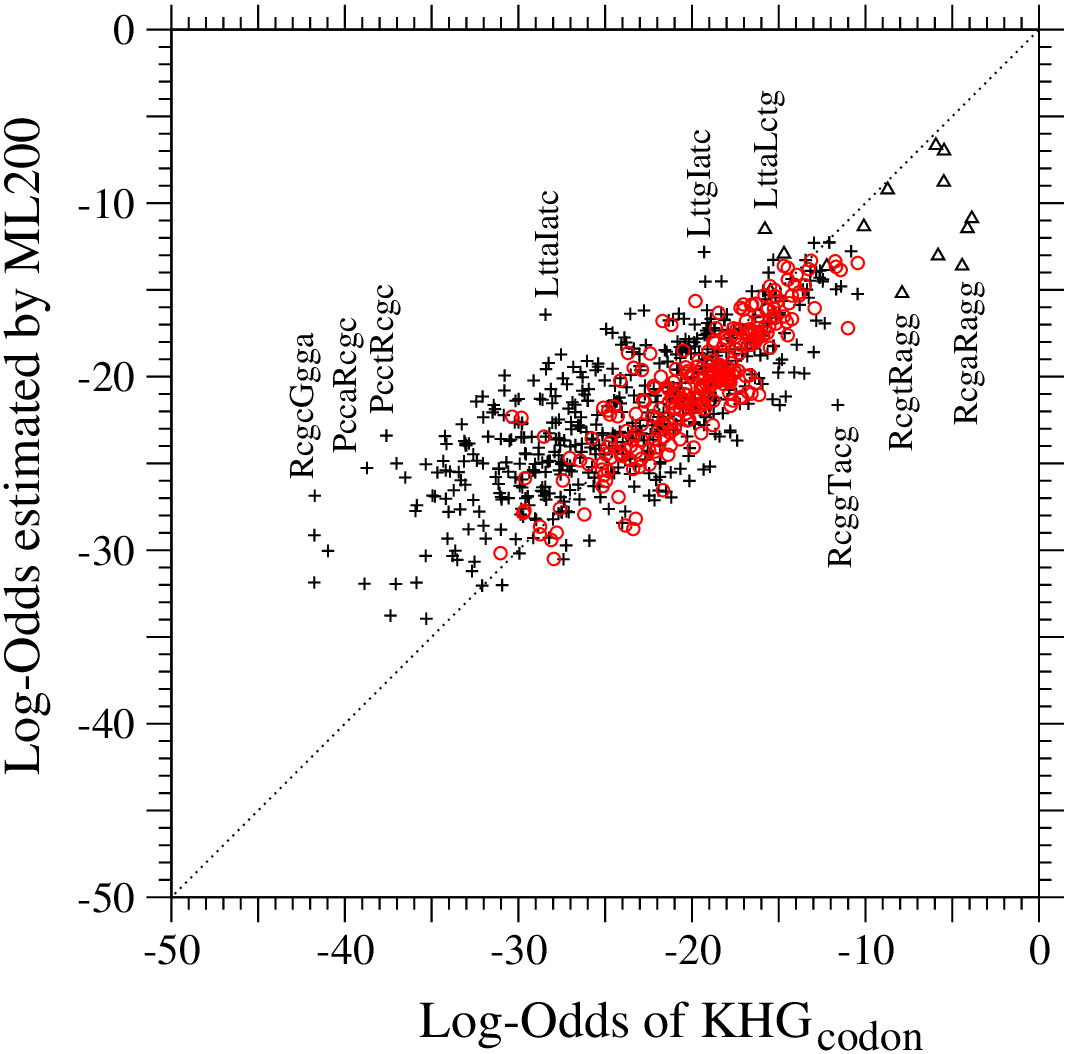}
}
} 
\caption{\label{fig: ML200_logodds_codon_1pam_KHG_2x}
Each element 
log-$O(\langle S \rangle(\hat{\tau},\hat{\sigma}))_{\mu \nu}$
of the log-odds matrix 
corresponding to double nucleotide changes 
in the ML-200 model fitted to the 1-PAM KHG codon substitution matrix
is plotted against the log-odds log-$O(S^{\script{KHG}}(\mbox{1 PAM}))_{\mu \nu}$
calculated from KHG.
Upper triangle, plus, and circle marks show the log-odds values for
synonymous pairs and one-, and two-step amino acid pairs,
respectively. 
The dotted line shows the line 
of equal values between the ordinate and the abscissa.
}
\end{figure*}

\FigureInLegends{\newpage}

\begin{figure*}[ht]
\FigureInLegends{
\centerline{
\includegraphics*[width=130mm,angle=0]{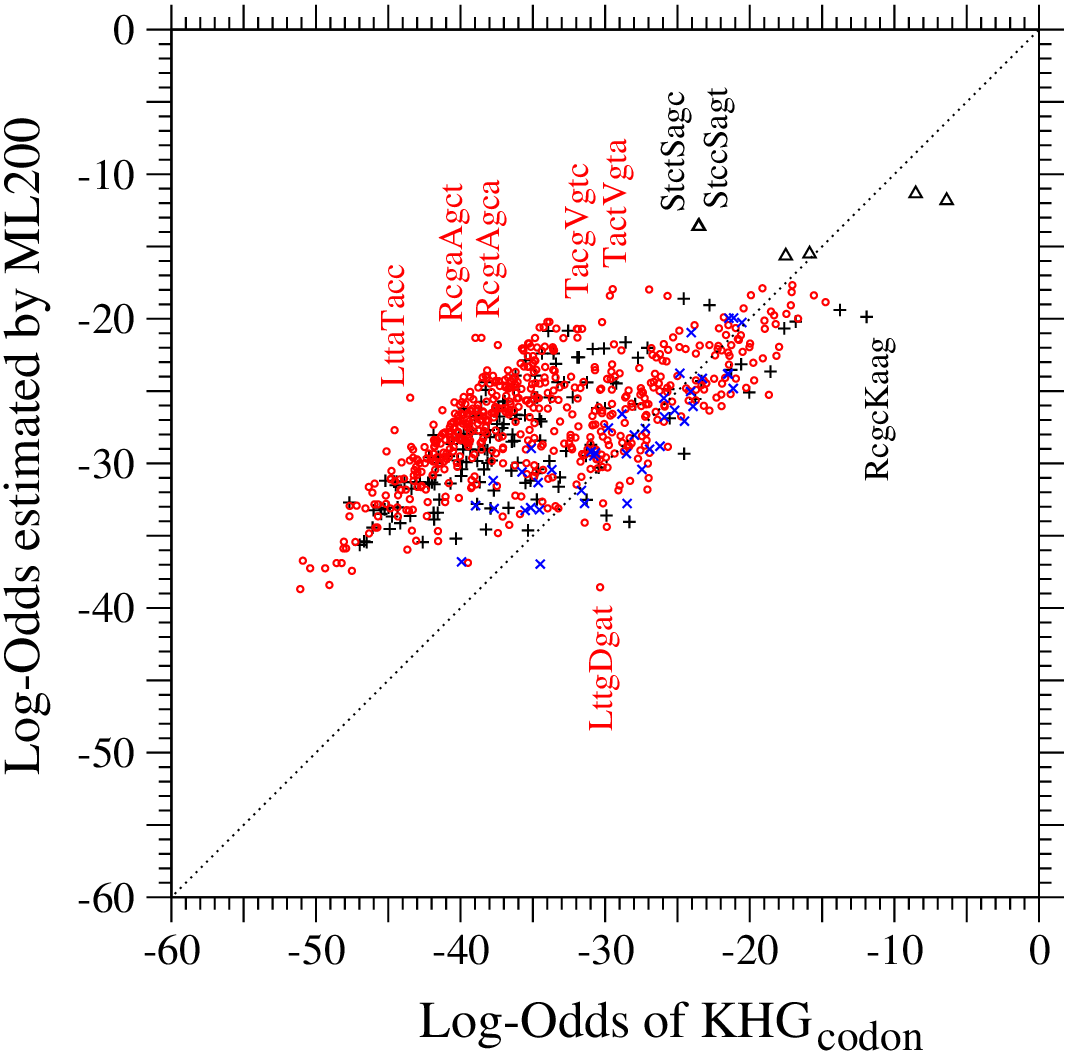}
}
} 
\caption{\label{fig: ML200_logodds_codon_1pam_KHG_3x}
Each element 
log-$O(\langle S \rangle(\hat{\tau},\hat{\sigma}))_{\mu \nu}$
of the log-odds matrix 
corresponding to triple nucleotide changes 
in the ML-200 model fitted to the 1-PAM codon KHG substitution matrix
is plotted against the log-odds log-$O(S^{\script{KHG}}(\mbox{1 PAM}))_{\mu \nu}$
calculated from KHG.
Upper triangle, plus, circle, and cross marks show the log-odds values for
synonymous pairs and one-, two-, and three-step amino acid pairs,
respectively. 
The dotted line shows the line 
of equal values between the ordinate and the abscissa.
}
\end{figure*}

\FigureInLegends{\newpage}

\begin{figure*}[ht]
\FigureInLegends{
\centerline{
\includegraphics*[width=130mm,angle=0]{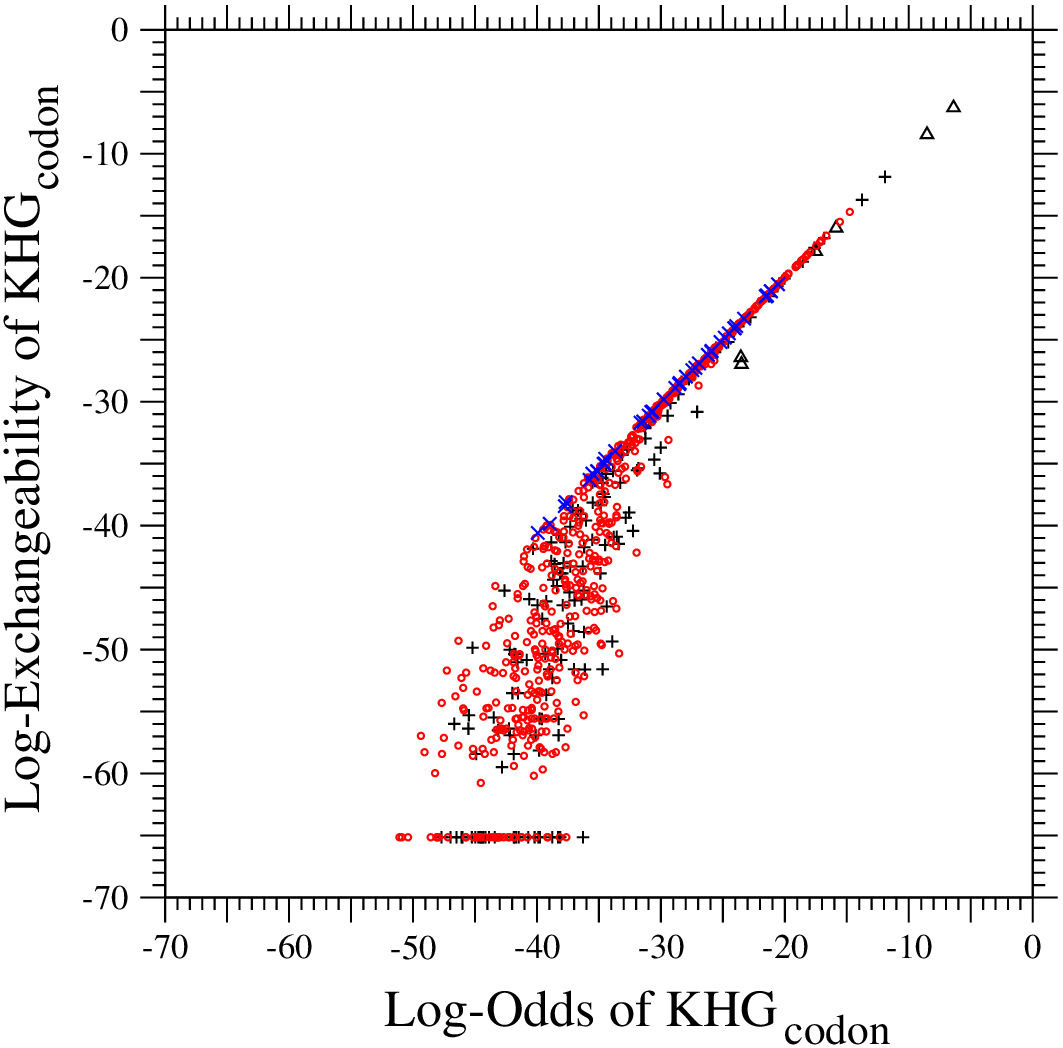}
}
} 
\caption{\label{fig: ML200_logodds_vs_exchange_scaled_1pam_observed_KHG_3x}
\label{fig: logodds_vs_exchange_scaled_1pam_observed_KHG_3x}
Codon log-exchangeabilities of codon substitutions corresponding to triple nucleotide changes in the 1-PAM KHG
are plotted against the log-odds log-$O(S^{\script{KHG}}(\mbox{1 PAM}))_{\mu \nu}$
calculated from KHG.
The log-exchangeability of the 1-PAM KHG 
is defined as $(10 / \log 10) \log [R^{\script{KHG}}_{\mu\nu} \cdot t_{\script{1-PAM}} / f_{\nu} ]$.
Upper triangle, plus, circle, and cross marks show the log-odds values for
synonymous pairs and one-, two-, and three-step amino acid pairs,
respectively. 
Log-exchangeabilities for the codon pairs whose instantaneous rate is estimated to be $0$ in KHG are
shown to be about $-65$ in this figure.
The dotted line shows the line 
of equal values between the ordinate and the abscissa.
}
\end{figure*}

} 

\FigurePanel{

\FigureInLegends{\newpage}

\begin{figure}[ht]
\FigureInLegends{
\textbf{A} \hspace*{47em} \textbf{B}

\centerline{
\includegraphics*[width=80mm,angle=0]{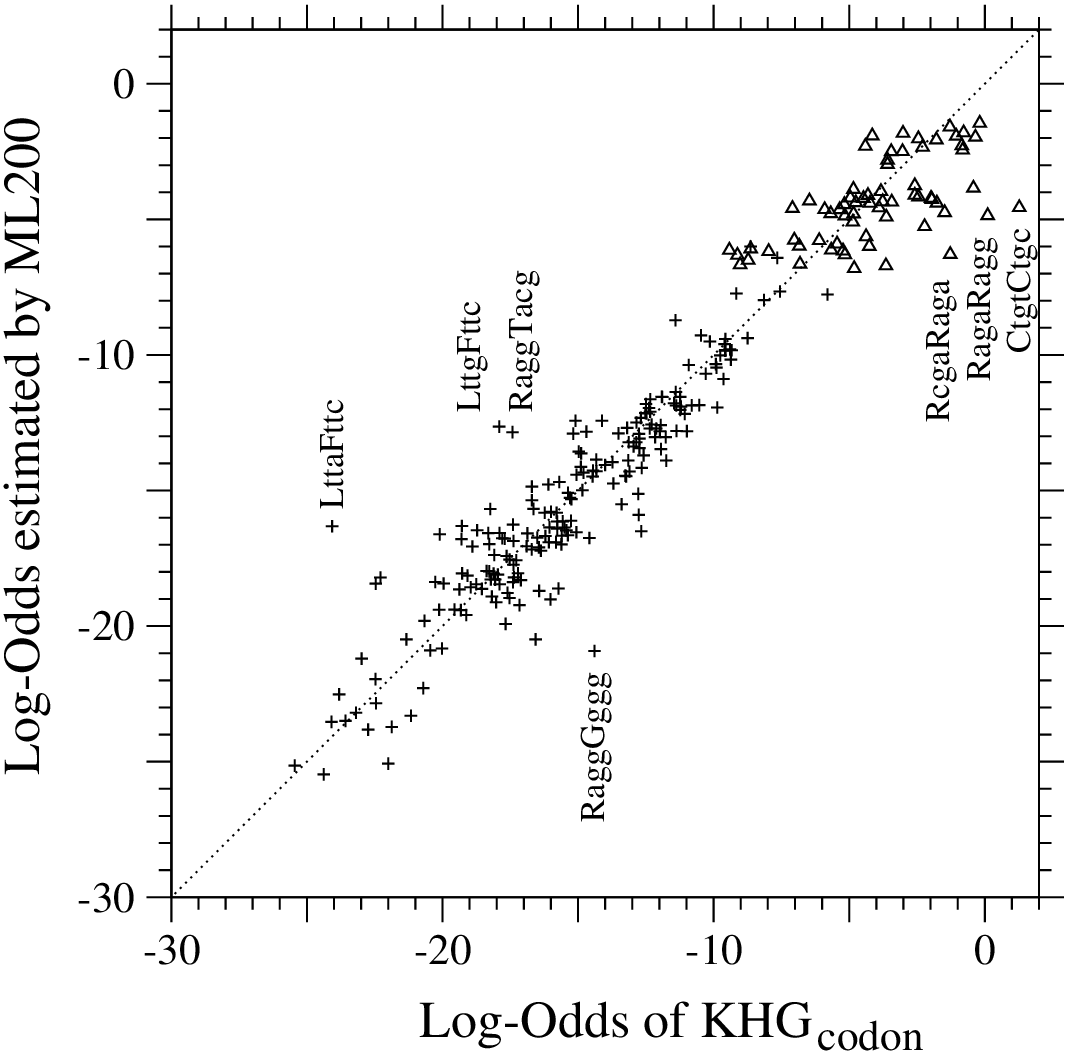}
\hspace*{5mm}
\includegraphics*[width=80mm,angle=0]{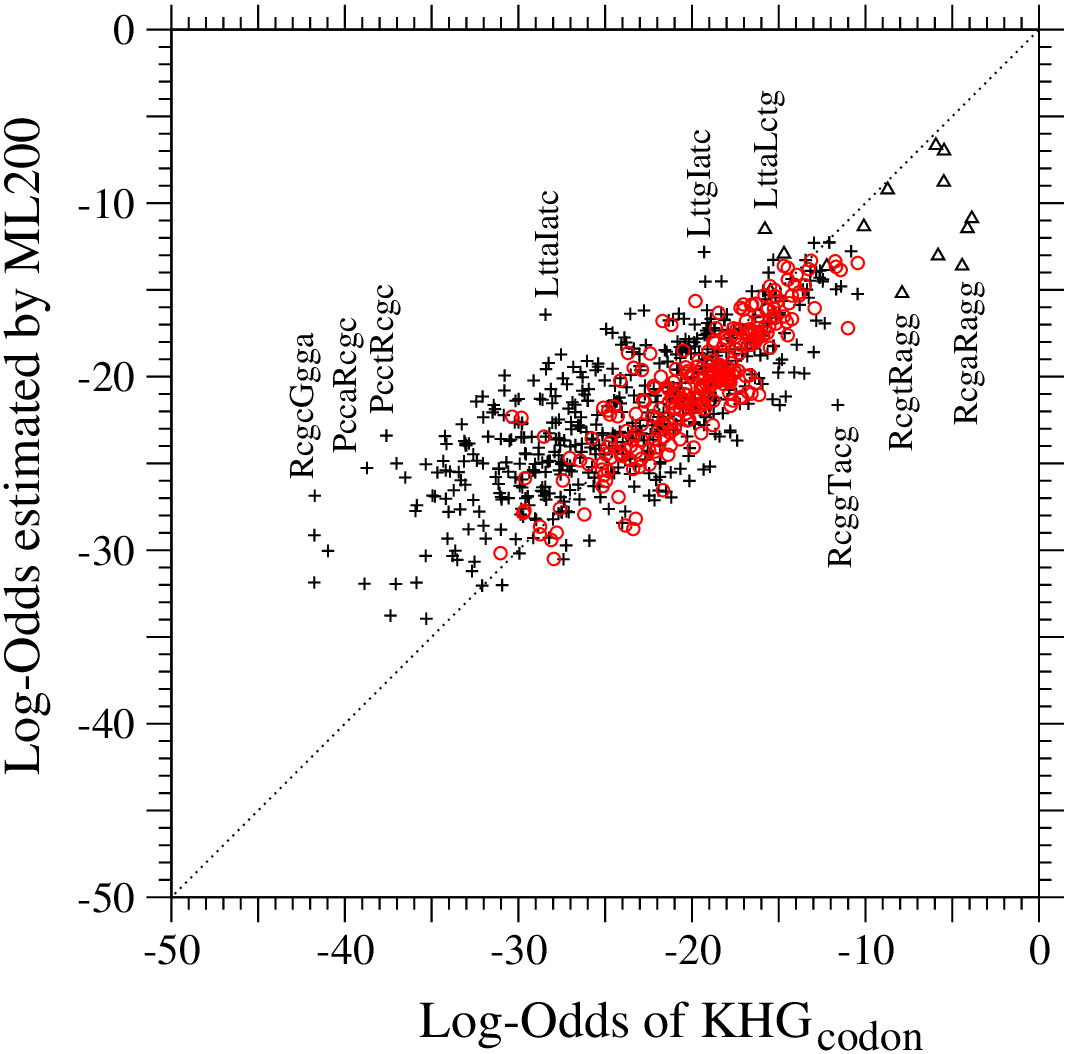}
}
\vspace*{1em}
\centerline{
\includegraphics*[width=80mm,angle=0]{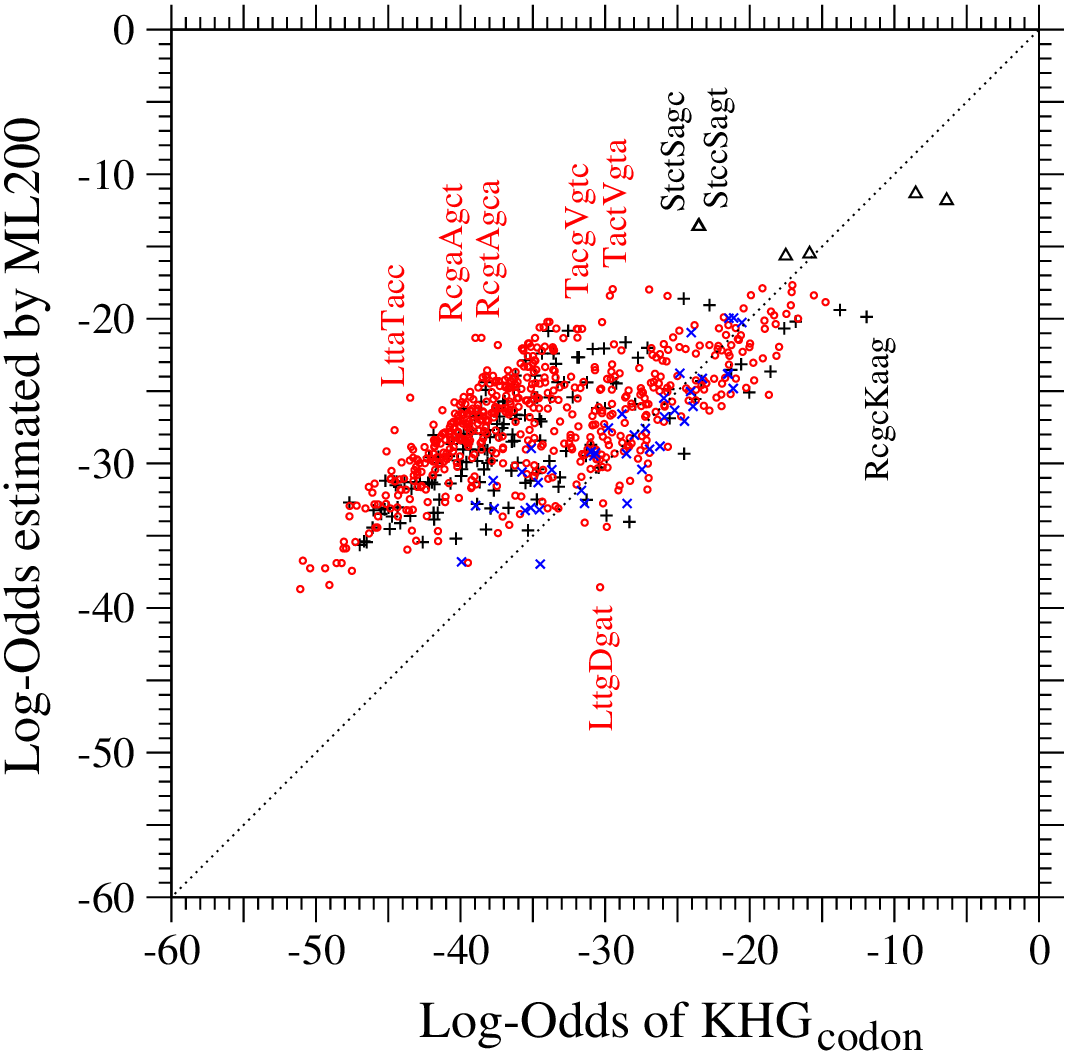}
\hspace*{5mm}
\includegraphics*[width=80mm,angle=0]{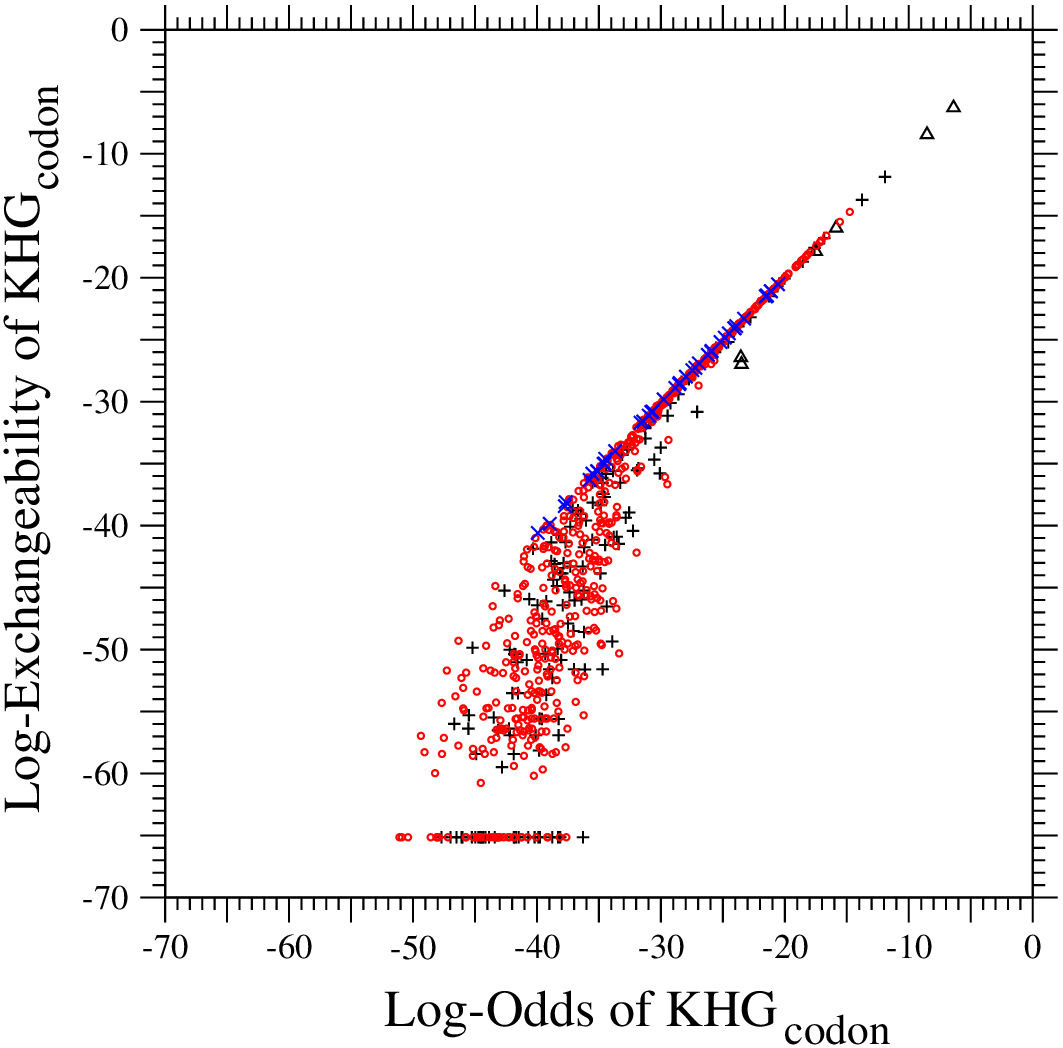}
}

\textbf{C} \hspace*{47em} \textbf{D}
} 
\caption{
\label{fig: ML200_logodds_codon_1pam_KHG}
\label{fig: ML200_logodds_codon_1pam_KHG_1x}
\label{fig: ML200_logodds_codon_1pam_KHG_2x}
\label{fig: ML200_logodds_codon_1pam_KHG_3x}
\label{fig: ML200_logodds_vs_exchange_scaled_1pam_observed_KHG_3x}
\label{fig: logodds_vs_exchange_scaled_1pam_observed_KHG_3x}
\BF{
The ML-200 model fitted to KHG.
} 
Each element 
log-$O(\langle S \rangle(\hat{\tau},\hat{\sigma}))_{\mu \nu}$
of the log-odds matrix 
corresponding to (A) single, (B) double, and (C) triple nucleotide changes 
in the ML-200 model fitted to the 1-PAM KHG codon substitution matrix
is plotted against the log-odds log-$O(S^{\script{KHG}}(\mbox{1 PAM}))_{\mu \nu}$
calculated from KHG.
In (D),
codon log-exchangeabilities of the 1-PAM KHG codon substitution matrix corresponding to triple nucleotide changes 
are plotted against the log-odds log-$O(S^{\script{KHG}}(\mbox{1 PAM}))_{\mu \nu}$
calculated from KHG.
The log-exchangeability of the 1-PAM KHG 
is defined as $(10 / \log 10) \log [R^{\script{KHG}}_{\mu\nu} \cdot t_{\script{1-PAM}} / f_{\nu} ]$.
Upper triangle, plus, circle, and cross marks show the log-odds values for
synonymous pairs and one-, two-, and three-step amino acid pairs,
respectively. 
Log-exchangeabilities for the codon pairs whose instantaneous rates are estimated to be $0$ in KHG are
shown to be about $-65$ in this figure.
The dotted line in each figure shows the line 
of equal values between the ordinate and the abscissa.
}
\end{figure}

} 

} 

\TextFig{

\FigureEach{
\FigureInLegends{\newpage}

\begin{figure*}[ht]
\FigureInLegends{
\centerline{
\includegraphics*[width=130mm,angle=0]{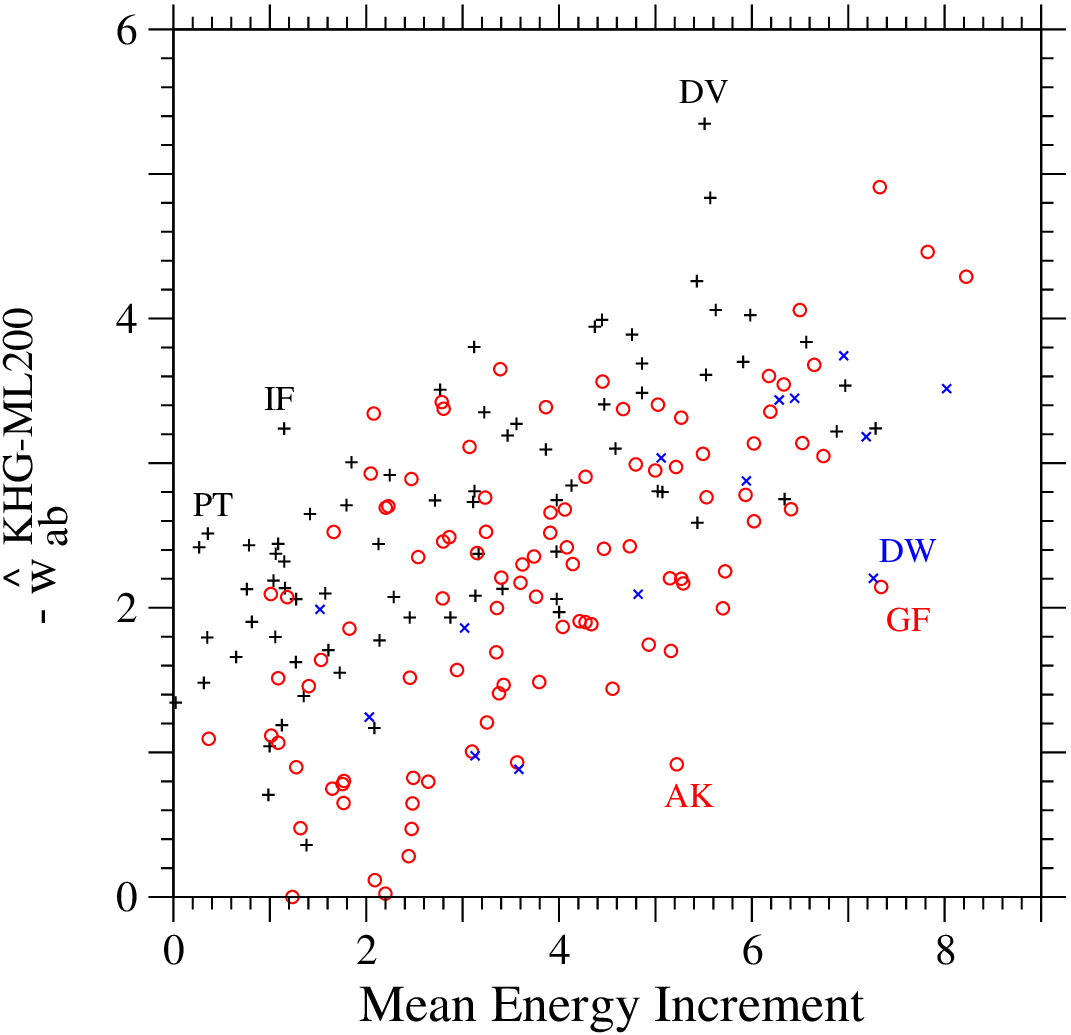}
}
} 
\caption{\label{fig: wij_KHG-ML199_vs_EI}
\label{fig: wij_KHG-ML200_vs_EI}
The ML estimate, $\hat{w}^{\script{KHG-ML200}}_{ab}$, 
of selective constraints on substitutions of the single step amino acid pair
in the ML-200 model fitted to the 1-PAM KHG matrix
is plotted against the mean energy increment,
($\Delta \hat{\varepsilon}^{\script{c}}_{ab} + \Delta \hat{\varepsilon}^{\script{v}}_{ab}$) 
defined in Supporting Information, Text S1.
due to an amino acid substitution.
Plus, circle, and cross marks show the values for
one-, two-, and three-step amino acid pairs,
respectively. 
The correlation coefficient between the axes is
equal to 0.71 for single-step amino acid pairs,
0.65 for multi-step amino acid pairs,
and 0.60 for all amino acid pairs.
}
\end{figure*}
} 

\FigurePanel{
\FigureInLegends{\newpage}

\begin{figure}[ht]
\FigureInLegends{
\textbf{A} \hspace*{47em} \textbf{B}

\centerline{
\includegraphics*[width=80mm,angle=0]{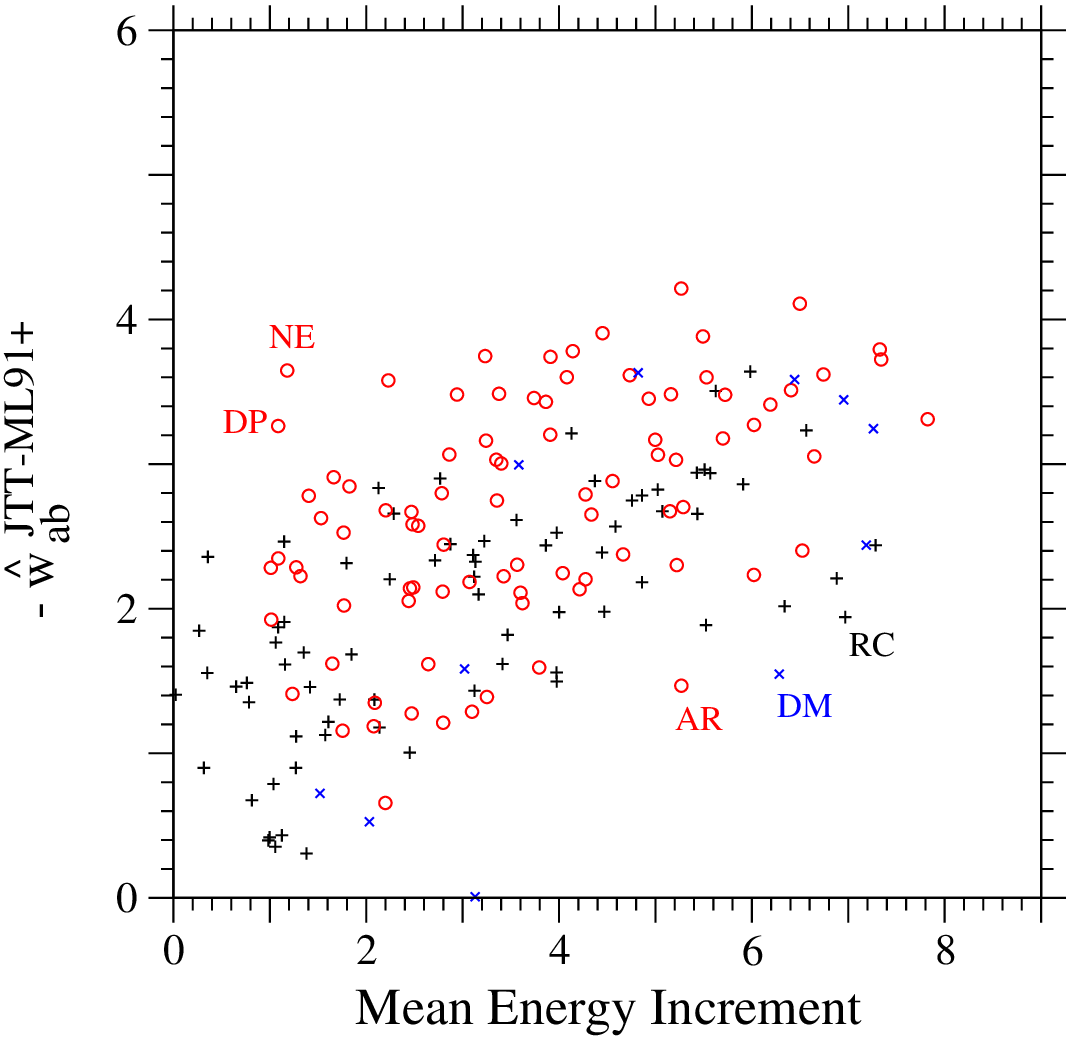}
\hspace*{5mm}
\includegraphics*[width=80mm,angle=0]{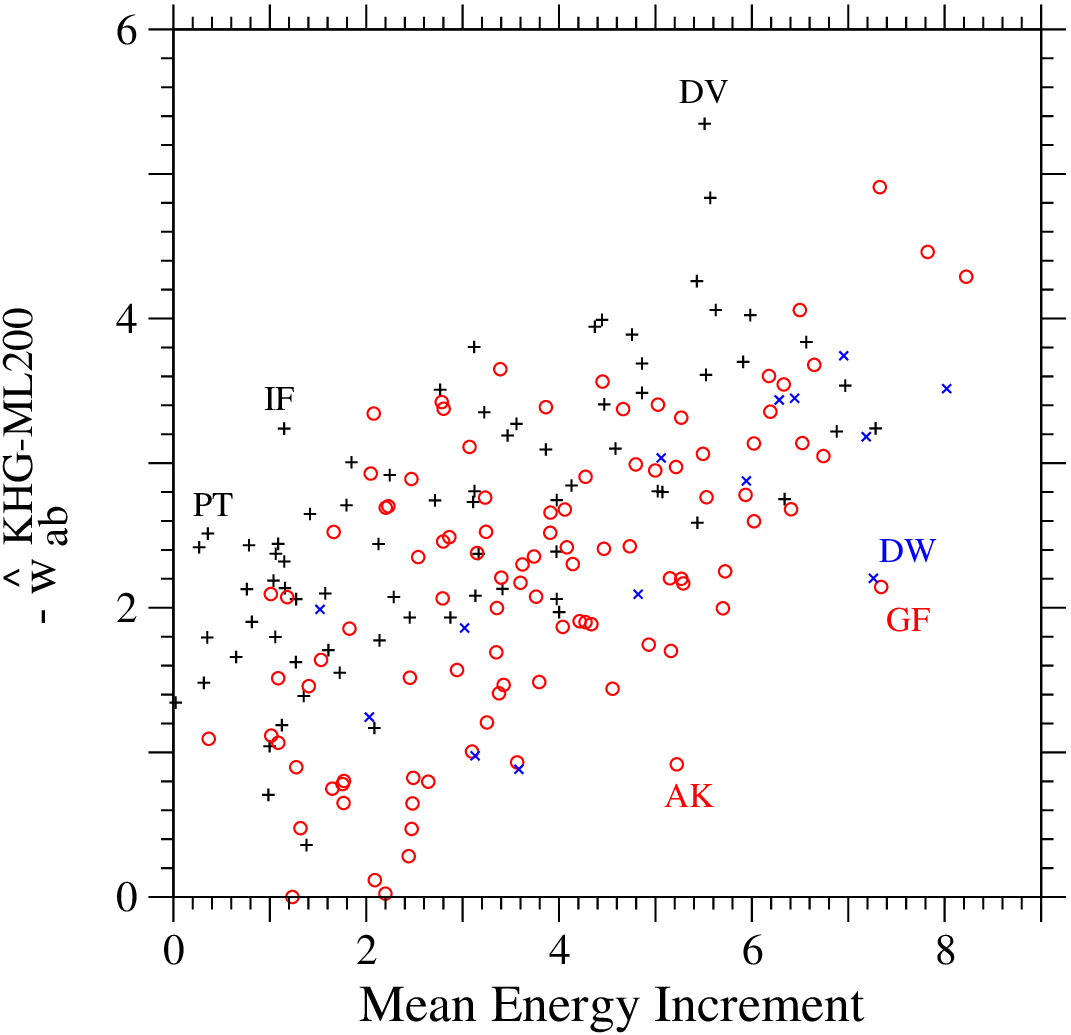}
}
} 
\caption{
\label{fig: wij_vs_energy_increments}
\label{fig: wij_JTT-ML91+_vs_EI}
\label{fig: wij_KHG-ML199_vs_EI}
\label{fig: wij_KHG-ML200_vs_EI}
\BF{
Selective constraint for each amino acid pair estimated from JTT and from KHG.
} 
The ML estimate, (A) $- \hat{w}^{\script{JTT-ML91+}}_{ab}$ 
in the ML-91+ model fitted to the 1-PAM JTT amino acid substitution matrix
and (B) $- \hat{w}^{\script{KHG-ML200}}_{ab}$
in the ML-200 model fitted to the 1-PAM KHG codon substitution matrix,
for each amino acid pair is plotted against the mean energy increment 
due to an amino acid substitution,
($\Delta \hat{\varepsilon}^{\script{c}}_{ab} + \Delta \hat{\varepsilon}^{\script{v}}_{ab}$) 
defined 
by \SEqs{4, S1-5, and S1-6}.
In (A),
the estimates $\hat{w}_{ab}$ for
the least exchangeable class of multi-step amino acid pairs
are not shown.
Plus, circle, and cross marks show the values for
one-, two-, and three-step amino acid pairs,
respectively. 
}
\end{figure}
} 

} 

\SupFig{

\FigureEach{

\FigureInLegends{\newpage}

\begin{figure*}[ht]
\FigureInLegends{
\centerline{
\includegraphics*[width=130mm,angle=0]{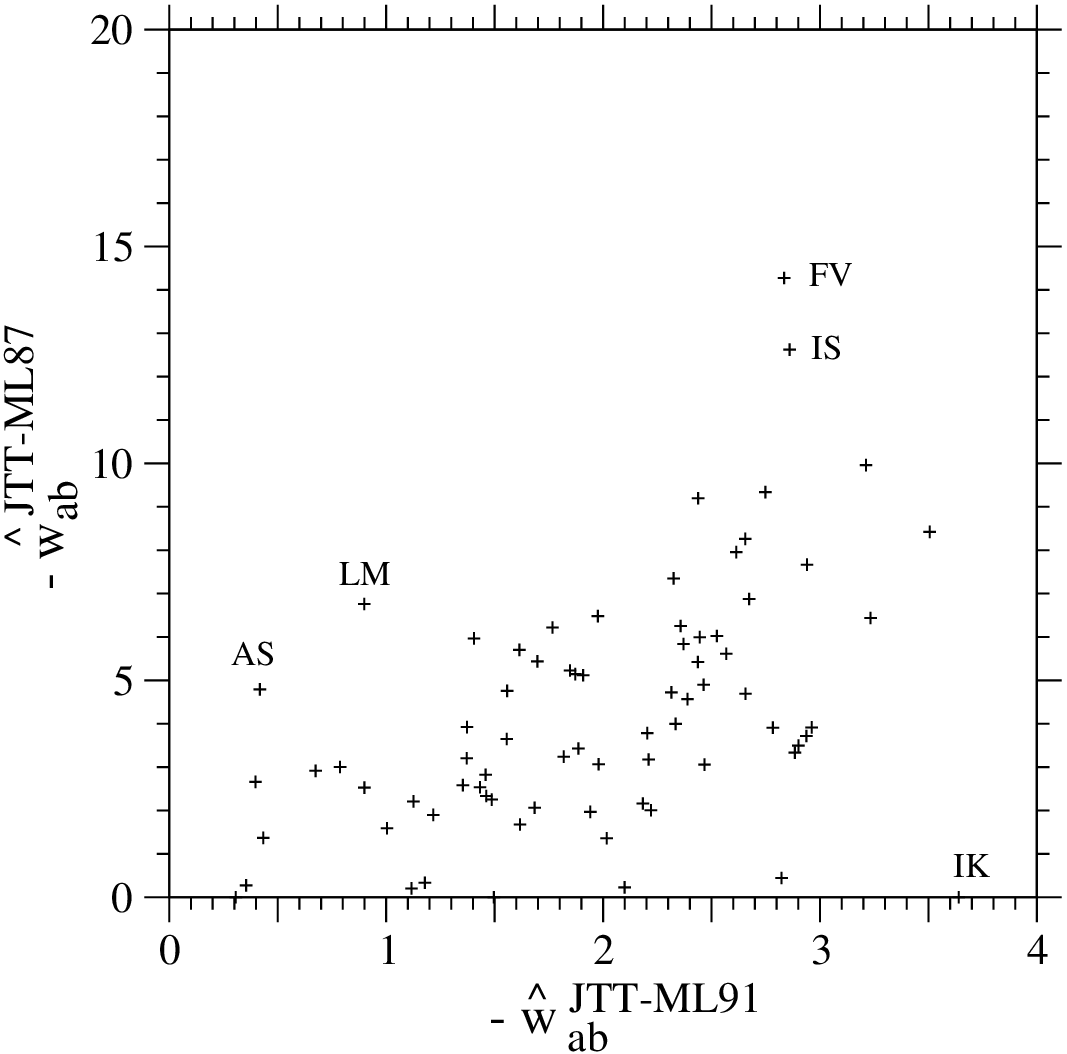}
}
} 
\caption{\label{fig: wij_JTT-ML87_vs_JTT-ML91}
\label{fig: wij1_JTT-ML87_vs_JTT-ML91}
\label{fig: wij1_ML87_vs_ML91}
The ML estimator, $\hat{w}_{ab}$, 
of selective constraints on
substitutions of the single step amino acid pair
in the ML-87 model fitted to the 1-PAM JTT matrix
is plotted against that in the ML-91 model.
}
\end{figure*}

} 

\FigureEach{

\FigureInLegends{\newpage}

\begin{figure*}[ht]
\FigureInLegends{
\centerline{
\includegraphics*[width=130mm,angle=0]{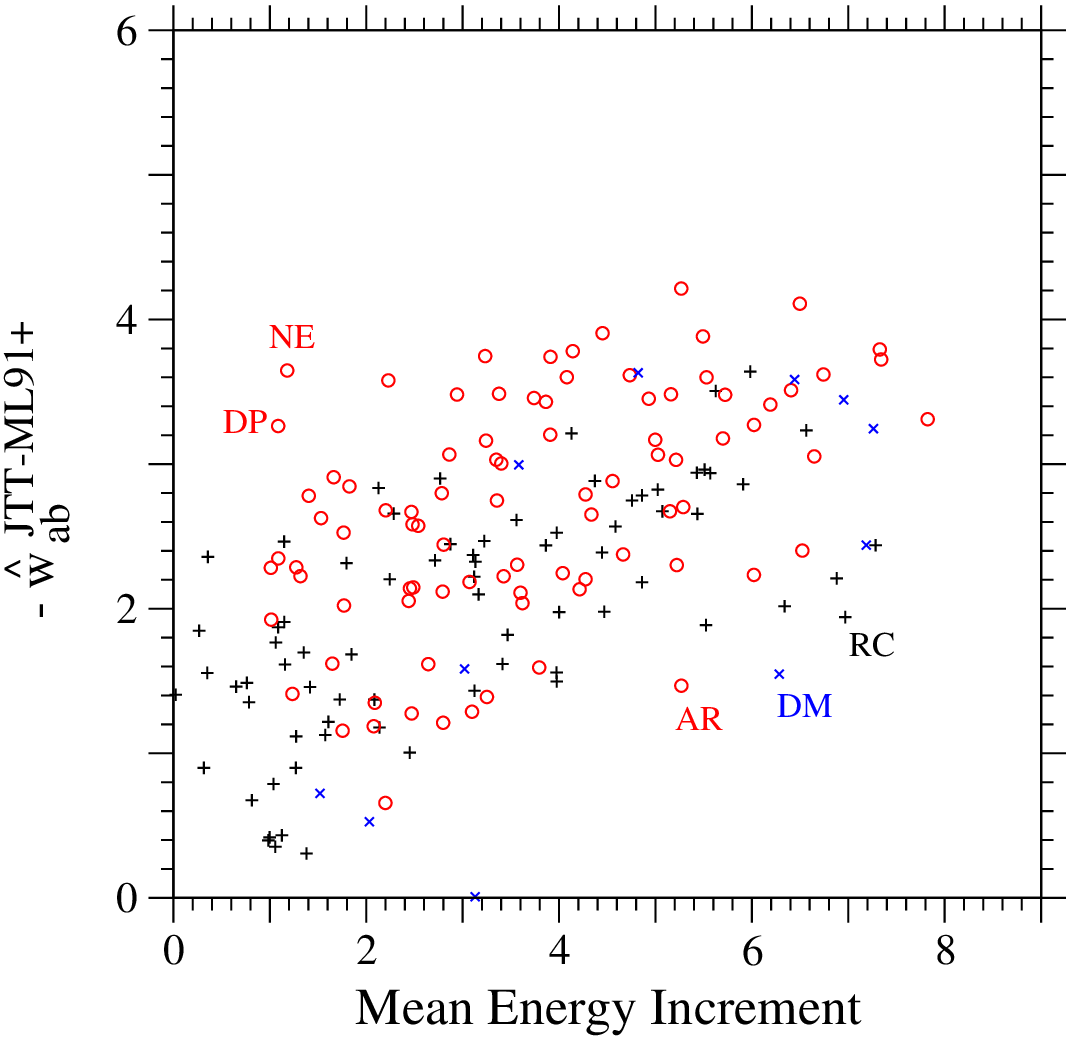}
}
} 
\caption{
\label{fig: wij_vs_energy_increments}
\label{fig: wij_JTT-ML91+_vs_EI}
The ML estimate, $\hat{w}^{\script{JTT-ML91+}}_{ab}$, for
selective constraints on substitutions of each amino acid pair
in the ML-91+ model fitted to the 1-PAM JTT matrix
is plotted against the mean energy increment
due to an amino acid substitution,
($\Delta \hat{\varepsilon}^{\script{c}}_{ab} + \Delta \hat{\varepsilon}^{\script{v}}_{ab}$) 
defined in Supporting Information, Text S1.
The estimates $\hat{w}_{ab}$ for
the least exchangeable class of multi-step amino acid pairs
are not shown.
Plus, circle, and cross marks show the values for
one-, two-, and three-step amino acid pairs,
respectively. 
The correlation coefficient for single step amino acid pairs is equal to.
0.66.
}
\end{figure*}

} 

\FigurePanel{

\FigureInLegends{\clearpage}

\FigureInLegends{\small}

\begin{figure}[ht]
\FigureInLegends{
\textbf{A} \hspace*{45em} \textbf{B}
\centerline{
\includegraphics*[width=75mm,angle=0]{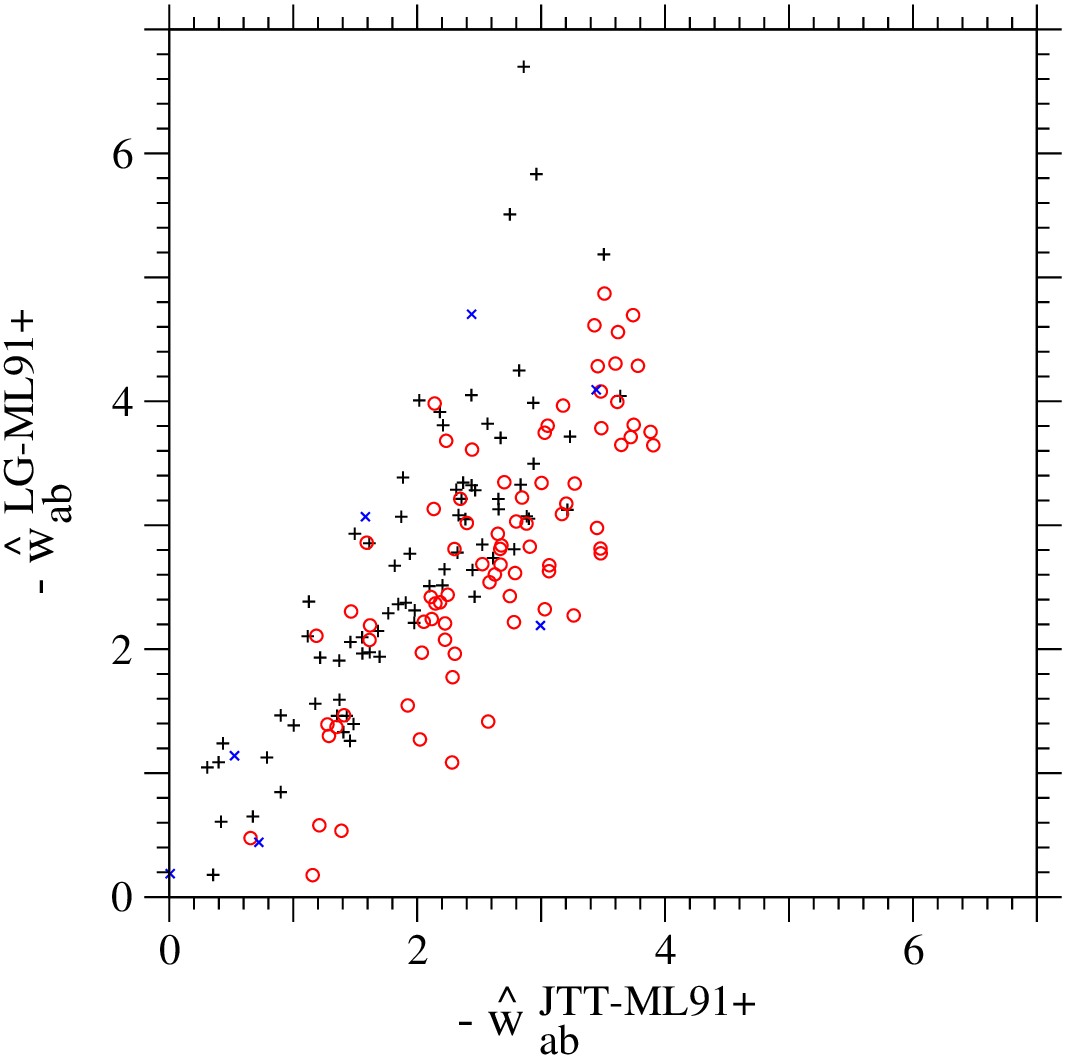}
\hspace*{5mm}
\includegraphics*[width=75mm,angle=0]{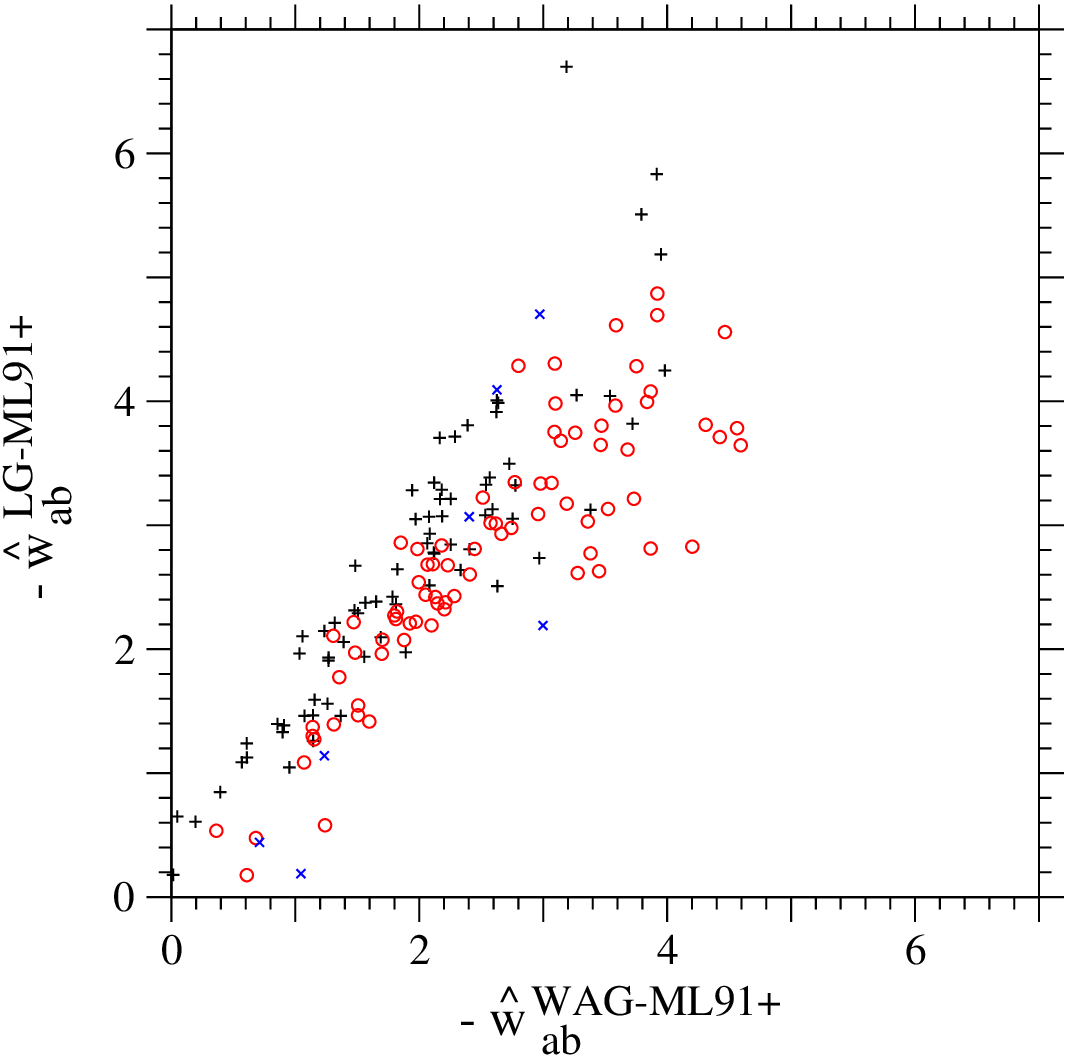}
}
\textbf{C} \hspace*{45em} \textbf{D}
\centerline{
\includegraphics*[width=75mm,angle=0]{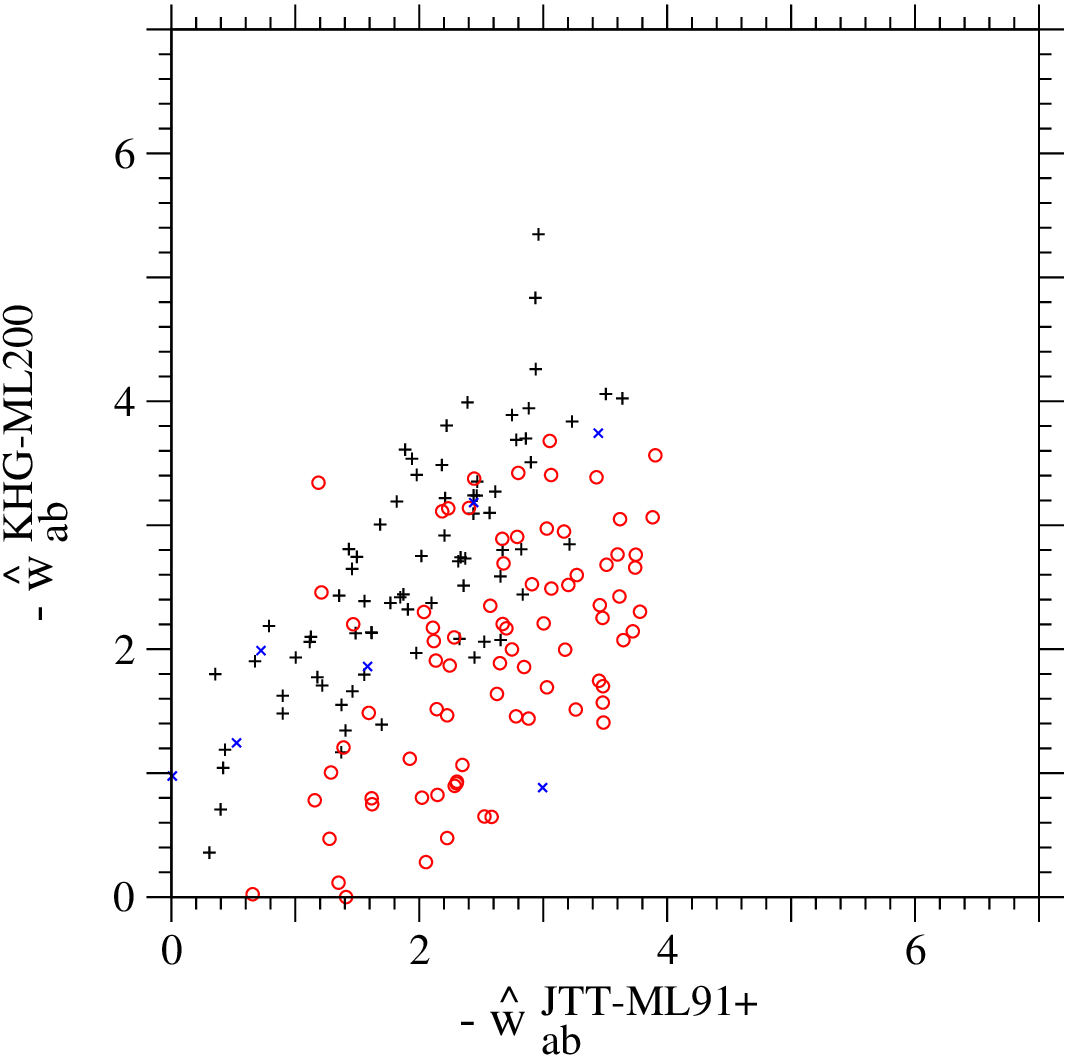}
\hspace*{5mm}
\includegraphics*[width=75mm,angle=0]{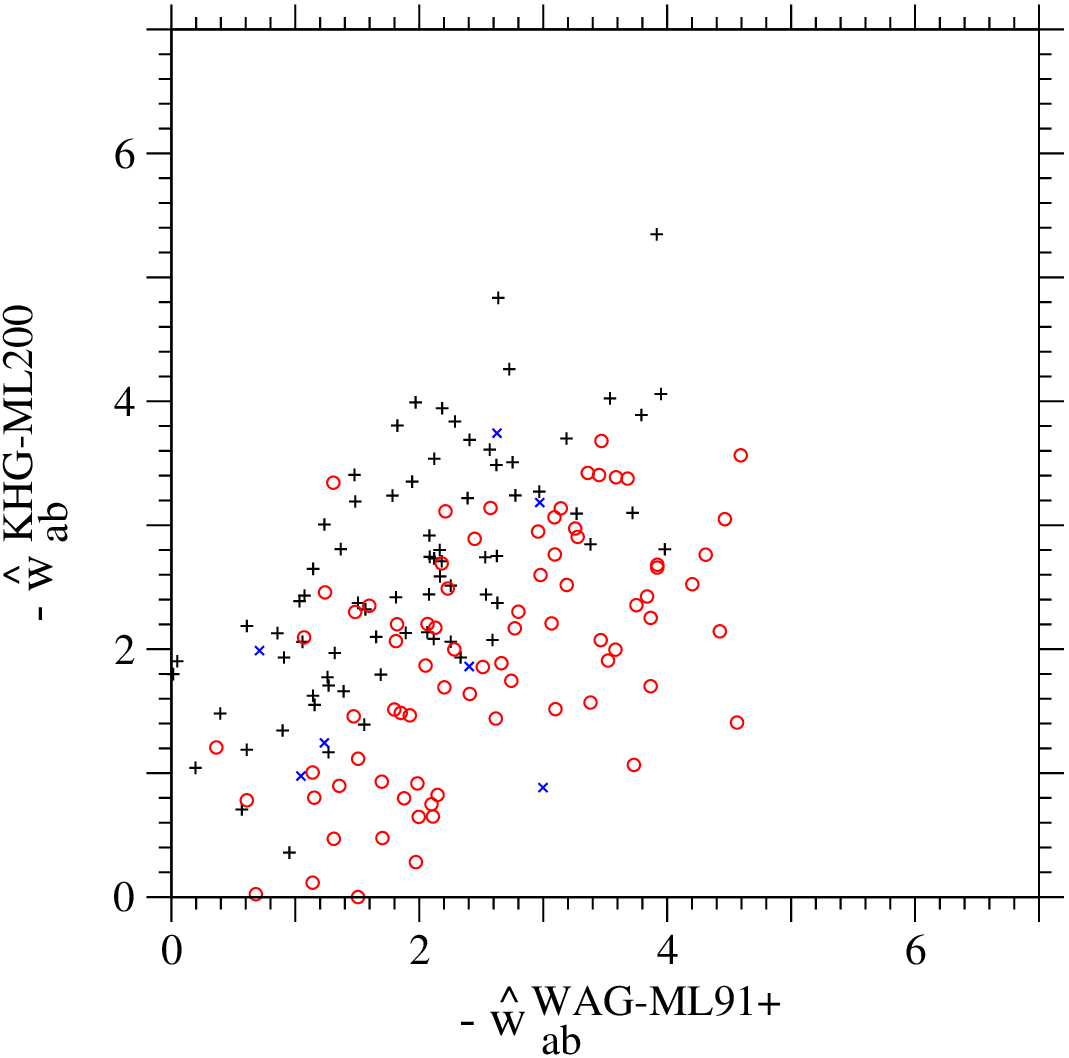}
}
\textbf{E} \hspace*{45em} \textbf{F}
\centerline{
\includegraphics*[width=75mm,angle=0]{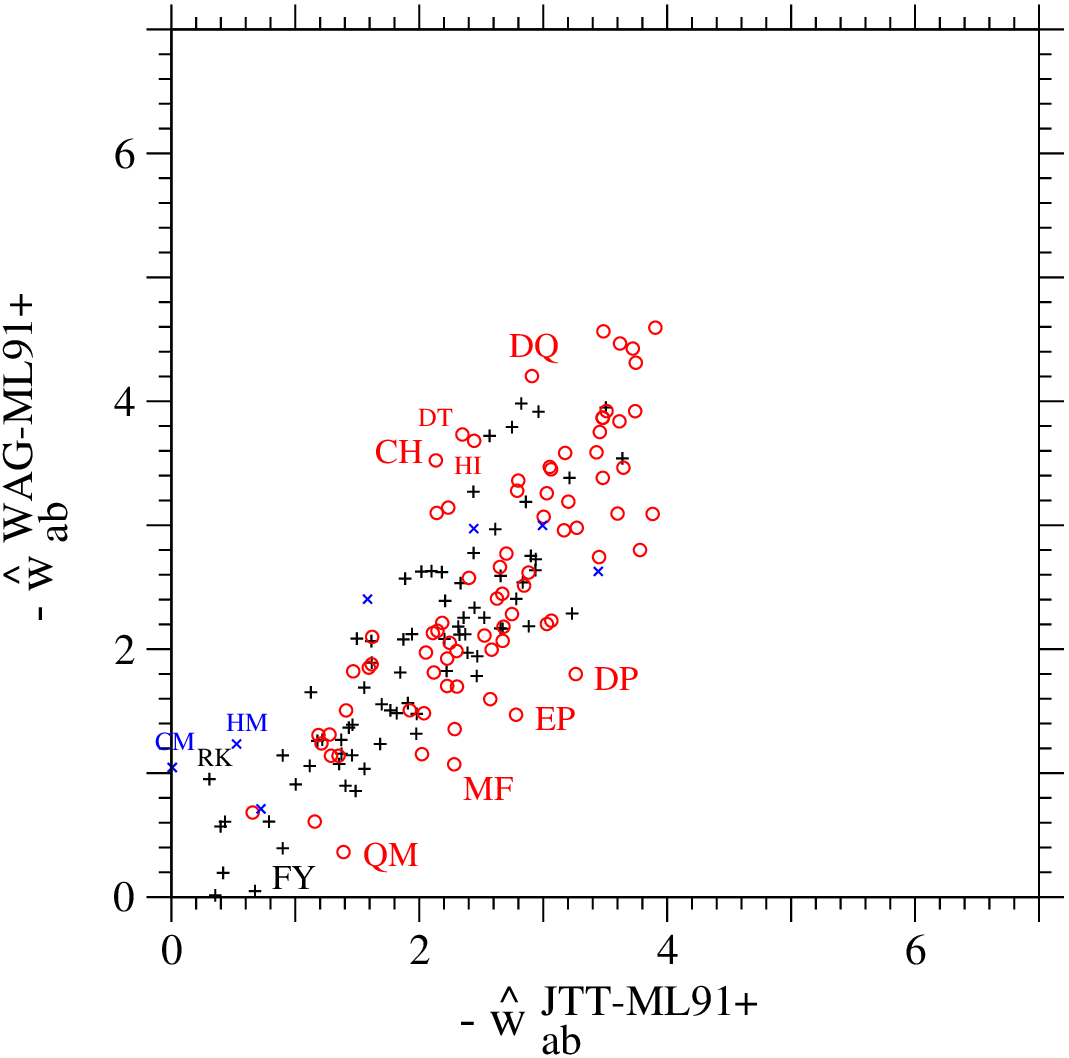}
\hspace*{5mm}
\includegraphics*[width=75mm,angle=0]{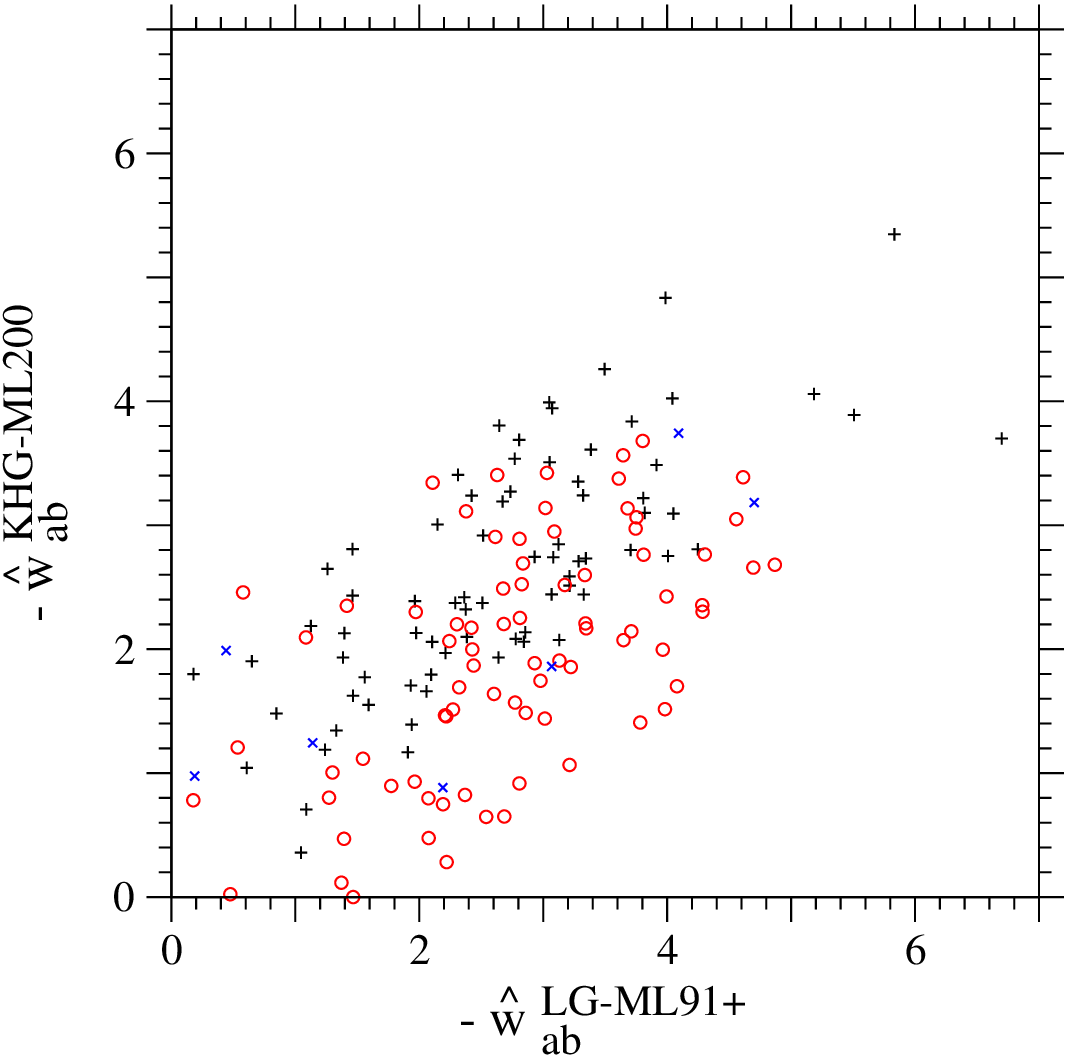}
}
} 
\caption{
\label{fig: wij_JTT_vs_WAG-ML91+}
\label{fig: wij_LG_vs_WAG-ML91+}
\label{fig: wij_JTT-ML91+_vs_KHG-ML200}
\label{fig: wij_LG-ML91+_vs_KHG-ML200}
\FigureInLegends{\small}
\BF{
Comparison between various estimates of
selective constraint for each amino acid pair 
} 
The ML estimates
of selective constraint on substitutions of each amino acid pair
are compared between the models fitted to various empirical substitution matrices.
The estimates $\hat{w}_{ab}$ for multi-step amino acid pairs
that belong to the least exchangeable class at least in one of the models
are not shown.
Plus, circle, and cross marks show the values for
one-, two-, and three-step amino acid pairs,
respectively. 
\FigureInLegends{\normalsize}
}
\end{figure}

\FigureInLegends{\normalsize}

\FigureInLegends{\clearpage}

\begin{figure}[ht]
\FigureInLegends{
\textbf{A} \hspace*{47em} \textbf{B}

\centerline{
\includegraphics*[width=80mm,angle=0]{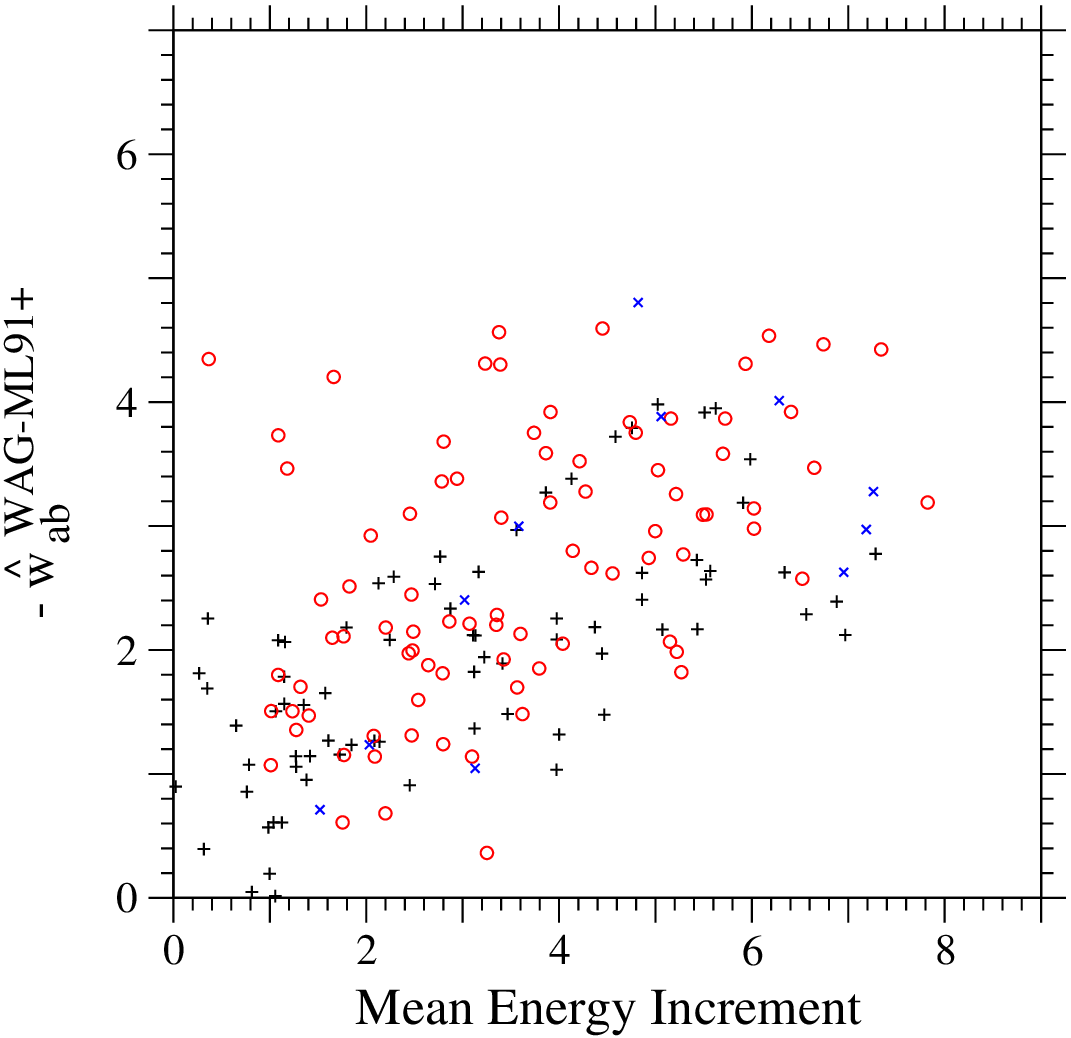}
\hspace*{5mm}
\includegraphics*[width=80mm,angle=0]{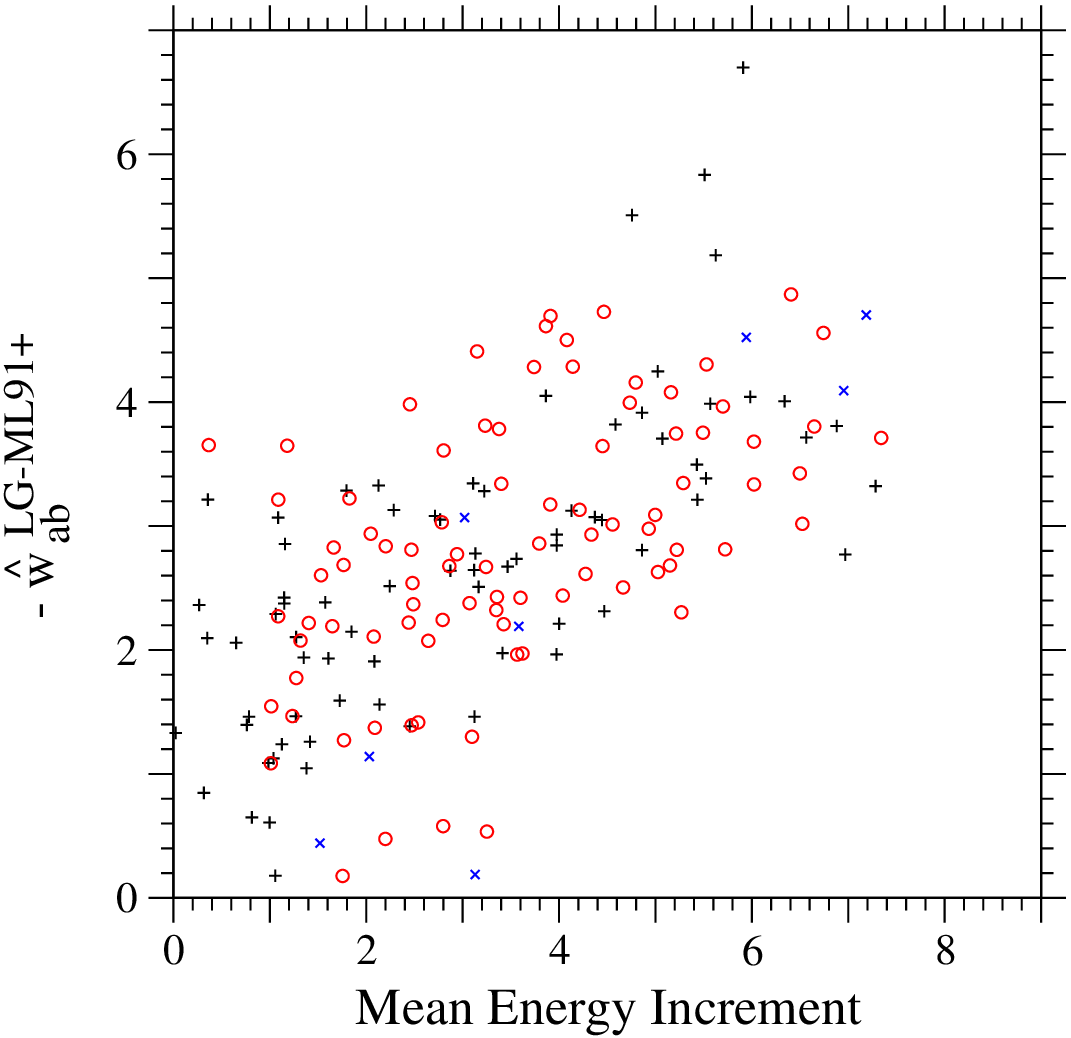}
}
} 
\caption{
\label{fig: wij_EI_vs_WAG-ML91+}
\label{fig: wij_EI_vs_LG-ML91+}
\BF{
Selective constraint for each amino acid pair 
estimated from WAG and from LG.
} 
The ML estimate, 
$- \hat{w}^{\script{WAG-ML91+}}_{ab}$ in (A) and
$- \hat{w}^{\script{LG-ML91+}}_{ab}$ in (B), 
of selective constraint on substitutions of each amino acid pair
in the ML-91+ models fitted to the 1-PAM matrices of WAG and LG
is plotted against the mean energy increment
due to an amino acid substitution,
($\Delta \hat{\varepsilon}^{\script{c}}_{ab} + \Delta \hat{\varepsilon}^{\script{v}}_{ab}$) 
defined 
by \SEqs{4, S1-5, and S1-6}.
The estimates $\hat{w}_{ab}$ for the least exchangeable class 
of multi-step amino acid pairs
are not shown.
Plus, circle, and cross marks show the values for
one-, two-, and three-step amino acid pairs,
respectively. 
}
\end{figure}

\FigureInLegends{\clearpage}

\begin{figure}[ht]
\FigureInLegends{
\textbf{A} \hspace*{47em} \textbf{B}

\centerline{
\includegraphics*[width=80mm,angle=0]{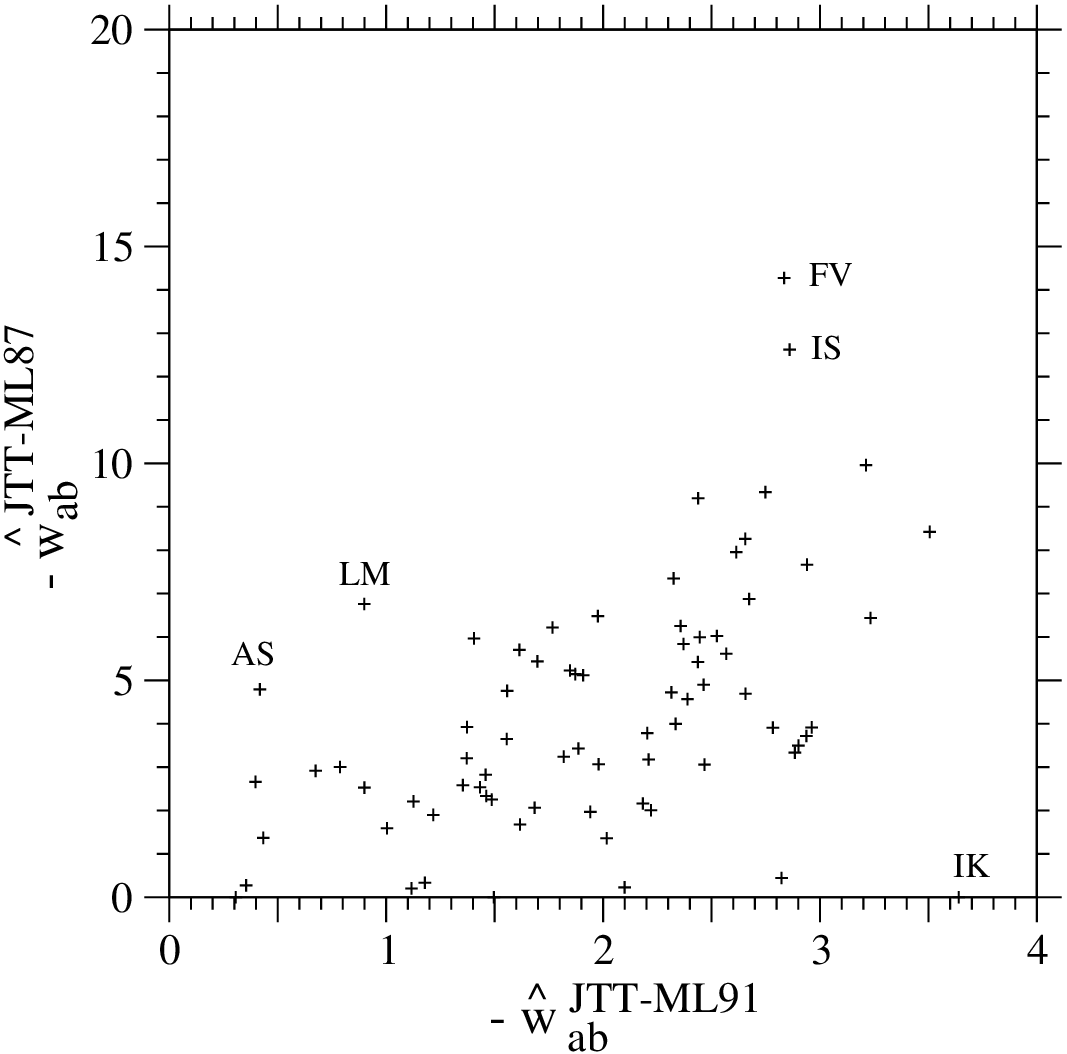}
\hspace*{5mm}
\includegraphics*[width=80mm,angle=0]{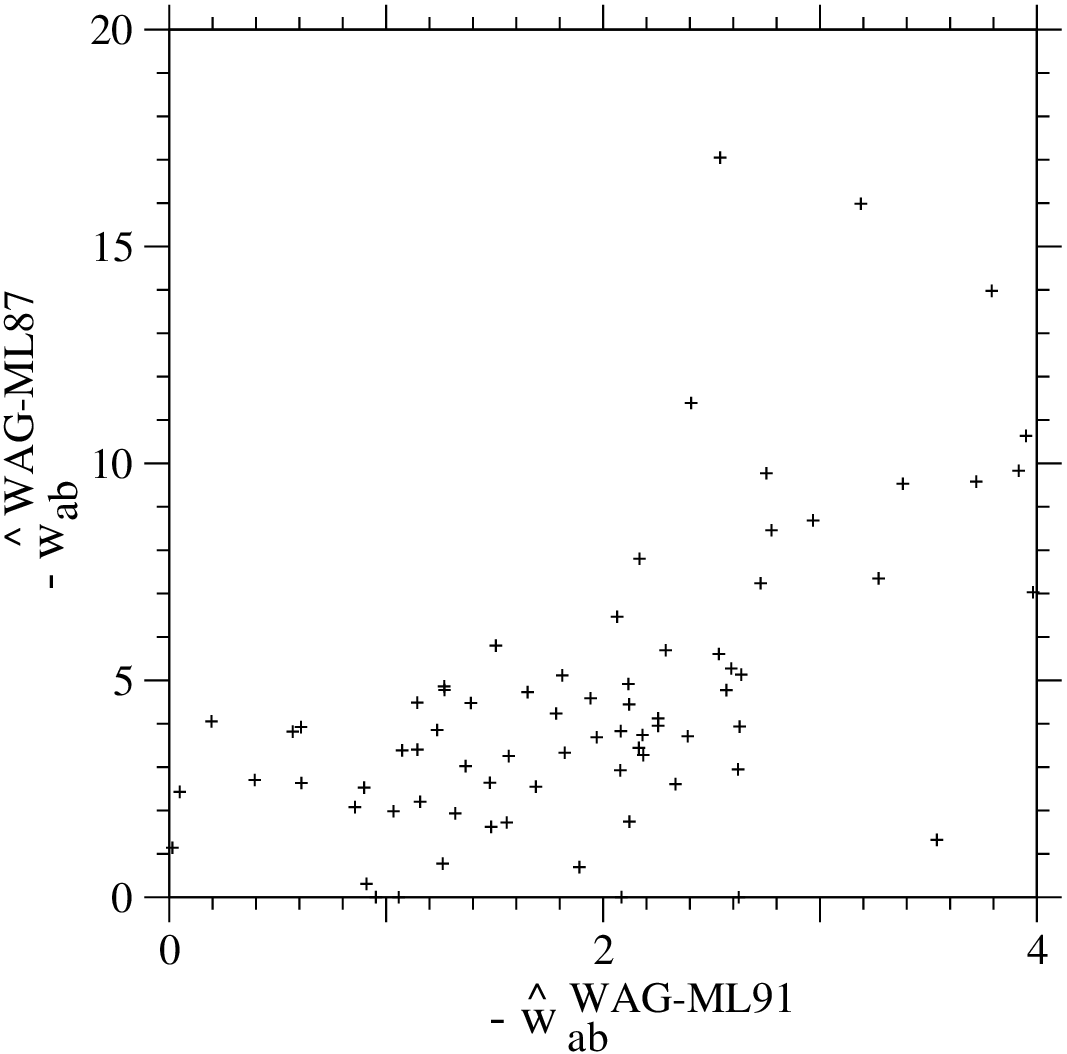}
}
} 
\caption{
\label{fig: wij_JTT-ML87_vs_JTT-ML91}
\label{fig: wij1_JTT-ML87_vs_JTT-ML91}
\label{fig: wij1_ML87_vs_ML91}
\label{sfig: wij_vs_energy_increments}
\label{fig: wij_WAG-ML87_vs_WAG-ML91}
\label{fig: wij1_WAG-ML87_vs_WAG-ML91}
\BF{
Comparison of the ML estimates of selective constraint for each amino acid pair 
between the ML-87 and the ML-91 models.
} 
The ML estimate
of selective constraint for each single step amino acid pair
in the ML-87 model fitted to 
(A) the 1-PAM JTT matrix or (B) the 1-PAM WAG matrix
is plotted against that in the ML-91 model.
}
\end{figure}

} 

} 

\SupFig{

\FigureEach{
\FigureInLegends{\newpage}

\begin{figure*}[ht]
\FigureInLegends{
\centerline{
\includegraphics*[width=130mm,angle=0]{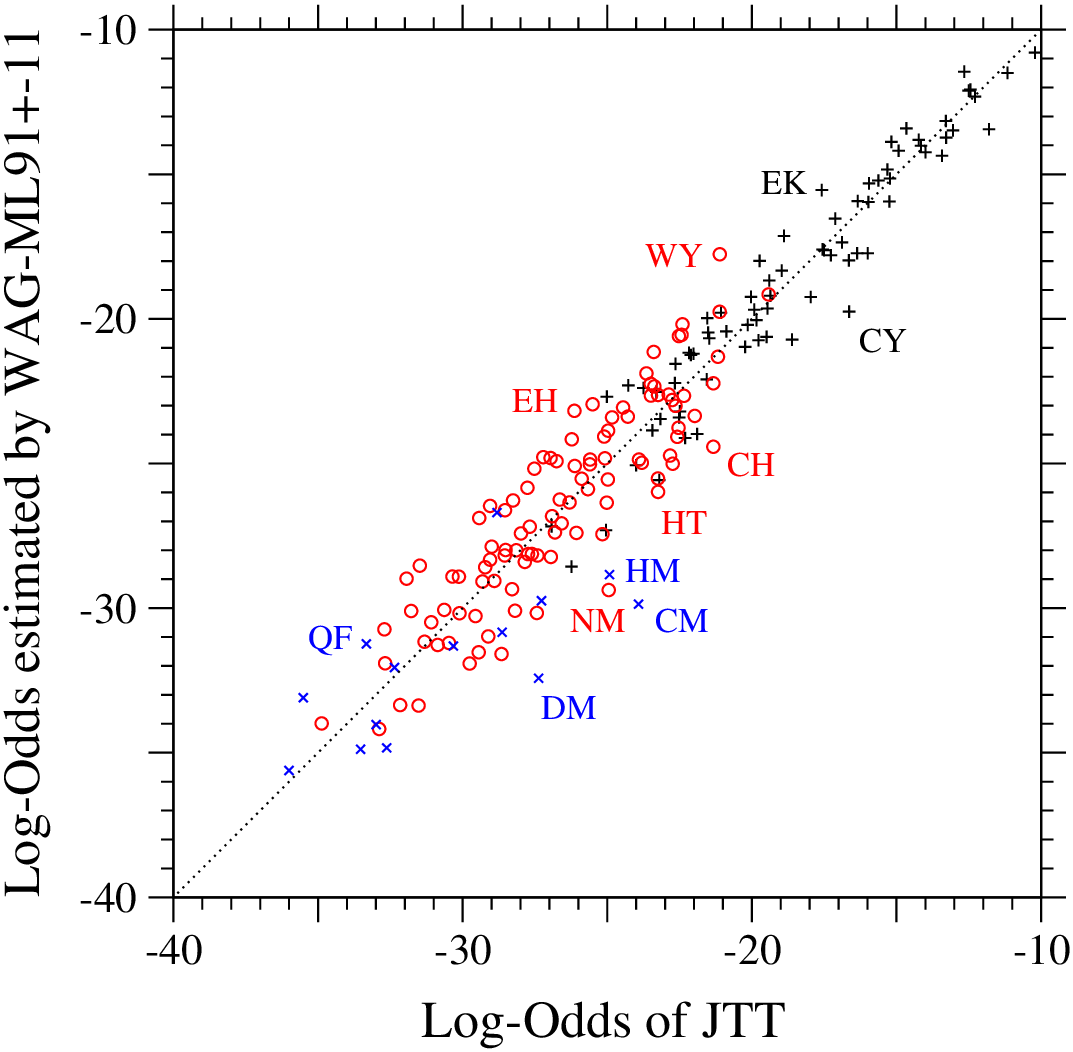}
}
} 
\caption{
\label{fig: WAG-ML-91+_log-odds_1pam_JTT}
\label{fig: WAG-ML91+_log-odds_1pam_JTT}
Each element 
log-$O(\langle S \rangle(\hat{\tau},\hat{\sigma}))_{ab}$
of the log-odds matrix 
of the WAG-ML91+-11 model fitted to the 1-PAM JTT matrix
is plotted against the log-odds log-$O(S^{\script{JTT}}(\mbox{1 PAM}))_{ab}$
calculated from the JTT.
Plus, circle, and cross marks show the log-odds values for
one-, two-, and three-step amino acid pairs,
respectively.  
The dotted line shows the line 
of equal values between the ordinate and the abscissa.
}
\end{figure*}

} 

} 

\TextFig{

\FigureEach{

\FigureInLegends{\newpage}

\begin{figure*}[ht]
\FigureInLegends{
\centerline{
\includegraphics*[width=130mm,angle=0]{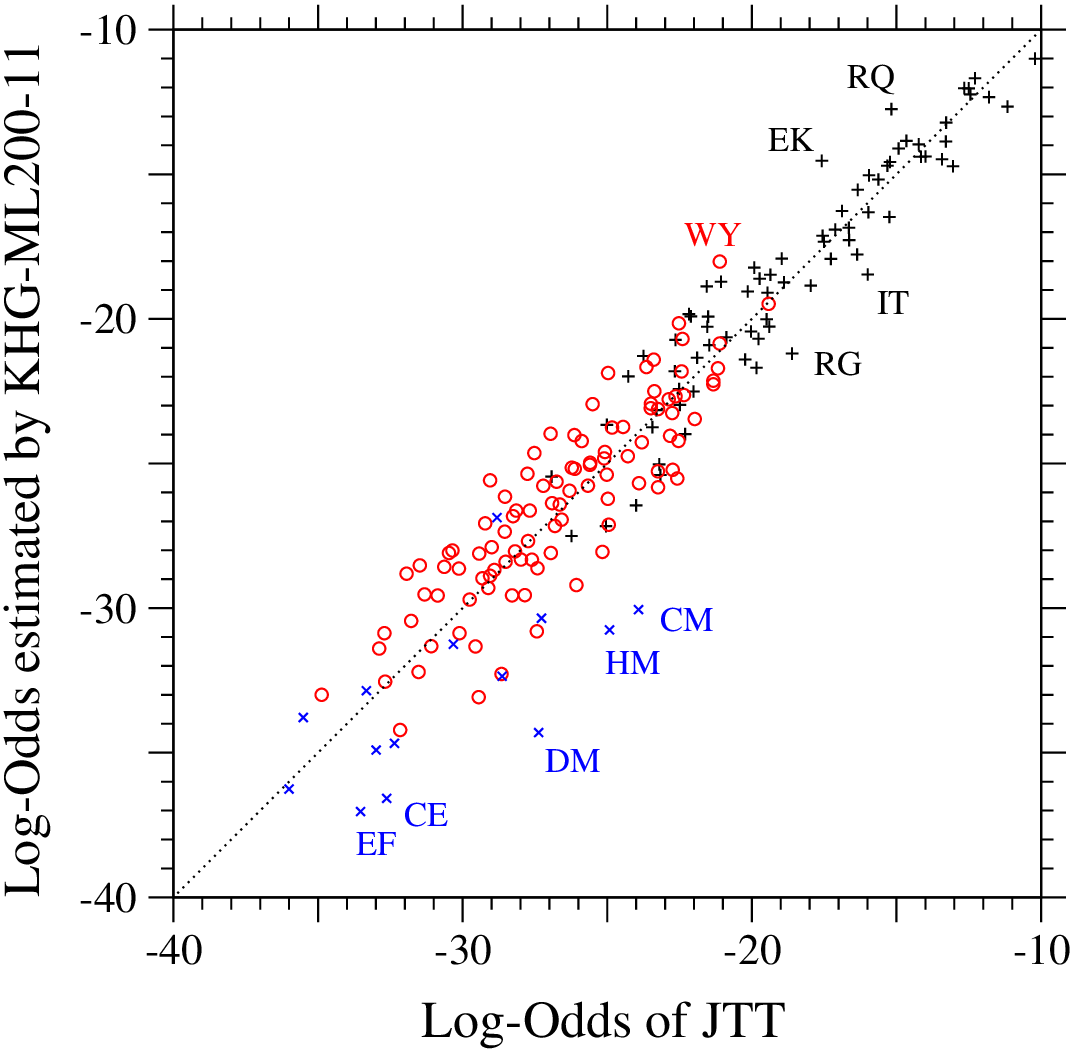}
}
} 
\caption{\label{fig: KHG-ML200_log-odds_1pam_JTT}
\label{fig: KHG-ML199_log-odds_1pam_JTT}
Each element 
log-$O(\langle S \rangle(\hat{\tau},\hat{\sigma}))_{ab}$
of the log-odds matrix 
of the KHG-ML200-11 model fitted to the 1-PAM JTT matrix
is plotted against the log-odds log-$O(S^{\script{JTT}}(\mbox{1 PAM}))_{ab}$
calculated from the JTT.
Plus, circle, and cross marks show the log-odds values for
one-, two-, and three-step amino acid pairs,
respectively.  
The dotted line shows the line 
of equal values between the ordinate and the abscissa.
}
\end{figure*}

} 

} 

\SupFig{

\FigureEach{

\FigureInLegends{\newpage}

\begin{figure*}[ht]
\FigureInLegends{
\centerline{
\includegraphics*[width=130mm,angle=0]{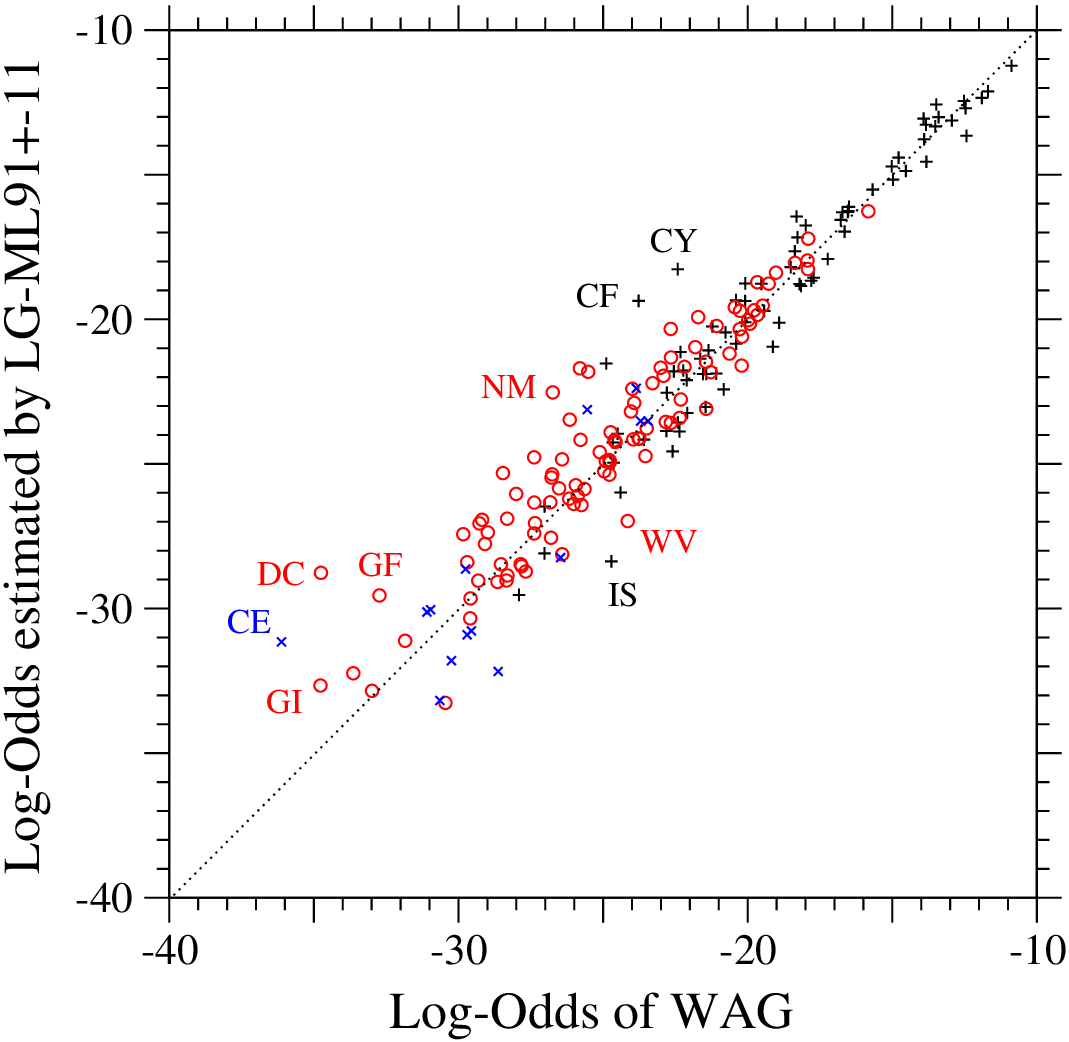}
}
} 
\caption{\label{fig: LG-ML-91+_log-odds_1pam_WAG}
\label{fig: LG-ML-91_log-odds_1pam_WAG}
Each element 
log-$O(\langle S \rangle(\hat{\tau},\hat{\sigma}))_{ab}$
of the log-odds matrix 
of the LG-ML91+-11 model fitted to the 1-PAM WAG matrix
is plotted against the log-odds log-$O(S^{\script{WAG}}(\mbox{1 PAM}))_{ab}$
calculated from WAG.
Plus, circle, and cross marks show the log-odds values for
one-, two-, and three-step amino acid pairs,
respectively.  
The dotted line shows the line 
of equal values between the ordinate and the abscissa.
}
\end{figure*}

} 

} 

\TextFig{

\FigureEach{

\FigureInLegends{\newpage}

\begin{figure*}[ht]
\FigureInLegends{
\centerline{
\includegraphics*[width=130mm,angle=0]{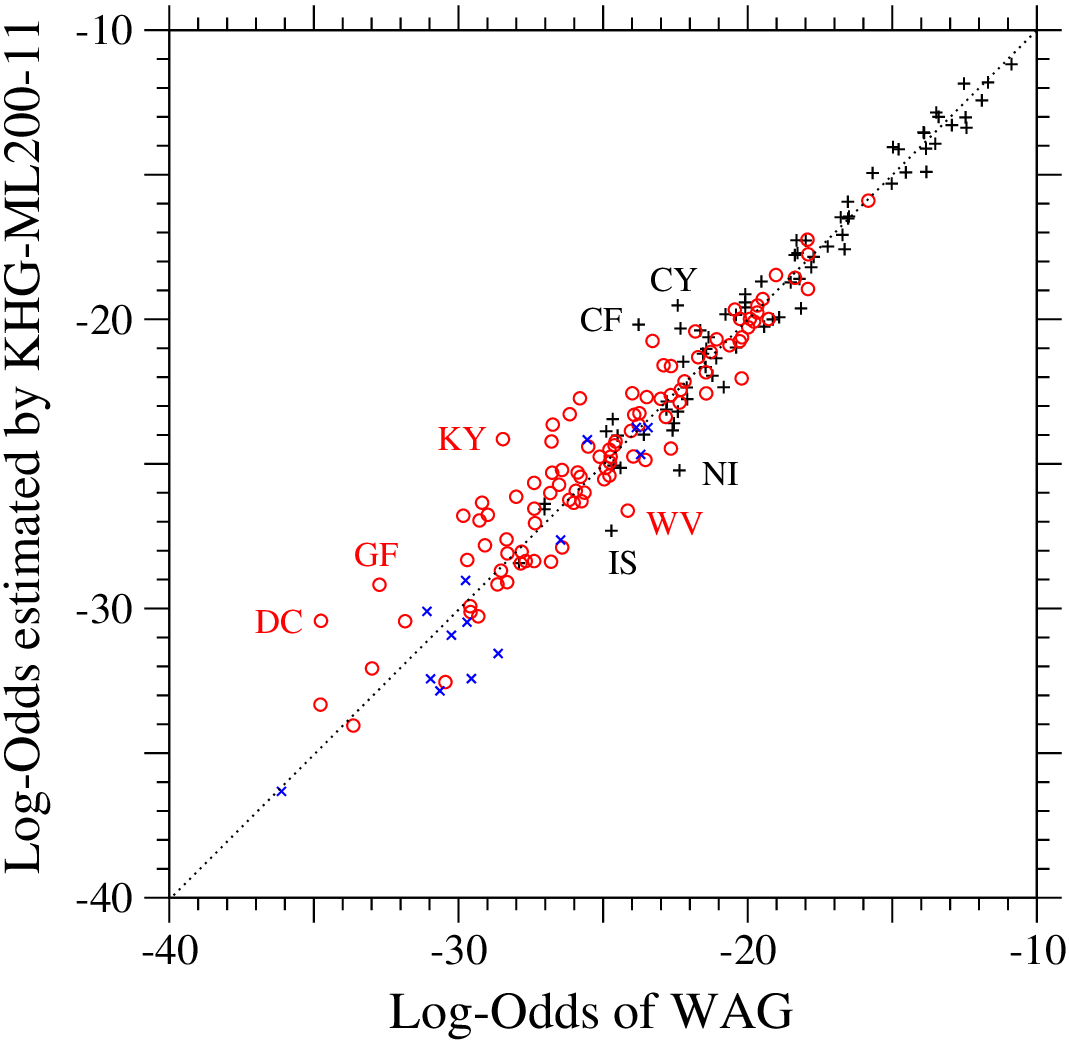}
}
} 
\caption{\label{fig: KHG-ML200_log-odds_1pam_WAG}
\label{fig: KHG-ML199_log-odds_1pam_WAG}
Each element 
log-$O(\langle S \rangle(\hat{\tau},\hat{\sigma}))_{ab}$
of the log-odds matrix 
of the KHG-ML200-11 model fitted to the 1-PAM WAG matrix
is plotted against the log-odds log-$O(S^{\script{WAG}}(\mbox{1 PAM}))_{ab}$
calculated from WAG.
Plus, circle, and cross marks show the log-odds values for
one-, two-, and three-step amino acid pairs,
respectively.  
The dotted line shows the line 
of equal values between the ordinate and the abscissa.
}
\end{figure*}

} 

} 

\SupFig{

\FigureEach{

\FigureInLegends{\newpage}

\begin{figure*}[ht]
\FigureInLegends{
\centerline{
\includegraphics*[width=130mm,angle=0]{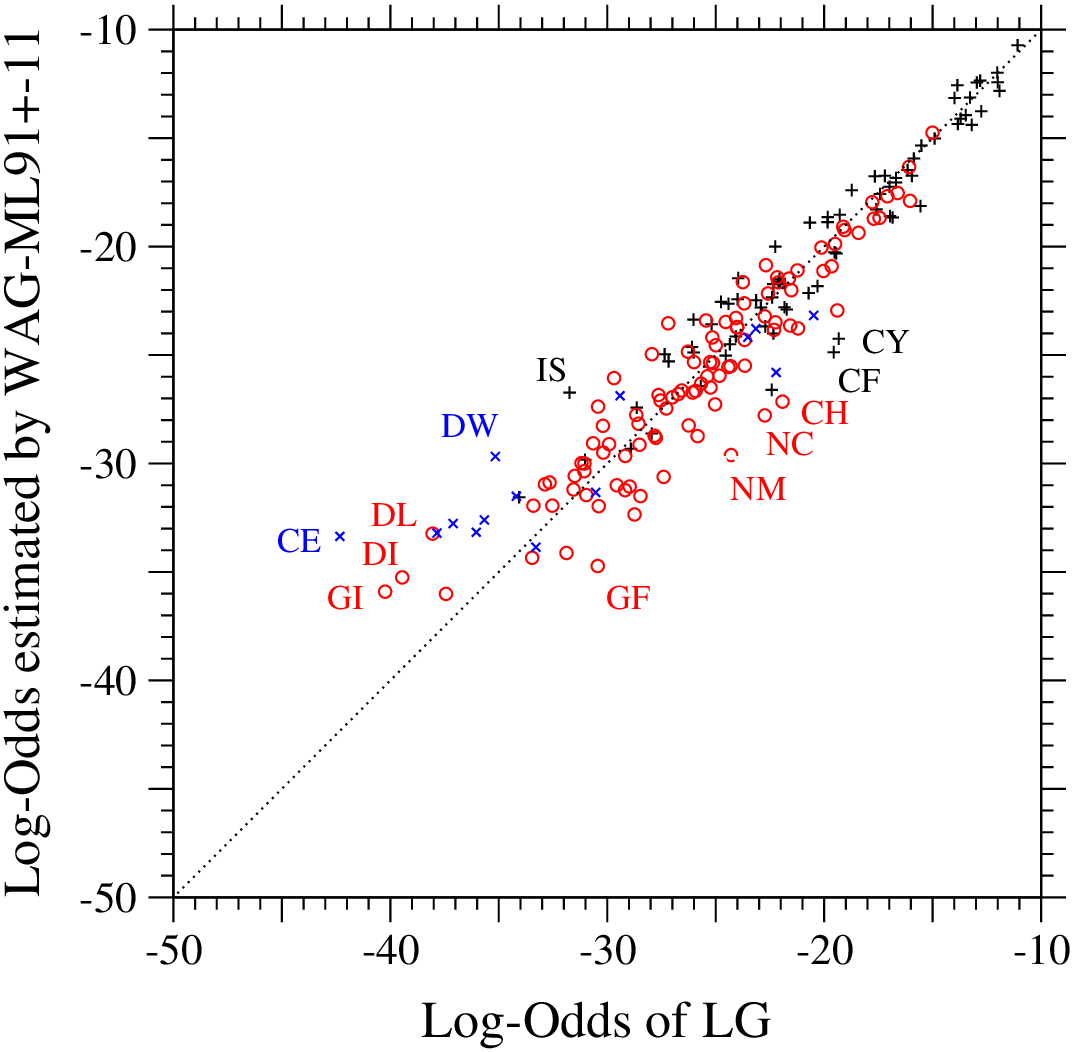}
}
} 
\caption{\label{fig: WAG-ML91+_log-odds_1pam_LG}
Each element 
log-$O(\langle S \rangle(\hat{\tau},\hat{\sigma}))_{ab}$
of the log-odds matrix 
of the WAG-ML91+-11 model fitted to the 1-PAM LG matrix
is plotted against the log-odds log-$O(S^{\script{LG}}(\mbox{1 PAM}))_{ab}$
calculated from LG.
Plus, circle, and cross marks show the log-odds values for
one-, two-, and three-step amino acid pairs,
respectively.  
The dotted line shows the line 
of equal values between the ordinate and the abscissa.
}
\end{figure*}

} 

} 

\TextFig{

\FigureEach{

\FigureInLegends{\newpage}

\begin{figure*}[ht]
\FigureInLegends{
\centerline{
\includegraphics*[width=130mm,angle=0]{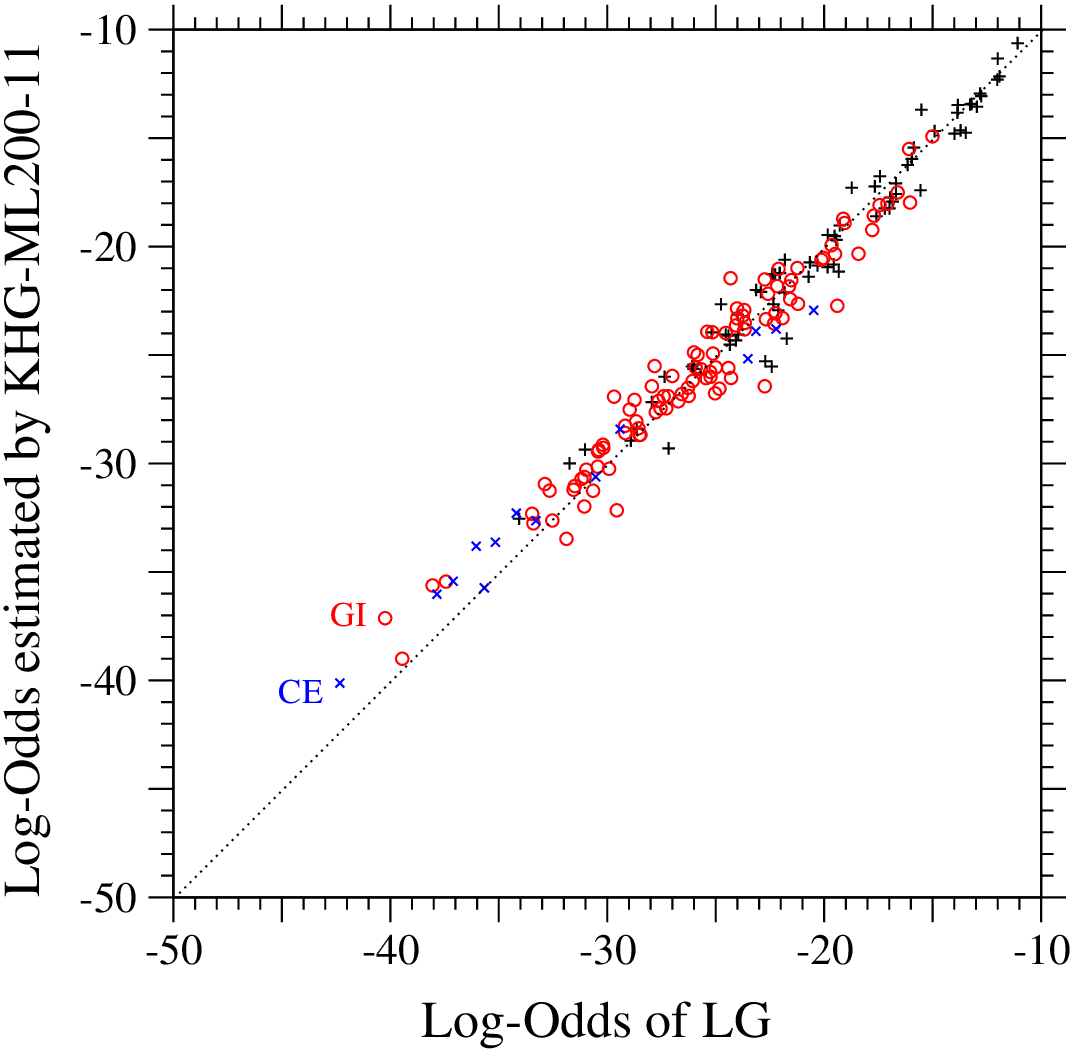}
}
} 
\caption{\label{fig: KHG-ML200_log-odds_1pam_LG}
\label{fig: KHG-ML199_log-odds_1pam_LG}
Each element 
log-$O(\langle S \rangle(\hat{\tau},\hat{\sigma}))_{ab}$
of the log-odds matrix 
of the KHG-ML200-11 model fitted to the 1-PAM LG matrix
is plotted against the log-odds log-$O(S^{\script{LG}}(\mbox{1 PAM}))_{ab}$
calculated from LG.
Plus, circle, and cross marks show the log-odds values for
one-, two-, and three-step amino acid pairs,
respectively.  
The dotted line shows the line 
of equal values between the ordinate and the abscissa.
}
\end{figure*}

} 

\FigurePanel{

\FigureInLegends{\newpage}

\begin{figure}[ht]
\FigureInLegends{
\textbf{A} \hspace*{47em} \textbf{B}

\centerline{
\includegraphics*[width=80mm,angle=0]{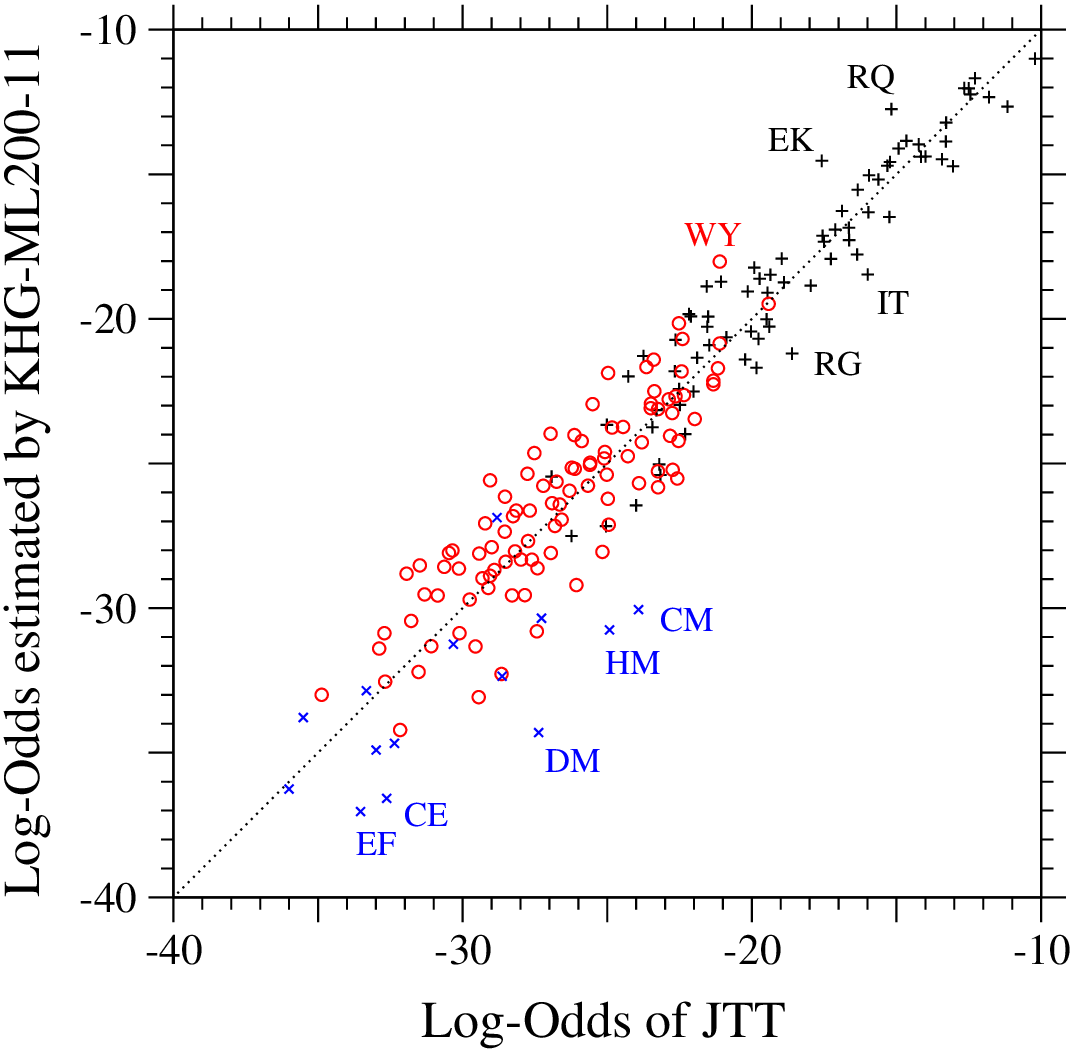}
\hspace*{5mm}
\includegraphics*[width=80mm,angle=0]{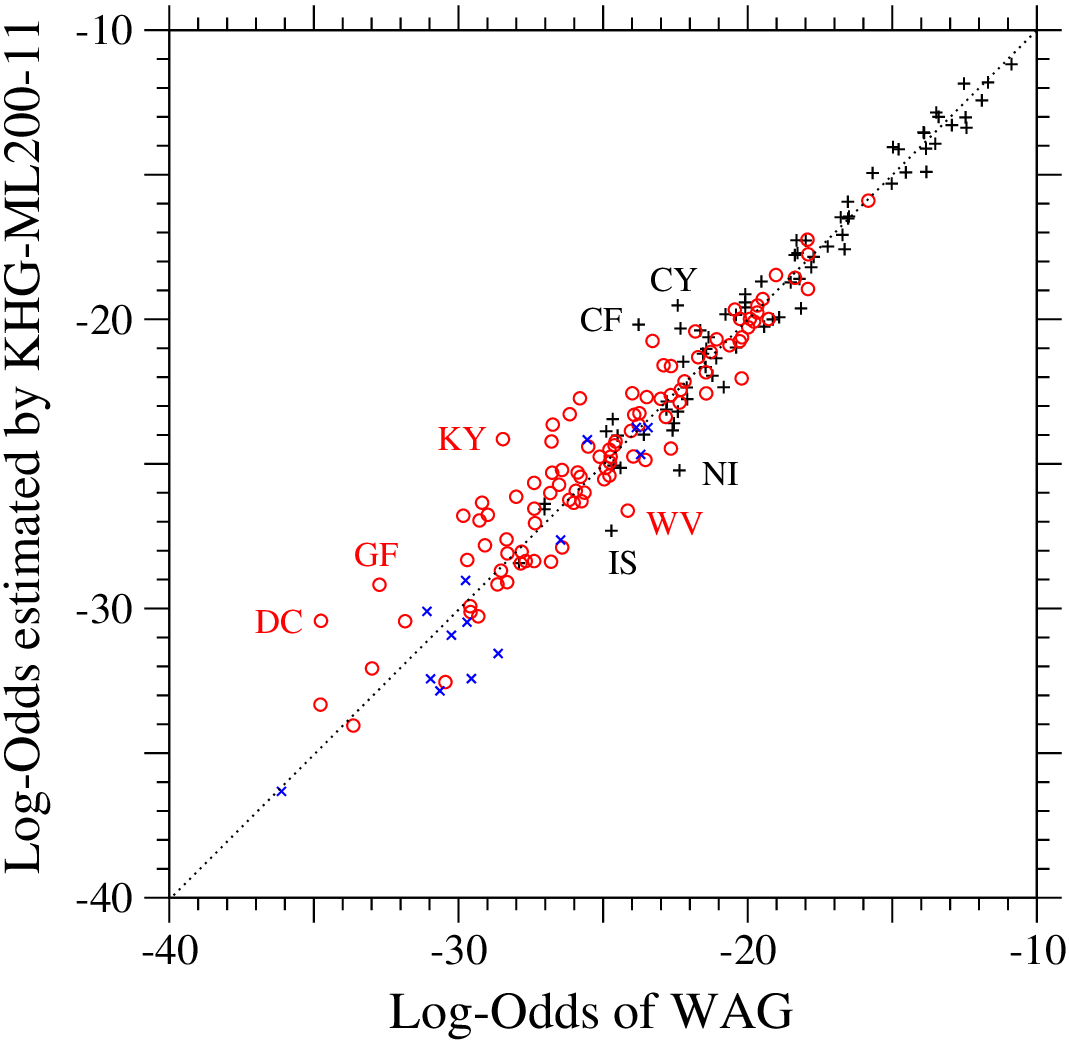}
}
\vspace*{1em}
\centerline{
\includegraphics*[width=80mm,angle=0]{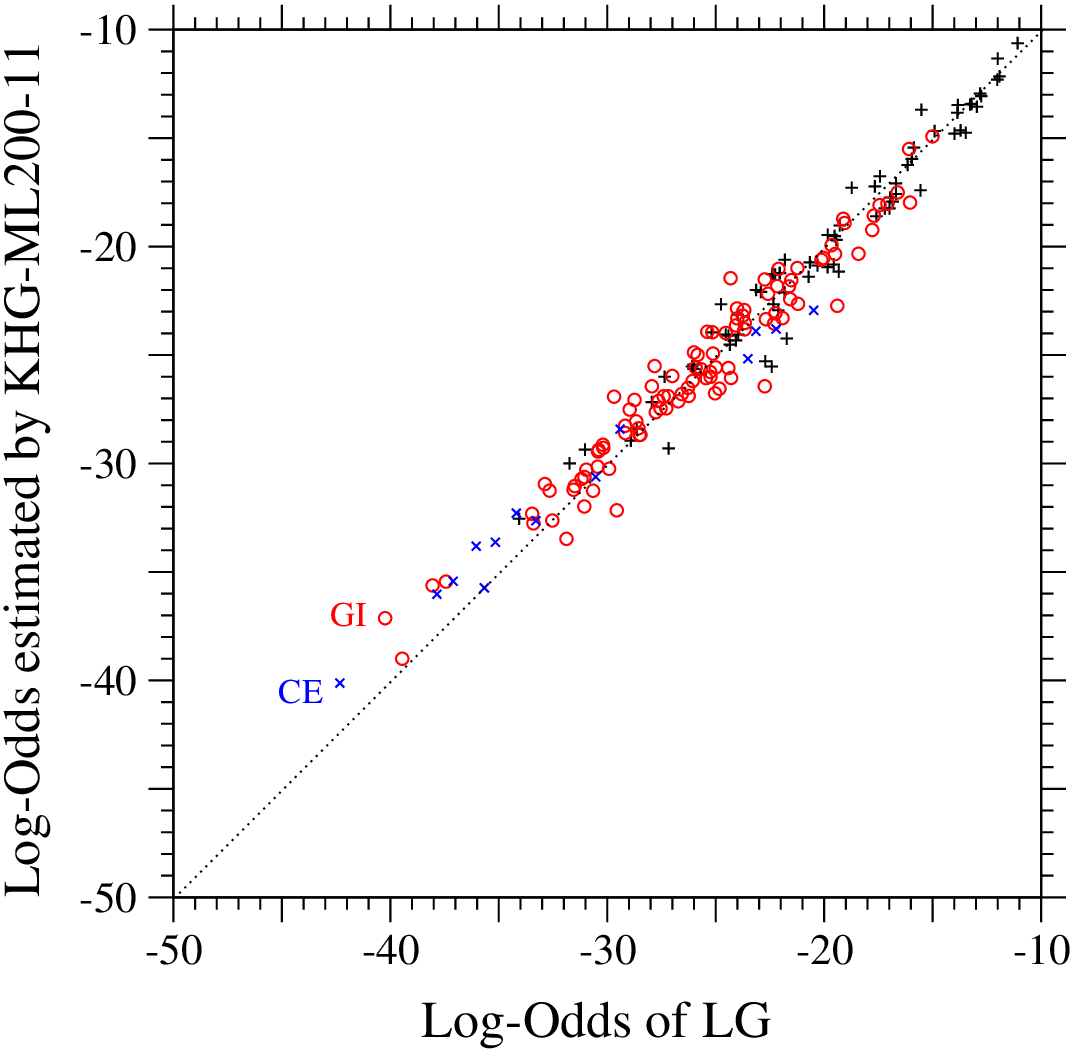}
\hspace*{5mm}
\includegraphics*[width=80mm,angle=0]{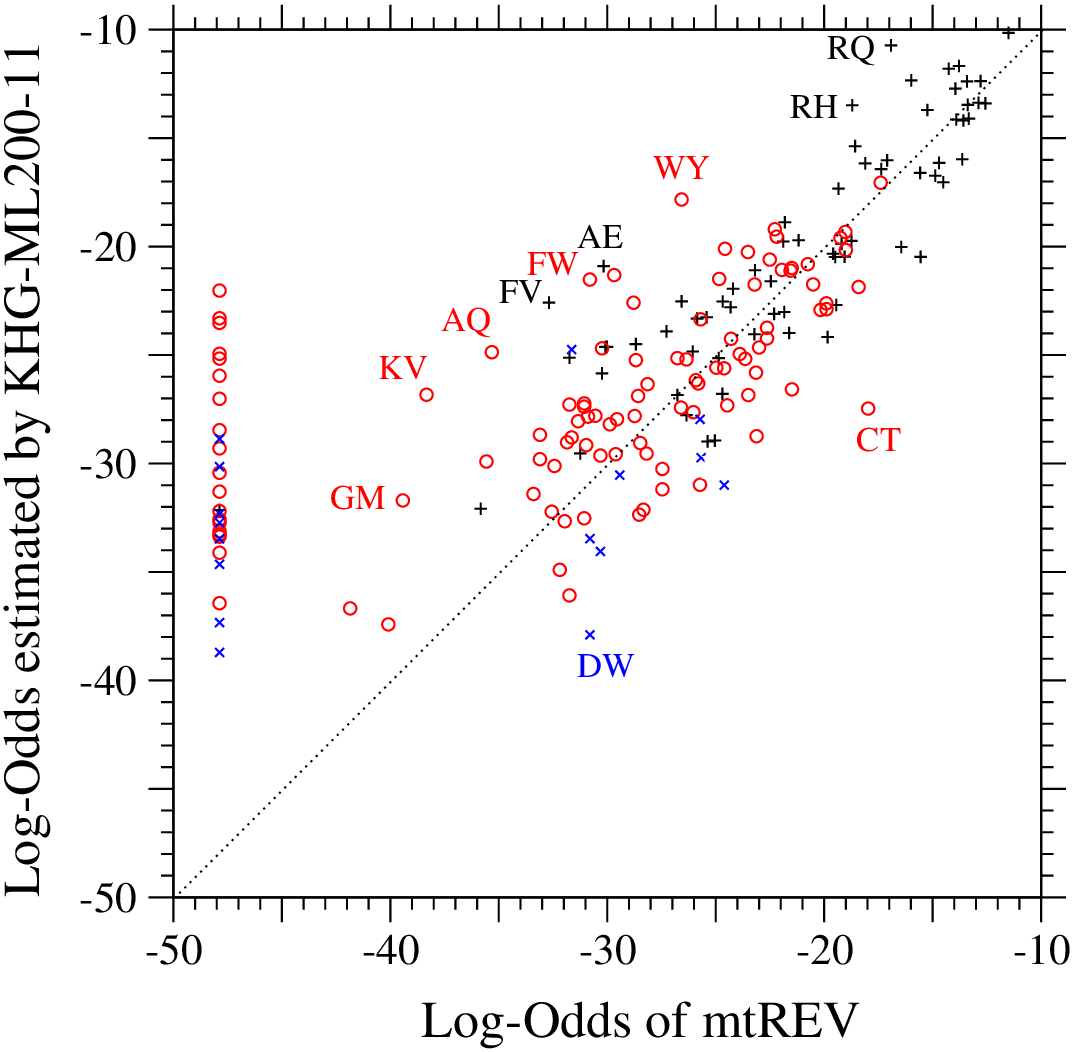}
}

\textbf{C} \hspace*{47em} \textbf{D}
} 
\caption{
\label{fig: KHG-ML200_log-odds_1pam_JTT}
\label{fig: KHG-ML199_log-odds_1pam_JTT}
\label{fig: KHG-ML200_log-odds_1pam_WAG}
\label{fig: KHG-ML199_log-odds_1pam_WAG}
\label{fig: KHG-ML200_log-odds_1pam_LG}
\label{fig: KHG-ML199_log-odds_1pam_LG}
\label{fig: KHG-ML200_log-odds_1pam_MtRevRmtx}
\label{fig: KHG-ML199_log-odds_1pam_MtRevRmtx}
\BF{
The KHG-ML200-11 model fitted to each of JTT, WAG, LG, and mtREV.
} 
Each element 
log-$O(\langle S \rangle(\hat{\tau},\hat{\sigma}))_{ab}$
of the log-odds matrices
of the KHG-ML200-11 model 
fitted to the 1-PAM matrices of 
(A) JTT, (B) WAG, (C) LG, and (D) mtREV
is plotted against the log-odds log-$O(S^{\script{LG}}(\mbox{1 PAM}))_{ab}$
calculated from the corresponding empirical substitution matrices.
Plus, circle, and cross marks show the log-odds values for
one-, two-, and three-step amino acid pairs,
respectively.  
The dotted line in each figure shows the line 
of equal values between the ordinate and the abscissa.
The log-odds elements of mtREV whose values are smaller than about $-47.8$ 
are all assumed to be $-47.8$; see the original paper\CITE{AH:96}.
}
\end{figure}

} 

} 

\SupFig{

\FigureEach{

\FigureInLegends{\newpage}

\begin{figure*}[ht]
\FigureInLegends{
\centerline{
\includegraphics*[width=130mm,angle=0]{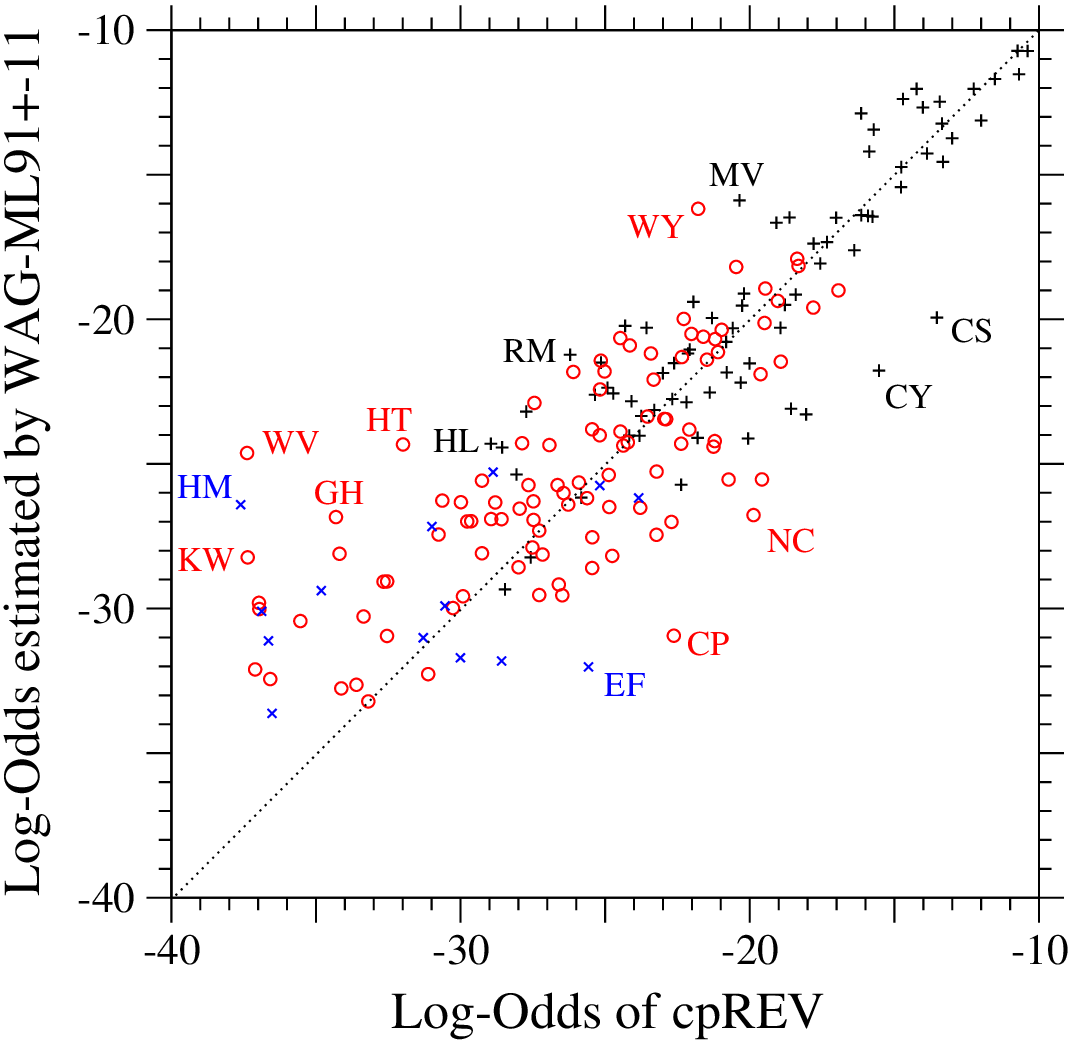}
}
} 
\caption{
\label{fig: WAG-ML91+_log-odds_1pam_CpRev}
\label{fig: WAG-ML-91+_log-odds_1pam_CpRev}
Each element 
log-$O(\langle S \rangle(\hat{\tau},\hat{\sigma}))_{ab}$
of the log-odds matrix 
of the WAG-ML91+-11 model fitted to the 1-PAM cpREV matrix
is plotted against the log-odds log-$O(S^{\script{cpREV}}(\mbox{1 PAM}))_{ab}$
calculated from cpREV.
Plus, circle, and cross marks show the log-odds values for
one-, two-, and three-step amino acid pairs,
respectively.  
The dotted line shows the line 
of equal values between the ordinate and the abscissa.
}
\end{figure*}

} 

\FigureEach{
\FigureInLegends{\newpage}

\begin{figure*}[ht]
\FigureInLegends{
\centerline{
\includegraphics*[width=130mm,angle=0]{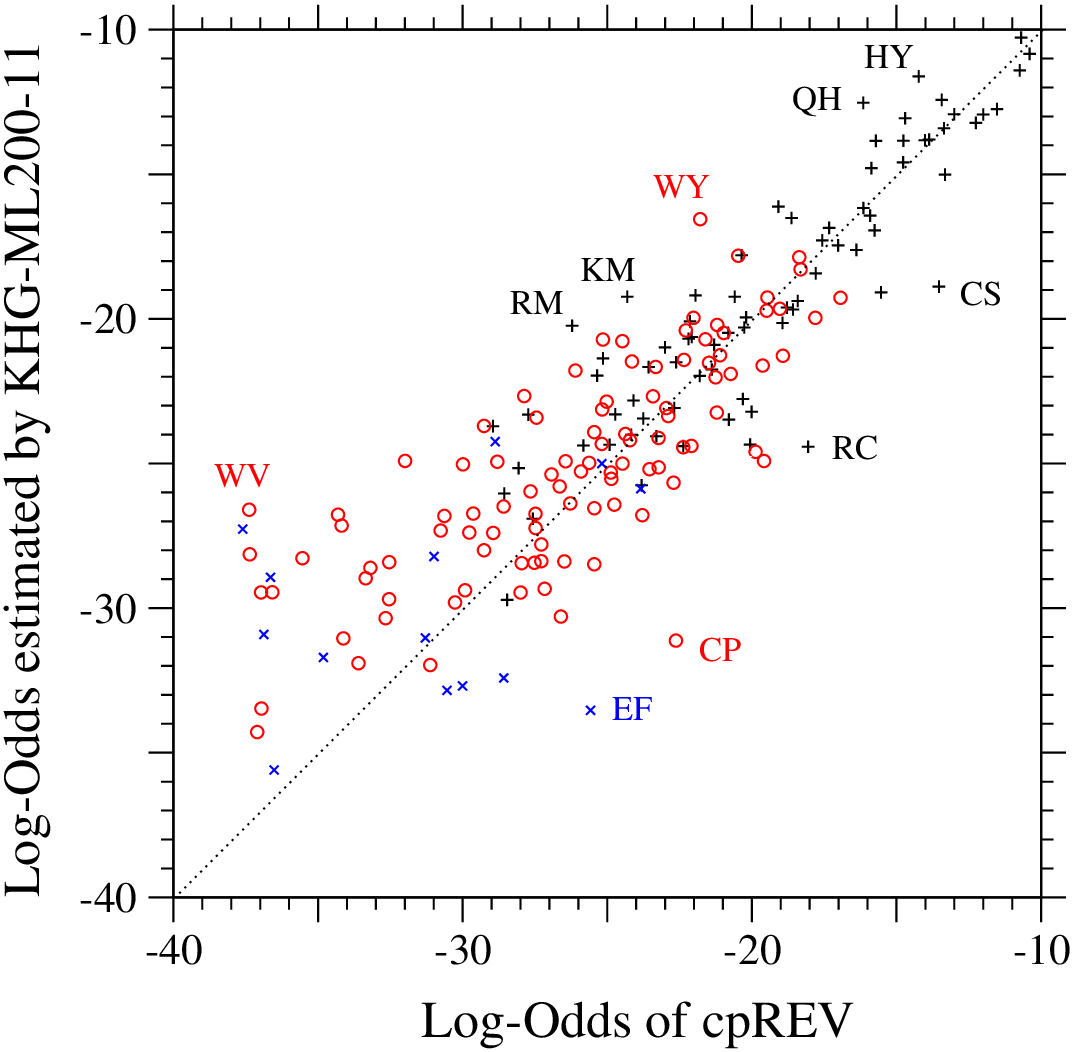}
}
} 
\caption{\label{fig: KHG-ML200_log-odds_1pam_CpRev}
\label{fig: KHG-ML199_log-odds_1pam_CpRev}
Each element 
log-$O(\langle S \rangle(\hat{\tau},\hat{\sigma}))_{ab}$
of the log-odds matrix 
of the KHG-ML200-11 model fitted to the 1-PAM cpREV matrix
is plotted against the log-odds log-$O(S^{\script{cpREV}}(\mbox{1 PAM}))_{ab}$
calculated from cpREV.
Plus, circle, and cross marks show the log-odds values for
one-, two-, and three-step amino acid pairs,
respectively.  
The dotted line shows the line 
of equal values between the ordinate and the abscissa.
}
\end{figure*}

} 

\FigureEach{
\FigureInLegends{\newpage}

\begin{figure*}[ht]
\FigureInLegends{
\centerline{
\includegraphics*[width=130mm,angle=0]{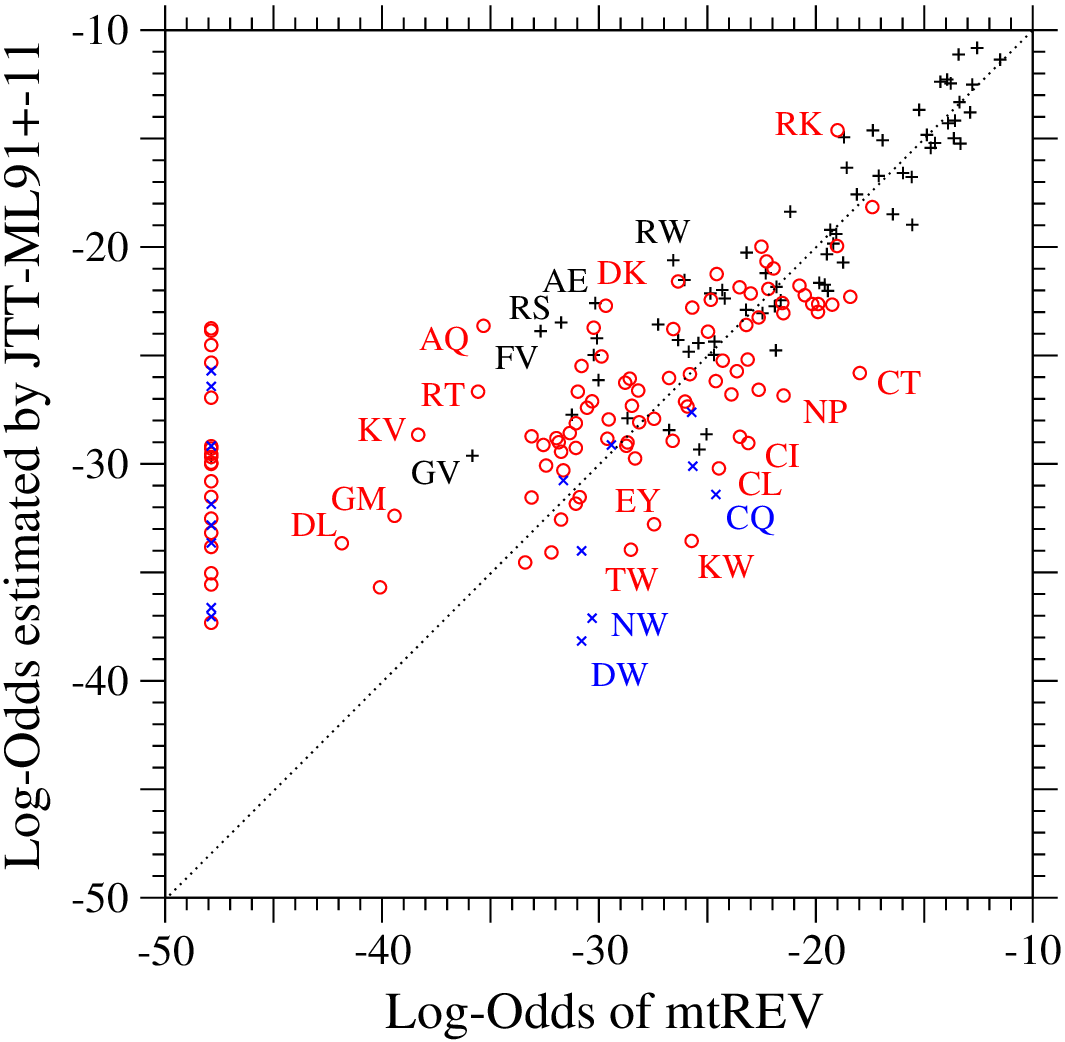}
}
} 
\caption{
\label{fig: JTT-ML-91+_log-odds_1pam_MtRevRmtx}
\label{fig: JTT-ML91+_log-odds_1pam_MtRevRmtx}
\label{fig: JTT-ML-91_log-odds_1pam_MtRevRmtx}
Each element 
log-$O(\langle S \rangle(\hat{\tau},\hat{\sigma}))_{ab}$
of the log-odds matrix 
of the JTT-ML91+-11 model fitted to the 1-PAM mtREV matrix
is plotted against the log-odds log-$O(S^{\script{mtREV}}(\mbox{1 PAM}))_{ab}$
calculated from mtREV.
Plus, circle, and cross marks show the log-odds values for
one-, two-, and three-step amino acid pairs,
respectively.  
The log-odds elements of mtREV whose values are smaller than about $-47.8$ 
are all assumed to be $-47.8$; see the original paper\CITE{AH:96}
The dotted line shows the line 
of equal values between the ordinate and the abscissa.
}
\end{figure*}

} 

\FigureEach{

\FigureInLegends{\newpage}

\begin{figure*}[ht]
\FigureInLegends{
\centerline{
\includegraphics*[width=130mm,angle=0]{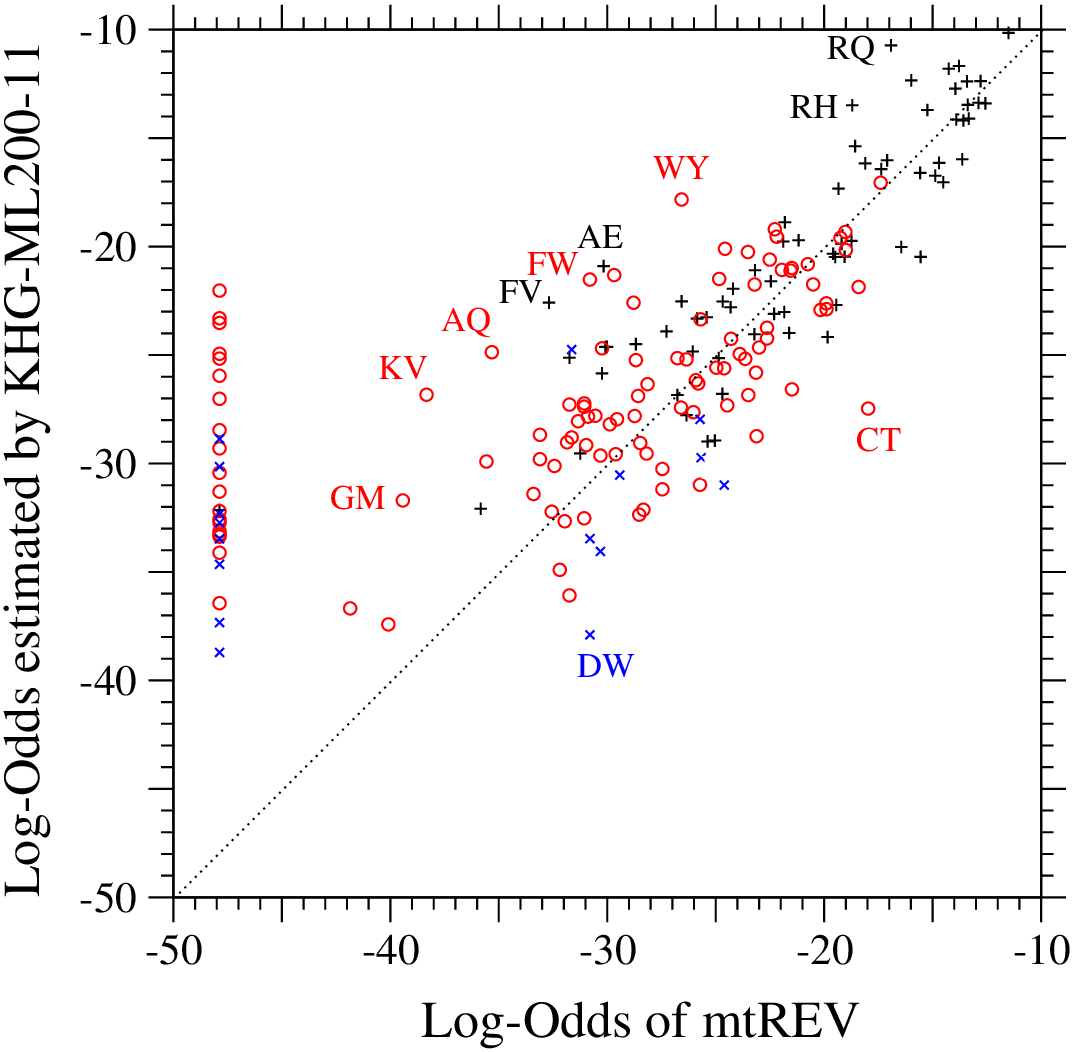}
}
} 
\caption{\label{fig: KHG-ML200_log-odds_1pam_MtRevRmtx}
\label{fig: KHG-ML199_log-odds_1pam_MtRevRmtx}
Each element 
log-$O(\langle S \rangle(\hat{\tau},\hat{\sigma}))_{ab}$
of the log-odds matrix 
of the KHG-ML200-11 model fitted to the 1-PAM mtREV matrix
is plotted against the log-odds log-$O(S^{\script{mtREV}}(\mbox{1 PAM}))_{ab}$
calculated from mtREV.
Plus, circle, and cross marks show the log-odds values for
one-, two-, and three-step amino acid pairs,
respectively.  
The log-odds elements of mtREV whose values are smaller than about $-47.8$ 
are all assumed to be $-47.8$; see the original paper\CITE{AH:96}
The dotted line shows the line 
of equal values between the ordinate and the abscissa.
}
\end{figure*}

} 

\FigurePanel{

\FigureInLegends{\newpage}
\FigureInLegends{\small}

\begin{figure}[ht]
\FigureInLegends{
\textbf{A} \hspace*{45em} \textbf{B}
\centerline{
\includegraphics*[width=75mm,angle=0]{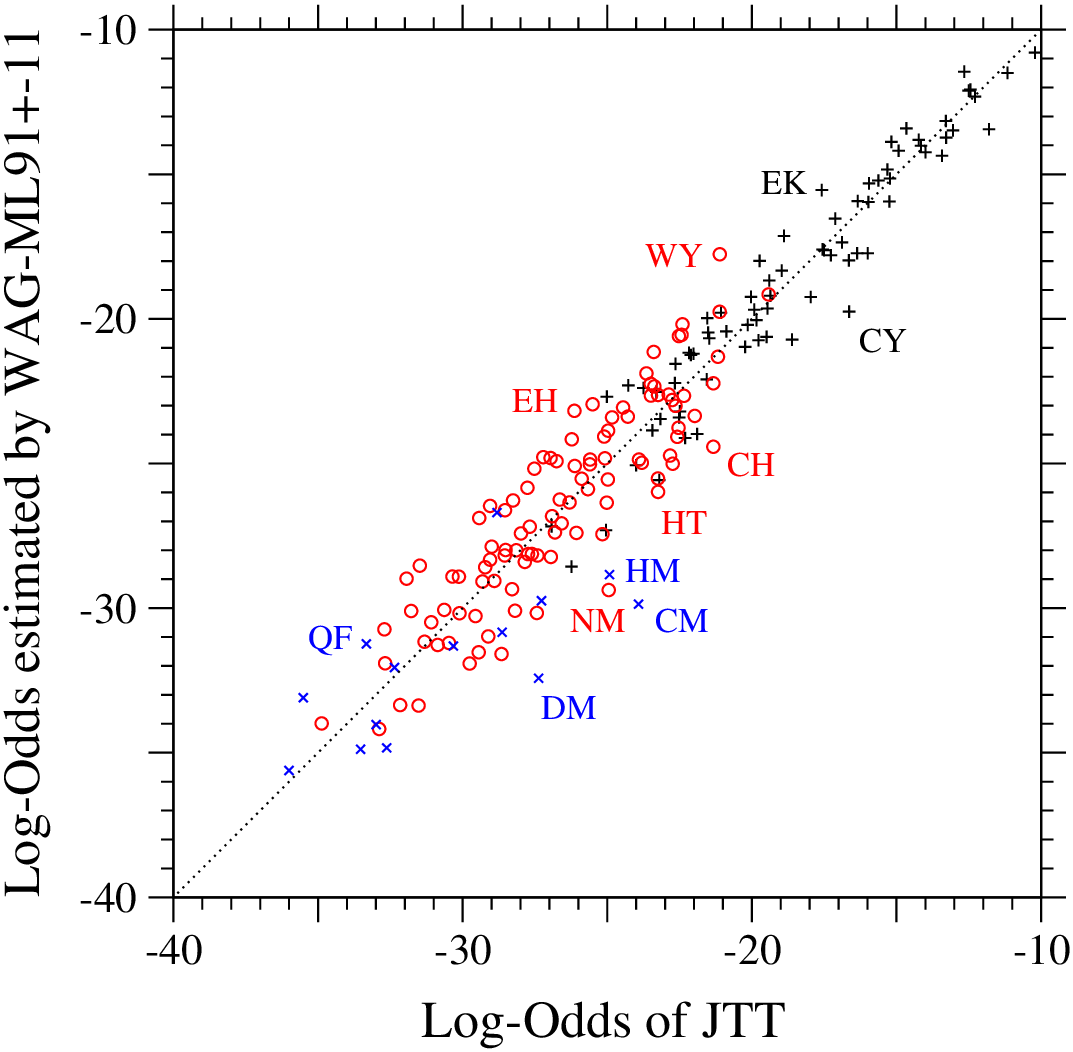}
\hspace*{5mm}
\includegraphics*[width=75mm,angle=0]{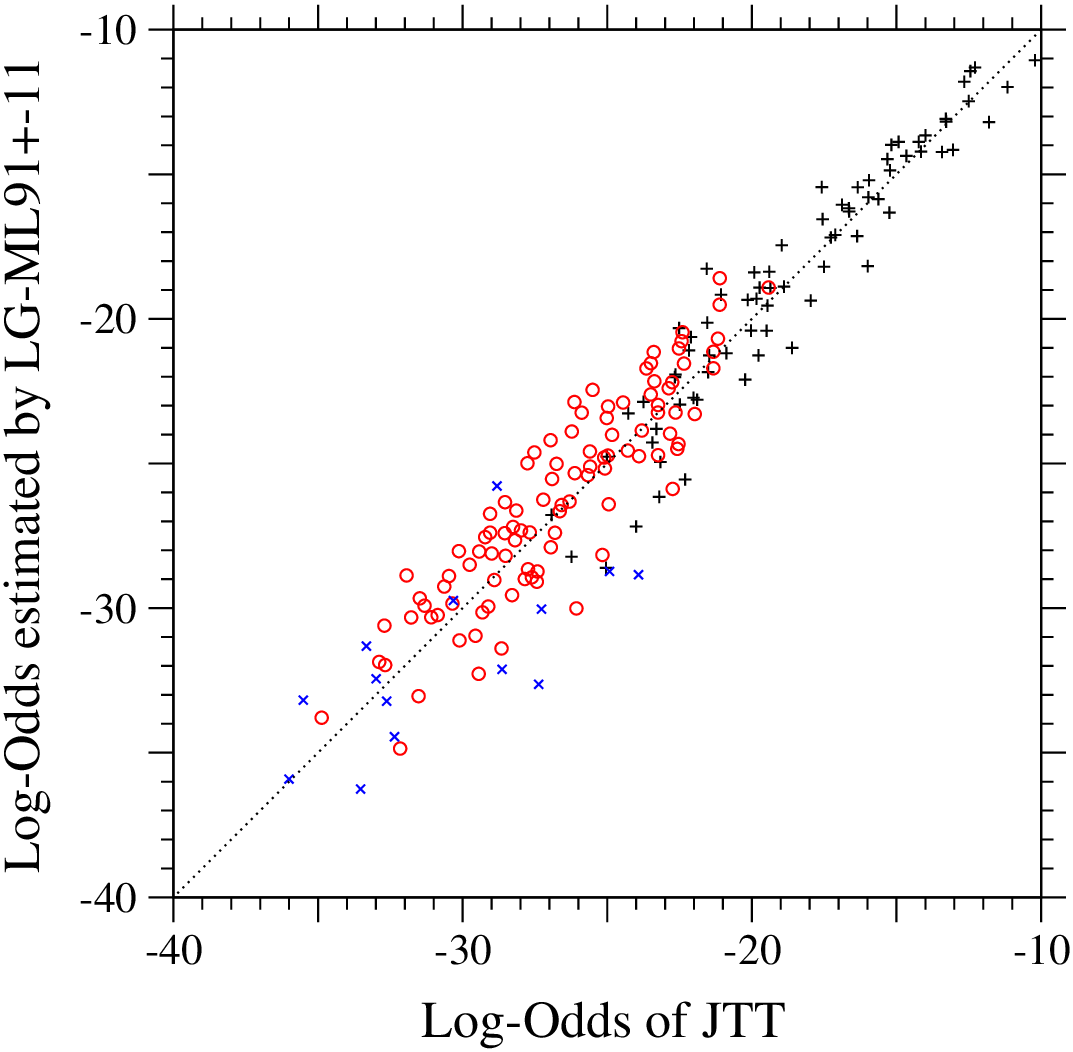}
}
\textbf{C} \hspace*{45em} \textbf{D}
\centerline{
\includegraphics*[width=75mm,angle=0]{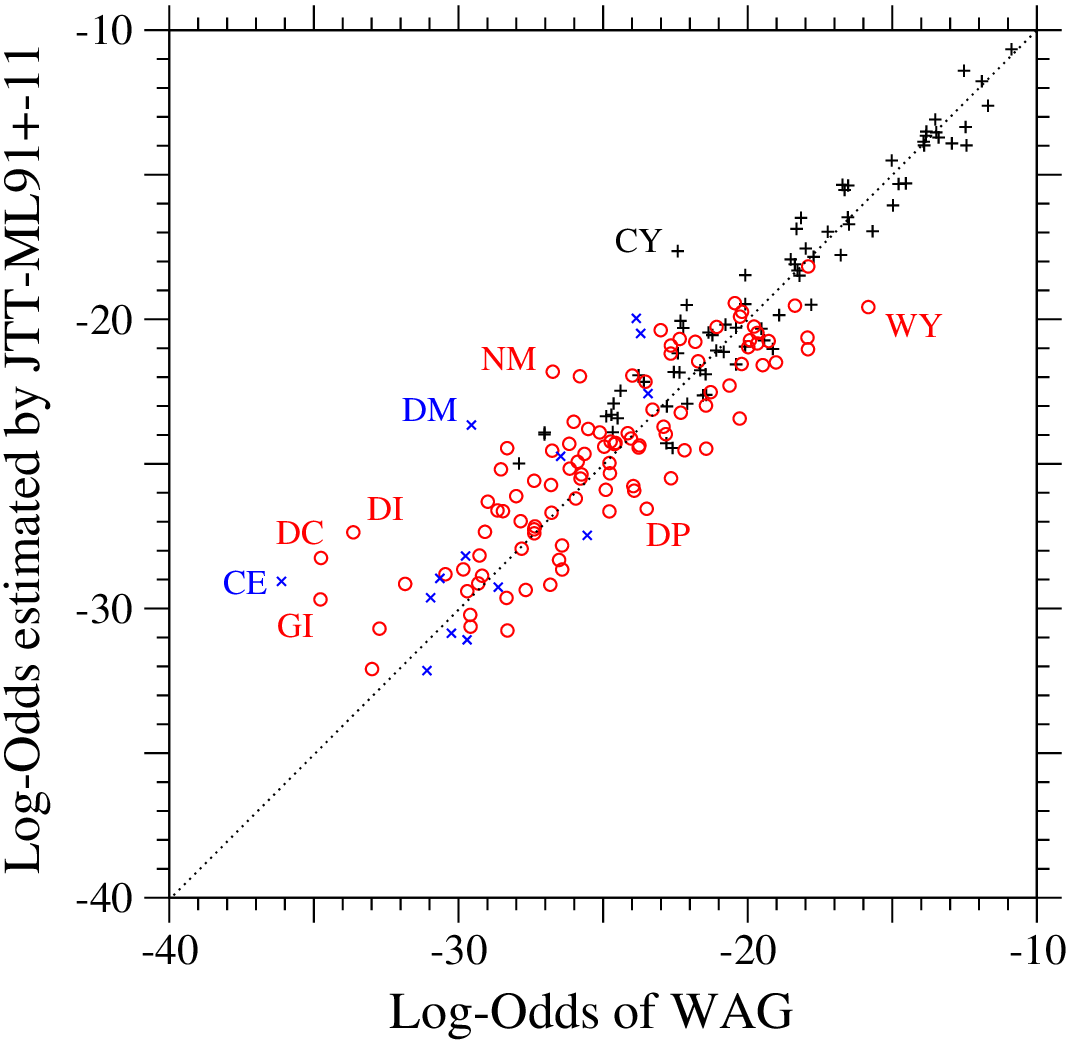}
\hspace*{5mm}
\includegraphics*[width=75mm,angle=0]{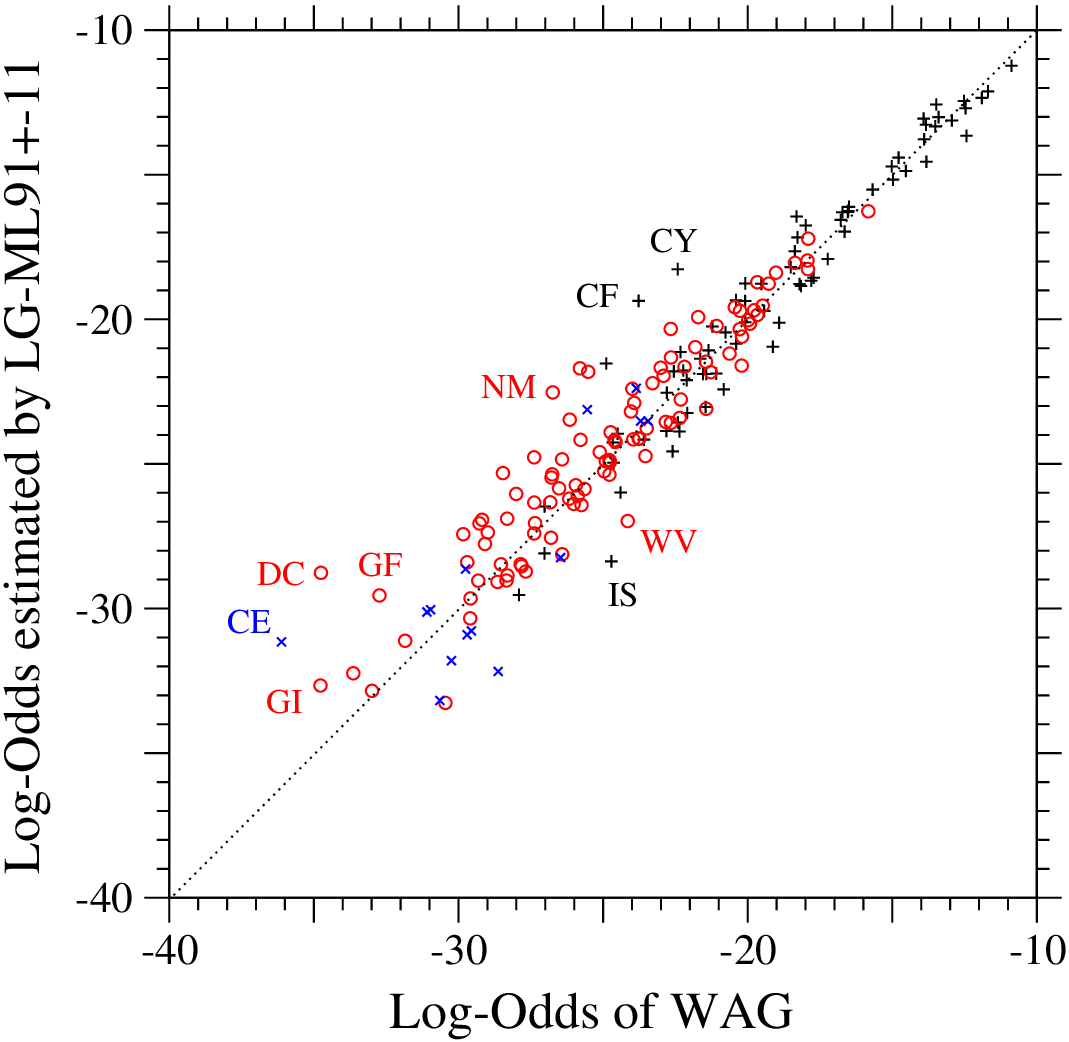}
}
\textbf{E} \hspace*{45em} \textbf{F}
\centerline{
\includegraphics*[width=75mm,angle=0]{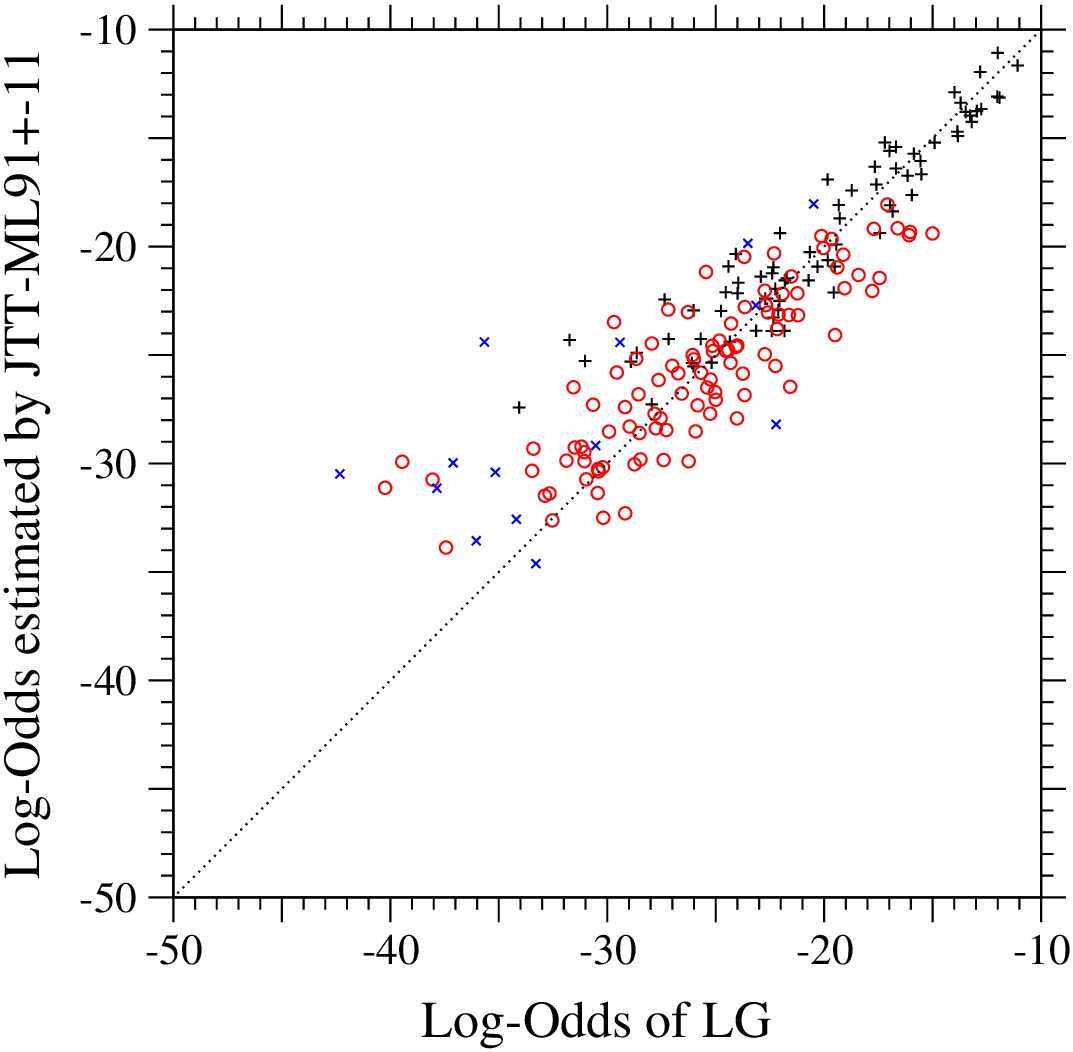}
\hspace*{5mm}
\includegraphics*[width=75mm,angle=0]{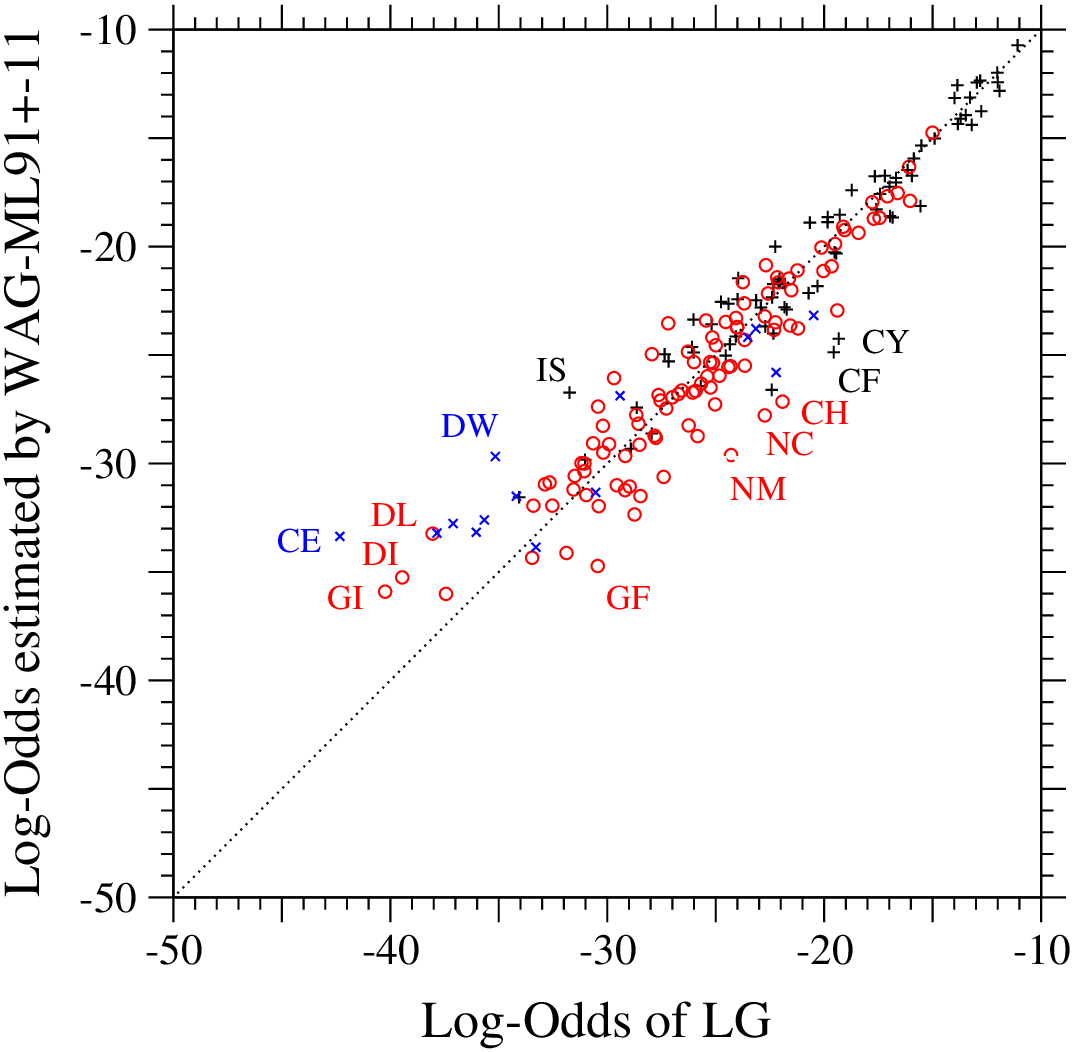}
}

} 
\caption{
\label{fig: X-ML91+_log-odds_1pam_JTT}
\label{fig: X-ML91+_log-odds_1pam_WAG}
\label{fig: X-ML91+_log-odds_1pam_LG}
\FigureInLegends{\small}
\BF{
Models fitted to each of JTT, WAG, and LG.
} 
Each element 
log-$O(\langle S \rangle(\hat{\tau},\hat{\sigma}))_{ab}$
of the log-odds matrix 
of the model fitted to each empirical substitution matrix
is plotted against the log-odds log-$O(S^{\script{obs}}(\mbox{1 PAM}))_{ab}$
calculated from the corresponding empirical substitution matrix.
Plus, circle, and cross marks show the log-odds values for
one-, two-, and three-step amino acid pairs,
respectively.  
The dotted line in each figure shows the line 
of equal values between the ordinate and the abscissa.
\FigureInLegends{\normalsize}
}
\end{figure}

\FigureInLegends{\normalsize}

} 

\FigurePanel{

\FigureInLegends{\newpage}

\FigureInLegends{\small}

\begin{figure}[ht]
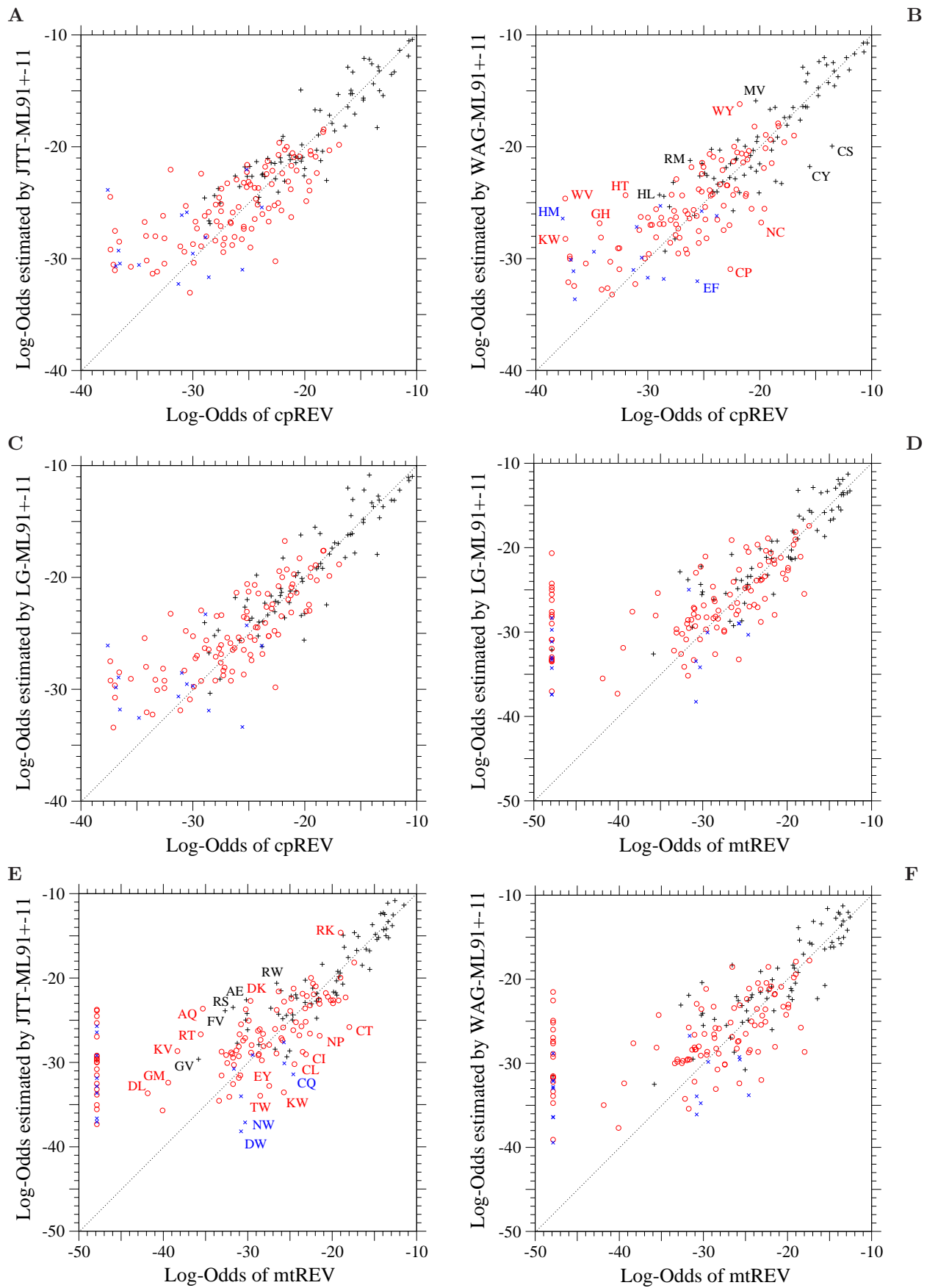

\FigureInLegends{
\textbf{A} \hspace*{45em} \textbf{B}
\centerline{
\includegraphics*[width=75mm,angle=0]{FIGS/JTT-ML91+mod_logodds_1pam_CpRev_JTT+kT+multi+5r+3f+sigma.eps}
\hspace*{5mm}
\includegraphics*[width=75mm,angle=0]{FIGS/WAG-ML91+mod_logodds_1pam_CpRev_WAG+kT+multi+5r+3f+sigma.eps}
}
\textbf{C} \hspace*{45em} \textbf{D}
\centerline{
\includegraphics*[width=75mm,angle=0]{FIGS/LG-ML91+mod_logodds_1pam_CpRev_LG+kT+multi+5r+3f+sigma.eps}
\hspace*{5mm}
\includegraphics*[width=75mm,angle=0]{FIGS/LG-ML91+mod_logodds_1pam_MtRevRmtx_LG+kT+multi+5r+3f+sigma.eps}
}
\textbf{E} \hspace*{45em} \textbf{F}
\centerline{
\includegraphics*[width=75mm,angle=0]{FIGS/JTT-ML91+mod_logodds_1pam_MtRevRmtx_JTT+kT+multi+5r+3f+sigma.eps}
\hspace*{5mm}
\includegraphics*[width=75mm,angle=0]{FIGS/WAG-ML91+mod_logodds_1pam_MtRevRmtx_WAG+kT+multi+5r+3f+sigma.eps}
}

} 
\caption{
\label{fig: X-ML91+_log-odds_1pam_CpRev}
\label{fig: X-ML91+_log-odds_1pam_MtRevRmtx}
\FigureInLegends{\small}
\BF{
Models fitted to each of cpREV and mtREV.
} 
Each element 
log-$O(\langle S \rangle(\hat{\tau},\hat{\sigma}))_{ab}$
of the log-odds matrix 
of the model fitted to each empirical substitution matrix
is plotted against the log-odds log-$O(S^{\script{obs}}(\mbox{1 PAM}))_{ab}$
calculated from the corresponding empirical substitution matrix.
Plus, circle, and cross marks show the log-odds values for
one-, two-, and three-step amino acid pairs,
respectively.  
The dotted line in each figure shows the line 
of equal values between the ordinate and the abscissa.
\FigureInLegends{\normalsize}
}
\end{figure}

\FigureInLegends{\normalsize}

} 

\FigurePanel{

\FigureInLegends{\newpage}

\begin{figure}[ht]
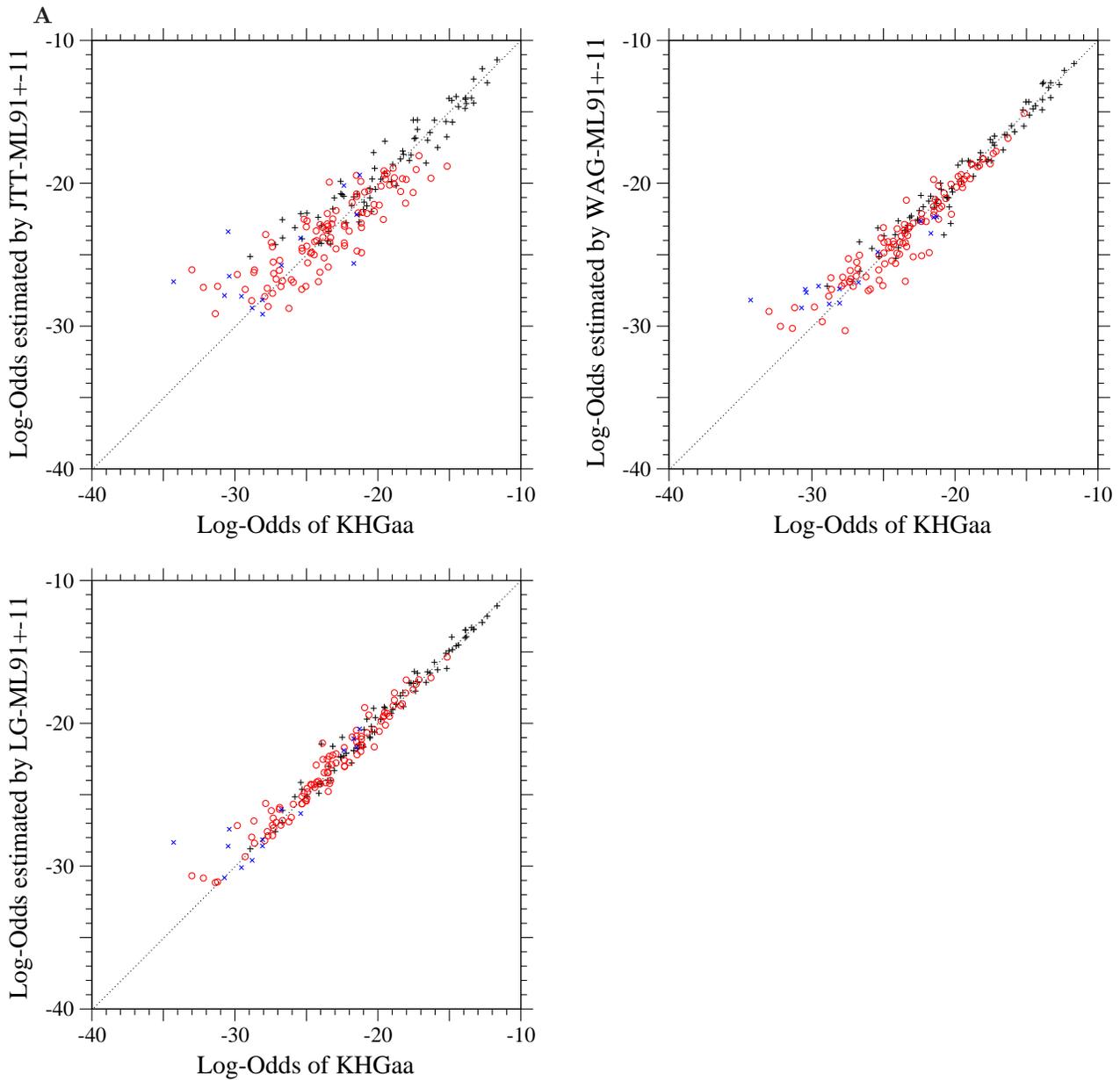

\FigureInLegends{
\textbf{A} \hspace*{47em} \textbf{B}

\centerline{
\includegraphics*[width=80mm,angle=0]{FIGS/JTT-ML91+mod_logodds_1pam_KHGasm_JTT+kT+multi+5r+3f+sigma.eps}
\hspace*{5mm}
\includegraphics*[width=80mm,angle=0]{FIGS/WAG-ML91+mod_logodds_1pam_KHGasm_WAG+kT+multi+5r+3f+sigma.eps}
}
\vspace*{1em}
\centerline{
\includegraphics*[width=80mm,angle=0]{FIGS/LG-ML91+mod_logodds_1pam_KHGasm_LG+kT+multi+5r+3f+sigma.eps}
\hspace*{5mm}
\hspace*{80mm}
}

\textbf{C} 
} 
\caption{
\label{fig: X-ML91+_log-odds_1pam_KHGasm}
\BF{
Models fitted to the KHG-derived amino acid substitution matrix.
} 
Each element 
log-$O(\langle S \rangle(\hat{\tau},\hat{\sigma}))_{ab}$
of the log-odds matrix 
of the model 
fitted to the 1-PAM KHG-derived amino acid substitution matrix (KHGaa)
is plotted against the log-odds log-$O(S^{\script{obs}}(\mbox{1 PAM}))_{ab}$
calculated from KHGaa.
Plus, circle, and cross marks show the log-odds values for
one-, two-, and three-step amino acid pairs,
respectively.  
The dotted line in each figure shows the line 
of equal values between the ordinate and the abscissa.
}
\end{figure}

} 

\FigurePanel{

\FigureInLegends{\newpage}

\begin{figure}[ht]
\FigureInLegends{
\textbf{A} \hspace*{47em} \textbf{B}

\centerline{
\includegraphics*[width=80mm,angle=0]{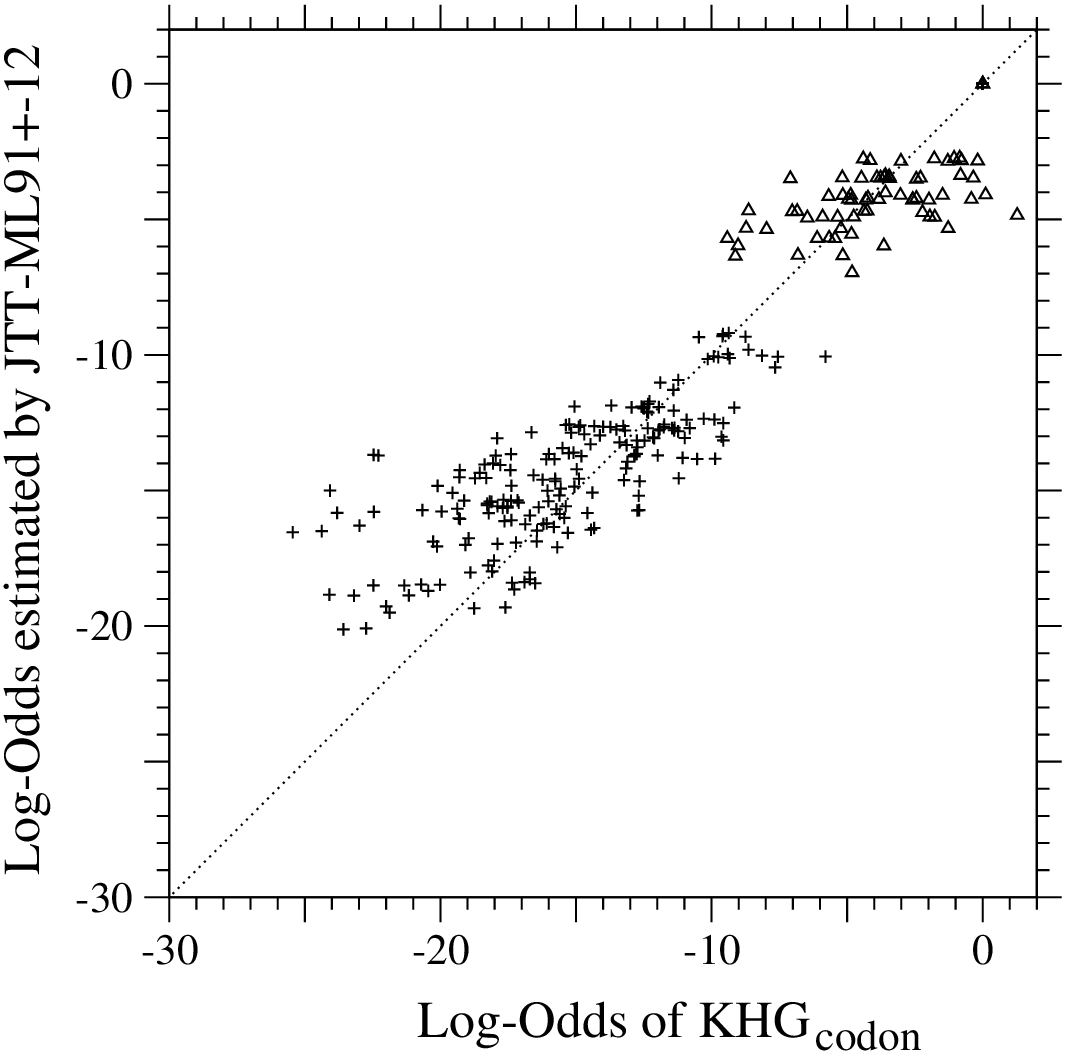}
\hspace*{5mm}
\includegraphics*[width=80mm,angle=0]{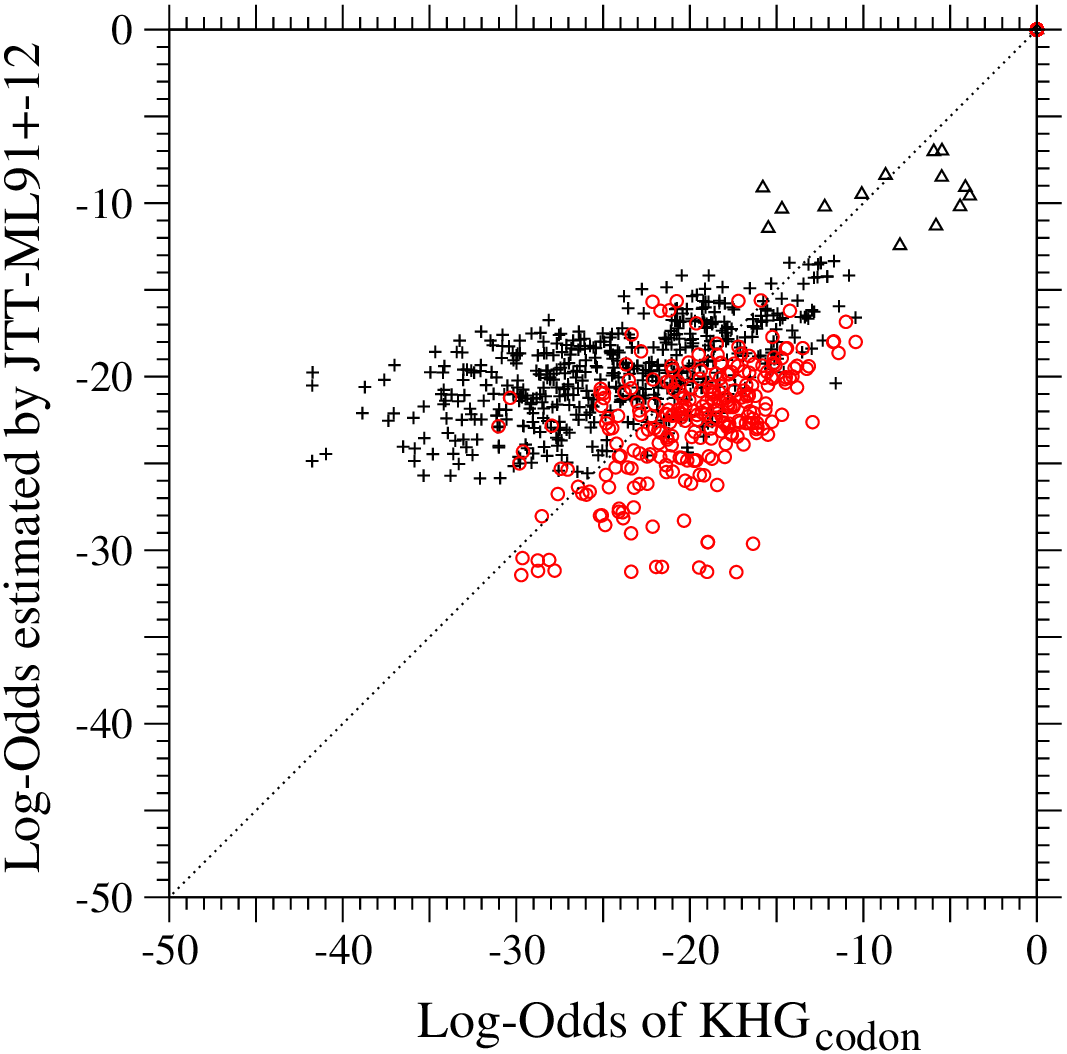}
}
\vspace*{1em}
\centerline{
\includegraphics*[width=80mm,angle=0]{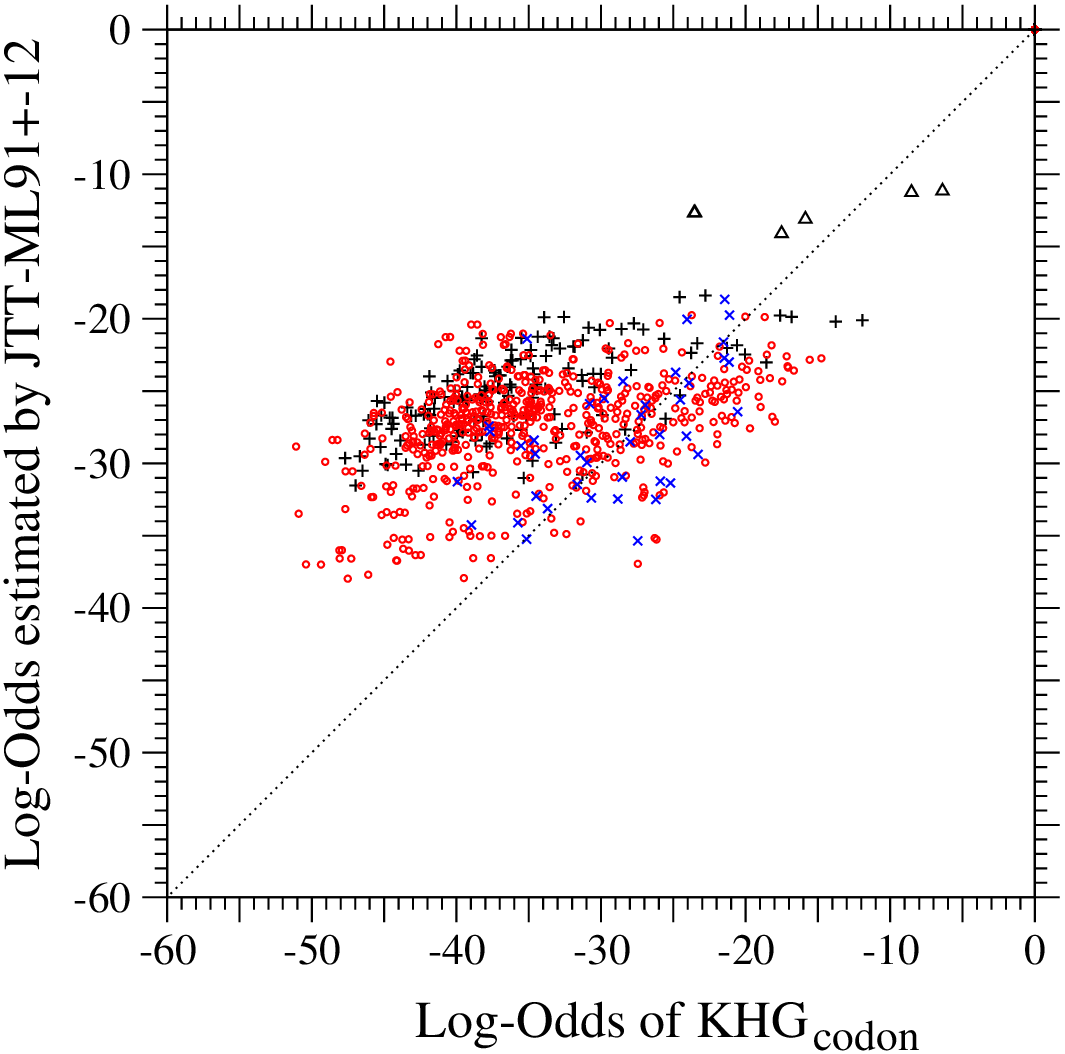}
\hspace*{5mm}
\hspace*{80mm}
}

\textbf{C} 
} 
\caption{
\label{fig: JTT-ML91+_log-odds_1pam_KHG}
\BF{
The JTT-ML91+-12 model fitted to the 1-PAM KHG codon substitution matrix.
} 
Each element 
log-$O(\langle S \rangle(\hat{\tau},\hat{\sigma}))_{\mu \nu}$
of the log-odds matrix 
corresponding to (A) single, (B) double, and (C) triple nucleotide changes 
in the JTT-ML91+-12 model fitted to the 1-PAM KHG codon substitution matrix
is plotted against the log-odds log-$O(S^{\script{KHG}}(\mbox{1 PAM}))_{\mu \nu}$
calculated from KHG.
Upper triangle, plus, circle, and cross marks show the log-odds values for
synonymous pairs and one-, two-, and three-step amino acid pairs,
respectively. 
The dotted line in each figure shows the line 
of equal values between the ordinate and the abscissa.
}
\end{figure}

} 

\FigurePanel{

\FigureInLegends{\newpage}

\begin{figure}[ht]
\FigureInLegends{
\textbf{A} \hspace*{47em} \textbf{B}

\centerline{
\includegraphics*[width=80mm,angle=0]{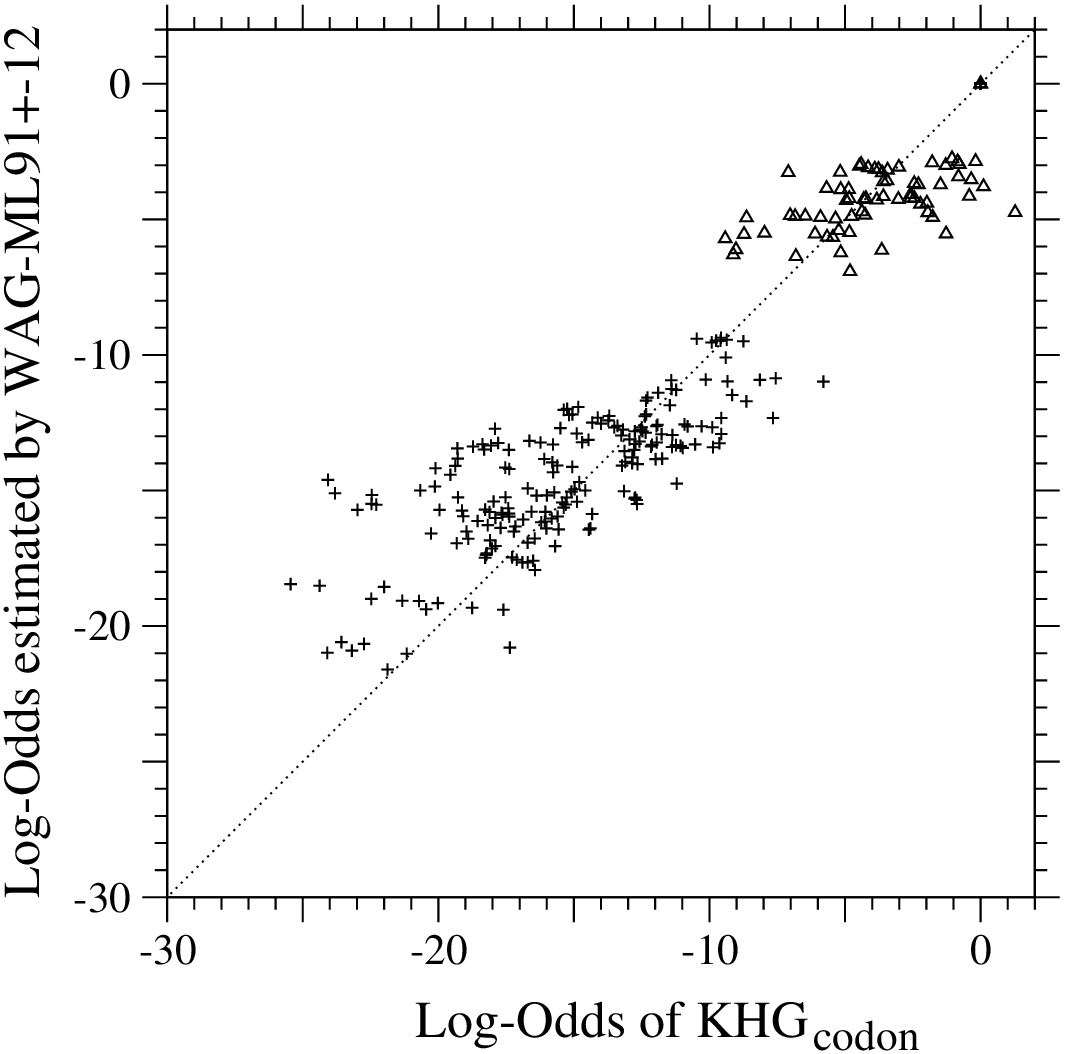}
\hspace*{5mm}
\includegraphics*[width=80mm,angle=0]{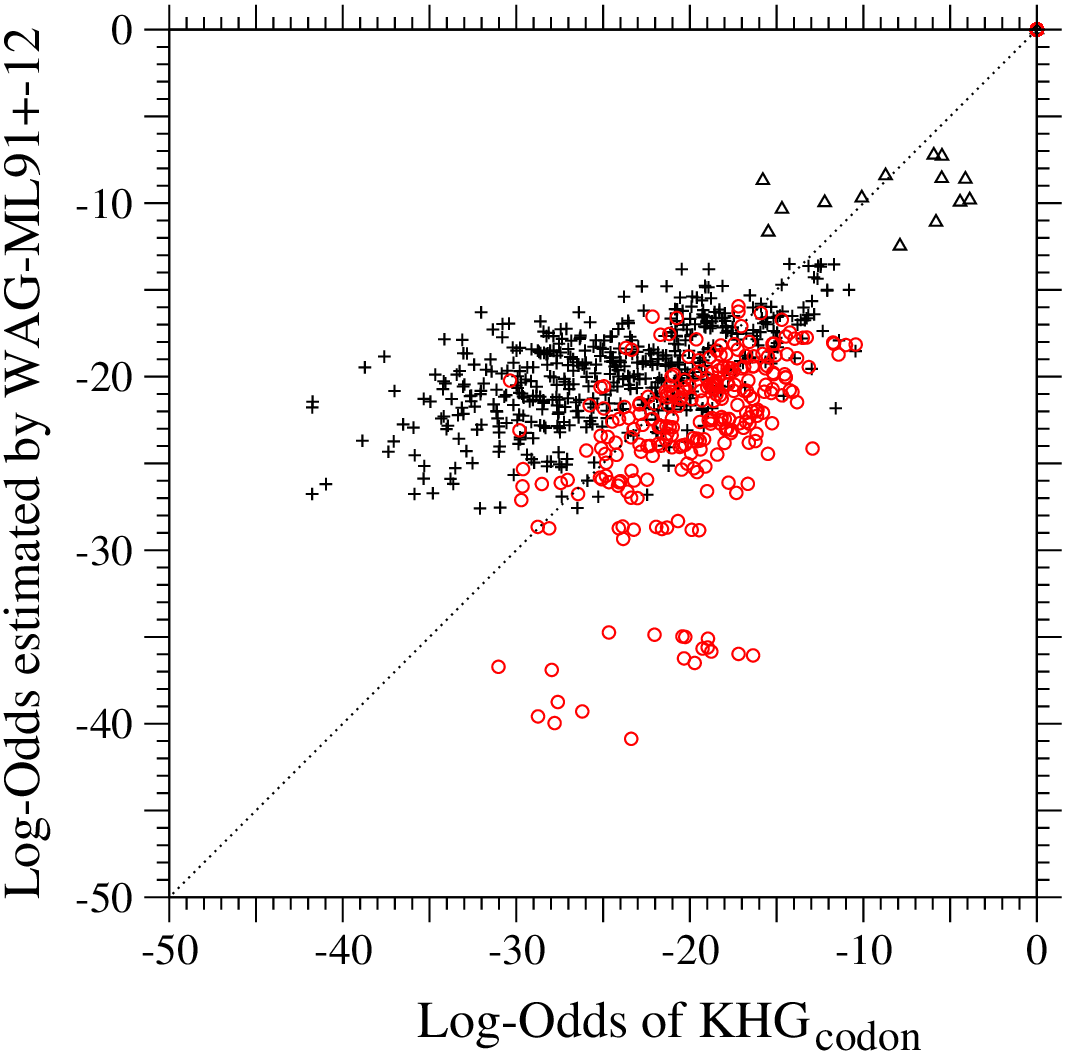}
}
\vspace*{1em}
\centerline{
\includegraphics*[width=80mm,angle=0]{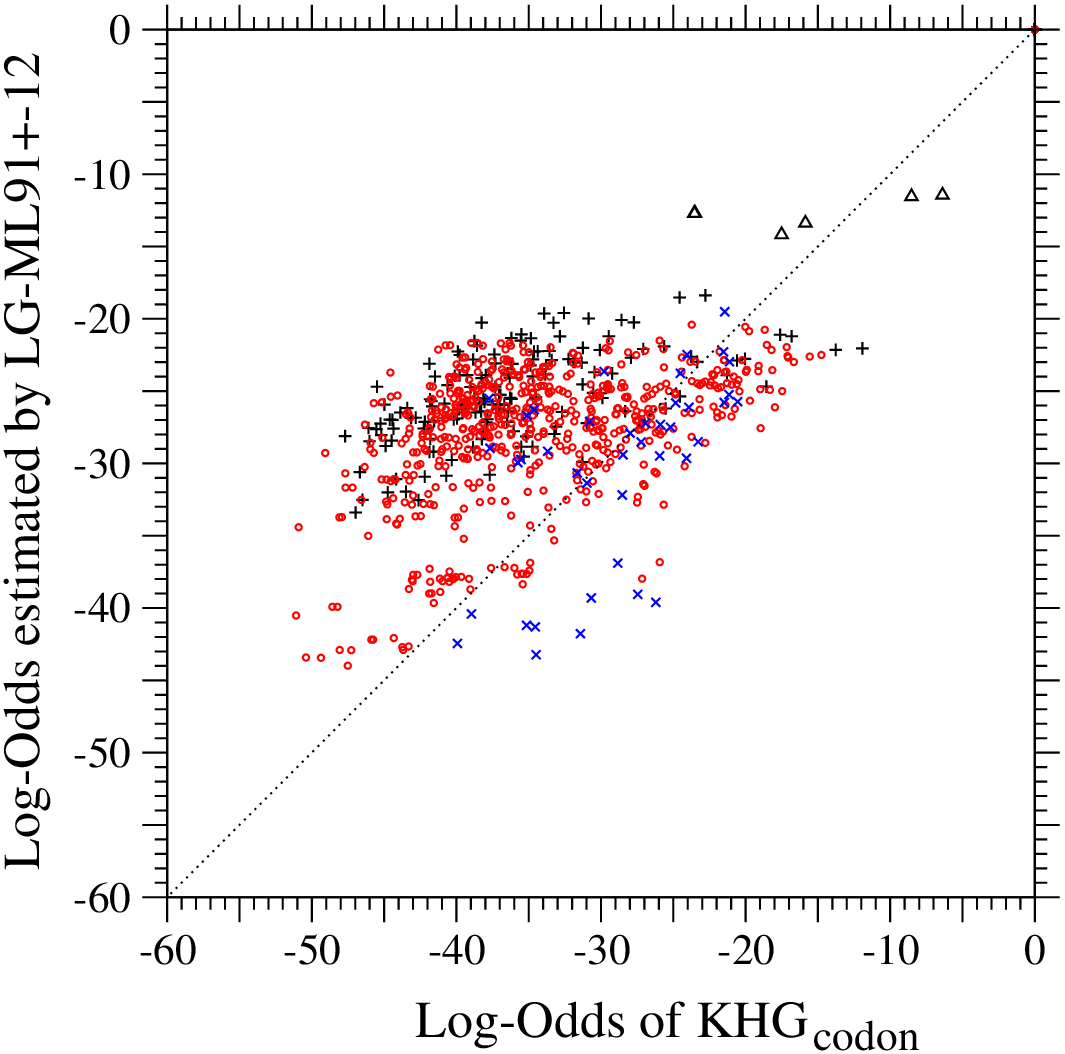}
\hspace*{5mm}
\hspace*{80mm}
}

\textbf{C} 
} 
\caption{
\label{fig: WAG-ML91+_log-odds_1pam_KHG}
\BF{
The WAG-ML91+-12 model fitted to the 1-PAM KHG codon substitution matrix.
} 
Each element 
log-$O(\langle S \rangle(\hat{\tau},\hat{\sigma}))_{\mu \nu}$
of the log-odds matrix 
corresponding to (A) single, (B) double, and (C) triple nucleotide changes 
in the WAG-ML91+-12 model fitted to the 1-PAM KHG codon substitution matrix
is plotted against the log-odds log-$O(S^{\script{KHG}}(\mbox{1 PAM}))_{\mu \nu}$
calculated from KHG.
Upper triangle, plus, circle, and cross marks show the log-odds values for
synonymous pairs and one-, two-, and three-step amino acid pairs,
respectively. 
The dotted line in each figure shows the line 
of equal values between the ordinate and the abscissa.
}
\end{figure}

} 

\FigurePanel{

\FigureInLegends{\newpage}

\begin{figure}[ht]
\FigureInLegends{
\textbf{A} \hspace*{47em} \textbf{B}

\centerline{
\includegraphics*[width=80mm,angle=0]{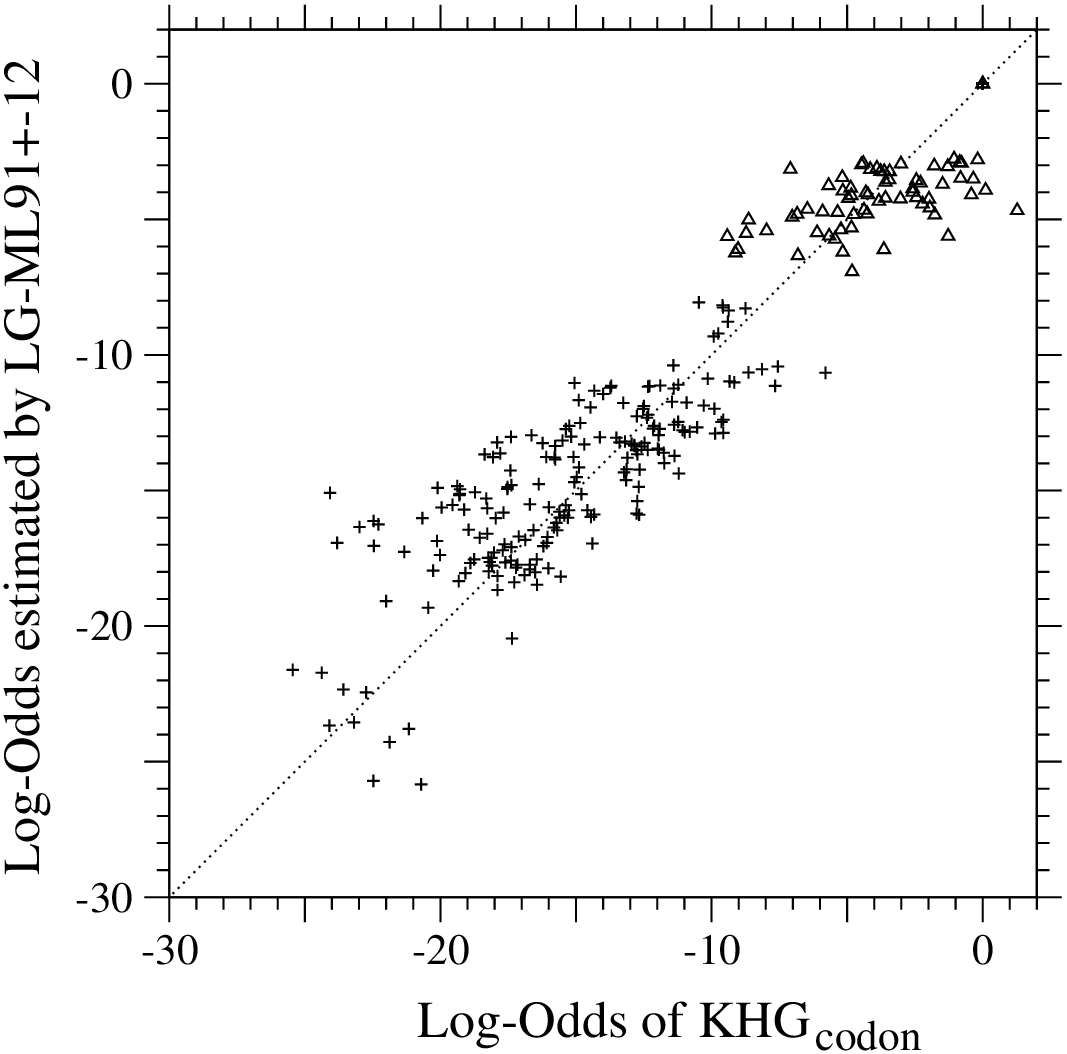}
\hspace*{5mm}
\includegraphics*[width=80mm,angle=0]{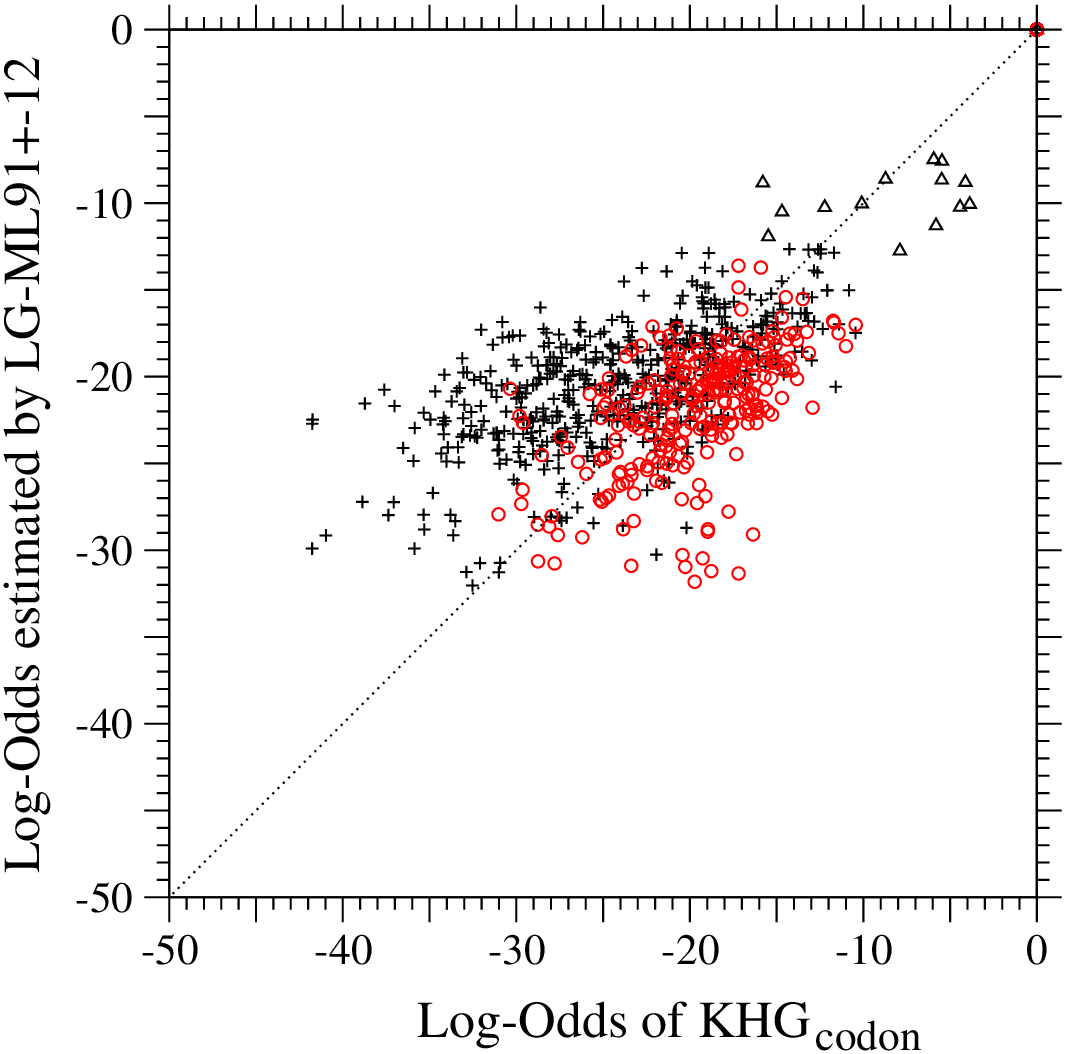}
}
\vspace*{1em}
\centerline{
\includegraphics*[width=80mm,angle=0]{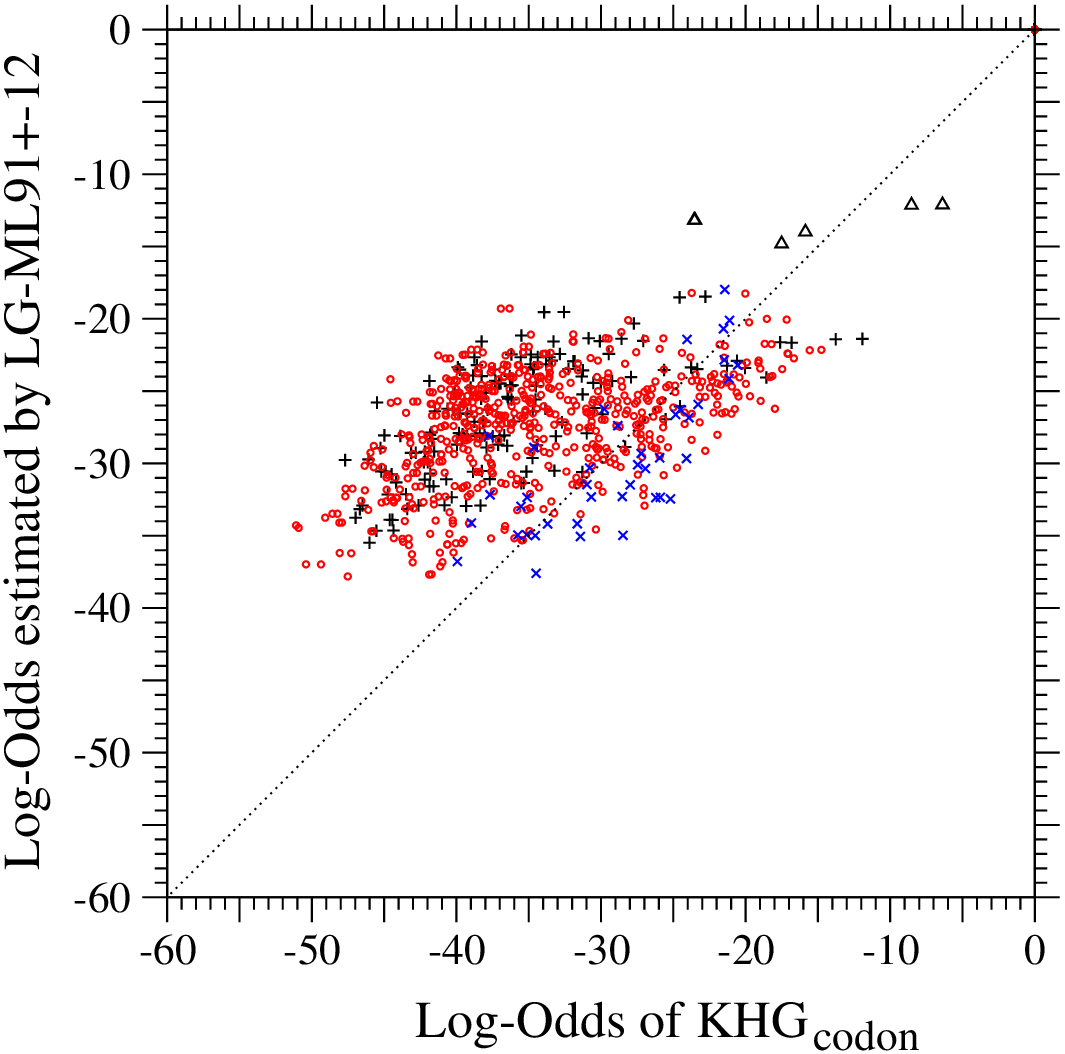}
\hspace*{5mm}
\hspace*{80mm}
}

\textbf{C} 
} 
\caption{
\label{fig: LG-ML91+_log-odds_1pam_KHG}
\BF{
The LG-ML91+-12 model fitted to the 1-PAM KHG codon substitution matrix.
} 
Each element 
log-$O(\langle S \rangle(\hat{\tau},\hat{\sigma}))_{\mu \nu}$
of the log-odds matrix 
corresponding to (A) single, (B) double, and (C) triple nucleotide changes 
in the LG-ML91+-12 model fitted to the 1-PAM KHG codon substitution matrix
is plotted against the log-odds log-$O(S^{\script{KHG}}(\mbox{1 PAM}))_{\mu \nu}$
calculated from KHG.
Upper triangle, plus, circle, and cross marks show the log-odds values for
synonymous pairs and one-, two-, and three-step amino acid pairs,
respectively. 
The dotted line in each figure shows the line 
of equal values between the ordinate and the abscissa.
}
\end{figure}

} 

} 

\SupFig{

\FigureInLegends{\newpage}

\FigureInLegends{\small}

\begin{figure*}[ht]
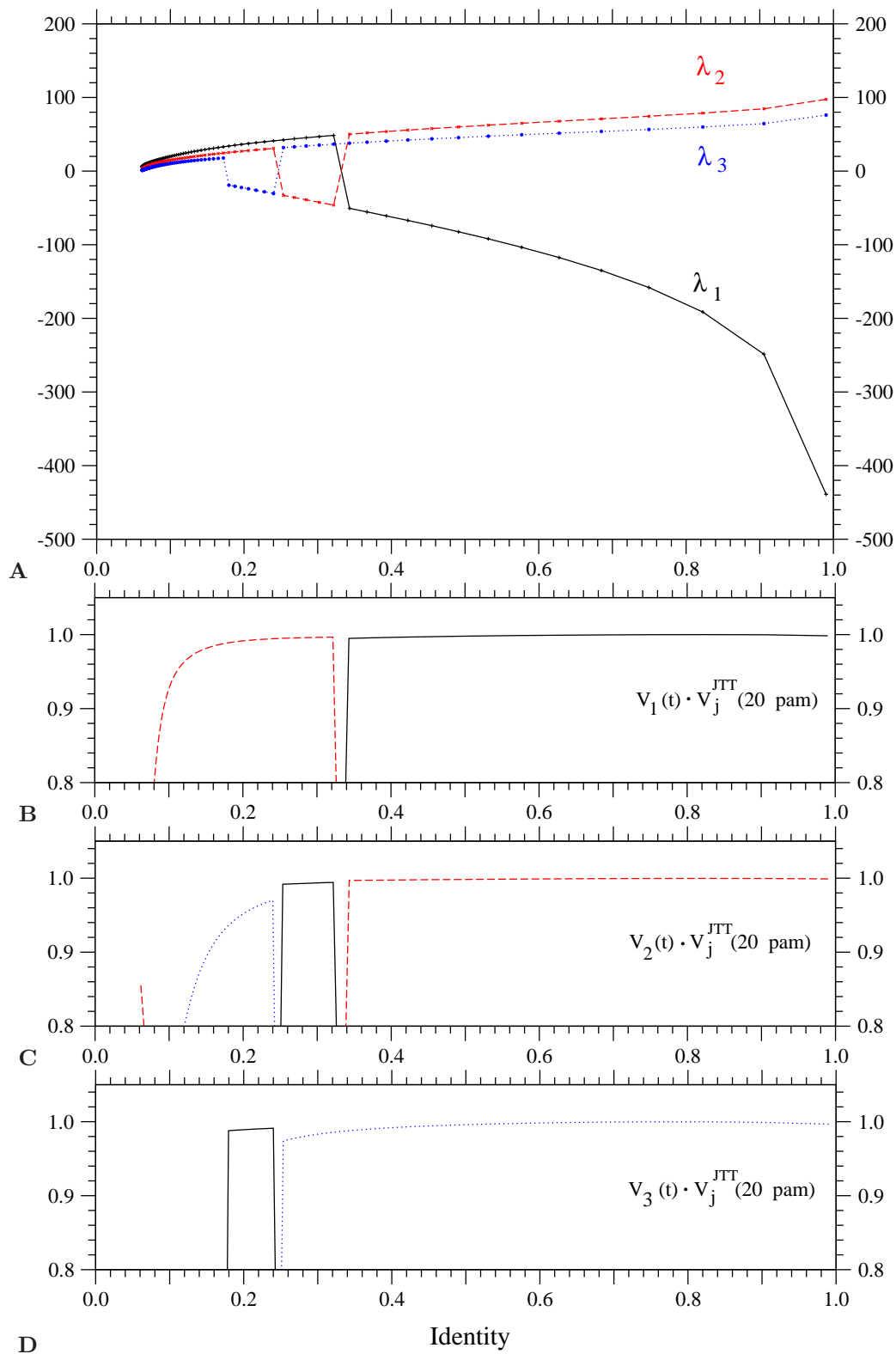

\FigureInLegends{
\centerline{
\textbf{A}
\includegraphics*[width=133mm,angle=0]{FIGS/logodds-eval_ML91+.eps}
}
\centerline{
\textbf{B}
\includegraphics*[width=130mm,angle=0]{FIGS/logodds-evec012vsJTT0_ML91+.eps}
}
\centerline{
\textbf{C}
\includegraphics*[width=130mm,angle=0]{FIGS/logodds-evec012vsJTT1_ML91+.eps}
}
\centerline{
\textbf{D}
\includegraphics*[width=130mm,angle=0]{FIGS/logodds-evec012vsJTT2_ML91+.eps}
}
} 
\caption{\label{fig: log-odds-ev_ML91+}
\label{fig: log-odds-ev_ML91}
\FigureInLegends{\small}
\BF{
Temporal changes of the eigenvalues and the eigenvectors
of the log-odds matrix 
log-$O(\langle S \rangle(t))$
calculated by the ML-91+ model fitted to JTT
as a function of sequence identity.
} 
In (A), 
the solid, the broken, and the dotted lines show 
the temporal changes of the first ($\lambda_1$),
the second ($\lambda_2$), and the third ($\lambda_3$) principal eigenvalues, 
respectively.
The inner products of the eigenvectors with the eigenvectors of the JTT 20-PAM log-odds matrix, 
$\VEC{V}_i(t) \cdot \VEC{V}_j^{\script{JTT}}(\mbox{20-PAM})$,
are shown 
in (B) for the first principal eigenvector ($i = 1$),
in (C) for the second principal eigenvector ($i = 2$), 
and in (D) for the third principal eigenvector ($i = 3$),
by solid lines for $j = 1$, by broken lines for $j = 2$, and 
by dotted lines for $j = 3$.
\FigureInLegends{\normalsize}
}
\end{figure*}
\FigureInLegends{\normalsize}

} 

\clearpage

\pagestyle{myheadings}

\renewcommand{\TableInLegends}[1]{#1}

\setcounter{table}{0}
\renewcommand{\thetable}{S\arabic{table}}
\newpage

\renewcommand{\TextTable}[1]{}
\renewcommand{\SupTable}[1]{#1}

\renewcommand{\FullTable}[1]{#1}

\TextTable{ 

\begin{table}[!ht]
\caption{\label{tbl: model_names}
\BF{
Brief description of models.
} 
}
\vspace*{2em}

\begin{tabular}{p{3cm}|p{12cm}}
\hline
Model name	
		&	Description	\\
\hline
No-Constraints-$n$	& No amino acid dependences of selective constraints; $\beta = 0$.
			 The suffix $n$ means the number of ML parameters.
		\\
\hline
EI-$n$		
		& $\hat{w}_{ab}^{\script{estimate}} \equiv \Delta \hat{\varepsilon}^{\script{c}}_{ab} + \Delta \hat{\varepsilon}^{\script{v}}_{ab}$
		based on the Energy-Increment-based (EI) method, which is described in
		Supporting Information, Text S1, is used 
		to estimate $w_{ab}$ in 
		\Eq{\ref{eq: estimation_of_fitness}}.
		The suffix $n$ means the number of ML parameters.
		\\
\hline
Miyata-$n$	& The amino acid pair distance $d_{ab}$ estimated by 
		Miyata et al.\CITE{MMY:79} 
		is used as $w_{ab}^{\script{estimate}} = - d_{ab}$ 
		to estimate $w_{ab}$ in \Eq{\ref{eq: estimation_of_fitness}}.
		The suffix $n$ means the number of ML parameters.
		\\
\hline
Grantham-$n$	& The amino acid distance $d_{ab}$ estimated by Grantham\CITE{G:74}
		is used as $w_{ab}^{\script{estimate}} = - d_{ab}$ 
		to estimate $w_{ab}$ in \Eq{\ref{eq: estimation_of_fitness}}.
		The suffix $n$ means the number of ML parameters.
		\\
\hline
ML-$n$		& Selective constraints $\{ w_{ab} \}$ are estimated 
		by maximizing the likelihood of JTT\CITE{JTT:92}, 
	WAG\CITE{WG:01}, or LG\CITE{LG:08}, 
	and called $\{ w_{ab}^{\script{JTT/WAG/LG-ML}n} \}$.
	The suffix $n$ means the number of ML parameters.
	In the ML-87, multiple nucleotide changes are 
	disallowed, and $\{ w_{ab} \}$ for all 75 single-step amino acid pairs
	are estimated.
	In the ML-91 and the ML-94, multiple nucleotide changes are allowed,
	and $\{ w_{ab} \}$ for all 75 single-step amino acid pairs  and for 
	6 groups of multiple-step amino acid pairs are estimated.
	In the ML-91, equal codon usage is assumed.
	In the ML-200 for codon substitution matrices,
	$\{ w_{ab} \}$ for all 190 amino acid pairs are estimated.
		\\
\hline
ML-$n+$		
	&
	First, the ML-$n$ is used to estimate parameters, and then
	$\{ w_{ab} \}$ for all multiple-step amino acid pairs are estimated
	by maximizing the likelihood with fixing all other 
	parameters to the values estimated by the ML-$n$.
		\\
\hline
JTT-ML91-$n$,	
WAG-ML91-$n$,
LG-ML91-$n$
		& Selective constraints 
	$\{ w_{ab}^{\script{JTT/WAG/LG-ML91}} \}$ estimated 
		by maximizing the likelihood of 
	JTT/WAG/LG\CITE{JTT:92,WG:01,LG:08}
		in the ML-91 model are used 
		as $\{ w_{ab}^{\script{estimate}} \}$
                in \Eq{\ref{eq: estimation_of_fitness}}.
	The suffix $n$ means the number of ML parameters.
		\\
\hline
JTT-ML91+-$n$,	
WAG-ML91+-$n$,
LG-ML91+-$n$
		& Selective constraints 
	$\{ w_{ab}^{\script{JTT/WAG/LG-ML91+}} \}$ estimated 
		by maximizing the likelihood of 
	JTT/WAG/LG\CITE{JTT:92,WG:01,LG:08}
		in the ML-91+ model are used 
		as $\{ w_{ab}^{\script{estimate}} \}$
                in \Eq{\ref{eq: estimation_of_fitness}}.
	The suffix $n$ means the number of ML parameters.
	The JTT/WAG/ LG-ML91+-0 models correspond to the JTT/WAG/LG-F models, respectively.
		\\
\hline
KHG-ML200-$n$
		& Selective constraints $\{ w_{ab}^{\script{KHG-ML200}} \}$ estimated 
	by maximizing the likelihood of 
	the KHG codon substitution matrix\CITE{KHG:07} 
		in the ML-200 model are used 
		as $\{ w_{ab}^{\script{estimate}} \}$
                in \Eq{\ref{eq: estimation_of_fitness}}.
	The suffix $n$ means the number of ML parameters.
	The KHG-ML200-0 models correspond to the KHG-F model.
		\\
\hline
\end{tabular}

\end{table}

} 

\SupTable{

\TableInLegends{
\newpage
} 
\begin{table}[ht]
\caption{\label{tbl: optimizations_no-selection_JTT_WAG_CpRev_MtRevRmtx}
\label{tbl: optimizations_no-selection_JTT_CpRev_MtRevRmtx}
\label{tbl: optimizations_no-selection}
ML estimates 
of the present models without 
selective constraints on amino acids for the 1-PAM substitution matrices of 
JTT, WAG, cpREV, and mtREV.
}
\vspace*{2em}
\TableInLegends{

\footnotesize
\begin{tabular}{ll|rr|rr|rr|rr}
\hline
	&	& \multicolumn{2}{c|}{JTT}	&\multicolumn{2}{c|}{WAG}	& \multicolumn{2}{c|}{cpREV}	& \multicolumn{2}{c}{mtREV}	\\
		\cline{3-10}
	&	& \multicolumn{2}{c|}{No-Constraints- $^a$}	& \multicolumn{2}{c|}{No-Constraints- $^a$}	& \multicolumn{2}{c|}{No-Constraints- $^a$}	& \multicolumn{2}{c}{No-Constraints- $^a$} \\
id no.	&parameter
		&1	
							&10	
		&1					&10	
		&1	
							&10	
		&1	
							&10	\\
\hline
0&
$-\hat{w}_0$		&(0.0)	& (0.0) & (0.0)	& (0.0)	& (0.0) & (0.0)	&(0.0)	& (0.0)	 	\\
1&
$1/\hat{\beta}$	&($\infty$) &($\infty$) &($\infty$) &($\infty$) & ($\infty$) & ($\infty$) &($\infty$) & ($\infty$) \\
2&
$\hat{m}_{[tc][ag]}$	&($\rightarrow 0$)	
											&$\rightarrow 0$	
			&($\rightarrow 0$)	&0.279	
			&($\rightarrow 0$)	
											&0.0455	
			&($\rightarrow 0$)	
											&0.0405	\\
3&
$\hat{m}_{tc|ag}/\hat{m}_{[tc][ag]}$ 	&2.16	
											&2.20	
					&1.61	&1.54	

					&2.17	
											&2.62	
					&2.32	
											&3.24	\\
4&
$\hat{m}_{ag}/\hat{m}_{tc|ag}$		&(1.0)	
											&1.28	
					&(1.0)	&1.36	
					&(1.0)	
											&1.50	
					&(1.0)	
											&1.47	\\
5&
$\hat{m}_{ta}/\hat{m}_{[tc][ag]}$	&(1.0)	
											&0.629	
					&(1.0)	&0.687	
					&(1.0)	
											&0.480	
					&(1.0)	
											&0.595	\\
6&
$\hat{m}_{tg}/\hat{m}_{[tc][ag]}$	&(1.0)	
											&0.708	
					&(1.0)	&0.622	
					&(1.0)	
										 	&0.775	
					&(1.0)	
										 	&0.373	\\
7&
$\hat{m}_{ca}/\hat{m}_{[tc][ag]}$	&(1.0)	
											&1.28	
					&(1.0)	&1.45	
					&(1.0)	
											&1.64	
					&(1.0)	
											&1.96	\\
8&
$\hat{f}^{\script{mut}}_{t+a}$		&(0.5)	
											&0.495	
					&(0.5)	&0.401	
					&(0.5)	
										 	&0.279	
					&(0.5)	
										 	&0.226	\\
9&
$\hat{f}^{\script{mut}}_t/\hat{f}^{\script{mut}}_{t+a}$	
					&(0.5)	
											&0.486	
					&(0.5)	&0.503	
					&(0.5)	
											&0.563	
					&(0.5)	
											&0.583	\\
10&
$\hat{f}^{\script{mut}}_c/\hat{f}^{\script{mut}}_{c+g}$	
					&(0.5)	
											&0.335	
					&(0.5)	&0.354	
					&(0.5)	
										 	&0.306	
					&(0.5)	
										 	&0.223	\\
14&
$\hat{\sigma}$				&($\rightarrow 0$)	
											&1.76	
					&($\rightarrow 0$)	&1.58	
					&($\rightarrow 0$)	
										 	&2.96	
					&($\rightarrow 0$)	
										 	&2.46	\\
\hline
&
$\hat{\tau} \hat{\sigma}$		&0.0137	
											&0.0228	
					&0.0136	&0.0206	
					&0.0139	
											&0.0296	
					&0.0149	
											&0.0296	\\
&
\#parameters				&21	
											&30	
					&21	&30	
					&21	
										 	&30	
					&21	
										 	&30	\\
&
$\hat{I}_{KL}(\hat{\VEC{\theta}}) \times 10^8 \ ^b$
					&$729533$	
																	&207260	
					&$1156393$	&233841	
					&$1014962$	
																	&249448	
					&$945289$	
																	&305500	\\

&
$\Delta \mbox{AIC} \ ^c$		&86428.1	&24595.5	
					&37917.6	&7719.1	
					&3478.0		&904.5	
					&2644.1		&901.0	\\
\FullTable{
\hline
&
Ratio of substitution rates 			&	&	&	&	&	&	&	&	\\
& \ \  per codon					&	&	&	&	&	&	&	&	\\
&
\hspace*{1em} the total base/codon 		&1.0	
												&1.30	
						&1.0	&1.47	
						&1.0	
												&1.40	
						&1.0	
												&1.35	\\
&
\hspace*{1em} transition/transversion           &1.13	
												&1.00	
						&0.848	&0.752	
						&1.11	
												&1.02	
						&1.24	
												&1.10	\\
&
\hspace*{1em} nonsynonymous/synonymous$^d$          &2.75	
												&4.15	
						&2.84	&5.77	
						&2.60	
												&4.91	
						&2.09	
												&3.30	\\
\hline
&
Ratio of substitution rates 	 		&	&	&	&	&	&	&	&	\\
& \ \  per codon for $\sigma \rightarrow 0$	&	&	&	&	&	&	&	&	\\
&
\hspace*{1em} the total base/codon    		&1.0	
												&1.0	
						&1.0	&1.21	
						&1.0	
												&1.04	
						&1.0	
												&1.02	\\
&
\hspace*{1em} transition/transversion           &1.13	
												&1.20	
						&0.848	&0.853	
						&1.11	
												&1.43	
						&1.24	
												&1.45	\\
&
\hspace*{1em} nonsynonymous/synonymous$^d$          &2.75	
												&2.83	
						&2.84	&4.26	
						&2.60	
												&3.19	
						&2.09	
												&2.08	\\
} 
\hline
\end{tabular}

\vspace*{1em}

\noindent
$^a$ In all models,
equal codon usage 
($\hat{f}^{\script{usage}}_{t} = \hat{f}^{\script{usage}}_{a} = \hat{f}^{\script{usage}}_{c} = \hat{f}^{\script{usage}}_{g} = 0.25$)
is assumed. 
If the value of a parameter is parenthesized, the parameter is not variable but fixed to the value specified.

\noindent
$^b$ $\hat{I}_{KL}(\hat{\VEC{\theta}}) = -(\ell(\hat{\VEC{\theta}})/N + 2.98607330)$ for JTT, 
						$- (\ell(\hat{\VEC{\theta}})/N + 2.97444860)$ for WAG,
						$- (\ell(\hat{\VEC{\theta}})/N + 2.95801048)$ for cpREV,
					and  $- (\ell(\hat{\VEC{\theta}})/N + 2.85313622)$ for mtREV; see text for details.

\noindent
$^c$ $\Delta \mbox{AIC} \equiv 2 N \hat{I}_{KL}(\hat{\VEC{\theta}}) + 2 \times $ \#parameters 
with $N \simeq 5919000$ for JTT, 
$N \approx 1637663$ for WAG,
$N \approx 169269$ for cpREV 
and $N \approx 137637$ for mtREV; see text for details.

$^d$  Note that these ratios are not the ratios of the rates per site but per codon; see text for details.

\normalsize
} 
\end{table}

} 

\TextTable{ 

\begin{table}[ht]
\caption{\label{tbl: optimizations_EI_AIC}
\label{tbl: optimizations_AIC}
\BF{
$\Delta$AIC values
of the present models 
without and with
the selective constraints on amino acids, 
} 
which are based on 
mean energy increments due to an amino acid substitution (EI), 
the Miyata's and the Grantham's physico-chemical distances,
for the 1-PAM amino acid substitution matrices of 
JTT, WAG, cpREV, and mtREV.
}
\vspace*{2em}

\footnotesize

\begin{tabular}{l|l|r|r|r|r}
\hline
	&	& \multicolumn{4}{c}{$\Delta \mbox{AIC} \ ^a$}  \\
	\hline
Model &\#parameters	& JTT    &WAG    &cpREV  &mtREV
							\\
	&(id no. $^b$)	&	&	& 	&	\\
\hline
No-Constraints-	&		&		&	&	&	\\
\ \ 1	&21($\beta=0$, 3)	&86428.1	&37917.6 &3478.0 &2644.1 \\
\ \ 10	&30($\beta=0$, 2-10,14)	&24595.6	&7719.1	&904.5	&901.0	\\
\ \ 13	&33($\beta=0$, 2-14)	&22913.6	&7141.5	& 874.9	& 798.8	\\
\hline
EI-	&		&		&	&	&	\\
\ \ 2	&22(1,3)	&77337.9	&35058.8&3186.0	&2396.6	\\
\ \ 2G	&22(1,14)	&24197.7	&5571.6	&974.0	&1066.8	\\
\ \ 3	&23(1,3,14)	&16463.7	&4995.0	&761.5	&776.4	\\
\ \ 4	&24(1-3,14)	&15808.7	&4443.6	&743.0	&753.9	\\
\ \ 8	&28(1-7,14)	&15715.0	&4327.8	&722.0	&728.2	\\
\ \ 7	&27(1-3,8-10,14)&15081.0	&4312.6	&650.7	&688.7	\\
\ \ 10	&30(1,3-10,14)	&15435.7	&4801.8	&670.7	&702.8	\\
\ \ 10M	&30(1-10)	&15270.7	&4250.4	&645.3	&674.3	\\
\ \ 11	&31(1-10,14)	&14999.0	&4202.5	&636.0	&674.3	\\
\ \ 10MU&30(1-3,8-14)	&13464.3	&3959.7	&578.9	&662.4	\\
\ \ 12	&32(1,3-13)	&72316.3	&33908.4&2939.7	&2215.0	\\
\ \ 13	&33(1,3-14)	&13819.7	&4554.2	&623.6 	&655.5	\\
\ \ 13M	&33(1-13)	&13436.2	&3822.4	&551.1	&623.3	\\
\ \ 14	&34(1-14)	&13151.9	&3748.0	&541.9	&614.8	\\
\hline
Miyata-	&		&		&	&	&	\\
\ \ 4	&24(1-3,14)	&16090.1	&4938.1	&750.3	&783.0	\\
\ \ 7	&27(1-3,8-10,14)&15767.2	&4715.4	&654.5	&701.6	\\
\ \ 10	&30(1,3-10,14)  &16446.1	&5124.9	&679.2	&708.5	\\
\ \ 11	&31(1-10,14)    &15536.8	&4429.5	&628.4	&658.4	\\
\ \ 13	&33(1,3-14)     &15058.2	&4943.1	&656.5	&682.3	\\
\ \ 14	&34(1-14)       &14338.5	&4254.0	&603.7	&613.6	\\
\hline
Grantham-	&		&		&	&	&	\\
\ \ 4	&24(1-3,14)	&20505.1	&5953.7	&916.4	&887.1	\\
\ \ 7	&27(1-3,8-10,14)&18898.2	&5814.0	&840.6	&832.9	\\
\ \ 10	&30(1,3-10,14)  &18744.5	&5749.0	&805.4	&799.8	\\
\ \ 11	&31(1-10,14)    &18680.9	&5579.7	&803.2	&796.5	\\
\ \ 13	&33(1,3-14)     &16784.9	&5512.9	&765.0	&741.0	\\
\ \ 14	&34(1-14)       &16729.7	&5477.1	&755.0	&739.5	\\
\hline
\end{tabular}

\vspace*{1em}

\noindent
$^a$ $\Delta \mbox{AIC} \equiv 2 N \hat{I}_{KL}(\hat{\VEC{\theta}}) + 2 \times $ \#parameters
with 
$N \simeq 5919000$ for JTT,
$N \approx 1637663$ for WAG,
$N \approx 169269$ for cpREV,
and $N \approx 137637$ for mtREV; see text for details.

$^b$
ML parameters in each model are specified by
the parameter id numbers in the parenthesis,
and other parameters are fixed at
$\mbox{id}_{0} = 0$, $\mbox{id}_{1} = \infty$, $\mbox{id}_{2} \rightarrow 0$,
$\mbox{id}_{3-7} = 1.0$, $\mbox{id}_{8-13} = 0.5$, and $\mbox{id}_{14} \rightarrow 0$.
Each id number corresponds to the parameter id number listed in 
\Table{\ref{tbl: optimizations_wij_JTT_WAG_LG_KHG}}.

\end{table}

} 

\SupTable{

\TableInLegends{
\newpage
} 
\begin{table}[ht]
\caption{\label{tbl: optimizations_EI_JTT_WAG_CpRev_MtRevRmtx}
\label{tbl: optimizations_EI-14_JTT}
\label{tbl: optimizations_EI-14_WAG}
\label{tbl: optimizations_EI-14_CpRev}
\label{tbl: optimizations_EI-14_MtRevRmtx}
\label{tbl: optimizations_EI-14_WAG_CpRev_MtRevRmtx}
\label{tbl: optimizations_EI_WAG_CpRev_MtRevRmtx}
ML estimates 
of the present models with 
the selective constraints based on 
mean energy increments due to an amino acid substitution (EI) 
for the 1-PAM substitution matrices of JTT, WAG, cpREV, and mtREV.
}
\vspace*{2em}
\TableInLegends{

\footnotesize

\begin{tabular}{l|rr|rr|rr|rr}
\hline
				&\multicolumn{2}{c|}{JTT}	&\multicolumn{2}{c|}{WAG}	&\multicolumn{2}{c|}{cpREV} 	&\multicolumn{2}{c}{mtREV}  \\
				\cline{2-9}
				&EI-10 $^a$	&EI-11 $^a$	&EI-10 $^a$	&EI-11 $^a$	&EI-10 $^a$	&EI-11 $^a$	&EI-10 $^a$	&EI-11 $^a$	\\
\hline
$-\hat{w}_0$				& (0.0)	& (0.0)	& (0.0)	& (0.0)	& (0.0)	& (0.0)	& (0.0)	& (0.0)	\\
$1/\hat{\beta}$			& 2.50	& 2.60	&1.78	&2.14	&2.15	&2.26	&2.14	&2.29	\\			
$\hat{m}_{[tc][ag]}$		&($\rightarrow 0$) &0.308	&($\rightarrow 0$) &0.916	&($\rightarrow 0$) &0.684	&($\rightarrow 0$) &0.737	\\
$\hat{m}_{tc|ag}/\hat{m}_{[tc][ag]}$
				& 2.51	& 2.22	&1.82	&1.58	&2.82	&2.24	&4.21	&3.06	\\
$\hat{m}_{ag}/\hat{m}_{tc|ag}$	& 1.01	& 1.01	&1.13	&1.10	&1.19	&1.14	&1.05	&1.01	\\
$\hat{m}_{ta}/\hat{m}_{[tc][ag]}$ &1.02	&1.07	&1.26	&1.22	&0.992	&1.14	&1.48	&1.44	\\
$\hat{m}_{tg}/\hat{m}_{[tc][ag]}$ &1.06	& 1.09	&0.985	&1.01	&1.34	&1.23	&0.792	&0.797	\\
$\hat{m}_{ca}/\hat{m}_{[tc][ag]}$ &0.937 & 0.891 &1.04	&0.949	&0.974	&0.925	&1.17	&1.08	\\
$\hat{f}^{\script{mut}}_{t+a}$	& 0.582 & 0.565 &0.516	&0.486	&0.376	&0.405	&0.359	&0.403	\\
$\hat{f}^{\script{mut}}_t/\hat{f}^{\script{mut}}_{t+a}$	
				& 0.522 & 0.525	&0.603	&0.575	&0.647	&0.642	&0.671	&0.646	\\
$\hat{f}^{\script{mut}}_c/\hat{f}^{\script{mut}}_{c+g}$	
				& 0.432	& 0.450	&0.495	&0.511	&0.450	&0.462	&0.388	&0.404	\\
$\hat{\sigma}$ 			&3.20	&0.918	&11.7	&0.998	&7.26	&0.969	&5.25	&0.339	\\
\hline
$\hat{\tau} \hat{\sigma}$	&0.0358	&0.0217	&0.0709	&0.0204	&0.0558	&0.0211	&0.0531	&0.0185	\\
\#parameters			&30	&31	& 30 	& 31	&30	&31	&30	&31	\\
$\hat{I}_{KL}(\hat{\VEC{\theta}}) \times 10^8 \ ^b$ 
				&$129885$	&$126178$	&144772	&126415			&180379	&169548			&233525	&222441	\\
$\Delta \mbox{AIC} \ ^c$	&15435.7	&14999.0	&4801.8	&4202.5			&670.7	&636.0			&702.8	&674.3	\\
\FullTable{
\hline
Ratio of substitution rates     &	&	&	&			&	&			&	&       \\
\ \ per codon 			&	&	&	&			&	&			&	&       \\
\hspace*{1em} the total base/codon
				&1.36	&1.35	&1.53	&1.54	&1.45	&1.48	&1.38	&1.44	\\
\hspace*{1em} transition/transversion           
				&1.09	&1.11	&0.803	&0.834	&1.08	&1.13	&1.34	&1.41	\\
\hspace*{1em} nonsynonymous/synonymous$^d$          
				&2.09	&2.13	&2.48	&2.82	&2.45	&2.65	&1.75	&1.92	\\
\hline
Ratio of substitution rates per codon & &	&	&	&	&			&	&\\
\ \ for $\sigma \rightarrow 0$			&	&	&	&			&	&			&	&       \\
\hspace*{1em} total base/codon  &1,0	&1.18	&1.0	&1.38	&1.0	&1.31	&1.0	&1.37	\\
\hspace*{1em} transition/transversion           
				&1.49	&1.28	&1.25	&0.944	&1.93	&1.36	&2.35	&1.56	\\
\hspace*{1em} nonsynonymous/synonymous$^d$          
				&1.12	&1.59	&0.945	&2.13	&1.15	&1.99	&0.767	&1.64	\\
\hline
Ratio of substitution rates per codon 	 	& 	&	&	&	&	&			&	&\\
\ \ for $w_{ab}=0$ and $\sigma \rightarrow 0$	&	&	&	&			&	&			&	&       \\
\hspace*{1em} total base/codon  &1.0	&1.28	&1.0	&1.59	&1.0	&1.48	&1.0	&1.59	\\
\hspace*{1em} transition/transversion           
				&1.31	&1.15	&0.983	&0.830	&1.51	&1.50	&2.15	&1.57	\\
\hspace*{1em} nonsynonymous/synonymous$^d$          
				&2.57	&3.83	&2.82	&6.53	&2.74	&1.16	&1.84	&4.51	\\
} 
\hline
\end{tabular}

\vspace*{1em}

\noindent
$^a$ In all models, 
equal codon usage (
$\hat{f}^{\script{usage}}_{t} = \hat{f}^{\script{usage}}_{a} = \hat{f}^{\script{usage}}_{c} = \hat{f}^{\script{usage}}_{g} = 0.25$
) is assumed.            
If the value of a parameter is parenthesized, the parameter is not variable but fixed to the value specified.

\noindent
$^b$ $\hat{I}_{KL}(\hat{\VEC{\theta}}) = $
				$-(\ell(\hat{\VEC{\theta}})/N + 2.98607330)$ for JTT, 
				$- (\ell(\hat{\VEC{\theta}})/N + 2.97444860)$ for WAG,
				$- (\ell(\hat{\VEC{\theta}})/N + 2.95801048)$ for cpREV,
				and  $- (\ell(\hat{\VEC{\theta}})/N + 2.85313622)$ for mtREV; see text for details.

\noindent
$^c$ $\Delta \mbox{AIC} \equiv 2 N \hat{I}_{KL}(\hat{\VEC{\theta}}) + 2 \times $ \#parameters
with 
$N \simeq 5919000$ for JTT,
$N \approx 1637663$ for WAG,
$N \approx 169269$ for cpREV,
and $N \approx 137637$ for mtREV; see text for details.

$^d$  Note that these ratios are not the ratios of the rates per site but per codon; see text for details.

} 
\end{table}

\TableInLegends{
\newpage
} 
\begin{table}[ht]
\caption{\label{tbl: optimizations_selection_JTT_WAG}
\label{tbl: optimizations_selection_JTT}
\label{tbl: optimizations_selection_WAG}
ML estimates 
of the present models with 
the selective constraints based on the Grantham's and the Miyata's amino acid distances 
for the 1-PAM substitution matrices of JTT and WAG.
}
\vspace*{2em}
\TableInLegends{

\footnotesize
\begin{tabular}{l|rr|rr|rr|rr}
\hline
		& \multicolumn{4}{c|}{JTT}	&\multicolumn{4}{c}{WAG}	\\
		\cline{2-9}
		& \multicolumn{2}{c|}{Grantham- $^a$} & \multicolumn{2}{c|}{Miyata- $^a$}
		& \multicolumn{2}{c|}{Grantham- $^a$} & \multicolumn{2}{c}{Miyata- $^a$}	\\
					&10 	
							&11	&10 	& 11  		
										&10	&11	&10	&11	\\
\hline
$-\hat{w}_0$					& (0.0)	
							&(0.0)	& (0.0)	& (0.0)	
										&(0.0)	&(0.0)	&(0.0)	&(0.0) 	\\
$1/\hat{\beta}$				& 82.0	
							&81.9	& 1.71	& 1.82	
										&58.9	&65.1	&1.28	&1.59 	\\
$\hat{m}_{[tc][ag]}$			&($\rightarrow 0$) 
							&0.0392	& ($\rightarrow 0$) & 0.617 
										&($\rightarrow 0$)	&0.353	&($\rightarrow 0$)	&1.33 	\\
$\hat{m}_{tc|ag}/\hat{m}_{[tc][ag]}$	& 2.12	
							&2.09	& 2.32	& 1.92	
										&1.49	&1.44	&1.64	&1.40 	\\
$\hat{m}_{ag}/\hat{m}_{tc|ag}$		& 1.08	
							&1.08	& 1.05	& 1.05	
										&1.18	&1.17	&1.15	&1.11 	\\
$\hat{m}_{ta}/\hat{m}_{[tc][ag]}$	& 0.864	
							&0.863	& 0.925	& 0.983	
										&0.987	&0.938	&1.02	&1.02 	\\
$\hat{m}_{tg}/\hat{m}_{[tc][ag]}$	& 0.961	
							&0.983	& 0.922	& 0.985	
										&0.816	&0.907	&0.813	&0.912 	\\
$\hat{m}_{ca}/\hat{m}_{[tc][ag]}$	& 1.16	
							&1.16	& 1.26	& 1.12	
										&1.39	&1.32	&1.55	&1.23 	\\
$\hat{f}^{\script{mut}}_{t+a}$		& 0.582	
							&0.581	&0.574	& 0.543	
										&0.528	&0.517	&0.499	&0.466 	\\
$\hat{f}^{\script{mut}}_t/\hat{f}^{\script{mut}}_{t+a}$	
					& 0.512	
							&0.513	&0.513	& 0.505	
										&0.573	&0.562	&0.575	&0.531 	\\
$\hat{f}^{\script{mut}}_c/\hat{f}^{\script{mut}}_{c+g}$	
					& 0.384	
							&0.385	&0.448	& 0.479	
										&0.412	&0.420	&0.513	&0.541 	\\
$\hat{\sigma}$				&2.80	
							&2.37	&2.98	& $0.00938$	
										&9.00	&2.97	&9.87	& $0.00118$ \\
\hline
$\hat{\tau} \hat{\sigma}$		&0.0330	
							&0.0306 & 0.0342 &0.0147	
										&0.0596	&0.0317	&0.0632	&0.0135 	\\
\#parameters				&30	
							&31	& 30	& 31	
										&30	& 31	& 30	&31	\\
$\hat{I}_{KL}(\hat{\VEC{\theta}}) \times 10^8 \ ^b$	
					&$157835$	
								&157281 &$138419$	&$130721$	
													&173694	&168463	&154639	&133347 \\
$\Delta \mbox{AIC} \ ^c$		& 18744.5	
								&18680.9	& 16446.1	& 15536.8	
															&5749.0	&5579.7	&5124.9	&4429.5 \\ 
\FullTable{
\hline
Ratio of substitution rates per codon          	&	
								&	&	&	
											&	&	&	&	\\
\hspace*{1em} the total base/the total codon	&1.35	&1.35	&1.35	&1.34	&1.51	&1.50	&1.51	&1.53	\\
\hspace*{1em} transition/transversion		&1.04	&1.04	&1.07	&1.10	&0.768	&0.779	& 0.791	& 0.812	\\
\hspace*{1em} nonsynonymous/synonymous$^d$	&2.21	&2.20	&2.14	&2.18	&2.54	&2.65	&2.53	&2.93	\\
\hline
Ratio of substitution rates per codon 		&	
								&	&	&      
											&	&	&	&	\\
\ \  for $\sigma \rightarrow 0$			&	&	&	&	&	&	&	&	\\
\hspace*{1em} the total base/the total codon	&1.0	&1.02	&1.0	&1.33	&1.0	&1.16	&1.0	&1.53	\\
\hspace*{1em} transition/transversion		&1.33	&1.31	&1.42	&1.10	&1.06	&0.951	&1.17	&0.813	\\
\hspace*{1em} nonsynonymous/synonymous$^d$	&1.22	&1.28	&1.17	&2.17	&1.04	&1.52	&1.02	&2.93	\\
Ratio of substitution rates per codon 		&	
								&	&	&       
											&	&	&	&	\\
\ \  for $w_{ab}=0$ and $\sigma \rightarrow 0$	&	&	&	&	&	&	&	&	\\
\hspace*{1em} the total base/the total codon	&1.0	&1.04	&1.0	&1.48	&1.0	&1.26	&1.0	&1.74	\\
\hspace*{1em} transition/transversion		&1.12	&1.10	&1.21	&0.990	&0.803	&0.771	&0.881	&0.736	\\
\hspace*{1em} nonsynonymous/synonymous$^d$	&2.67	&2.81	&2.63	&5.24	&2.97	&4.20	&2.92	&8.49	\\
} 
\hline
\end{tabular}

\vspace*{1em}

\noindent
$^a$ In all models, 
equal codon usage
($\hat{f}^{\script{usage}}_{t} = \hat{f}^{\script{usage}}_{a} = \hat{f}^{\script{usage}}_{c} = \hat{f}^{\script{usage}}_{g} = 0.25$)
is assumed.
If the value of a parameter is parenthesized, the parameter is not variable but fixed to the value specified.

\noindent
$^b$ $\hat{I}_{KL}(\hat{\VEC{\theta}}) = -(\ell(\hat{\VEC{\theta}})/N + 2.98607330)$ 
for JTT, and 
$- (\ell(\hat{\VEC{\theta}})/N + 2.97444860)$ for WAG; see text for details.

\noindent
$^c$ $\Delta \mbox{AIC} \equiv 2 N \hat{I}_{KL}(\hat{\VEC{\theta}}) + 2 \times $ \#parameters 
with $N = 5919000$
for JTT, and 
$N \approx 1637663$ for WAG; see text for details.

$^d$  Note that these ratios are not the ratios of the rates per site but per codon; see text for details.

\normalsize
} 
\end{table}

} 

\TextTable{ 

\newpage
\begin{table}[ht]
\caption{\label{tbl: optimizations_wij_JTT_WAG_LG_KHG}
\label{tbl: optimizations_wij_JTT_WAG}
\label{tbl: optimizations_wij_JTT_WAG_LG}
\BF{
ML estimates and $\Delta$AIC values
of the present models
for the 1-PAM amino acid substitution matrices of JTT, WAG, and LG, 
and the 1-PAM codon substitution matrix of KHG.
} 
}
\vspace*{2em}

\footnotesize

\begin{tabular}{ll|rrr|rrr|rr|r}
\hline
	&
						& \multicolumn{3}{c|}{JTT} 
						& \multicolumn{3}{c|}{WAG}
						& \multicolumn{2}{c|}{LG}
						& \multicolumn{1}{c}{KHG}	\\
		&	&	&	&	&	&	&	&	&	&(codon)	\\
\hline
id	& parameter
						& ML--87$^a$ 
							& ML--91$^a$ & ML--94 
						& ML--87$^a$ 
							& ML--91$^a$ & ML--94 
							& ML--91$^a$ & ML--94 
							& ML--200	\\
no.		&	&	&	&	&	&	&	&	&	&		\\
\hline
0& $-\hat{w}_0$				& N/A 	
							& N/A	& N/A
						& N/A
							& N/A	& N/A
							& N/A	& N/A
							& N/A				\\
1& $1/\hat{\beta}$				& N/A   	
							& N/A  	& N/A  
						& N/A  
							& N/A  	& N/A  
							& N/A  	& N/A  
							& N/A  				\\
2&
$\hat{m}_{[tc][ag]}$			
						&($\rightarrow 0$)	
									&0.637	& 0.662 
						&($\rightarrow 0$)
									&1.28	& 1.29	
									&1.08	& 1.19	
									&0.939	\\
3&
$\hat{m}_{tc|ag}/\hat{m}_{[tc][ag]}$	
						&0.0919	
							&1.57	&1.59	
						&0.746	
							&1.70	& 1.69 
							&1.85	& 1.81 
							&0.843	\\
4&
$\hat{m}_{ag}/\hat{m}_{tc|ag}$		
						&1.77	
							&1.14	& 1.15	
						&1.98	
							&1.32	&1.31	
							&1.23	&1.21	
							&0.945	\\
5&
$\hat{m}_{ta}/\hat{m}_{[tc][ag]}$	
						&0.0293	
							&0.729	&0.730	
						&0.0477	
							&0.791	&0.784	
							&0.676	&0.682	
							&1.52	\\
6&
$\hat{m}_{tg}/\hat{m}_{[tc][ag]}$	
						&3.21	
							&0.940	&0.950	
						&3.64	
							&1.04	&1.01	
							&1.07	&1.07	
							&0.554	\\
7&
$\hat{m}_{ca}/\hat{m}_{[tc][ag]}$	
						&0.719	
							&1.19	&1.18	
						&0.110	
							&1.23	&1.23	
							&1.28	&1.25	
							&0.573	\\
8&
$\hat{f}^{\script{mut}}_{t+a}$		
						&0.408	
							&0.459	&0.446 	
						&0.372	
							&0.367	&0.392 	
							&0.388	&0.403 	
							&0.497	\\
9&
$\hat{f}^{\script{mut}}_t/\hat{f}^{\script{mut}}_{t+a}$	
					
						&0.113	
							&0.501	&0.522	
					
						&0.234	
							&0.587	&0.513	
							&0.450	&0.439	
							&0.513	\\
10&
$\hat{f}^{\script{mut}}_c/\hat{f}^{\script{mut}}_{c+g}$	
					
						&0.698	
							&0.429	&0.436 	
					
						&0.425	
							&0.479	&0.471 	
							&0.427	&0.383 	
							&0.470	\\
11&
$\hat{f}^{\script{usage}}_{t+a}$	
						&0.0682	
							&(0.5)	&0.483	
						&0.0669	
							&(0.5)	&0.221	
							&(0.5)	&0.447	
							&NA	\\
12&
$\hat{f}^{\script{usage}}_t/\hat{f}^{\script{usage}}_{t+a}$
					
						&0.461	
							&(0.5)	&0.491	
					
						&0.330	
							&(0.5)	&0.429	
							&(0.5)	&0.555	
							&NA	\\
13&
$\hat{f}^{\script{usage}}_c/\hat{f}^{\script{usage}}_{c+g}$
					
						&0.386	
							&(0.5)	&0.558	
					
						&0.310	
							&(0.5)	&0.306	
							&(0.5)	&0.249	
							&NA	\\
14&
$\hat{\sigma}$				
						&27.3	
							&0.738	&0.740	
						&43.3	
							&0.905	&0.840	
							&0.415	&0.395	
							&$\rightarrow 0$	\\
\hline
\multicolumn{2}{l|}	
{\hspace*{1em} $\hat{\tau} \hat{\sigma}$}		
						&0.334	
							&0.0243	&0.0246	
						&0.317	
							&0.0223	&0.0207	 
							&0.0246	&0.0240	 
							&0.0240	\\
\multicolumn{2}{l|}	
{\hspace*{1em} \#parameters}				
						&107	
							&111	&114	
						&107	
							&111	&114	
							&111	&114	
							&261	\\
\multicolumn{2}{l|}	
{\hspace*{1em} $\hat{I}_{KL}(\hat{\VEC{\theta}}) \times 10^8 \ ^b$}
					&$15695$
							&638	& $613$	
					&$35319$
							&1903	& $1438$ 
							&2771	&2335	
							&269946	\\
\multicolumn{2}{l|}	
{\hspace*{1em} $\Delta \mbox{AIC} \ ^c$}			&2072.0
							&297.5	&300.6	
						&1370.8
							&284.3	&275.1	
								&782.5	&700.4	
							& \mbox{unknown}	\\
\hline					
\multicolumn{2}{l|}	
{Ratio of substitution rates} 	 	&	&	&	&	&	&	&	&	& \\
\multicolumn{2}{l|}	
{\hspace*{1em} per codon}		&       &       &       &       &       &       &       &       & \\
\multicolumn{2}{l|}	
{\hspace*{1em} the total base/codon}    		
							&1.28	
								&1.35	&1.35	  
							&1.38	
								&1.53	&1.52	  
								&1.38	&1.39	
								&1.29 \ \ \	\\
		&	&	&	&	&	&	&	&	&	&(1.29)$^d$	\\
\multicolumn{2}{l|}	
{\hspace*{1em} transition/transversion}           
							&0.464	
								&1.08	&1.08	  
							&0.482	
								&0.932	&0.806	  
								&1.18	&1.20	
								&0.764 \ \ \ \\
		&	&	&	&	&	&	&	&	&	&(0.765)$^d$	\\
\multicolumn{2}{l|}	
{\hspace*{1em} nonsynonymous/synonymous$^e$ }        
							&1.13
								&1.37	&1.34	  
							&1.57	
								&2.07	&2.40	  
								&1.05     &1.20	
								&0.726 \ \ \ \\
		&	&	&	&	&	&	&	&	&	&(0.723)$^d$	\\
\hline
\multicolumn{2}{l|}	
{Ratio of substitution rates }			&	&	&	&	&	&	&	&	& \\
\multicolumn{2}{l|}	
{\hspace*{1em} per codon for $\sigma \rightarrow 0$} &       &       &       &       &       &       &       &       & \\
\multicolumn{2}{l|}	
{\hspace*{1em} total base/codon}    		
							&1.0	
								&1.22	&1.22	  
							&1.0	
								&1.38	&1.40	  
								&1.31	&1.33	
								&1.29	\\
\multicolumn{2}{l|}	
{\hspace*{1em} transition/transversion}           
							&0.101	
								&1.21	&1.22	  
							&0.647	
								&1.11	&0.932	  
								&1.31	&1.35	
								&0.764	\\
\multicolumn{2}{l|}	
{\hspace*{1em} nonsynonymous/synonymous$^e$}          
							&0.0644
								&1.04	&1.02	  
							&0.138	
								&1.50	&1.79	  
								&0.853	&0.889	
								&0.726	\\
\hline
\multicolumn{2}{l|}	
{Ratio of substitution rates per}			 	&	&	&	&	&	&	&	&	& \\
\multicolumn{2}{l|}	
{\hspace*{1em} codon for $w_{ab}=0$ and $\sigma \rightarrow 0$}	&       &       &       &       &       &       &       &       & \\
\multicolumn{2}{l|}	
{\hspace*{1em} total base/codon}                  
                                                        &1.0    
                                                                &1.45	&1.46     
                                                        &1.0
                                                                &1.72	&1.74     
								&1.67	&1.71	
								&1.51	\\
\multicolumn{2}{l|}	
{\hspace*{1em} transition/transversion}           
                                                        &0.0605 
                                                                &0.829	&0.831    
                                                        &0.499
                                                                &0.933	&0.849    
								&0.992	&0.981	
								&0.427	\\
\multicolumn{2}{l|}	
{\hspace*{1em} nonsynonymous/synonymous$^e$}      
                                                        &11.3   
                                                                &5.58	&5.74     
                                                        &11.1
                                                                &8.68	&11.1     
								&7.45	&8.46	
								&6.81	\\
\hline
\end{tabular}

\vspace*{1em}

\noindent
$^a$ If the value of a parameter is parenthesized, the parameter is not variable but fixed to the value specified.

\noindent
$^b$ $\hat{I}_{KL}(\hat{\VEC{\theta}}) = -(\ell(\hat{\VEC{\theta}})/N + 2.98607330)$ for JTT,
$-(\ell(\hat{\VEC{\theta}})/N + 2.97444860)$ for WAG,
$-(\ell(\hat{\VEC{\theta}})/N + 2.96853414)$ for LG,
and
$-(\ell(\hat{\VEC{\theta}})/N + 4.19073314)$ for KHG; see text for details. 

\noindent
$^c$ $\Delta \mbox{AIC} \equiv 2 N \hat{I}_{KL}(\hat{\VEC{\theta}}) + 2 \times $ \#parameters 
with $N \simeq 5919000$ for JTT, 
$N \approx 1637663$ for WAG,
$N \approx 10114373$ for LG,
and
the value of $N$ is unknown for KHG; see text for details.

\noindent
$^d$  The value in the parenthesis corresponds to the one for the KHG codon substitution probability matrix.

\noindent
$^e$  Note that these ratios are not the ratios of the rates per site but per codon; see text for details.

\normalfont
\end{table}

\newpage
\begin{table}[ht]
\caption{\label{tbl: correlation_of_wij_between_EI_JTT_WAG_LG_KHG}
\label{tbl: correlation_of_wij}
\BF{
Correlations of $\hat{w}_{ab}$ between various estimates;
} 
the lower half shows the correlation coefficients of $\hat{w}_{ab}$ 
for 75 single-step amino acid pairs
and the upper half does those of $\hat{w}_{ab}$ 
for 86 multi-step amino acid pairs
by excluding 29 amino acid pairs of the least exchangeable category in 
the JTT-ML91, the WAG-ML91 or the LG-ML91.
}
\vspace*{2em}

\begin{tabular}{l|cccccc}
\hline
Model		& EI	& JTT-ML91+	& WAG-ML91+	& LG-ML91+	&  \multicolumn{2}{c}{KHG-ML200} 	\\
\hline
EI		&	&0.45           &0.51           &0.59           &0.55 &(0.65)$^a$	\\
JTT-ML91+	&0.66   &               &0.80           &0.80           &0.51		\\
WAG-ML91+	&0.68   &0.87           &               &0.86           &0.55		\\
LG-ML91+	&0.71   &0.82           &0.90           &               &0.58		\\
KHG-ML200	&0.71   &0.77           &0.69           &0.74		&		\\
\hline
\end{tabular}

\vspace*{1em}
$^a$ The value in the parenthesis is the correlation coefficient
for which the $\hat{w}_{ab}$ for all multi-step amino acid pairs are taken into account.
The correlation coefficient of $\hat{w}_{ab}$ for all amino acid pairs 
between the EI and the KHG-ML200 is equal to 0.60.

\end{table}

} 

\SupTable{

} 

\TextTable{ 

\newpage
\begin{table}[th]
\caption{\label{tbl: optimizations_wij_AIC}
\label{tbl: optimizations_ML91+mod_AIC}
\BF{
$\Delta$AIC values 
of the present models with 
the respective selective constraints on amino acids,
} 
$\hat{w}^{\script{JTT-ML91+}}$, $\hat{w}^{\script{WAG-ML91+}}$,
$\hat{w}^{\script{LG-ML91+}}$, and $\hat{w}^{\script{KHG-M200}}$,
for the various 1-PAM substitution matrices.
}
\vspace*{2em}

\footnotesize

\begin{tabular}{l|l|rrrrr|r|r}
\hline
	& \#parameters	& \multicolumn{5}{c}{$\Delta \mbox{AIC} \ ^b$}  & \multicolumn{2}{|c}{$\hat{I}_{KL}(\hat{\VEC{\theta}}) \times 10^8 \ ^c$}    \\
		\hline
Model name	& \#parameters	 	&JTT	&WAG	&LG	&cpREV	 &mtREV	&KHG & KHG	\\
	& (id no. $^a$)		&	&	&	&	&	&(amino acid)	&(codon)	\\
\hline
JTT-ML91+- &			&	&	&	&	&	&	&	\\
\ \ 0	&20			&	&2657.5	&20807.0 &461.7  &426.0	&	\\
\ \ 1	&21(14)			&	&2065.1	&20382.6 &433.9  &424.4	&	\\
\ \ 4	&24(1-3,14)		&	&1773.7	&16148.3 &439.2  &401.9	&	\\
\ \ 7	&27(1-3,8-10,14)	&	&1257.8	&12330.2 &303.4  &295.5	&	\\
\ \ 11	&31(1-10,14)	&	&1152.9	&12140.0 &291.5  &286.5	& 40931	\\
\ \ 12	&32(0-10,14)	&	&	&	&	&	&	& 473668	\\
\hline
WAG-ML91+- &			&	&	&	&	&	&	&	\\
\ \ 0	&20			&9095.4	&	&10537.3 &316.2  &535.1	& 	\\
\ \ 1	&21(14)			&8928.9	&	&9196.3  &317.1  &532.8	&	\\
\ \ 4	&24(1-3,14)		&6274.9	&	&6354.9	 &281.4  &414.0	&	\\
\ \ 7	&27(1-3,8-10,14)	&3658.3 &	&5294.9	 &261.6  &383.6	&	\\
\ \ 11	&31(1-10,14)	&3299.2 &	&4813.3	 &259.1  &365.1	& 12789 \\
\ \ 12	&32(0-10,14)	&	&	&	&	&	&	& 496804	\\
\hline
LG-ML91+- &			&	&	&	&	&	&	&	\\
\ \ 0	&20			&13669.8 &1806.0 &	 &487.1  &593.4	&	\\
\ \ 1	&21(14)			&12176.2 &1188.8 &	 &421.4  &558.0	&	\\
\ \ 4	&24(1-3,14)		&6325.7	 &811.6	 &	 &340.6  &391.6	&	\\
\ \ 7	&27(1-3,8-10,14)	&3983.0	 &636.0	 &	 &267.0  &329.8	&	\\
\ \ 11	&31(1-10,14)	&3878.5	 &574.7	 &	 &267.1  &314.9	& 5732 	\\
\ \ 12	&32(0-10,14)	&	&	&	&	&	&	& 436557	\\
\hline
KHG-ML200- &			&	&	&	&	&	&	&	\\
\ \ 0	&20			&15063.5 &953.4 &12568.9 &403.6 &593.6	&	&	\\
\ \ 1	&21(14)			&15078.6 &955.4 &12570.9 &405.6 &595.6	&	&	\\
\ \ 4	&24(1-3,14)		&6398.0	 &540.7	&5683.3	&297.4	&399.3	&	&	\\
\ \ 7	&27(1-3,8-10,14)	&4611.5	 &533.4	&3804.2	&259.9	&358.0	&	&	\\
\ \ 11	&31(1-10,14)	&4429.9  &518.7	&3006.1	&251.7	&334.1	&	&	\\
\hline
\end{tabular}

\vspace*{1em}

$^a$
Parameter id numbers in the parenthesis mean 
ML parameters in each model
and other parameters except for $\beta = 1$ and $w_0 = 0$ are fixed              
to the value of the corresponding parameter
listed in the column of the ML-91 or the ML-200 in \Table{\ref{tbl: optimizations_wij_JTT_WAG}};
each id number 
corresponds to the parameter id number listed in   
\Table{\ref{tbl: optimizations_wij_JTT_WAG}}. 

\noindent
$^b$ $\Delta \mbox{AIC} \equiv 2 N \hat{I}_{KL}(\hat{\VEC{\theta}}) + 2 \times $ \#parameters
with 
$N \simeq 5919000$ for JTT,
$N \approx 1637663$ for WAG,
$N \approx 10114373$ for LG,
$N \approx 169269$ for cpREV,
and
$N \approx 137637$ for mtREV; see text for details.

\noindent
$^c$ $\hat{I}_{KL}(\hat{\VEC{\theta}}) = $
$- (\ell(\hat{\VEC{\theta}})/N + 2.97009788)$ for the KHG-derived amino acid substitution probability matrix,
and $- (\ell(\hat{\VEC{\theta}})/N + 4.19073314)$ for the KHG codon substitution probability
matrix; see text for details.

\noindent

\end{table}

} 

\TextTable{

\setlength{\topmargin}{-3.0cm}

\newpage
\begin{table}[ht]
\caption{
\label{tbl: optimizations_wij_WAG_JTT_CpRev_MtRevRmtx_X-ML91+}
\label{tbl: optimizations_wij_WAG_JTT_CpRev_MtRevRmtx_X-ML94+}
\label{tbl: optimizations_wij_JTT_WAG_LG_X-ML91+}
\label{tbl: optimizations_wij_JTT_WAG_LG_X-ML}
\label{tbl: optimizations_wij_JTT_WAG_LG_X-ML_KHG-ML}
\BF{
ML estimates
of the present models with 
the respective selective constraints
for the 1-PAM amino acid substitution matrices of JTT, WAG, and LG.
} 
}
\vspace*{2em}

\small

\begin{tabular}{l|rrr|rrr|rrr}
\hline
		&\multicolumn{3}{c}{JTT} &\multicolumn{3}{|c}{WAG} 	  &\multicolumn{3}{|c}{LG}		\\
		\hline
	 	&WAG- $^a$	&LG- $^a$	&KHG- $^a$    &JTT- $^a$ 	& LG- $^a$ 	&KHG- $^a$    &JTT- $^a$ 	&WAG- $^a$  	&KHG- $^a$  \\
		\hline
	 	&\multicolumn{2}{c|}{ML91+-11} 		&{\footnotesize{ML200-11}}	&\multicolumn{2}{|c|}{ML91+-11}		&{\footnotesize{ML200-11}}	&\multicolumn{2}{|c|}{ML91+-11} &{\footnotesize{ML200-11}}	\\
\hline
$- \hat{w}_0$					&(0.0)	& (0.0)	&(0.0)	&(0.0)	&(0.0)	&(0.0)	&(0.0)	&(0.0)	& (0.0)	\\
$1/\hat{\beta}$				&1.08	&1.32	&1.07	&1.04 	&1.28	&1.01	&0.830	&0.798	&0.757	\\
$\hat{m}_{[tc][ag]}$			&0.429	&0.304	&0.257	&1.29	&0.921	&0.648	&1.45	&1.543	&0.577	\\
$\hat{m}_{tc|ag}/\hat{m}_{[tc][ag]}$	&2.36	&2.42	&1.26	&1.19	&1.71	&0.850	&1.16	&1.82	&0.783	\\
$\hat{m}_{ag}/\hat{m}_{tc|ag}$		&1.22 	&1.16	&0.915	&1.26	&1.27	&1.00	&1.20	&1.26	&0.869	\\
$\hat{m}_{ta}/\hat{m}_{[tc][ag]}$	&0.649	&0.654	&1.32	&0.814	&0.802	&1.54	&0.668	&0.634	&1.59	\\
$\hat{m}_{tg}/\hat{m}_{[tc][ag]}$	&1.13	&1.01	&0.622	&0.862	&0.947	&0.568	&0.988	&1.20	&0.524	\\
$\hat{m}_{ca}/\hat{m}_{[tc][ag]}$	&1.18	&1.31	&0.605	&1.27	&1.33	&0.597	&1.24	&1.20	&0.446	\\
$\hat{f}^{\script{mut}}_{t+a}$		&0.481	&0.507	&0.578	&0.351	&0.405	&0.512	&0.333	&0.335	&0.534	\\
$\hat{f}^{\script{mut}}_t/\hat{f}^{\script{mut}}_{t+a}$	
					&0.527	&0.488	&0.490	&0.548	&0.527	&0.519	&0.462	&0.518	&0.463	\\
$\hat{f}^{\script{mut}}_c/\hat{f}^{\script{mut}}_{c+g}$	
					&0.429	&0.390	&0.413	&0.461	&0.435	&0.463	&0.455	&0.468	&0.446	\\
$\hat{\sigma}$				&1.09	&1.28	&0.604	&0.893  &0.751	&$\rightarrow 0$	&0.886	&0.718	&$\rightarrow 0$	\\
\hline
$\hat{\tau} \hat{\sigma}$		&0.0263	&0.0310	&0.0363	&0.0220	&0.0230	&0.0275	&0.0246	&0.0231	&0.0444	\\
\#parameters				&31	&31	&31	&31	&31	&31	&31	&31	&31	\\
$\hat{I}_{KL}(\hat{\VEC{\theta}}) \times 10^8 \ ^b$	
					&$27346$&$32239$&36897	&$33306$&$15653$&13945	&$59707$&$23488$&14554	\\
$\Delta \mbox{AIC} \ ^c$		&3299.2	&3878.5	&4429.9	&1152.9  &574.7	&518.7	&12140.0&4813.3 &3006.1	\\
\hline
Ratio of substitution	 		&	&	&	&	&	&	&	&	&	\\
\ rates per codon			&	&	&	&	&	&	&	&	&	\\
\ the total base/codon    		&1.35	&1.32 	&1.19	& 1.51  &1.45	&1.19	&1.47 	&1.49	&1.12	\\
\ transition/transversion           	&1.23 	&1.25	&1.02	& 0.815 &0.959	&0.753	&0.902	&1.08	&0.789	\\
\ non-/synonymous$^d$          		&1.49 &1.17 &0.612 & 2.07  &1.59 &0.577	&1.56 &1.60 &0.293	\\
\hline
For $\sigma \rightarrow 0$		&	&	&	&	&	&	&	&	&	\\
\ the total base/codon    		&1.19	&1.13	&1.09	& 1.37  &1.33	&1.19	&1.34 	&1.39	&1.12	\\
\ transition/transversion           	&1.51	&1.57	&1.06	& 0.923 &1.10	&0.753	&1.03	&1.29	&0.789	\\
\ non-/synonymous$^d$  	        	&1.03 &0.755 &0.449 & 1.54  &1.19 &0.577 &1.14 &1.20 &0.293	\\
\hline
For $w_{ab}=0$ and $\sigma \rightarrow 0$ &	&	&	&	&	&	&	&	&	\\
\ the total base/codon    		&1.38	&1.29	&1.18	& 1.66  &1.60	&1.38	&1.68	&1.80	&1.34	\\
\ transition/transversion           	&1.27	&1.28	&0.642	& 0.645 &0.926	&0.440	&0.622 	&0.989	&0.390	\\
\ non-/synonymous$^d$  	        	&4.67 &3.99 &3.71	& 8.62  &7.02	&5.35	&8.79 &9.49	&5.23	\\
\hline
\end{tabular}

\vspace*{1em}

\footnotesize

\noindent
$^a$ In all models, 
equal codon usage 
($\hat{f}^{\script{usage}}_{t} = \hat{f}^{\script{usage}}_{a} = \hat{f}^{\script{usage}}_{c} = \hat{f}^{\script{usage}}_{g} = 0.25$)
is assumed.                          
If the value of a parameter is parenthesized, the parameter is not variable but fixed to the value specified.

\noindent
$^b$ $\hat{I}_{KL}(\hat{\VEC{\theta}}) = $
$- (\ell(\hat{\VEC{\theta}})/N + 2.98607330)$ for JTT, 
$- (\ell(\hat{\VEC{\theta}})/N + 2.97444860)$ for WAG,
and
$- (\ell(\hat{\VEC{\theta}})/N + 2.96853414)$ for LG.

\noindent
$^c$ $\Delta \mbox{AIC} \equiv 2 N \hat{I}_{KL}(\hat{\VEC{\theta}}) + 2 \times $ \#parameters
with 
$N \simeq 5919000$ for JTT,
$N \approx 1637663$ for WAG,
and
$N \approx 10114373$ for LG;
see text for details.

\noindent
$^d$  Note that these ratios are not the ratios of the rates per site but per codon; see text for details.

\normalsize
\end{table}

\newpage
\begin{table}[ht]
\caption{
\label{tbl: optimizations_wij_CpRev_MtRevRmtx_X-ML91+}
\label{tbl: optimizations_wij_CpRev_MtRevRmtx_X-ML}
\label{tbl: optimizations_wij_CpRev_MtRevRmtx_X-ML_KHG-ML}
\BF{
ML estimates
of the present models with 
the respective selective constraints
for the 1-PAM amino acid substitution matrices of cpREV and mtREV.
} 
}
\vspace*{2em}

\small

\begin{tabular}{l|rrrr|rrrr}
\hline
			&\multicolumn{4}{c}{cpREV} 	&\multicolumn{4}{|c}{mtREV}  \\
		\hline
	 	&JTT- $^a$ 	&WAG- $^a$	&LG- $^a$ 	&KHG- $^a$    & JTT- $^a$ 	&WAG- $^a$ 	&LG- $^a$  	&KHG- $^a$    \\
		\hline
	 	&\multicolumn{3}{c|}{ML91+-11} 	&ML200-11	&\multicolumn{3}{|c|}{ML91+-11}	&ML200-11	\\
\hline
$- \hat{w}_0$					&(0.0)	&(0.0)	&(0.0)	&(0.0)	&(0.0)	&(0.0)	& (0.0)	&(0.0)	\\
$1/\hat{\beta}$				&0.940	&0.977	&1.18	&1.02	&0.690	&0.845	&0.977	&0.752	\\
$\hat{m}_{[tc][ag]}$			&0.865	&0.917	&0.611	&0.521	&0.564	&0.524	&0.321	&0.228	\\
$\hat{m}_{tc|ag}/\hat{m}_{[tc][ag]}$	&1.50	&2.23	&2.353	&1.14	&2.01	&3.43	&3.82	&1.64	\\
$\hat{m}_{ag}/\hat{m}_{tc|ag}$		&1.28	&1.30	&1.24	&0.973	&1.06	&1.13	&1.08	&0.752	\\
$\hat{m}_{ta}/\hat{m}_{[tc][ag]}$	&0.746	&0.705	&0.733	&1.61	&0.681	&0.595	&0.638	&2.00	\\
$\hat{m}_{tg}/\hat{m}_{[tc][ag]}$	&1.17	&1.37	&1.25	&0.747	&0.792	&0.893	&0.839	&0.411	\\
$\hat{m}_{ca}/\hat{m}_{[tc][ag]}$	&1.23	&1.17	&1.26	&0.566	&1.65	&1.67	&1.76	&0.623	\\
$\hat{f}^{\script{mut}}_{t+a}$		&0.283	&0.306	&0.328	&0.442	&0.262	&0.270	&0.287	&0.426	\\
$\hat{f}^{\script{mut}}_t/\hat{f}^{\script{mut}}_{t+a}$	
					&0.611	&0.654	&0.609	&0.597	&0.601	&0.652	&0.598	&0.631	\\
$\hat{f}^{\script{mut}}_c/\hat{f}^{\script{mut}}_{c+g}$	
					&0.425	&0.446	&0.393	&0.425	&0.349	&0.304	&0.260	&0.332	\\
$\hat{\sigma}$				&1.93	&1.43	&1.75	&0.158	&3.48	&2.18	&3.37	&2.89	\\
\hline
$\hat{\tau} \hat{\sigma}$		&0.0325	&0.0285	&0.0339	&0.0288	&0.0603	&0.0445	&0.0653	&0.0923	\\
\#parameters				&31	&31	&31	&31	&31	&31	&31	&31	\\
$\hat{I}_{KL}(\hat{\VEC{\theta}}) \times 10^8 \ ^b$	
					&$67803$	&$58229$	&$60586$	&56032	&$81541$	&$110126$	&$91860$	&98837	\\
$\Delta \mbox{AIC} \ ^c$		&291.5	&259.1	&267.1	&251.7	&286.5	&365.1	&314.9	&334.1	\\
\hline
Ratio of substitution	 		&	&	&	&	&	&	&	&	\\
\ rates per codon			&	&	&	&	&	&	&	&	\\
\ the total base/codon    		& 1.45	& 1.46	&1.41	&1.20	& 1.36	& 1.37	&1.33	&1.23	\\
\ transition/transversion           	& 1.05	& 1.20	&1.25	&1.05	& 1.44	& 1.65	&1.74	&1.45	\\
\ non-/synonymous$^d$          		& 1.74 & 1.80 &1.38	&0.631	& 0.908 & 1.04 &0.772 &0.403	\\
\hline
For $\sigma \rightarrow 0$		&	&	&	&	&	&	&	&	\\
\ the total base/codon    		& 1.21	& 1.26	&1.20	&1.16	& 1.11	& 1.15	&1.09	&1.05	\\
\ transition/transversion           	& 1.42	& 1.66	&1.77	&1.07	& 2.52	& 2.73	&3.31	&1.96	\\
\ non-/synonymous$^d$  	        	& 1.03 & 1.10 &0.794 &0.573 & 0.387 & 0.515 &0.312 &0.163	\\
\hline
For $w_{ab}=0$ and $\sigma \rightarrow 0$ &	&	&	&	&	&	&	&	\\
\ the total base/codon    		&1.45	&1.55	&1.44	&1.33	&1.31	&1.37	&1.26	&1.16	\\
\ transition/transversion           	&0.797	&1.20	&1.25	&0.569	&1.06	&1.78	&1.98	&0.883	\\
\ non-/synonymous$^d$  	        	&6.06 &6.33	&5.14	&4.97	&3.40	&3.09	&2.58	&3.02	\\
\hline
\end{tabular}

\vspace*{1em}

\footnotesize

\noindent
$^a$ In all models, 
equal codon usage 
($\hat{f}^{\script{usage}}_{t} = \hat{f}^{\script{usage}}_{a} = \hat{f}^{\script{usage}}_{c} = \hat{f}^{\script{usage}}_{g} = 0.25$)
is assumed.                          
If the value of a parameter is parenthesized, the parameter is not variable but fixed to the value specified.

\noindent
$^b$ $\hat{I}_{KL}(\hat{\VEC{\theta}}) = $
$- (\ell(\hat{\VEC{\theta}})/N + 2.95801048)$ for cpREV, 
and
$- (\ell(\hat{\VEC{\theta}})/N + 2.85313622)$ for mtREV; see text for details.

\noindent
$^c$ $\Delta \mbox{AIC} \equiv 2 N \hat{I}_{KL}(\hat{\VEC{\theta}}) + 2 \times $ \#parameters
with 
$N \approx 169269$ for cpREV,
and $N \approx 137637$ for mtREV; see text for details.

$^d$  Note that these ratios are not the ratios of the rates per site but per codon; see text for details.

\normalsize

\end{table}

\newpage
\begin{table}[ht]
\caption{
\label{tbl: optimizations_wij_KHGasm_KHG_X-ML91+}
\label{tbl: optimizations_wij_KHGasm_KHG_X-ML}
\BF{
ML estimates
of the present models with 
the respective selective constraints
for the 1-PAM KHG-derived amino acid and KHG codon substitution matrices.
} 
}
\vspace*{2em}

\small

\begin{tabular}{l|rrr|rrr}
\hline
			&\multicolumn{3}{c}{KHG (amino acid)} 	&\multicolumn{3}{|c}{KHG (codon)}  \\
		\hline
	 	&JTT- $^a$\ \ \ \ 	&WAG- $^a$\ \ \ \	&LG- $^a$\ \ \ \ 	&JTT- $^a$\ \ \ \ 	&WAG- $^a$\ \ \ \	&LG- $^a$\ \ \ \ 	\\
		\hline
	 	&\multicolumn{3}{c}{ML91+-11} 	&\multicolumn{3}{|c}{ML91+-12}	\\
\hline
$- \hat{w}_0$					&(0.0)	&(0.0)	&(0.0)	&1.29	&1.50	&1.11	\\
$1/\hat{\beta}$				&0.952	&0.912	&1.22	&1.72	&2.02	&1.91	\\
$\hat{m}_{[tc][ag]}$			&1.545	&1.68	&1.33	&1.23	&1.21	&1.15	\\
$\hat{m}_{tc|ag}/\hat{m}_{[tc][ag]}$	&1.19	&1.73	&1.69	&0.992	&1.07	&1.09	\\
$\hat{m}_{ag}/\hat{m}_{tc|ag}$		&1.24	&1.28	&1.22	&1.09	&1.12	&1.10	\\
$\hat{m}_{ta}/\hat{m}_{[tc][ag]}$	&0.689	&0.682	&0.748	&1.26	&1.25	&1.25	\\
$\hat{m}_{tg}/\hat{m}_{[tc][ag]}$	&0.855	&1.07	&0.943	&0.646	&0.662	&0.671	\\
$\hat{m}_{ca}/\hat{m}_{[tc][ag]}$	&1.32	&1.26	&1.31	&0.815	&0.806	&0.813	\\
$\hat{f}^{\script{mut}}_{t+a}$		&0.317	&0.334	&0.377	&0.480	&0.484	&0.488	\\
$\hat{f}^{\script{mut}}_t/\hat{f}^{\script{mut}}_{t+a}$	
					&0.533	&0.579	&0.512	&0.499	&0.499	&0.493	\\
$\hat{f}^{\script{mut}}_c/\hat{f}^{\script{mut}}_{c+g}$	
					&0.460	&0.480	&0.441	&0.464	&0.459	&0.459	\\
$\hat{\sigma}$				&2.64	&2.25	&1.30	&$\rightarrow 0$	&0.0496	&$\rightarrow 0$	\\
\hline
$\hat{\tau} \hat{\sigma}$		&0.0308	&0.0286	&0.0247	&0.0240	&0.0247	&0.0240	\\
\#parameters				&31	&31	&31	&32	&32	&32	\\
$\hat{I}_{KL}(\hat{\VEC{\theta}}) \times 10^8 \ ^b$	
					&40931	&12789	&5732	&473668	&496804	&436557	\\
\hline
Ratio of substitution	 		&	&	&	&	&	&	\\
\ rates	per codon			&	&	&	&	&	&	\\
\ the total base/codon    		& 1.64	& 1.66	&1.59	& 1.29	& 1.29	&1.29	\\
\ transition/transversion           	& 0.772	& 0.859	&0.891	& 0.759	& 0.765	&0.767	\\
\ non-/synonymous$^c$          		& 2.56 & 2.61	&2.03	& 0.728	& 0.727	&0.724	\\
\hline
For $\sigma \rightarrow 0$		&	&	&	&	&	&	\\
\ the total base/codon    		& 1.39	& 1.45	&1.43	& 1.29	& 1.28	&1.29	\\
\ transition/transversion           	& 0.977	& 1.15	&1.08	& 0.759	& 0.770	&0.767	\\
\ non-/synonymous$^c$  	        	& 1.48 & 1.54 	&1.36	& 0.728	& 0.704	&0.724	\\
\hline
For $w_{ab}=0$ and $\sigma \rightarrow 0$		&	&	&	&	\\
\ the total base/codon    		&1.71	&1.83	&1.75	&1.65	&1.65	&1.64	\\
\ transition/transversion           	&0.637	&0.926	&0.892	&0.51	&0.552	&0.561	\\
\ non-/synonymous$^c$  	        	&9.41 	&10.3 	&8.86	&8.16	&8.07	&7.77	\\
\hline
\end{tabular}

\vspace*{1em}

\footnotesize

\noindent
$^a$ In all models, codon frequencies are taken to be equal to the observed ones.
If the value of a parameter is parenthesized, the parameter is not variable but fixed to the value specified.

\noindent
$^b$ $\hat{I}_{KL}(\hat{\VEC{\theta}}) = $
$- (\ell(\hat{\VEC{\theta}})/N + 2.97009788)$ for the KHG-derived amino acid substitution probability matrix,
and $- (\ell(\hat{\VEC{\theta}})/N + 4.19073314)$ for the KHG codon substitution probability 
matrix; see text for details.

$^d$  Note that these ratios are not the ratios of the rates per site but per codon; see text for details.

\normalsize
\end{table}

} 

\TextTable{

} 

\TextTable{

\newpage
\begin{table}[ht]
\caption{\label{tbl: LK_of_mt_tree}
\BF{
Log-likelihoods of a phylogenetic tree\CITE{AH:96} 
} 
of the concatenated sequences of 12 protein-coding sequences 
encoded on the same strand of mitochondrial DNA 
from 20 vertebrate species with 2 races from human.
}
\vspace*{2em}

\begin{tabular}{l|rrrrrr}
\hline
Codon Substitution 
	&\#p$^b$ & $\ell +$ \hspace*{3em} &$\mbox{AIC} -$ \hspace*{2em}	&$\hat{\sigma}$ & $\hat{m}_{[tc][ag]}$
	& $\hat{m}_{tc|ag}/\hat{m}_{[tc][ag]}$
		\\
Model$^a$	&	&$\Red{116898.6}$	&$\Red{233917.3}$	&		&
	&
		\\
\hline
 LG-1-F$^c$	&60      
					&$\Red{-1293.8}$	&$\Red{2587.6}$		\\
 KHGaa-1-F$^{cd}$	&60      
					&$\Red{-1293.0}$	&$\Red{2586.1}$		\\
 WAG-1-F$^c$	&60      
					&$\Red{-1108.1}$	&$\Red{2216.1}$		\\
 JTT-1-F$^c$    &60      
					&$\Red{-836.4}$       &$\Red{1672.8}$		\\
 mtREV-1-F$^c$  &60      
					&$0.0$		&$0.0$			\\
		\\
 No-Constraints-1-F$^e$	&60     
					&$\Red{-1731.0}$	&$\Red{3462.1}$
					&$(2.46)$     	&$(0.040)$	&$(3.24)$ 	\\
 WAG-ML91+-1-F$^e$	&60	
					&$\Red{1021.4}$	&$\Red{-2042.7}$	
					&$(2.18)$	&$(0.524)$	&$(3.43)$	\\
 JTT-ML91+-1-F$^e$	&60	
					&$\Red{1237.7}$	&$\Red{-2475.5}$	
					&$(3.48)$	&$(0.564)$	&$(2.01)$	\\
 LG-ML91+-1-F$^e$	&60	
					&$\Red{1382.2}$	&$\Red{-2764.4}$
					&$(3.37)$	&$(0.321)$	&$(3.82)$	\\
 EI-1-F$^e$		&60	
					&$\Red{1395.8}$	&$\Red{-2791.6}$
					&$(0.339)$	&$(0.737)$	&$(3.06)$	\\
 KHG-ML200-1-F$^e$	&60	
					&$\Red{1676.9}$	&$\Red{-3353.9}$
					&$(2.89)$	&$(0.228)$	&$(1.64)$	\\
		\\
 No-Constraints-11-F	&70      
					&$\Red{772.2}$	&$\Red{-1524.4}$	
					&$0.906$	&$0.273$	&$3.37$	\\
 EI-12-F	        &71      
					&$\Red{1966.6}$	&$\Red{-3911.2}$
					&$0.326$	&$0.549$	&$3.60$	\\
 WAG-ML91+-12-F         &71      
					&$\Red{2268.3}$	&$\Red{-4514.5}$	
					&$1.84$     	&$0.471$ 	&$4.16 $	\\
 JTT-ML91+-12-F         &71      
					&$\Red{2275.1}$	&$\Red{-4528.1}$
					&$3.57 $	&$0.506 $	&$2.91 $	\\
 KHG-ML200-12-F         &71      
					&$\Red{2355.7}$	&$\Red{-4689.4}$	
					&$0.469 $	&$0.226 $	&$2.50 $	\\
 LG-ML91+-12-F          &71      
					&$\Red{2510.0}$	&$\Red{-4997.9}$	
					&$1.26 $ 	&$0.357 $	&$4.32 $	\\
			\\
 No-Constraints-11-F-dG4  &71      
					&$\Red{2495.4}$	&$\Red{-4968.9}$
					&$0.000$	&$0.182 $	&$3.62 $	\\
 EI-12-F-dG4            &72      
					&$\Red{3742.4}$	&$\Red{-7460.7}$
					&$0.000$	&$0.392 $	&$3.95 $	\\
 JTT-ML91+-12-F-dG4     &72      
					&$\Red{4156.9}$	&$\Red{-8289.8}$
					&$0.064 $	&$0.385 $	&$3.11 $	\\
 KHG-ML200-12-F-dG4     &72      
					&$\Red{4190.0}$	&$\Red{-8356.0}$
					&$0.000$	&$0.147 $	&$2.60 $	\\
 WAG-ML91+-12-F-dG4     &72      
					&$\Red{4196.4}$	&$\Red{-8368.7}$
					&$0.042 $	&$0.342 $	&$4.61 $	\\
 LG-ML91+-12-F-dG4      &72      
					&$\Red{4412.6}$	&$\Red{-8801.1}$
					&$0.029 $	&$0.253 $	&$4.83 $	\\
\hline
\end{tabular}

\vspace*{1em}

$^a$ In all models named with a suffix "F", 
codon frequencies are taken to be equal to those in coding sequences.
A suffix "dG4" means the discrete approximation of the $\Gamma$ distribution
with 4 categories\CITE{Y:94} for rate variation.
The parameter $w_0$ in \Eq{\ref{eq: estimation_of_fitness}} is optimized in all models.

$^b$ The number of parameters; the value for the mtREV-1-F
is not quite correct, because mtREV was estimated 
from the almost same set of protein sequences\CITE{AH:96}.

$^c$ The exchangeabilties of nonsynonymous and synonymous codon pairs are equal to 
$\exp w_0$ multiplied by 
those of the corresponding amino acid pairs 
and all equal to the mean amino acid exchangeability
in the empirical amino acid substitution matrix specified, respectively.

$^d$ KHGaa means the amino acid substitution matrix derived from KHG.

$^e$ All parameters except $w_0$ and codon frequencies are fixed to those ML estimates
of each model fitted to mtREV.

\end{table}

} 

\end{document}